\documentclass[a4paper,fleqn,usenatbib]{mnras}
\usepackage[T1]{fontenc}
\usepackage{ae,aecompl}

\usepackage{graphicx}	
\usepackage{amsmath}	
\usepackage{amssymb}	
\usepackage[utf8]{inputenc}
\usepackage{epsfig}
\usepackage{url}
\usepackage{capt-of}
\usepackage{multicol}
\usepackage{epstopdf}
\usepackage{caption}
\usepackage{textcomp}
\usepackage[usenames,dvipsnames]{color}
\usepackage{float,lscape}
\usepackage{pdfpages}
\usepackage{pdflscape}

\DeclareMathOperator\erf{erf}

\title[The Tully-Fisher Relation of COLD GASS Galaxies]{The Tully-Fisher Relation of COLD GASS Galaxies}

\author[Tiley et al.]{Alfred L.\ Tiley,$^{1}$
Martin Bureau,$^{1}$ Am\'{e}lie Saintonge,$^{2}$ Selcuk Topal,$^{1}$
\newauthor Timothy A.\ Davis$^{3,4}$ and
Kazufumi Torii$^{5}$
\\
$^{1}$Sub-department of Astrophysics, Department of Physics, University of Oxford, Denys Wilkinson Building, \\  \hspace{0.13cm}Keble Road, Oxford OX1 3RH, UK\\
$^{2}$Department of Physics and Astronomy, University College London, Gower Place, London WC1E 6BT, UK\\
$^{3}$Centre for Astrophysics Research, University of Hertfordshire, Hatfield, Herts AL10 9AB, UK\\
$^{4}$School of Physics \& Astronomy, Cardiff University, Queens Buildings, The Parade, Cardiff CF24 3AA, UK\\
$^{5}$Department of Physics, Nagoya University, Chikusa-ku, Nagoya, Aichi 464-8601, Japan
}

\date{Accepted XXX. Received YYY; in original form ZZZ}

\pubyear{2016}

\begin{document}
\label{firstpage}
\pagerange{\pageref{firstpage}--\pageref{lastpage}}
\maketitle

\begin{abstract}
We present the stellar mass ($M_{*}$) and {\it Wide-Field Infrared Survey Explorer} ({\it WISE}) absolute
Band~1 magnitude ($M_{W1}$) Tully-Fisher relations (TFRs) of subsets of galaxies from the CO Legacy Database for the Galex Arecibo SDSS Survey (COLD GASS). We examine the benefits and drawbacks of several commonly used fitting functions in the context of measuring CO(1-0) line widths (and thus rotation velocities), favouring the Gaussian Double Peak function. We find the $M_{W1}$ and $M_{*}$ TFR, for a carefully selected sub-sample, to be $M_{W1} = (-7.1\pm0.6) \left[\log{\left(\frac{W_{50}/\sin{i}}{\text{km~s}^{-1}}\right)}-2.58\right] - 23.83\pm0.09$
  and
  $\log{(M_{*}/M_{\odot})} = (3.3\pm0.3) \left[\log{\left(\frac{W_{50}/\sin{i}}{\text{km~s}^{-1}}\right)}-2.58\right] + 10.51\pm0.04$, respectively, where $W_{50}$ is the width of a galaxy's CO(1-0) integrated profile at $50\%$ of its maximum and the inclination $i$ is derived from the galaxy axial ratio measured on the SDSS $r$-band image. We find no evidence for any significant offset between the TFRs of COLD GASS galaxies and those of comparison samples of similar redshifts and morphologies. The slope of the COLD GASS $M_{*}$ TFR agrees with the relation of \citet{Pizagno:2005aa}. However, we measure a comparatively shallower slope for the COLD GASS $M_{W1}$ TFR as compared to the relation of \citet{TullyPierce2000}. We attribute this to the fact that the COLD GASS sample comprises galaxies of various (late-type) morphologies. Nevertheless, our work provides a robust reference point with which to compare future CO TFR studies.
\end{abstract}

\begin{keywords}
galaxies: general 
\end{keywords}



\section{Introduction}
\label{sec:intro}

\subsection{Background}
\label{subsec:background}

The Tully-Fisher relation \citep[TFR;][]{Tully:1977aa}, one of the best studied galaxy
scaling relations, can be derived easily by considering circular
motion under gravity from a spherical mass distribution:
$M(r) \propto v^2(r)r$, where $M$ is the mass of the body enclosed
within a radial distance $r$ from its centre and $v$ is the rotational
velocity at $r$. Defining the
dynamical mass-to-light ratio $M/L$ and surface mass density
$\Sigma \equiv M/\pi r^2$ of the body, one then obtains the TFR:
\begin{equation} \label{eq:TFR}
  L \propto \frac{v^4}{\Sigma(M/L)}\,\,\,.
\end{equation}

Discovered observationally, the TFR was
initially used as a galaxy distance indicator and, once coupled with
the galaxy systemic velocity, as a tool to measure the Hubble constant \citep[e.g. ][]{Aaronson:1983, Bottinelli:1988, Giovanelli:1997, Sakai:2000,Tutui:2001}. Indeed, measuring a characteristic rotation velocity
and an apparent luminosity, the TFR-predicted absolute luminosity then
yields a distance estimate \citep[e.g.][]{Sandage:1976, Bottinelli:1980, Teerikorpi:1987, Fouque:1990, Bureau:1996, Ekholm:2000}. This of course
assumes that one knows the mass-to-light ratio and surface density of
the object studied, or more commonly that one compares populations of
galaxies assumed to have identical $M/L$ and $\Sigma$ (e.g.\ galaxies
of a given morphological type at a given epoch).

The TFR implies a tight relationship between luminous mass (traced by
$L$) and dynamical or total mass (traced by $v$), evidence that the
growths over time of the luminous and dark matter present in galaxies
are closely connected. Most importantly, given a reliable measure of
distance, the TFR can be turned around and used to probe the
mass-to-light ratio and surface density of galaxies. Indeed, many studies have used the zero-point, slope and scatter of the TFR to constrain cosmological models and models of galaxy formation \citep[e.g.][]{Cole:1989,Eisenstein:1996,Willick:1997,Steinmetz:1999,vandenBosch:2000,Yang:2003,Gnedin:2007,McGaugh:2012}. In addition, since a galaxy's
redshift is itself a good distance indicator beyond the local
Universe, the TFR is a powerful tool to directly probe the evolution of $M/L$
and $\Sigma$ over cosmological distances. Our goal here is thus to
allow such studies by providing a local TFR benchmark, using molecular gas (CO) as the kinematic tracer.

\subsection{Immediate Motivation}
\label{subsec:motivation}

The TFR was originally studied using blue optical restframe bands as
proxies for the stellar mass of galaxies \citep[see
e.g.][]{Tully:1977aa,McGaugh:1998aa,Ziegler:2002aa}. However, these
tend to trace young stellar populations and are considerably affected
by extinction, resulting in significant intrinsic scatter in the
relation \citep[e.g.][]{Verheijen:2001aa}. In recent years and in the
present work, near-infrared bands have been utilised instead
\citep[see
e.g.][]{Conselice:2005aa,Theureau:2007aa,Pizagno:2007aa,Puech:2008aa},
as these suffer little from extinction and better trace the bulk of
the stellar mass \citep[e.g.][]{Dickinson:2003aa}, resulting in
smaller intrinsic scatter.

Whilst there is a large body of work on the TFR using atomic hydrogen
(H{\small I}) 21~cm observations
\citep[e.g.][]{Tully:1977aa,Sprayberry:1995aa,Bell:2001aa}, there are
several advantages to using observations of carbon monoxide (CO), that
traces the cold molecular gas in galaxies. First, atomic hydrogen in galaxies is currently only routinely
detected in the local Universe. The Square Kilometre Array (SKA)
precursors Australian Square Kilometre Array Pathfinder (ASKAP) and
Karoo Array Telescope (MeerKat) can detect H{\small I} to moderate
redshifts \citep[$z\lesssim0.4$;][]{Meyer:2009,Holwerda:2010,deBlok:2011,Duffy:2012}, but the SKA itself will be
required to routinely detect galaxies at $z>1$ \citep{Abdalla:2015, Yahya:2015}. Multiple transitions of CO are however already routinely
detected in the bulk of the star-forming galaxy population at
$z\approx1$--$3$ \citep[e.g.][]{Daddi:2010, Magdis:2012, Magnelli:2012, Combes:2013, Freundlich:2013,Tacconi:2010,Tacconi:2013,Genzel:2015}, and in
star-bursting galaxies up to $z\approx7$ \citep[e.g.][]{Walter:2004, Riechers:2008a, Riechers:2008b, Riechers:2009, Wang:2011, Wagg:2014}. CO thus allows to extend TFR
studies probing the mass-to-light ratio and surface density of
galaxies to the earliest precursors of today's galaxies.

Second, previous work has
shown that the H{\small I} discs of galaxies, that are typically more
extended spatially than the molecular gas, can also be kinematically
unrelaxed, with e.g.\ large-scale warps
\citep[e.g.][]{Verheijen:2001aa}. Molecular gas is generally more
dynamically relaxed and suffers less from such problems. More
importantly, the atomic hydrogen in early-type galaxies is often
significantly disturbed, with much of the gas lying in tidal features
or nearby dwarf galaxies \citep[see e.g.][]{Morganti:2006aa, Serra:2012}, thus confusing
low-resolution observations such as those obtained with single-dish
telescopes. High-resolution interferometric observations are thus
necessary to identify those early-type galaxies with a regular
H{\small I} distribution appropriate to derive reliable TFRs \citep[see e.g.][]{denHeijer:2015}. On the other hand,
\citet{Davis:2011aa} clearly showed that CO single-dish observations
easily yield robust and unbiased TFRs for early-type galaxies. CO
observations thus offer the attractive possibility to derive TFRs more
accurate than those currently available, and this across the entire
Hubble sequence.

Our goal here is therefore to establish a benchmark CO TFR of local
galaxies, as a pre-requisite to extend the relation to higher
redshifts. There are clearly tracers other than CO that can be used at
large redshifts, primarily optical ionised gas emission lines such as
H$\alpha$, [O{\small II}] and [O{\small III}], and these should also
be pursued to provide independent probes of the evolution of the TFR
\citep[see
e.g.][]{Cresci:2009aa,Gnerucci:2011aa,Miller:2011aa,Miller:2012aa}. However,
it is known that ionised gas discs at $z\approx1$--$3$ are turbulent
\citep[see e.g.][]{Schreiber:2006, Swinbank:2012}, and great
care must be taken when measuring and interpreting their rotational
motions \citep[e.g.][]{Wisnioski:2015,Stott:2016,Tiley:2016}. With the advent of
the Atacama Large Millimeter/submillimeter Array
(ALMA)\footnote{http://www.almascience.org/}, CO emission tracing
dynamically cold gas is more easily detectable in high-redshift
galaxies than ever before, and embarking on CO TFR studies across the
Hubble sequence and redshift is particularly timely.

In this paper, we thus take a step toward establishing a local
benchmark for the CO TFR, using the CO(1-0) line as a kinematic
tracer. In future work our TFRs will be compared to those of a local sample
(Torii et al., in prep.) and a sample of $z\lesssim0.3$ galaxies (Topal et al., in prep.).

The sample, photometric data and kinematic data used in this paper are
described in \S~\ref{sec:data}. TFRs are
derived in \S~\ref{sec:TFRs} (the rotation velocity measure adopted
is defined in \S~\ref{subsec:measurew50} and extensively tested in Appendix~\ref{sec:velocity}). The results are discussed in
\S~\ref{sec:discussion}. We summarise and conclude briefly in
\S~\ref{sec:conclusions}.

\section{DATA}
\label{sec:data}

\subsection{CO Velocity Widths}
\label{subsec:CO}

As discussed at length in \S~\ref{subsec:measurew50} and Appendix~\ref{sec:velocity}, we adopt as our TFR
velocity measure the width of the integrated CO(1-0) line profile.
The \textit{GALEX} Arecibo SDSS Survey \citep[GASS;][]{Catinella:2009aa}
aimed to observe the neutral hydrogen content of a sample of
$\sim1000$ galaxies with the Arecibo 305~m telescope. More
specifically, GASS galaxies were selected to have redshifts
$0.025<z<0.05$, stellar masses $10^{10}<M_{*}/M_{\odot}<10^{11.5}$
with a flat distribution in $\log(M_{*}/M_{\odot})$, and positions
within the overlap region of the Sloan Digitized Sky Survey \citep[SDSS;][]{York:2000},
Arecibo Legacy Fast ALFA Survey \citep[ALFALFA;][]{Giovanelli:2005} and \textit{GALEX} Medium
Imaging Survey \citep[MIS;][]{Martin:2005,Morrissey:2005}. No other selection criterion was applied. A follow-up survey (Catinella, in prep.) has extended the GASS sample to probe down to stellar masses of $10^9 M_{\odot}$. The sample selection and survey strategy are identical to those of the original GASS survey, but for the difference that the lower mass galaxies are selected in the redshift interval $0.01<z<0.02$.

The CO Legacy Database for GASS (COLD GASS; Saintonge et al. 2011) adds information about the molecular gas contents of a randomly-selected sample of 500 GASS galaxies over the full mass range $10^9<M_{\ast}/M_{\odot}<10^{11.5}$. Most galaxies ($\approx80\%$) have
angular diameters small enough to be observed with a single pointing
of the Institut de Radioastronomie Millim\'{e}trique (IRAM) 30~m
telescope. For larger galaxies, an extra pointing offset from the
first was added. Fully reduced and baseline-subtracted integrated
CO(1-0) spectra of all massive COLD GASS galaxies ($M_{\ast}>10^{10} M_{\odot}$) are publicly available\footnote{http://www.mpa-garching.mpg.de/COLD\_GASS/} (binned
to $11.5$~km~s$^{-1}$ channels), and spectra for the lower mass galaxies have been made available to us ahead of publication. Once combined with other multi-band
observations, GASS and COLD GASS thus allow to measure the fraction of
the galaxies' baryonic mass contained in atomic gas, molecular gas and
stars.

\citeauthor{Saintonge:2011aa} define a ``secure'' detection in COLD GASS as
one with a signal-to-noise ratio $S/N>5$, where $S/N$ is calculated as
the ratio of the integrated line flux to its formal error. Considering the total data set available to us for this work, there are $260$ securely detected galaxies. These form the basis of the current sample.

\subsection{Near-infrared Luminosities}
\label{subsec:NIR}

Many previous TFR studies have used $K$-band magnitudes to probe the
bulk of the stellar mass in their sample galaxies. To that end,
$K_{\text{s}}$-band magnitudes from the Two Micron All Sky Survey \citep[2MASS;][]{Skrutskie:2006} were also considered here, but at the distances of the
COLD GASS galaxies the depth of 2MASS is insufficient to accurately
recover $K_{\rm{s}}$-band magnitudes. Other deeper surveys are available,
such as the UK Infrared Telescope Deep Sky Survey \citep[UKIDSS;][]{Lawrence:2007}, but these tend to have more limited sky coverage.

For this work, the {\it Wide-Field Infrared Survey Explorer} \citep[{\it WISE};][]{Wright:2010} Band~1 ($\approx3.4$~$\mu$m) magnitude ($W1$) of each
sample galaxy was thus adopted as a proxy for its total stellar
mass. This quantity is available for 222 of the 260 secure COLD GASS CO
detections. Specifically, the magnitude used was
the \textit{w1gmag} parameter from the All\textit{WISE} Source
Catalog\footnote{http://wise2.ipac.caltech.edu/docs/release/allwise/}.
This parameter is the $W1$ magnitude measured in an elliptical
aperture, with a size, shape and orientation based on that reported in the 2MASS Extended Source Catalog (XSC\footnote{http://irsa.ipac.caltech.edu/applications/2MASS/PubGalPS/}). To account for the larger {\it WISE} beam, the aperture is scaled-up accordingly. Those galaxies without an associated \textit{w1gmag} value are excluded from this work. 

It is unclear why 38 of the 260 galaxies securely detected do not have an associated \textit{w1gmag} value. The 38 galaxies in question are present in the XSC, and the majority of them are also deemed to be extended sources by {\it WISE} and are present in the All\textit{WISE} Source Catalog. However, the entries in the latter are not explicitly linked to the corresponding XSC objects despite their close proximity on the sky (typically $\sim1''$). The 38 galaxies do each have a {\it WISE} Band~1 magnitude measured in a $13''.75$ radius aperture ($\approx 23$ kpc at $z \approx 0.03$, the typical redshift of the COLD GASS galaxies), \textit{w1mag\_4} from the All\textit{WISE} Source
Catalog. To ensure we do not bias our sample by excluding the 38 galaxies in question, we use the available \textit{w1mag\_4} values to conduct a Kolmogorov-Smirnov (K-S) two-sample test between the full 260 securely detected galaxies and the 222 galaxies that remain after the exclusion of the 38 galaxies mentioned. We define a null hypothesis that the \textit{w1mag\_4} values of the latter are drawn from the same continuous distribution as those of the former, rejecting the null hypothesis if the $p$-value $p < 0.05$. The test produced a $p$-value $p = 0.68$, so we cannot reject the null hypothesis. We therefore proceed with our analysis, confident that the exclusion of those 38 galaxies without an associated \textit{w1gmag} value does not significantly bias the magnitude distribution of our remaining sample.

The bulk of the stellar mass is effectively probed at the wavelength
of $W1$, but this band is not so far red as to significantly suffer
from contamination due to dust emission. In addition,
\citet{Lagattuta:2013aa} found that the $(K-W1)$ colours of a sample
of $568$ late-type galaxies drawn from the 2MASS Tully-Fisher all-sky
galaxy catalog \citep{Masters:2008aa} are such that $(K-W1)\approx0$
with a scatter of only $\approx0.2$~dex. They did find a weak trend
towards bluer $(K-W1)$ colours at $W1\geq10.75$, but despite this it
is clear that both the $K$ and $W1$ bands, with similar wavelength
ranges, are tracing the same stellar populations. It is thus
acceptable to directly compare TFRs derived using either passband, at
least in the regime where differences in depths between surveys is not
an issue.

For each sample galaxy, the absolute {\it WISE} Band~1 magnitude
($M_{\text{W1}}$) was calculated as follows:
\begin{equation}
  M_{\text{W1}} = W1 - A_{\text{3.4$\mu$m}} - \mu\,\,\,,
\end{equation} 
where $A_{\text{3.4$\mu$m}}$ is the extinction in $W1$, calculated by
first adopting the reddening value $E(B-V)$ from the dust maps of
\citet{Schlafly:2011aa} taken from the NASA/IPAC Infrared Science
Archive\footnote{http://irsa.ipac.caltech.edu/applications/DUST/}, and
converting this to the corresponding $W1$-band extinction assuming
\begin{equation}
  A_{\text{3.4$\mu$m}} = R_{\text{3.4$\mu$m}}\, E(B-V)\,\,\,,
\end{equation}
where $R_{\text{3.4$\mu$m}}=0.18\pm0.1$ was adopted as measured in
\citet{Yuan:2013aa} using the stellar ``standard pair'' technique. The
distance modulus $\mu$ is taken from the NASA/IPAC Extragalactic
Database\footnote{http://ned.ipac.caltech.edu/} (NED), and is derived
from the galaxy's redshift adjusted for local deviations from the
Hubble flow due to the Shapley Cluster, Virgo Cluster and Great
Attractor.

Both a reddening and a distance modulus value were available for each of the 222 remaining galaxies. 

\subsection{Inclination Estimates}
\label{subsec:inclinations}

To account for projection effects, each sample galaxy's inclination
$i$ was calculated as
\begin{equation}
\cos(i)=\left(\frac{q^{2}-q_{0}^{2}}{1-q_{0}^{2}}\right)^{1/2}\,\,\,,
\end{equation} 
where $q$ is the observed ratio of the semi-minor to the semi-major
axis of the galaxy in SDSS $r$-band imaging, available for $217$ of the $222$ remaining sample galaxies. Those galaxies without an associated $q$ value are excluded from this work. The (intrinsic) axial
ratio of an edge-on galaxy, $q_{0}$, is morphology
dependent. We adopt the same prescription as \citet{Davis:2011aa},
whereby galaxies are divided into early types with $q_{0}=0.34$ and
late types with $q_{0}=0.2$. See \S~\ref{subsec:morphology} for a description of the morphological classifications used.

\subsection{Stellar Masses}
\label{subsec:stellarmass}

The stellar masses of our sample galaxies were taken from the Max
Planck Institute for Astrophysics-Johns Hopkins University Data
Release~7 (MPA-JHU DR7) derived data
catalogue\footnote{http://www.mpa-garching.mpg.de/SDSS/DR7/}, and were
determined using SDSS photometry via the spectral energy distribution
(SED)-fitting method described by \citet{Salim:2007aa}, assuming a
\citet{Chabrier:2003aa} initial mass function (IMF). In this method, each
galaxy's SED is compared to model SEDs from the library of
\citet{Bruzual:2003aa} to determine a stellar mass probability
distribution. The stellar mass and its uncertainty are then taken as
the median and half the difference between the $16^{\text{th}}$ and
$84^{\text{th}}$ percentile of this distribution, respectively, and
are available for $216$ of the $217$ remaining sample galaxies. The single galaxy without an associated stellar mass value is excluded from this work.

\subsection{AGN Candidates}
\label{subsec:AGN}

As the emission from an active galactic nucleus (AGN) can contaminate
and occasionally far outweigh that of the stellar body of a galaxy, it
is imperative to exclude from our sample galaxies hosting a
substantial AGN. AGN-hosting galaxies would otherwise be
systematically offset from the underlying TFR and would systematically
bias our fits.

We used publicly available classifications from the \textit{emissionLinesPort} table of the SDSS Data Release 10\footnote{\url{http://skyserver.sdss.org/dr10}} to exclude AGN-candidates from our sample. Each SDSS galaxy was classified by a fit to its spectrum using adaptations of the Gas AND Absorption Line Fitting \citep[GANDALF;][]{Sarzi:2006} and penalised PiXel Fitting \citep[pPXF;][]{Cappellari:2004} routines to extract several emission lines. These were then used to place each galaxy on a Baldwin, Phillips \& Terlevich \citep[BPT;][]{Baldwin:1981} diagram. Based on their position on the diagram, galaxies were divided into different categories that depend on the likelihood of the galaxy hosting an AGN. The classifications themselves are based on the work of \citet{Kauffmann:2003}, \citet{Kewley:2001} and \citet{Schawinski:2007}. We thus excluded $9$ galaxies classified as ``Seyfert", further reducing our adopted sample to $207$ galaxies.   

\subsection{Morphological Classes}
\label{subsec:morphology}

Our COLD GASS sample was cross-referenced with the Galaxy Zoo~1
catalog (GZ1; see \citealt{Lintott:2008aa,Lintott:2011aa}), providing
crowd-sourced classifications of $\sim600,000$ galaxies in the
SDSS. Each galaxy is classified as either a spiral, an elliptical or
``uncertain''. Of our remaining sample of $207$ galaxies drawn from
COLD GASS, $143$ are deemed to be spirals, $6$ ellipticals and the
remaining $58$ are uncertain. For the purposes of this work, and in particular the inclination correction described in \S~\ref{subsec:inclinations}, we equate those galaxies classified as spiral and uncertain to late-types, and those classified as elliptical to early-types. 

We do not initially exclude any galaxy based on its GZ1 morphological classification. However, the exclusion (or inclusion) of those galaxies deemed elliptical or uncertain is discussed further in \S~\ref{subsec:subsample}, where we present the details of a more restricted sub-sample. 

After applying all the criteria described in this section, we thus proceed with a final working sample of $207$ COLD GASS galaxies.   

%
%
\section{COLD GASS TULLY-FISHER RELATIONS}
\label{sec:TFRs}

\subsection{Measuring $W_{50}$}
\label{subsec:measurew50}
As a characteristic velocity measure, we adopt $W_{50}$, the width of
the CO(1-0) integrated profile at $50\%$ of its maximum, that should
be roughly equal to twice the maximum rotation velocity of the galaxy
$V_{\text{max}}$. In the presence of non-negligible noise, a fit to
the profile is usually superior to a direct measurement of $W_{50}$,
but the choice of the fitting function is not trivial and several
different functions have been used in the past.

While it is expected that the characteristic velocity measure will
vary depending on the function chosen, this is acceptable as long as
different galaxy samples being compared are measured in the same
manner. This is because the systematic effect introduced by any given
function will cancel out when the difference between two (or more)
samples is calculated. However, it is clearly preferable for the
measured characteristic velocity to be independent of the
signal-to-noise and amplitude-to-noise ratio ($A/N$) of the data, and
the measurements should not be systematically biased as the width
and/or shape of the profile vary. Despite previous work on the subject
\citep[e.g.][]{Saintonge:2007, Obreschkow:2009a, Obreschkow:2009b, Westmeier:2014aa}, as
the signal-to-noise ratio of CO data is generally low \citep[certainly lower
than that of typical H{\small I} spectra of nearby galaxies; e.g][]{Lavezzi:1998aa},
we deemed it prudent to compare the results of several different fitting
functions. 

This comparison is described in detail in Appendix~\ref{sec:velocity}, where we conclude that the Gaussian Double Peak function (Equation~\ref{eq:dgpeak}) is the most accurate and minimises potential biases as a function of $A/N$, inclination (apparent width) and rotation velocity (intrinsic width). We therefore deem it the most appropriate function with which to measure the CO(1-0) line widths of the COLD GASS galaxies, and adopt it throughout this work.

The CO velocity width ($W_{50}$) of every sample galaxy was thus measured by fitting the
Gaussian Double Peak function to its observed integrated CO(1-0)
spectrum from COLD GASS, this using the Levenberg-Marquardt algorithm
to minimise the reduced $\chi^2$, given by
\begin{equation}
  \chi_{\text{red}}^{2}\equiv\left(\sum\limits_{i}{\frac{[F(v_{i})-f(v_{i})]^{2}}{\sigma_{\text{rms}}^{2}}}\right)/\ \text{DOF}\,\,\,,
\end{equation} 
where $F(v_{i})$ is the observed CO(1-0) spectrum flux density in velocity bin
$v_{i}$, $f(v_{i})$ is the model flux density in that same velocity bin from
the Gaussian Double Peak function (i.e. the integral of the function across the bin divided by the bin width), $\sigma_{\text{rms}}$ is the rms noise
of the spectrum measured in a spectral range devoid of signal, DOF is
the number of degrees of freedom associated with the fit, and the sum
is taken over all velocity bins $v_{i}$ of the spectrum. The best fits
overlaid on the observed spectra are shown in
Appendix~\ref{sec:galfits} for all galaxies in our final sub-sample (see \S~\ref{subsec:subsample}).

Despite the fact that the Gaussian Double Peak function proved to be
the most robust function for fitting the CO profiles, this function
reduces to a single Gaussian in the limit where the half-width of the central parabola $w\to0$ (see Equation~\ref{eq:dgpeak}). As $w>0$ was the only
condition imposed on $w$ during the Gaussian Double Peak function fitting
process, there are cases where the best-fit value of $w$ is less than
the spectra's $11.5$~km~s$^{-1}$ velocity bin width. In these cases,
it is thus clear that a single Gaussian would be a more natural (and
perhaps better) fit.

With this in mind, a strategy was adopted whereby each galaxy spectrum
was fit with both the Gaussian Double Peak function and a standard
Gaussian (see Equation~\ref{eq:gaussian}). The value of $|1-\chi_{\text{red}}^{2}|$ for each function
was then calculated and the two values compared. The fit with the
smallest value was then adopted in each case. 

One further caveat was
necessary, since in some cases the best fit for the Gaussian Double
Peak function resulted in the peak of the central quadratic
$A_{\text{C}}$ (i.e.\ the central flux) being significantly greater
than the peak values of the flanking Gaussians $A_{\text{G}}$. In
these cases the central quadratic has a convex rather than a concave or flat
shape (as may be expected for respectively a double-horned or boxy-shaped profile). In these cases, specifically those for which $A_{\text{C}} > (3/2)A_{\text{G}}$, it was thus again
deemed more appropriate to adopt the best-fit Gaussian rather than the
Gaussian Double Peak function. 

Finally, for each spectrum {\sc mpfit} was used to explore the parameter space to find the combination of parameters that best fit the data. The best fit that {\sc mpfit} returns can be sensitive to the user-supplied initial guesses as local minima in the $\chi^{2}$ space can be mistaken for the global minimum by the fitting process. To reduce the chance of converging on a local minimum, and therefore not finding the true best-fit parameters, {\sc mpfit} was run several times for each spectrum, each time with a different set of initial guesses. The run with the smallest $|1-\chi_{\rm{red}}^{2}|$ value was then adopted as the best fit. The velocity width of the Gaussian Double Peak function at $50\%$ of the peak, $W_{50}$, was then calculated in an analytical manner:

\begin{equation}
W_{50}=2(w+\sigma\sqrt{2 \times \ln(2)})\,\,\,.
\end{equation} 

\noindent The same applies to the pure Gaussian function with $w=0$. 

The velocity width uncertainty, $\Delta W_{50}$, was then estimated by generating 150 realisations of the best-fit model, each with random Gaussian noise $\sigma_{\text{rms}}$ added. Each realisation was fit as described above and $\Delta W_{50}$ taken as the standard deviation of the velocity width distribution.

\subsection{Fitting the Tully-Fisher Relations}
\label{subsec:fitting}

Both a forward and a reverse straight line fit was made to each of the resulting Tully-Fisher relations using a Levenberg-Marquardt minimisation technique, again with {\sc mpfit}. The familiar forward fit minimises 

\begin{equation}
\chi_{\text{for}}^{2}\equiv\underset{i}{\Sigma}\left(\frac{1}{\sigma_{i}^{2}}\right)\{y_{i}-[m(x_{i}-x_{0})+b]\}~^{2}\,\,\,,
\end{equation}
where $x_{i}$ and $y_{i}$ are respectively the velocity and flux datum, $x_{0}$ is a ``pivot" point chosen to minimise the uncertainty in the straight line intercept $b$ (in practice we set $x_{0}$ to the median value of the $x_{i}$), $m$ is the straight line gradient, and the sum is over all sample galaxies. $\sigma_{i}$ is defined as 

\begin{equation}
  \sigma_{i}^{2}\equiv\sigma_{y,i}^{2}+m^{2}\sigma_{x,i}^{2}+\sigma_{\text{int}}^{2}\,\,\,,
\end{equation}
where $\sigma_{y,i}$ and $\sigma_{x,i}$ are respectively the uncertainty of an
individual data point in $y$ and $x$, and
$\sigma_{\text{int}}$ is a measure of the intrinsic scatter in the
Tully-Fisher relation, determined by
adjusting its value such that $\chi_{\text{for}}^{2}/\text{DOF} = 1$.

The total scatter in the relation is then

\begin{equation}
\sigma_{\rm tot}^{2}=\frac{\chi_{\text{for}}^{2}}{\underset{i}{\Sigma}\ (1/\sigma_{i}^{2})}\,\,\,.
\end{equation}

For the reverse fit, the figure of merit to minimise is

\begin{equation}
\chi_{\text{rev}}^{2}=\underset{i}{\Sigma}\left(\frac{1}{\zeta_{i}^{2}}\right)\left[x_{i}-(My_{i}+B+x_{0})\right]^{2},
\end{equation}
where similarly $M$ and $B$ are respectively the gradient and intercept of the straight line and

\begin{equation}
\zeta_{i}^{2}\equiv\sigma_{x,i}^{2}+M^{2}\sigma_{y,i}^{2}+\zeta_{\rm int}^{2}\,\,\,.
\end{equation}

\noindent $\zeta_{\rm int}$ is again the intrinsic scatter, determined such that $\chi_{\text{rev}}^{2}/\text{DOF} = 1$. The total scatter is then

\begin{equation}
\zeta_{\rm tot}^{2}=\frac{\chi_{\text{rev}}^{2}}{\underset{i}{\Sigma}\ (1/\zeta_{i}^{2})}\,\,\,.
\end{equation}

The reverse fit parameters can be directly compared to the forward fit parameters by defining the equivalent slope, intercept, intrinsic scatter and total scatter as $m^{'}\equiv 1/M$, $b^{'}\equiv -B/M$, $\sigma_{\rm int}^{'}\equiv M\zeta_{\rm int}$ and $\sigma_{\rm tot}^{'}\equiv M\zeta_{\rm tot}$, respectively \citep{Williams:2010aa}.

\subsection{Defining a Sub-sample}
\label{subsec:subsample}

\begin{figure*}
\centering
\begin{minipage}[]{1\textwidth}
\centering
\hspace{-0.85in}\includegraphics[width=0.75\textwidth, trim= 10 5 75 10, clip=True]{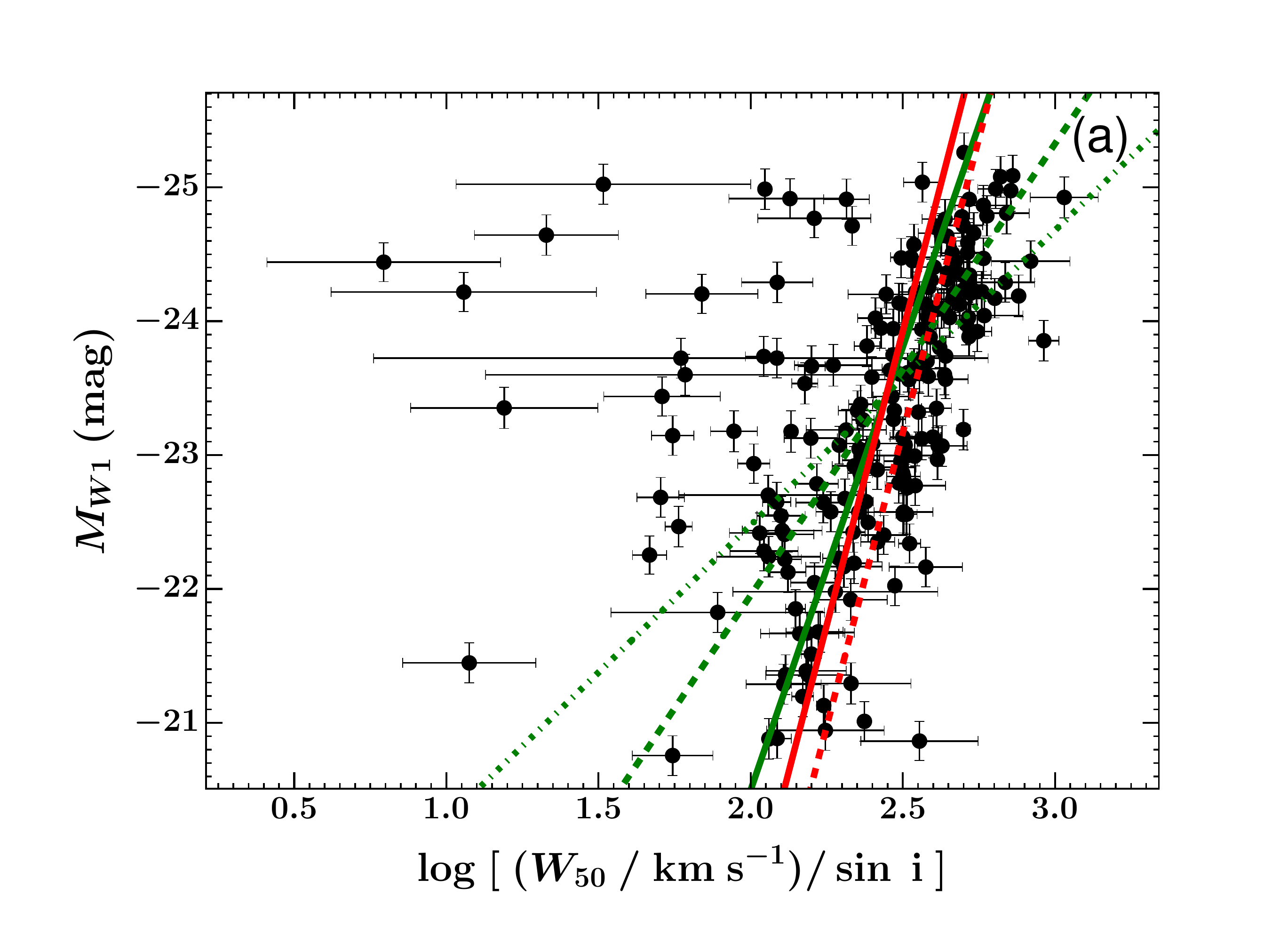}
\end{minipage}
\begin{minipage}[]{1\textwidth}
\centering
\hspace{-0.85in}\includegraphics[width=0.75\textwidth,trim= 10 5 75 50, clip=True]{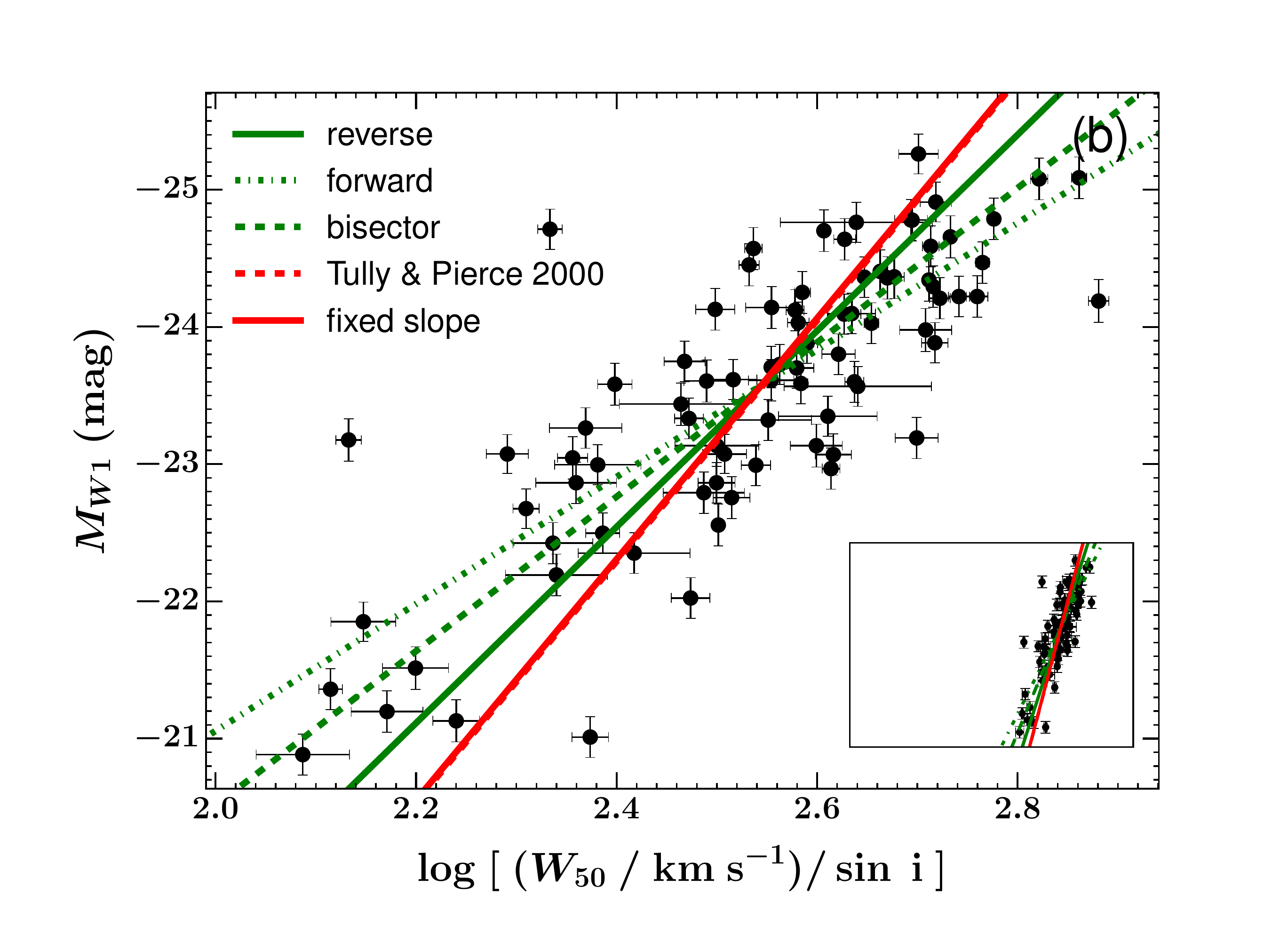}
\end{minipage}
\caption{COLD GASS $W1$-band Tully-Fisher relations. (a) and (b)
show the TFR for the initial COLD GASS sample and our final
sub-sample, respectively. The x-axis is the width of the
integrated CO(1-0) profile at $50\%$ of the peak, corrected for
the effect of inclination. The y-axis is the absolute {\it WISE}
Band~1 magnitude $W1$ ($\approx3.4$~$\mu$m). The green dot-dashed,
solid and dashed lines show the forward fit, reverse fit, and the
bisector of the two, respectively. The dashed red line is the
$K$-band TFR of \citet*{TullyPierce2000}. The solid red line
shows the best fit when the gradient is constrained to that of
\citeauthor{TullyPierce2000}. To demonstrate the reduction in scatter between the initial and final sub-sample, the embedded panel in (b) shows the TFR for the final sub-sample but over the same axis ranges as (a).}
\label{fig:W1}
\end{figure*}

\begin{figure*}
\centering
\begin{minipage}[]{1\textwidth}
\centering
\hspace{-0.85in}\includegraphics[width=0.75\textwidth, trim= 10 5 75 10, clip=True]{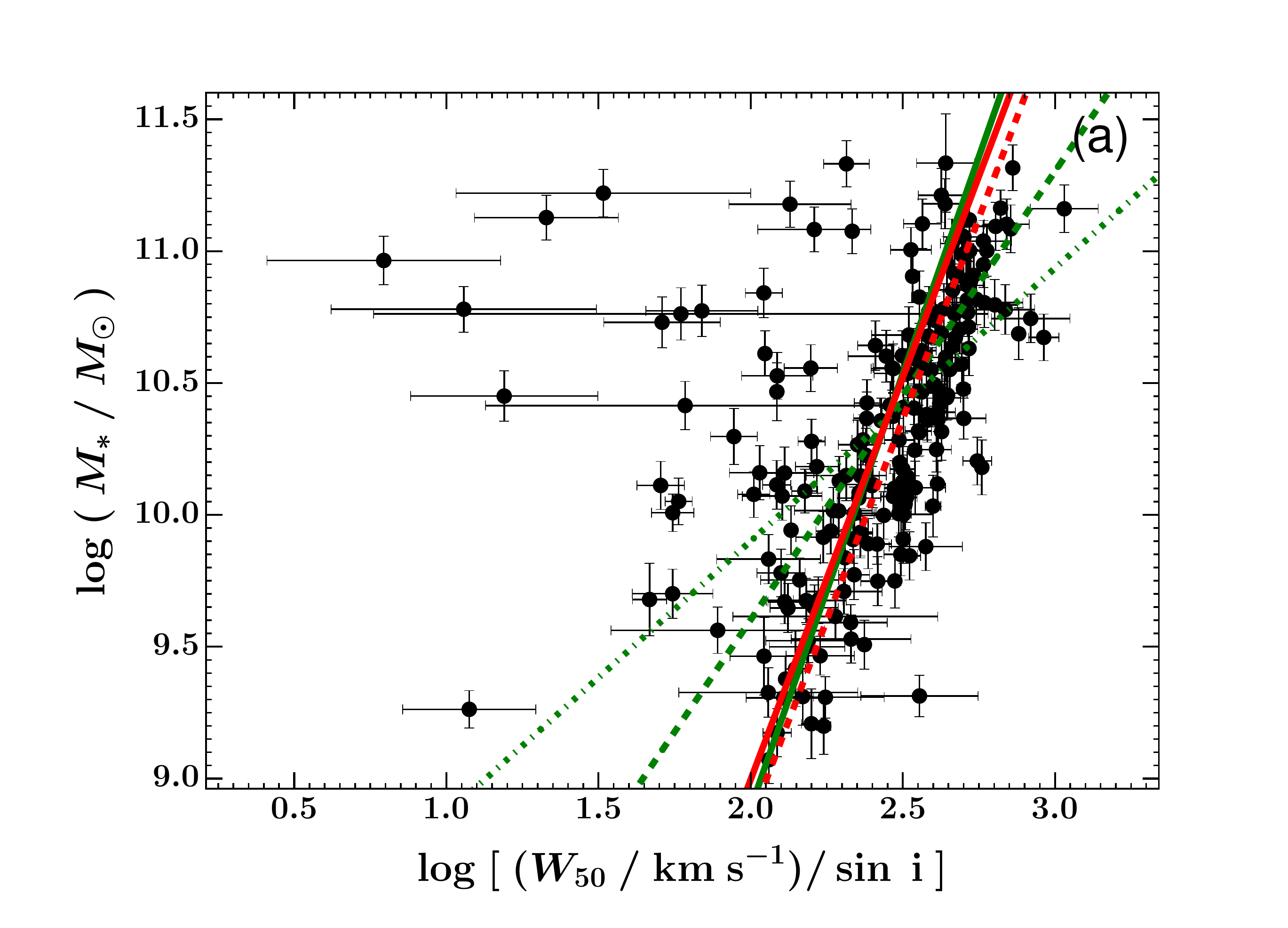}
\end{minipage}
\begin{minipage}[]{1\textwidth}
\centering
\hspace{-0.85in}\includegraphics[width=0.75\textwidth, trim= 10 5 75 50, clip=True]{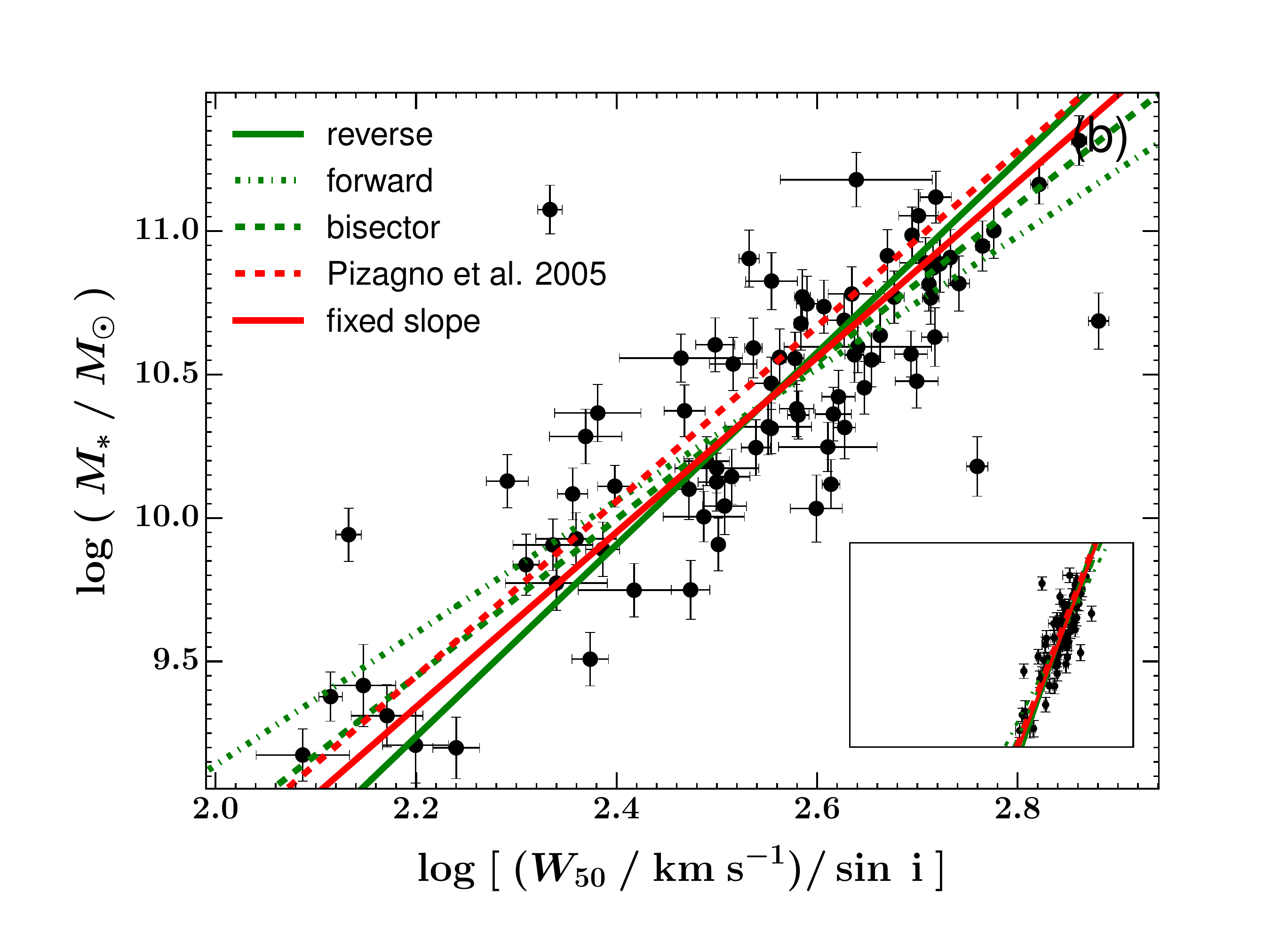}
\end{minipage}
  \caption{As Figure~\ref{fig:W1}, but for the COLD GASS stellar mass
    Tully-Fisher relations. The dashed red line is the stellar mass TFR
    of Pizagno et al.\ (2005). The solid red line shows the best fit
    when the gradient is constrained to that of \citeauthor{Pizagno:2005aa}}
  \label{fig:Mstar}
\end{figure*}

As described in \S~\ref{sec:data}, an initial sample was drawn from
the COLD GASS data that contains all galaxies with a secure CO(1-0)
detection, {\it WISE} $W1$ photometry, an SDSS $r$-band axial ratio, a stellar mass via multi-band
photometry, no AGN, and a GZ morphological classification for a total of $207$ galaxies. 
The $W1$-band and stellar mass Tully-Fisher relations for this initial
sample are shown in Figure~\ref{fig:W1}(a) and \ref{fig:Mstar}(a),
respectively, and the fit parameters are listed in Table~\ref{tab:W1}
and \ref{tab:Mstar}, respectively. Clearly, for both the $W1$-band and
stellar mass TFRs, the intrinsic and total scatters are very large for
this initial sample.

\begin{table*}
  \caption{Parameters of the $W1$-band Tully-Fisher relations.}
  \centering
  \begin{tabular}{lllllrrr}
    \hline
    Sample     & Pivot & Fit & Slope & Intercept & Intrinsic Scatter & Total Scatter & Offset \\
     &  & &  & (mag) & (mag) & (mag) & (mag) \\
    \hline
    Initial   & 2.49 & Forward  & $-2.2\phantom{0}\pm0.3$ & $-23.56\pm0.07$ & $0.92\pm0.05$ & $0.95\pm0.02$ & - \\
              & & Reverse  & $-6.6\phantom{0}\pm0.7$ & $-23.7\phantom{0}\pm0.1$ & $1.5\phantom{0}\pm0.1\phantom{0}$ & $1.62\pm0.01$ & - \\
              & & Bisector & $-3.4\phantom{0}\pm0.3$ & $-23.60\pm0.01$ & - & - & - \\
              & & Fixed    & $-8.78$                 & $-23.8\phantom{0}\pm0.1\phantom{0}$ & $2.0\phantom{0}\pm0.1\phantom{0}$ & $2.18\pm0.01$ & $-0.8\pm0.2$ \\
    Sub-sample & 2.58 & Forward  & $-4.6\phantom{0}\pm0.4$ & $-23.74\pm0.07$ & $0.59\pm0.05$ & $0.61\pm0.01$ & - \\
              & & Reverse  & $-7.1\phantom{0}\pm0.6$ & $-23.83\pm0.09$ & $0.73\pm0.06$ & $0.76\pm0.01$ & - \\
              & & Bisector & $-5.6\phantom{0}\pm0.3$ & $-23.77\pm0.01$ & - & - & - \\
              & & Fixed    & $-8.78$                 & $-23.9\phantom{0}\pm0.1\phantom{0}$ & $0.94\pm0.08$ & $0.96\pm0.01$ & $0.0\pm0.1$ \\
    \hline
  \end{tabular}
  \label{tab:W1}
\end{table*}

\begin{table*}
  \caption{Parameters of the stellar mass Tully-Fisher relations.}
  \centering
  \begin{tabular}{lllllrrr}
    \hline
    Sample & Pivot & Fit & Slope & Intercept & Intrinsic Scatter & Total Scatter & Offset \\
    &  &  &  & (dex) & (dex) & (dex)  & (dex) \\
    \hline
    Initial   & 2.49 & Forward  & $1.0\phantom{0}\pm0.1$ & $10.41\pm0.03$ & $0.44\pm0.03$ & $0.47\phantom{0}\pm0.01\phantom{0}$ & - \\
              & & Reverse  & $3.3\phantom{0}\pm0.4$ & $10.50\pm0.06$ & $0.77\pm0.05$ & $0.824\pm0.004$ & - \\
              & & Bisector & $1.7\phantom{0}\pm0.1$ & $10.44\pm0.01$ & - & - & - \\
              & & Fixed    & $3.05$                 & $10.49\pm0.05$ & $0.73\pm0.05$ & $0.765\pm0.003$ & $0.16\pm0.06$ \\
    Sub-sample & 2.58 & Forward  & $2.3\phantom{0}\pm0.2$ & $10.48\pm0.03$ & $0.27\pm0.02$ & $0.287\pm0.004$ & - \\
              & & Reverse  & $3.3\phantom{0}\pm0.3$ & $10.51\pm0.04$ & $0.32\pm0.03$ & $0.344\pm0.005$ & - \\
              & & Bisector & $2.7\phantom{0}\pm0.2$ & $10.49\pm0.01$ & - & - & - \\
              & & Fixed    & $3.05$                 & $10.50\pm0.03$ & $0.30\pm0.03$ & $0.318\pm0.003$ & $-0.11\pm0.04$ \\
    \hline
  \end{tabular}
  \label{tab:Mstar}
\end{table*}

There are several factors contributing to the observed
scatter. Firstly, as shown in Figure~\ref{fig:incselfconsistency}, the
measured velocity widths of nearly face-on galaxies are unreliable, in
the sense that they do not accurately reproduce the intrinsic widths
and their fractional errors are dependent on those widths. We thus
removed from the initial sample galaxies with an inclination
$i<30^{\circ}$, resulting in a sub-sample of $180$ galaxies.

Secondly, one must gauge whether the CO extends to sufficiently large
radii to sample the flat part of the galaxies' rotation curves. This
is essential, as otherwise the measured velocity widths will not be
representative of the total dynamical masses, introducing additional
scatter in the relations. Although not necessary, empirical studies
have shown that the kinematic tracer of galaxies with a boxy or
double-horned integrated profile usually extends beyond the
turn-over of the rotation curve \citep[e.g.][]{Davis:2011aa}. A Gaussian profile may imply that
insufficient CO is present in the outer parts to robustly recover the
flat part of the rotation curve, or that the rotation curve itself is
not flat. Of course, a Gaussian profile can also arise, despite sufficient CO present in the outer parts of a galaxy, if there is a particularly high concentration of gas in the central regions of the galaxy \citep[e.g.][]{Wiklind:1997,Lavezzi:1997aa}, but this is a risk worth taking. Lastly, given the finite width of the velocity channels,
rejecting profiles that do not appear boxy or double-horned may also exclude
profiles that are intrinsically boxy or double-horned but simply have
small line widths, an effect that would preferentially affect the
low-mass end of the TFRs. However, this drawback is small compared to
the benefits of significantly reduced scatter.

This approach is consistent with the understanding that the flat part
of a galaxy's rotation curve is accurately recovered provided that the
``flaring parameter'' of the integrated profile (i.e.\ the ratio of
the line width at $20\%$ of the peak flux to that at $50\%$) is less
than $\approx1.2$ \citep{Lavezzi:1997aa,Lavezzi:1998aa}. In other words, both methods
select integrated profiles with steep edges. We therefore reduced our
sample further by requiring that the integrated CO(1-0) profile of
each galaxy be either boxy or double-horned. In practice, this amounts
to excluding galaxies whose profile was best fit by a single Gaussian
function rather than the Gaussian Double Peak function, resulting in a
sub-sample of $94$ galaxies.

Thirdly, we consider how well our line profile fits recover the true
velocity widths at small amplitude-to-noise ratios $A/N$. As illustrated in
Figure~\ref{fig:fracdiff}, $A/N\geq1.5$ is
generally required to recover the true velocity width to $10\%$ or
better. Given the limited size
of the COLD GASS sample and the typical quality of the CO data, this
threshold also ensures a final sub-sample with a sufficiently large
number of galaxies for robust TFR fits. Applying this $A/N$ cut results in a
sub-sample of $88$ galaxies.

Fourthly, to have confidence in the accuracy of the velocity widths derived from the profile fits, we remove those galaxies with a fractional uncertainty $\Delta W_{50}\ /\ W_{50} > 20\%$. This further reduces our sub-sample to $84$ galaxies.

Lastly, we consider the morphologies of our sample galaxies, as
galaxies of different morphological types have different TFR
zero-points (presumably due to different surface mass densities and
mass-to-light ratios; see Eq.~\ref{eq:TFR}). In particular, there is a
clear offset between early- and late-type galaxies \citep[see
e.g.][]{Williams:2010aa,Davis:2011aa}. In fact, the work of \citet{Davis:2016} suggests that very massive early-type galaxies are again offset from the TFR of early-types of lower masses. There are also
possible variations of the TFR slope among late-type galaxies
\citep[e.g.][]{Lagattuta:2013aa}. 

\begin{figure}
  \includegraphics[scale=0.47, trim = 15 0 25 30, clip=True]{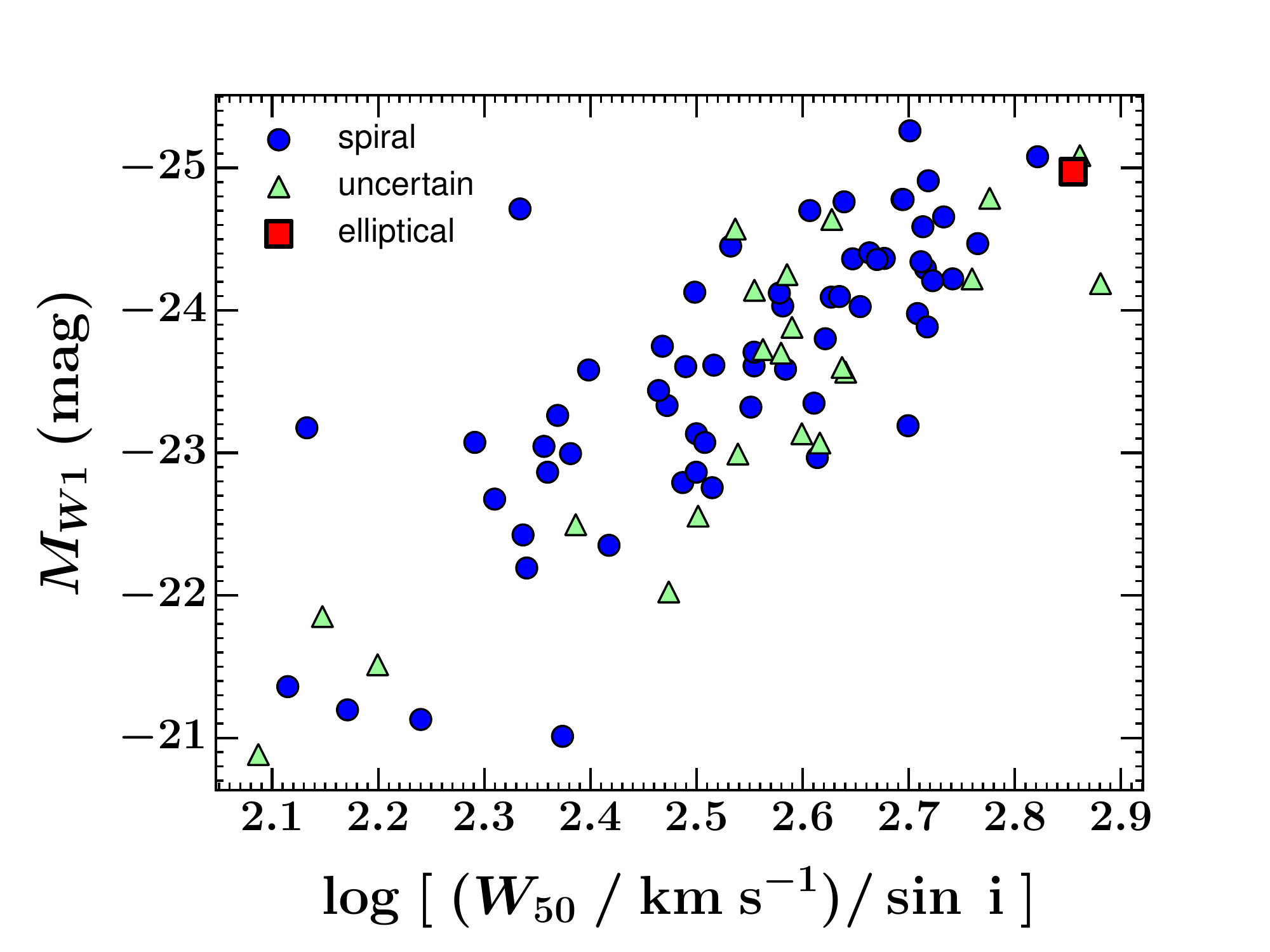}
  \caption{As Figure~\ref{fig:W1}, but for an intermediate (although
    nearly final) COLD GASS sub-sample (as described in
    \S~\ref{subsec:subsample}). Error bars are omitted for clarity and
    the galaxies are identified by their GZ1 morphological type (see
    \S~\ref{subsec:morphology}): blue circles for spirals
    (late-types), red squares for ellipticals (early-types), and green
    triangles for uncertain types. It is clear that those galaxies deemed uncertain follow broadly the same relation as, and do not exhibit increased scatter in comparison to, those deemed spirals.}
  \label{fig:morph}
\end{figure}

Figure~\ref{fig:morph} shows the TFR
of our galaxy sub-sample after all the aforementioned cuts have been
applied, and where all galaxies have been labeled according to their
morphological type (see \S~\ref{subsec:morphology}). Error bars have
been omitted for clarity. As there is only one early-type galaxy,
it was removed from the sub-sample. As the galaxies of uncertain
morphological type are essentially indistinguishable from the
late-types, with a similar slope, zero-point and scatter, there is no reason to exclude them
and we instead assimilate them to the late-type galaxies.

Overall, this thus leads to a final sub-sample comprising $83$
late-type galaxies, that we use for both $W1$-band and stellar mass
TFR fits. The resulting $W1$-band and
stellar mass Tully-Fisher relations for this final sub-sample are
shown in Figure~\ref{fig:W1}(b) and \ref{fig:Mstar}(b), respectively,
while the fit parameters are listed in Table~\ref{tab:W1} and
\ref{tab:Mstar}, respectively. As desired, the observed scatters of
our final sub-sample are significantly reduced compared to those of
the initial sample, this for both the $W1$-band and stellar mass TFR.

\subsection{The Tully-Fisher Relations}
\label{subsec:TFRs}

As described above, Figures~\ref{fig:W1} and \ref{fig:Mstar} show the
COLD GASS $W1$-band and stellar mass Tully-Fisher relation,
respectively, for both the initial and final sub-samples, with the
forward, reverse and bisector fits overlaid. Tables~\ref{tab:W1} and
\ref{tab:Mstar} list the corresponding fit parameters. The reverse fit
parameters have been adjusted to be directly comparable to the forward
fit parameters, as described in \S~\ref{subsec:fitting}.

A comparison relation is also displayed on each plot: for the $W1$-band TFR, the $K$-band
relation of \citet{TullyPierce2000}, who used H{\small I} integrated profiles to measure the galaxy rotation velocities; for the stellar mass TFR, the stellar mass relation of \citet{Pizagno:2005aa}, who used long-slit observations of H$\alpha$ emission to measure the ionised gas rotation. Based on the work of \citet{Madau:2014}, we apply a small offset in $\log(M_{*}/M_{\odot})$ of $- 0.034$ dex to the latter, to convert from a Kroupa \citep{Kroupa:2001} to a Chabrier IMF. A fit to the COLD GASS data with the slope
fixed to that of the comparison relation is also shown in each case,
allowing to gauge any offset between the comparison and COLD GASS
samples. We recall that with $(K-W1)\approx0\pm0.2$ for late-type
galaxies, a $W1$--$K$ comparison is justified
\citep{Lagattuta:2013aa}.

As expected, the reverse fits are always much steeper than the forward
fits, and indeed for both initial and final sub-samples and $W1$-band
and stellar mass TFRs, the slopes of the reverse fits are much closer
to those of the comparison samples. 

It is well established that a significant bias in the slope is introduced by using a forward fit \citep[e.g.][]{Schechter:1980,Teerikorpi:1987,Sandage:1988,TullyPierce2000,TC:2012,Sorce:2013}, stemming from the selection criteria imposed on galaxy samples used in TFR studies. This bias is explained in detail by \citet{WIllick:1994aa}. 

To understand why this approach is problematic, first consider the measured TFR of a complete (i.e. independent of any selection) sample of galaxies, with some Gaussian intrinsic scatter in magnitude (and indeed line width). For a given line width, a galaxy may be scattered either above or below the TFR if it is intrinsically brighter or dimmer than the corresponding mean magnitude for that line width (i.e. that predicted by the true, unbiased TFR). Now consider measuring the TFR for the same sample but for the common case where a limiting magnitude is imposed, such that galaxies dimmer than the limiting value are excluded from the analysis. At the faint end of the TFR, those galaxies that are intrinsically brighter than the mean will be preferentially included in the analysis, whilst those that are dimmer will be excluded. This acts to flatten the slope of the measured TFR, as at the small line width end the sample is biased brighter than the true underlying distribution. This bias is compounded by the fact that the uncertainties in the line width measurements are typically at least comparable (and often larger) than those in the luminosity measurements.  

Since in this scenario there is no selection via line width, the reverse fit, which minimises the residuals in line width, avoids this bias. \citet{WIllick:1994aa} points out that in practice some bias will still remain for the reverse fit when considering the TFR in a waveband other than that used to select the sample. This is particularly relevant to this work, since the GASS sample (and therefore COLD GASS) was not selected via a limiting $W1$ magnitude but was rather selected to be flat in $\log(M_{*}/M_{\odot})$ above a minimum stellar mass value (see \S~\ref{subsec:CO}). Nevertheless the bias in the reverse fit is much reduced compared to that of the forward fit. \citet{Willick:1995}, for example, find a reduction of a factor of 6 in the bias between the forward and reverse fit when examining the TFRs of several samples of spiral galaxies selected in the $I$ and $r$ bands. In light of this, the parameters in this work derived from
a forward fit should be treated with a degree of caution.

As desired following the exclusion of potential sources of scatter
(see \S~\ref{subsec:subsample}), for both the $W1$-band and the
stellar mass TFR there is a much greater scatter in the initial sample
than in the final sub-sample. In addition, the intrinsic and total
scatters are always larger in the reverse fit than in the forward fit,
reflecting the larger scatter in $W_{50}$ than in $M_{W1}$ and
$M_{*}$.

\section{Discussion}
\label{sec:discussion}

\subsection{Slope}
\label{subsec:slope}

Treating the forward fits with suspicion, it is most sensible to
compare the results of previous studies to the reverse fits only. The
unconstrained reverse fit to the $W1$ TFR is shallower than
the relation found by \citet{TullyPierce2000} for both the initial and final sub-sample, even allowing for the
uncertainties (although Tully \& Pierce do not quote an uncertainty on
their slope). However, the slope of both sub-sample's stellar mass
TFR agrees with that of the \citet{Pizagno:2005aa} TFR, after allowing for uncertainties (\citealt{Pizagno:2005aa} quote an uncertainty of $\pm0.12$ on their slope). The stellar
masses are model dependent, however, so the significance of this
result is uncertain, particularly given the comparatively shallower
slope of the $W1$ relation.

Several factors could of course affect the slope of the COLD GASS
TFRs, in particular a potential Malmquist bias and the fact that
galaxies of various primarily late-type morphologies were amalgamated
together. The former factor is likely to be most acute for the stellar
mass TFRs, as the COLD GASS sample was stellar mass-selected, but it is
unlikely to be important for the final sub-sample as our various
selection criteria, particular the integrated profile shape criterion (see \S~\ref{subsec:subsample}), have largely washed out any abrupt
stellar mass threshold. Of course the same criterion may also preferentially exclude those galaxies with small line widths. However, this effect is likely to be small in comparison to the Malmquist bias introduced via the GASS (and thus COLD GASS) selection function. The latter factor is
undoubtedly present to some extent, but it is hard to quantify without
better morphologies. Indeed, as our samples contain a variety of
(mainly late-type) galaxy morphologies, and the slope of the TFR
varies with galaxy type, the measured slopes are effectively averages
of mutliple slopes for different galaxy morphologies.

Lastly, we must also consider whether the very use of CO as a kinematic tracer may bias the slope of the TFR, such that the resultant slope is shallower than that of the H{\small I} TFR of the same sample. Indeed, the line widths of \citet{TullyPierce2000} are measured from H{\small I} observations, whereas the COLD GASS line widths are measured from CO(1-0). In Appendix~\ref{sec:HIcomparison}, for subsets of COLD GASS galaxies with both CO(1-0) and H{\small I} data, we compare the values of $W_{50}$ as derived from the width of both the CO(1-0) and H{\small I} integrated profiles (as described in \S~\ref{subsec:measurew50}). We find that the slope of the TFRs constructed using CO(1-0) are either comparable to or slightly shallower than those of the H{\small I} TFRs of the same galaxies. However, the slopes of both the CO(1-0) and H{\small I} TFRs agree within uncertainties. We may therefore cautiously attribute {\it some} of the difference in slope between the COLD GASS $W1$ TFR and the comparison relation to the use of CO(1-0) (rather than H{\small I}) as a kinematic tracer, but it should be stressed that, on the basis of the comparison in Appendix~\ref{sec:HIcomparison}, this is not likely to be the driving factor of the difference. Furthermore, as shown again in Appendix~\ref{sec:HIcomparison}, much is to be gained from using CO rather than H{\small I}, as it lead to much smaller intrinsic (and thus total) scatter for the sub-sample.       

\subsection{Inclinations}
\label{subsec:i}

The uncertainties on the measured inclinations $i$ contribute greatly
to the uncertainties on the inclination-corrected velocity widths
$W_{50}/\sin{i}$, particularly at small inclinations (and even with an
$i>30^{\circ}$ threshold). The robustness of any constructed TFR is thus
highly dependent on the accuracy of the inclination measurements. The
scatter of the TFR is likely to increase with decreasing accuracy,
with possibly a smaller systematic effect affecting the measured
slope.

The axial ratio method used in this work is appropriate for the
relatively shallow ground-based optical imaging used, but it is not
particularly refined. First, it naively assumes that galaxies can be
grouped into categories sharing a unique edge-on (intrinsic) axial
ratio (here $0.2$ and $0.34$ for late types and early types,
respectively). The morphologies of galaxies are in reality much more
diverse, and the intrinsic axial ratio is likely to vary within any
defined category. This is particularly relevant to our work, as the
inclination measurements are ultimately dependent on the relatively crude GZ1
morphological classifications. \citet{Davis:2011aa} in fact showed
that the scatter in the TFR of a sample of early-type galaxies is
reduced when one uses a measure of inclination derived from the
intrinsically very flat dust features visible in high resolution ({\it
  Hubble Space Telescope}) imaging, rather than stellar light as used
here. While the magnitude of the effect was certainly amplified by the
use of early-type galaxies, it nevertheless illustrates the point.

Having said that, as our inclinations are based on rather shallow
ground-based imaging, it may also be that they are systematically
underestimated (the galaxies appearing rounder than they really
should), leading to over-estimated inclination corrections to the
velocity widths. As this effect is likely to be more acute in smaller,
lower mass galaxies, however, it would lead to steeper rather than
shallower TFRs.

\subsection{Offset}
\label{subsec:offset}

We measure a small offset between the stellar mass TFR for both the initial and final sub-samples presented in this work and that of \citet{Pizagno:2005aa} ($+0.16\pm0.06$ and $-0.11\pm0.04$~dex, respectively). Considering the $W1$-band TFRs, we measure a large negative offset ($-0.8\pm0.2$~mag) between the initial sample's TFR and that of \citet{TullyPierce2000}. However, this offset disappears when considering the (more reliable) $W1$-band TFR for the final sub-sample.  
Importantly, in all cases, the offset between the COLD GASS and the comparison
sample TFR (at fixed slope) is much smaller than the intrinsic
scatter. There is thus no evidence for any significant offset between
the COLD GASS and comparison samples, as expected given the similar
redshifts and morphologies of the samples' galaxies.

\subsection{Scatter}
\label{subsec:scatter}

The intrinsic and total scatters of the $W1$-band Tully-Fisher
relation for our final sub-sample, for the forward unconstrained fit,
were found to be $0.59\pm0.05$ and $0.61\pm0.01$~mag, respectively. These values
are slightly larger than those of previous near-infrared TFR
studies. \citet{TullyPierce2000} found a $K'$-band total rms scatter of
$0.44$~mag for local spiral galaxies, whilst \citet{Verheijen:2001aa}
found a $0.32$~mag total scatter for the same passband. However,
\citet{Pizagno:2007aa} found an intrinsic scatter of
$0.42$--$0.46$~mag across the $g$, $r$, $i$ and $z$ bands. 

Considering the stellar mass TFR of the final sub-sample, the
intrinsic and total scatters of the forward unconstrained fit were
found to be $0.27\pm0.02$ and $0.287\pm0.004$~dex, respectively. As with the
$W1$-band relation, this is larger than previous TFR studies in the
local universe. \citet{Bell:2001aa} found a total
scatter of just $0.13$~dex, whilst \citet{Pizagno:2005aa} similarly
found an intrinsic scatter of $0.16$~dex.

The reasons why we measure a slightly higher intrinsic scatter than
previous studies are unclear. In Appendix~\ref{sec:HIcomparison} we show that the use of line widths derived from CO(1-0) integrated profiles leads to increased scatter in the TFR, compared to the same TFR constructed using H{\small I} integrated profiles. However, this difference in scatter disappears when we apply the criteria used to select the sub-sample described in \S~\ref{subsec:subsample}. This implies that the increased scatter is due to the inclusion of CO(1-0) profiles that do not display a boxy or double-horned shape. Once these systems are removed, the intrinsic and total scatter of the CO(1-0) TFRs are actually less than those of the H{\small I} TFRs for the same galaxies.

As discussed above, the mix of several
different late-type morphologies may affect the slope and scatter
(through different mass surface densities $\Sigma$ and mass-to-light ratios $M/L$; see \S~\ref{subsec:background}), while our inclinations may
be underestimated. While we have taken great care to estimate and
propagate uncertainties on our measurements, it may also be that our
observational errors are underestimated (leading to an overestimate of
the intrinsic scatter). The uncertainties on the stellar masses and
inclinations are particularly hard to reliably estimate.

\subsection{Sample Selection}
\label{subsec:sample}

The selection criteria of GASS, and thus COLD GASS, present two
potential problems when using galaxies drawn from these samples to
build TFRs. The first, discussed at length above, is the problem of
fitting a single TFR (slope and zero-point) to a sample of galaxies
with differing morphologies. The second is that both the GASS and COLD
GASS samples are chosen to be flat in $\log{M_{*}}$. This means that,
compared to e.g.\ a volume- or flux-limited sample, galaxies of large
masses will be over-represented. This could result in an inferred TFR
with a shallower than expected slope, depending on how heavily
weighted high-mass galaxies are in comparison to low-mass
galaxies. 

The sub-sample selection, as described in \S~\ref{subsec:subsample}, leads to a dramatic and significant reduction in the scatter of the TFRs. The main driver in this reduction is the exclusion of galaxies with profiles that do not appear boxy or double-horned. This ensures that we include in our analysis only those galaxies with sufficient CO in the outer parts to properly sample the flat parts of the rotation curve. This is at the expense of the possible rejection of profiles that are intrinsically boxy or double-horned but simply appear Gaussian due to their small line width and the finite width of the velocity channels. This cut by profile shape preferentially affects the low-mass end of the TFRs, and could therefore bias their resultant slopes. However, the effect is likely to be small (and opposite to the Malmquist bias) and the benefits of the significantly reduced  scatter far outweigh the drawbacks.

\section{Conclusions \& Future Work}
\label{sec:conclusions}

In an effort to firmly establish the CO Tully-Fisher relation (TFR) as
a useful tool to probe the evolution of galaxies over cosmic time, we
first tested the self-consistency and robustness of four functions
appropriate to fit the integrated line profiles of galaxies,
particularly in the low signal-to-noise ratio regime characteristic
of molecular gas observations. The Gaussian Double Peak function was
deemed to be the most self-consistent and to suffer the least from
possible systematic biases as a function of the amplitude-over-noise ratio $A/N$,
the galaxy inclination $i$ and the intrinsic flat circular velocity $V_{\text{c,flat}}$.

We then constructed the {\it WISE} $W1$-band and stellar mass TFRs
relations of galaxies drawn from the COLD GASS sample, both for an
initial sample of all galaxies with available data, and for a
restricted sub-sample of galaxies with high quality measurements (thus
decreasing the scatter). The rotation of the galaxies was determined
by fitting the Gaussian Double Peak function to the integrated CO(1-0)
line profile of each galaxy, and then measuring the width at $50\%$ of
the peak of the resultant fit ($W_{50}$). The $W1$ magnitudes were
drawn directly from the {\it WISE} catalogue, and the stellar mass for
each galaxy was determined via SED fitting of SDSS photometry.

The TFRs obtained from unconstrained forward fits have shallower
slopes than those expected from previous studies. Considering only the
more robust reverse fits, however, the best-fit TFRs for the final
COLD GASS sub-sample are
\begin{equation}
  M_{W1} = (-7.1\pm0.6) \left[\log{\left(\frac{W_{50}/\sin{i}}{\text{km~s}^{-1}}\right)}-2.58\right] - 23.83\pm0.09
\end{equation}

\noindent and

\begin{equation}
  \log{(M_{*}/M_{\odot})} = (3.3\pm0.3) \left[\log{\left(\frac{W_{50}/\sin{i}}{\text{km~s}^{-1}}\right)}-2.58\right] + 10.51\pm0.04\,\,\,.
\end{equation}
The unconstrained reverse fit slope of the COLD GASS sub-sample
$W1$-band TFR is still marginally shallower than that of
\citet{TullyPierce2000}, but the slope of the stellar mass TFR agrees within the uncertainties with the relation of \citet{Pizagno:2005aa}. The intrinsic
scatter (from forward fitting) is $0.59\pm0.05$~mag and $0.27\pm0.02$~dex for the
$W1$-band and stellar mass sub-sample TFR, respectively.

Fixing the slopes to those of the relations from the comparison
samples, small offsets are found with respect to the comparison
samples, that are however less than the intrinsic scatters. The COLD
GASS samples therefore agree with the comparison samples, although
they have slightly larger scatters than expected. Possible causes of
the increased scatters were discussed and include the method adopted to
measure inclinations, and fitting a single TFR to samples of galaxies
with various primarily late-type morphologies. Importantly, we showed that for a subset of COLD GASS galaxies in the final sub-sample with both CO(1-0) and H{\small I} data, the intrinsic and total scatters of the CO(1-0) TFRs were less than those of the same TFRs constructed using H{\small I} integrated profiles. 

The COLD GASS initial sample and final sub-sample contain a number of
galaxies comparable to those in previous CO TFR studies. Our work thus
provides a robust local benchmark to be used for comparison with
future CO work. In particular, Torii et al. (in prep.) will build
a local reference sample based on observations of very nearby
galaxies with the NANTEN2 telescope, and utilising identical fitting
methods to ours. Topal et al. (in prep.) will compare the TFRs
presented here with those measured using a sample of galaxies at $z\lesssim0.3$, including luminous infrared galaxies
(LIRGs) and galaxies from the Evolution of Gas in Normal Galaxies
(EGNoG) survey.

With the dawn of ALMA (and NOEMA), it is now possible to relatively
rapidly measure the CO emission of galaxies to large redshifts, when
the first objects were forming and slowly settling into the discs we
see today. In particular, significant samples of galaxies observed in
CO are now being built to probe the epoch of peak global star
formation activity ($1\lesssim z\lesssim3$), when turbulent gas-rich
galaxies were building the bulk of their stellar mass. Our work
therefore provide a robust reference point with which to compare
future TFR studies of those objects.

\section*{Acknowledgements}
We thank the anonymous referee for their constructive review of this work. AT acknowledges support from an STFC Studentship. MB acknowledges support from STFC rolling grants `Astrophysics at Oxford’ ST/H002456/1 and ST/K00106X/1. 
ST was supported by the Republic of Turkey, Ministry of National
Education, and The Philip Wetton Graduate Scholarship at Christ
Church. TAD acknowledges support from a Science and Technology Facilities Council Ernest Rutherford Fellowship. We thank the COLD GASS team for kindly providing us with additional data for this work.




\bibliographystyle{mnras}
\bibliography{TILEY_COLDGASS.bib} 




\appendix

\section{Velocity Measure:  $W_{50}$}
\label{sec:velocity}

\subsection{Fitting Functions}
\label{subsec:functions}

As stated in \S~\ref{subsec:measurew50}, as a characteristic velocity measure we adopt $W_{50}$, the width of the CO(1-0) integrated profile at 50\% of its maximum. Given the non-negligible noise in the COLD GASS spectra, a fit to the profile is preferred to a direct measurement of $W_{50}$. However, the choice of the fitting function is not trivial and several different functions have been used in the past. In this section we examine the relative merits of several of these functions for measuring $W_{50}$. Our goal is to ascertain which function is the most accurate and most importantly minimises potential biases as a function of $A/N$, inclination
(apparent width) and rotation velocity (intrinsic width) (all related to the total $S/N$).

The functions compared include the standard single Gaussian,
\begin{equation} \label{eq:gaussian}
  f(v) = A\,{\rm e}^{\frac{-(v-v_{0})^{2}}{2\sigma^{2}}}\,\,\,,
\end{equation} 
where $v$ is the velocity, $A>0$ is the amplitude of the peak (maximum
flux), $v_{0}$ is the velocity of the peak (and mean velocity; taken
to be within $\pm500$~km~s$^{-1}$ of the known systemic velocity), and
$\sigma>11.5$~km~s$^{-1}$ (the velocity bin width of the COLD GASS
spectra) is the root mean square (rms) velocity (i.e.\ the velocity
width) of the profile.

We also test a symmetric (with respect to the central velocity)
Gaussian Double Peak function, composed of a parabolic function
surrounded by two equidistant and identical (but mirrored)
half-Gaussians forming the low and high velocity edges of the profile:
\begin{equation} \label{eq:dgpeak}
  f(v)=
    \begin{cases}
      A_{\text{G}}\times\,{\rm e}^{\frac{-[v-(v_{0}-w)]^{2}}{2\sigma^{2}}} & v<v_{0}-w \\
      A_{\text{C}}+a\,(v-v_{0})^{2} & v_{0}-w\leq v\leq v_{0}+w\,\,\,, \\
      A_{\text{G}}\times\,{\rm e}^{\frac{-[v-(v_{0}+w)]^{2}}{2\sigma^{2}}} & v>v_{0}+w \\
    \end{cases}
\end{equation}
where $v_{0}$ is the central (mean) velocity (again taken to be within
$\pm500$~km~s$^{-1}$ of the known systemic velocity), $w>0$ is the
half-width of the central parabola, $\sigma>0$~km~s$^{-1}$ is the width of the profile edges, $A_{\text{G}}>0$ is the peak flux
of the two half-Gaussians (centred at $v_{0}\pm\,w$), $A_{\text{C}}>0$
is the flux at the central velocity, and
$a=(A_{\text{G}}-A_{\text{C}})\,/\,w^{2}$.

A variation of the Gaussian Double Peak function was also tested, the
Exponential Double Peak function used by \citet{Crocker:2011aa}, where
the two half-Gaussian edges are replaced by exponentials:
\begin{equation}
  f(v)=
  \begin{cases}
      A_{\text{G}}\times\,{\rm e}^{\frac{[v-(v_{0}-w)]}{\sigma}} & v<v_{0}-w \\
      A_{\text{C}}+a\,(v-v_{0})^{2} & v_{0}-w\leq v\leq v_{0}+w\,\,\,, \\
      A_{\text{G}}\times\,{\rm e}^{\frac{-[v-(v_{0}+w)]}{\sigma}} & v>v_{0}+w \\
    \end{cases}
\end{equation}
where $A_{\text{G}}>0$ is the peak flux of the two exponentials, but
all the parameters otherwise have the same meaning as for the Gaussian
Double Peak function.

Finally, a fourth function was also tested, the generalised Busy
Function adopted and discussed in detail by
\citet{Westmeier:2014aa}. Briefly, it is the product of two error
functions $\erf(x)$ and a polynomial. The resulting shape is similar
to that of the double-peaked functions above, i.e.\ typically a peak
(from the error functions) on either side of a central dip (from the
polynomial). The generalised form is described by
\begin{equation}
  \begin{array}{r@{}l}
    B(v)=\frac{a}{4} &{} \times\,\{\erf\,[\,b_{1}(v-\gamma_{1})\,]+1\} \\
                     &{} \times\,\{\erf\,[\,b_{2}(\gamma_{2}-v)\,]+1\} \\
                     &{} \times\,(c\,|v-v_{\text{p}}|^{n}+1)\,\,\,, \\
  \end{array}
\end{equation}
where $b_{1}>0$ and $b_{2}>0$ are respectively the slope of the left
and right error function, whilst $\gamma_{1}$ and $\gamma_{2}$
describe their respective width. $n$, $c>0$ and $v_{\text{p}}$ are
respectively the order, slope and offset of the polynomial. We adopted
$n=2$, yielding a parabola as for the double-peaked functions
above. $a>0$ determines the normalisation of the profile. 

\subsection{Tests}
\label{subsec:tests}

\begin{figure}
\begin{minipage}[]{1\textwidth}
\includegraphics[scale=0.21, trim= 5 50 5 0, clip=true]{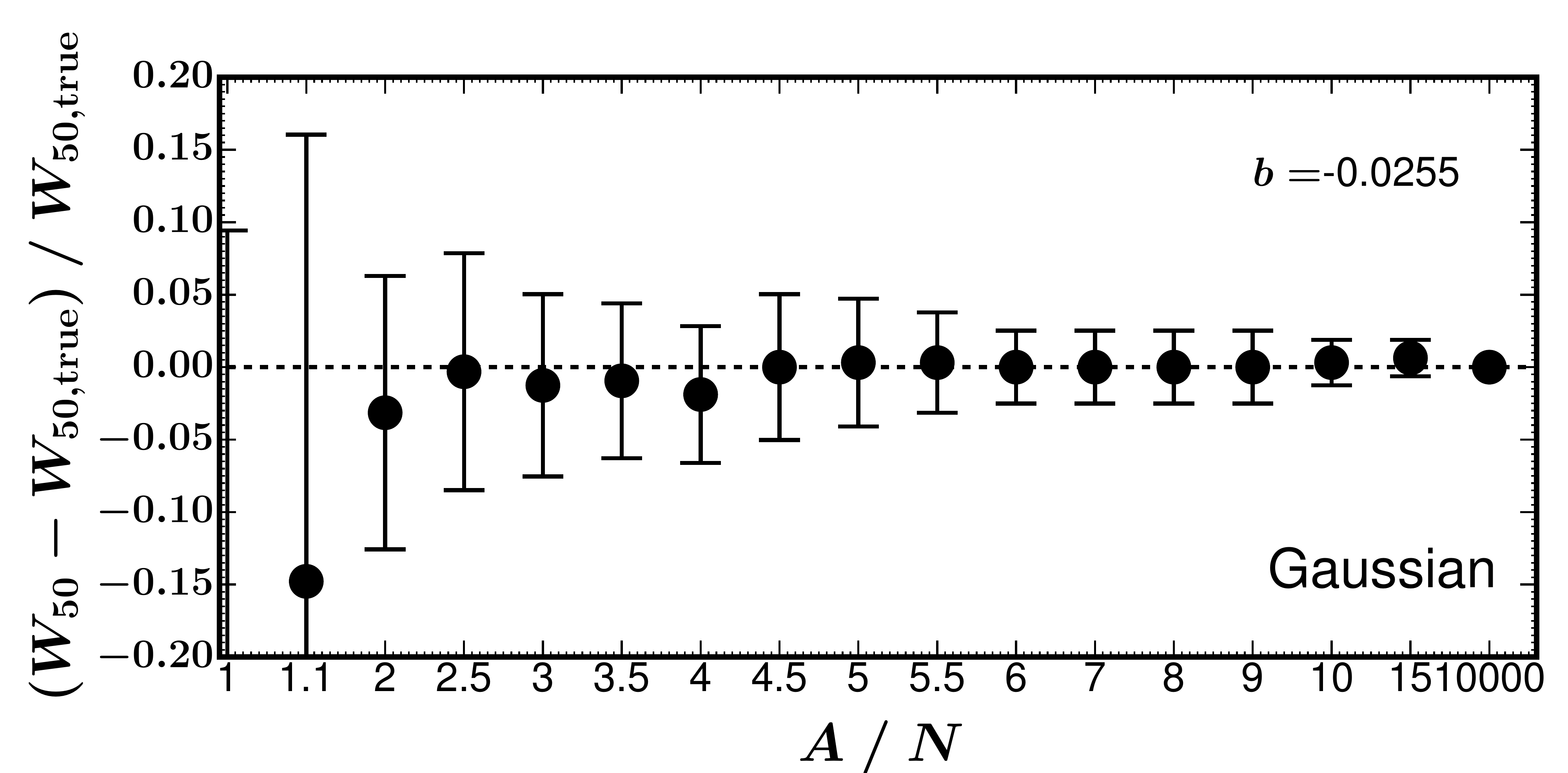}
\end{minipage}
\begin{minipage}[]{1\textwidth}
\includegraphics[scale=0.21, trim= 5 50 5 0, clip=true]{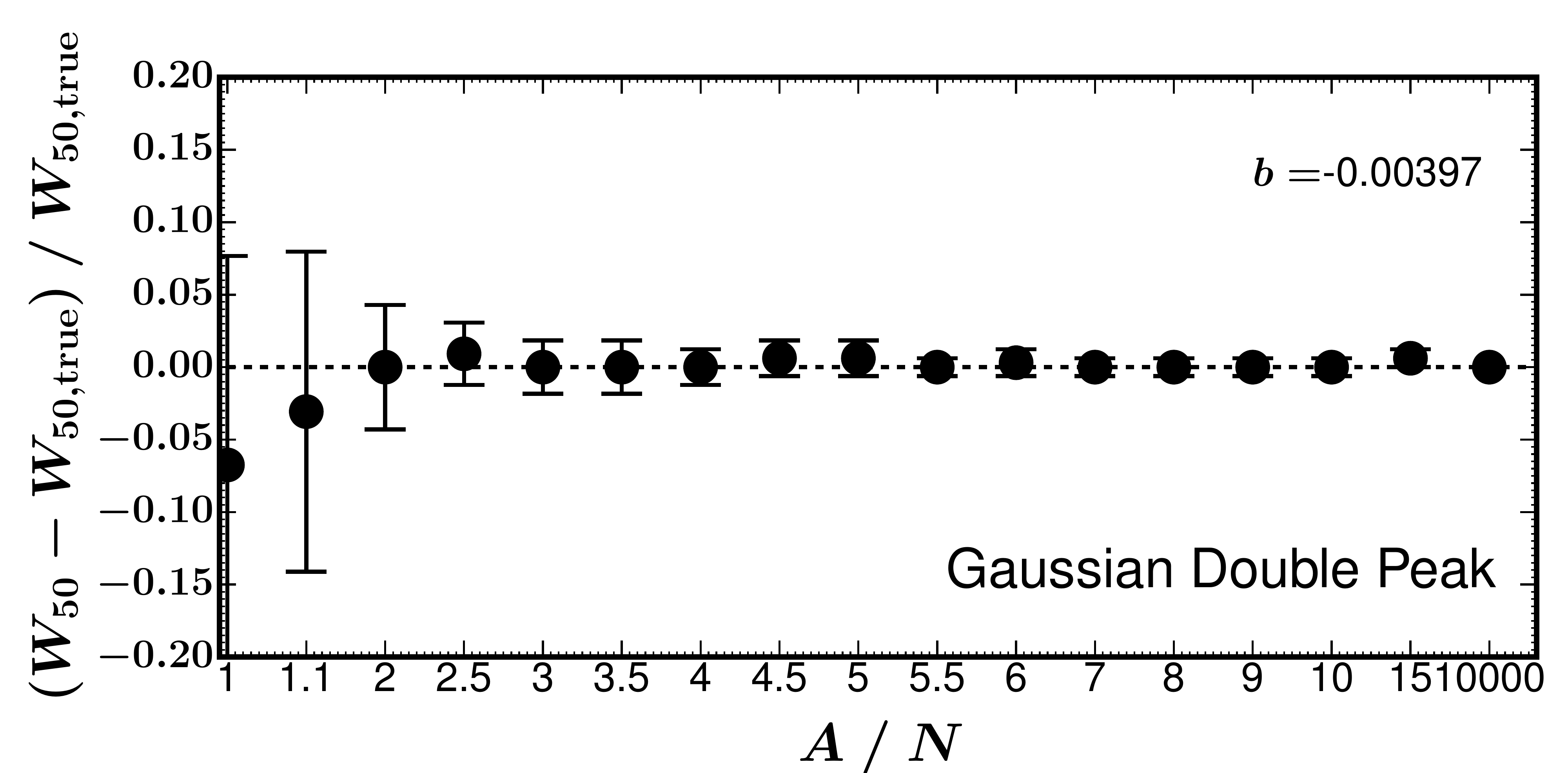}
\end{minipage}
\begin{minipage}[]{1\textwidth}
\includegraphics[scale=0.21, trim= 5 50 5 0, clip=true]{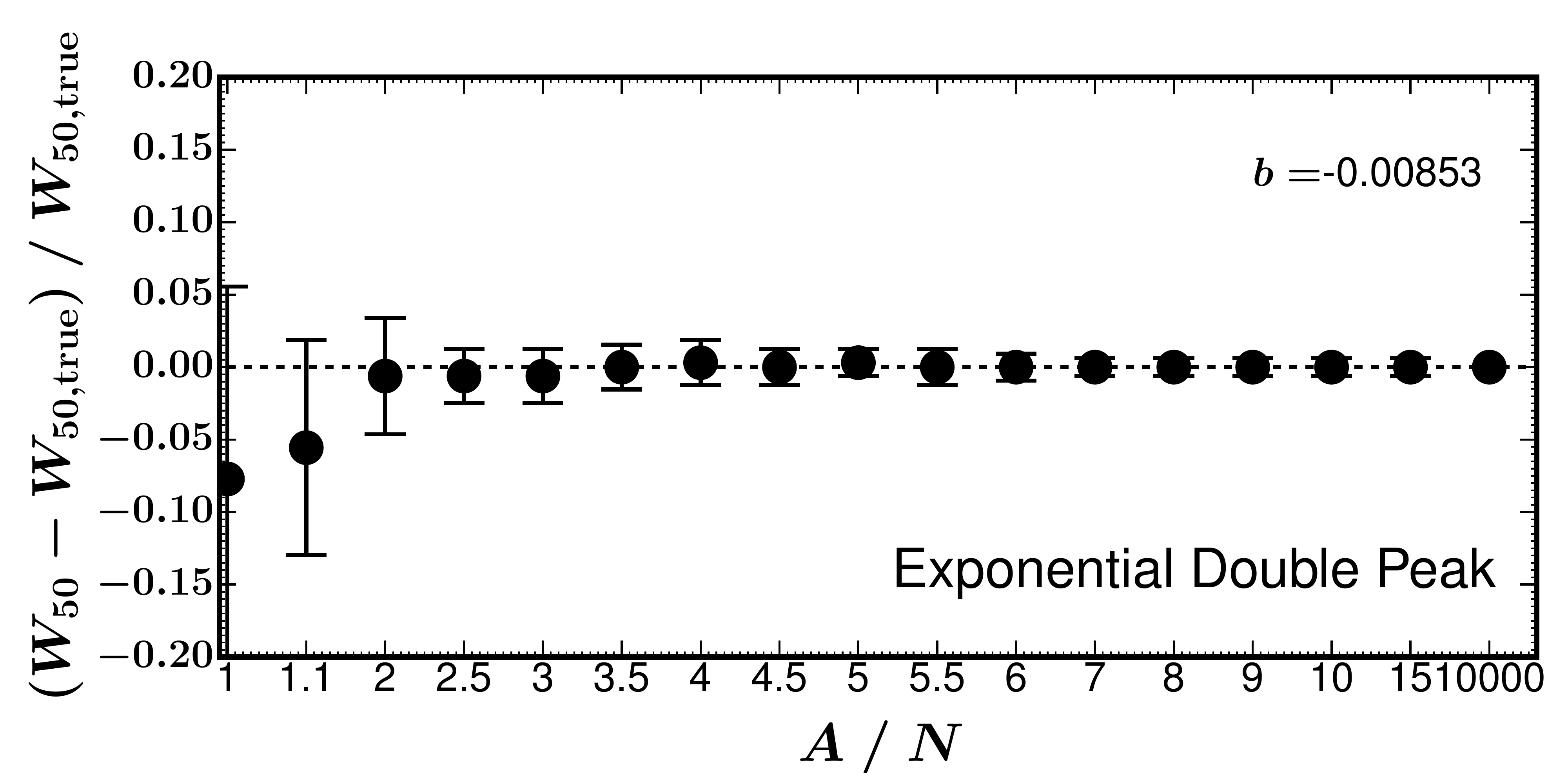}
\end{minipage}
\begin{minipage}[]{1\textwidth}
\includegraphics[scale=0.21, trim= 5 0 5 0, clip=true]{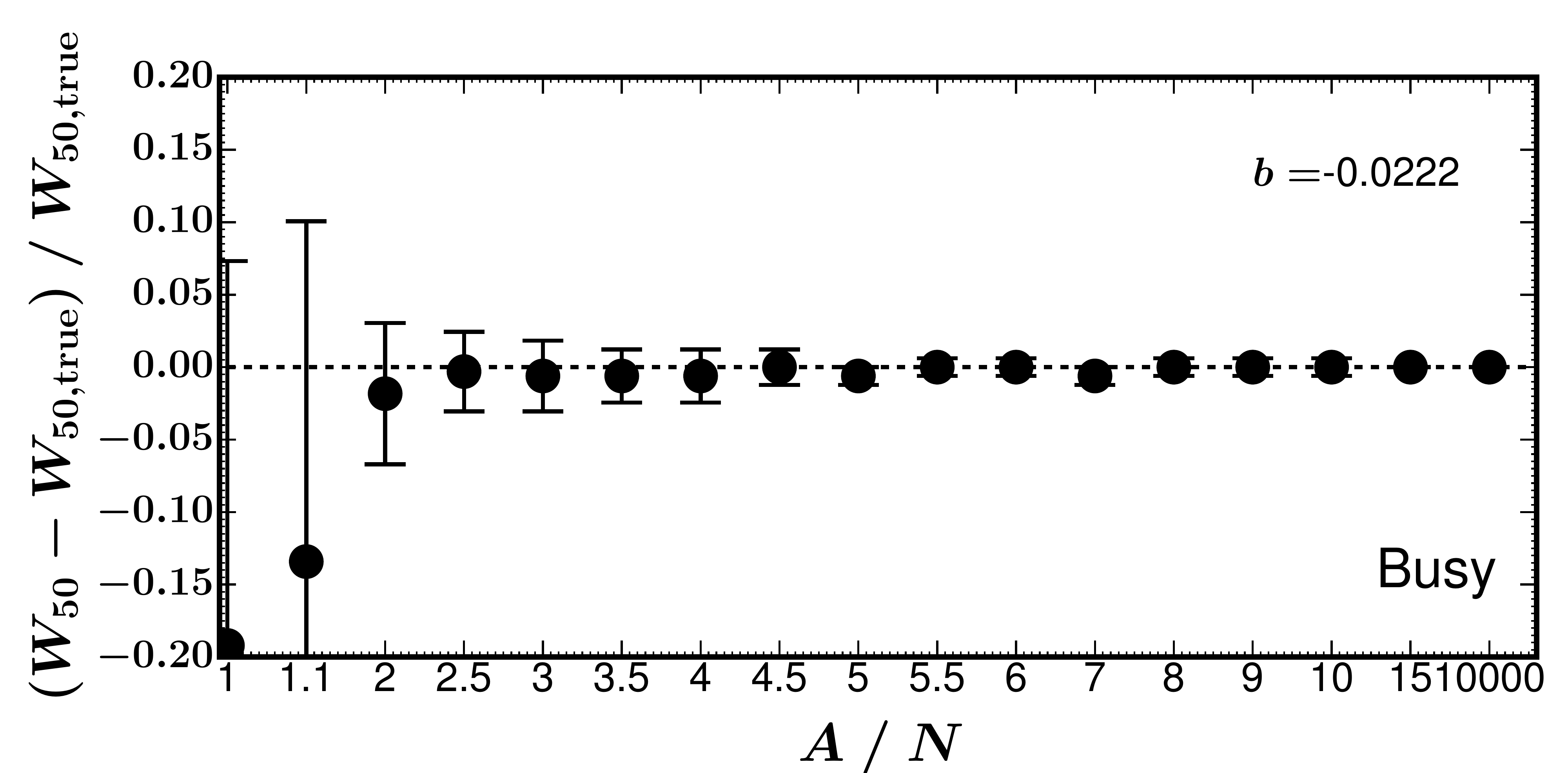}
\end{minipage}
\caption{Fractional difference between $W_{50}$ measured at $A/N=10000$ ($W_{50, \rm{true}}$, effectively noiseless) and $W_{50}$ measured for various values of $A/N$, this for an example case where the velocity $V_{\text{c,flat}}=210 \rm{\ km\ s^{-1}}$ and inclination $i = 70^\circ$. The 4 panels show how consistent the measured $W_{50}$ is for each of the four functions discussed in \S~\ref{subsec:functions} (i.e. Gaussian, Gaussian Double Peak, Exponential Double Peak, and Busy function). We quantify this with a bias measure $b$ defined in \S~\ref{subsec:tests} and shown in the top-right corner of each panel.}%
\label{fig:fracdiff}
\end{figure}

\begin{figure*}
\begin{minipage}[]{1\textwidth}
\includegraphics[scale=0.52, trim= 30 85 10 10, clip=true]{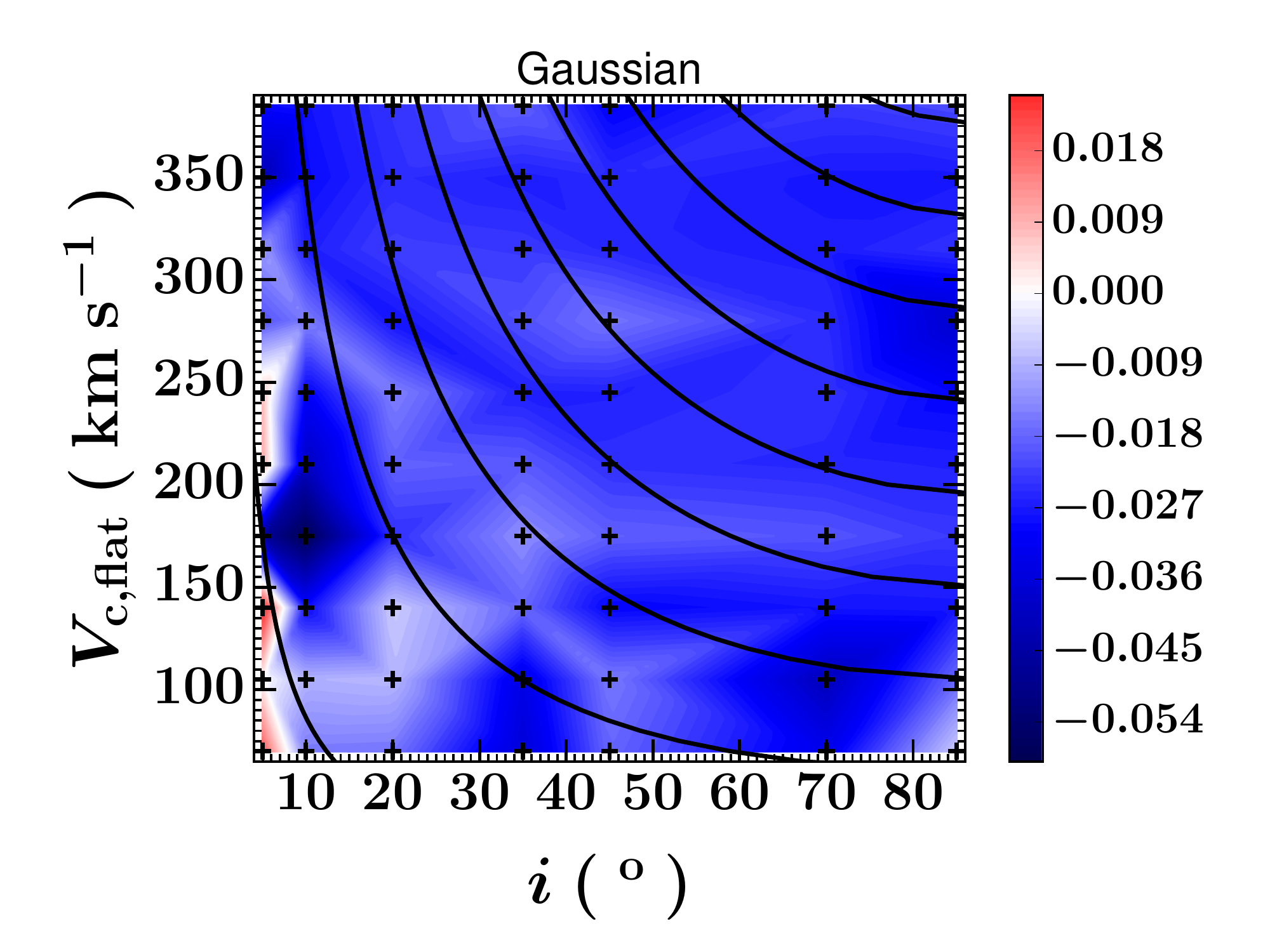}\includegraphics[scale=0.52, trim= 115 85 10 10, clip=true]{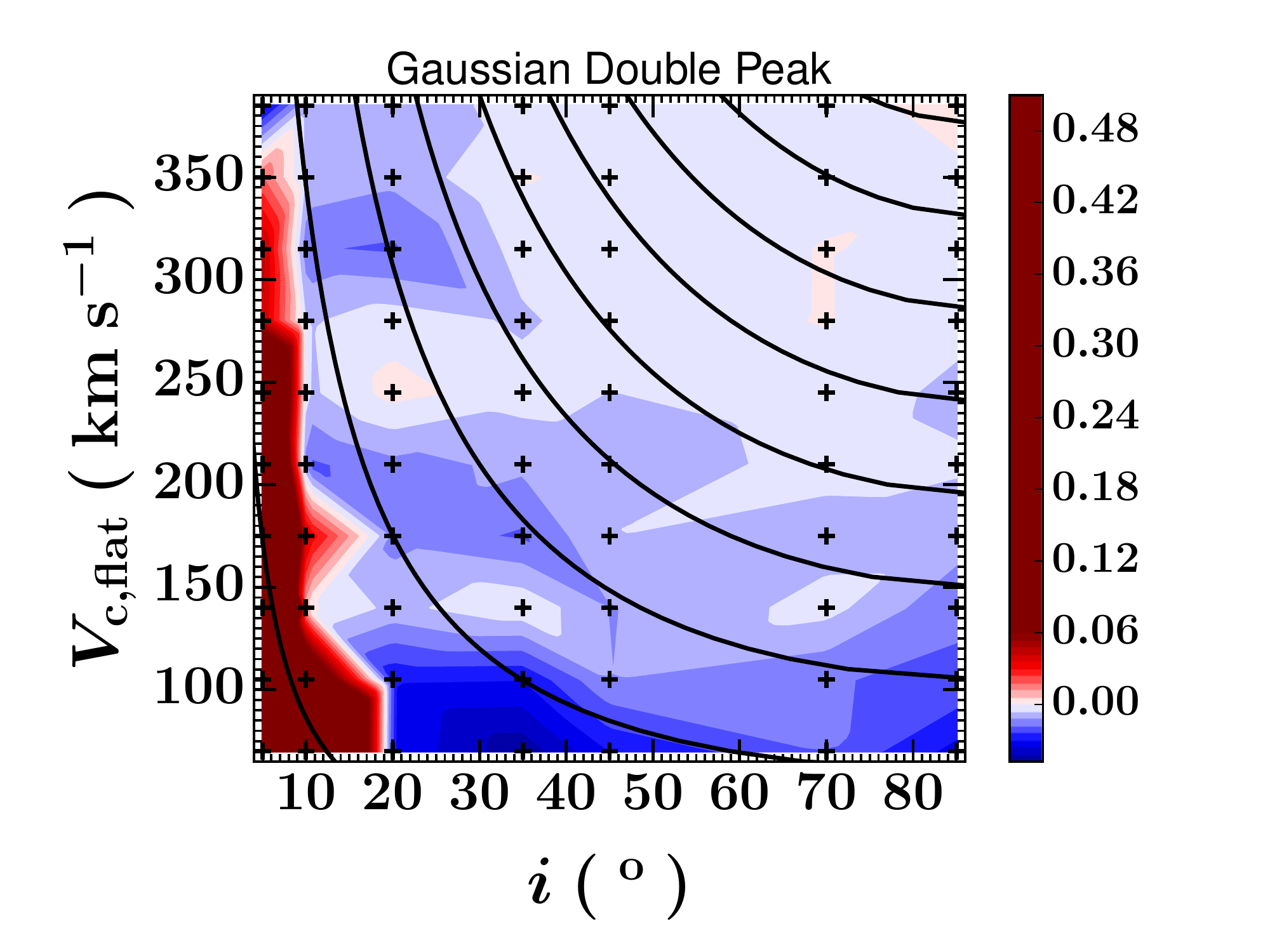}
\end{minipage}
\begin{minipage}[]{1\textwidth}
\includegraphics[scale=0.52, trim= 30 10 10 10, clip=true]{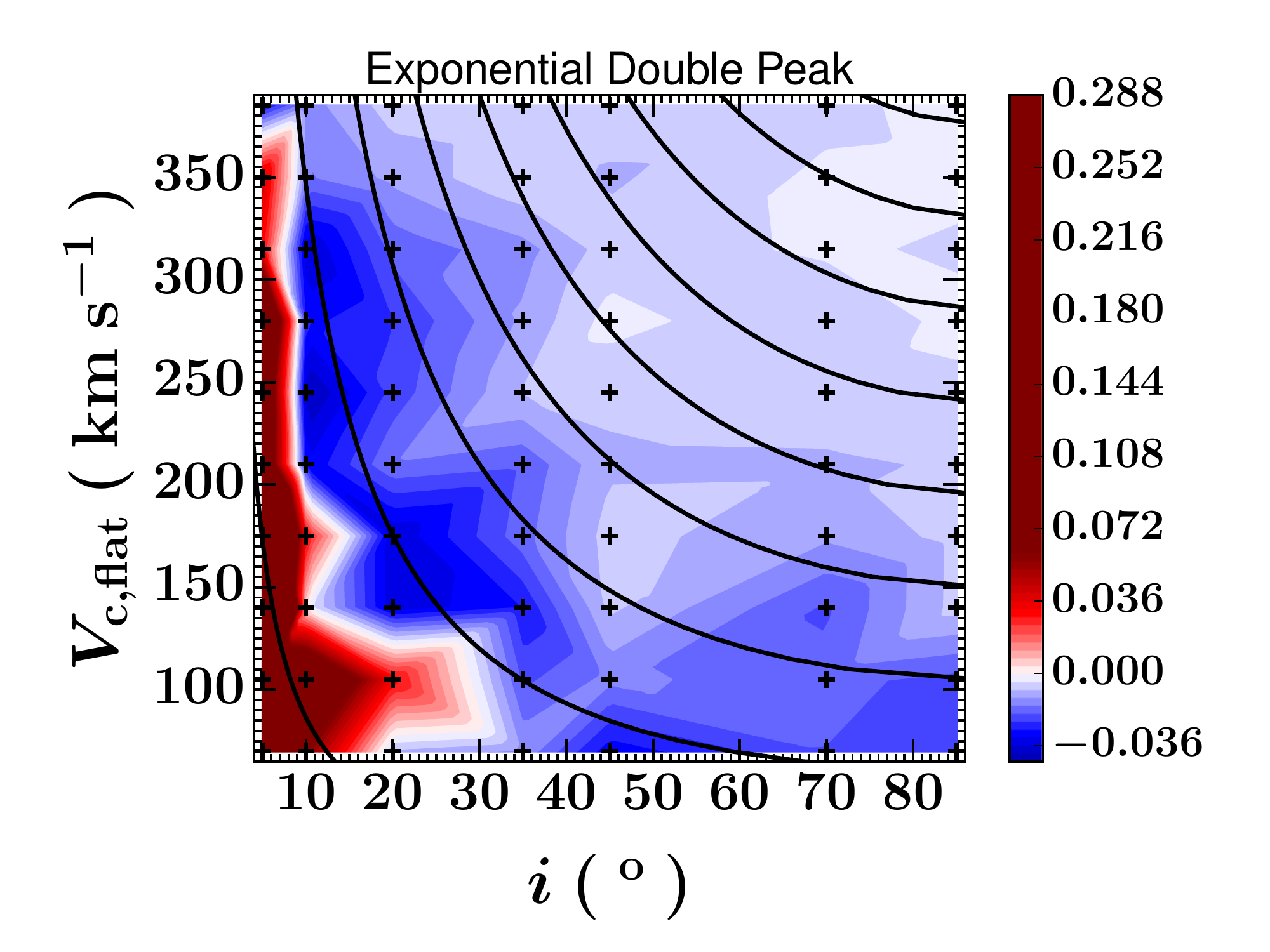}\includegraphics[scale=0.52, trim= 115 10 10 10, clip=true]{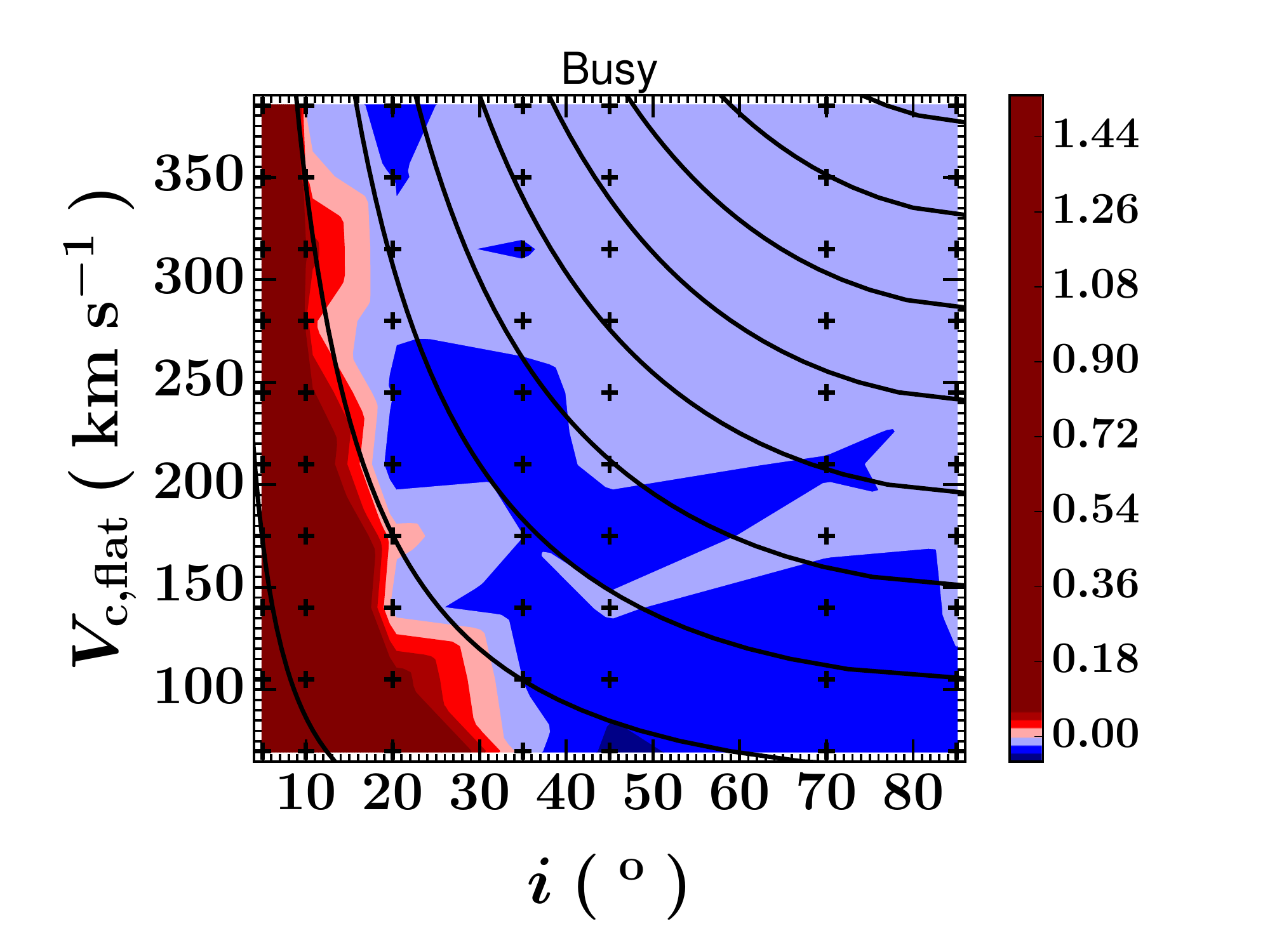}
\end{minipage}
\caption{The bias $b$ (as defined in the text) of the four analytical functions considered, as a function of the circular velocity $V_{\text{c,flat}}$ and the inclination $i$. For each function, the common colour scale shows whether the fit is positively (red) or negatively (blue) biased. The data points (black crosses) indicate
$V_{\text{c,flat}}$ -- $i$ pairs where a measurement was made. Black lines show curves of constant $V_{\text{c,flat}}\sin{i}$.}%
\label{fig:biasmaps}
\end{figure*}

To test which fitting function is most appropriate, a library of
noiseless integrated spectra was generated using the KINematic Molecular Simulation ({\sc
  KinMS}\footnote{purl.org/KinMS}) routine of \citet{Davis:2013}. Since the overall $S/N$ is a function of both $A/N$ and
the profile width, while we are ultimately only interested in
recovering the profile width for TFR studies, we have decided to probe
the effects of these two parameters separately.

The shape of a galaxy's integrated CO emission profile is primarily dependent on the physical properties of the galaxy, and to a lesser extent on the nature of the telescope used to observe it. Broadly, the width of the profile depends on the galaxy's projected circular velocity, and therefore its dynamical mass \citep{Casertano:1980}, whilst the breadth of the flanks of the profile depends on the intrinsic turbulence of the gas. The integral of the profile is proportional to the galaxy's total molecular gas mass \citep[e.g.][]{Maloney:1988}. The overall shape of the profile also depends on the distribution of the gas within the galaxy: the radial density profile will determine to what extent the CO emission samples the flat part of the galaxy's rotation curve - if the gas extends out to sufficient radii, the intrinsic profile will display a double-horned or boxy shape \citep[e.g.][]{Davis:2011aa}. Additionally, the central flux of the profile (and thus whether it is intrinsically double-horned or boxy) depends on the gas concentration within the disc \citep[e.g.][see \S~\ref{subsec:subsample}]{Wiklind:1997,Lavezzi:1997aa}. Whether or not the {\it observed} profile appears Gaussian, boxy or double-horned also depends on the width of the velocity channels and the beam size of the telescope used to observe the galaxy; if the velocity channels are too broad or the beam size is smaller than the radial size of the galaxy, then the integrated emission profile may appear Gaussian despite being intrinsically boxy or double-horned (in the latter case it is not a true integrated profile).          

Bearing all this in mind, the model spectra were created by assuming an edge-on exponential disc
of molecular gas with a scalelength of $2\arcsec$, a realistic
circular velocity curve that peaks at $3\arcsec$ and then remains
flat, and a fixed molecular gas velocity dispersion of
$12$~km~s$^{-1}$. A Gaussian single-dish telescope response with a
$22\arcsec$ beam (full width at half maximum) was used to integrate
the flux spatially, matching that of the IRAM $30$~m telescope. This
also ensures that our spectra contain essentially all the flux of the
modeled discs, and the resulting spectra are intrinsically double-horned
or boxy shaped.

Model integrated spectra with flat circular velocities
$V_{\text{c,flat}}$ ranging from $70$ to $385$~km~s$^{-1}$, and with inclinations ranging from $5^\circ$ to $85^\circ$, were
generated and then binned to $11.5$~km~s$^{-1}$ per channel to match
the COLD GASS spectra. Each of the
resulting spectra was then degraded by adding random Gaussian noise
such that the desired $A/N$ was reached, where the amplitude $A$ is
defined here as the peak of the spectrum before noise was added, and
the noise $N$ is defined as the root-mean-square (rms) of the spectrum
in an area devoid of emission (thus equal to the dispersion of the
Gaussian used to generate the noise).

For each input circular velocity $V_{\text{c,flat}}$, inclination $i$ and
amplitude-to-noise ratio $A/N$, $150$ realisations of the resulting
model spectrum were generated. Each of these realisations was then fit
with the four different functions described in
\S~\ref{subsec:functions}, and the width of the best fitting profile at
$50\%$ of the peak ($W_{50}$) was calculated. The fits were carried out
using the {\sc Python} package {\sc mpfit}\footnote{https://code.google.com/p/astrolibpy/source/browse/mpfit/mpfit.py} (\citealt{Markwardt:2009}; translated into {\sc Python} by Mark River and updated by Sergey Koposov), that
employs a Levenberg-Marquardt minimisation algorithm. The adopted
$W_{50}$ and its uncertainty for each combination of
$V_{\text{c,flat}}$, $i$ and $A/N$ were then taken respectively as the
median and median absolute deviation (with respect to the median itself) of
the $150$ associated measurements. 

Figure~\ref{fig:fracdiff} shows the fractional difference
between the true width ($W_{50,\rm{true}}$), defined as $W_{50}$ measured at $A/N = 10,000$ (i.e. for an effectively noiseless spectrum), and that measured as a function of $A/N$, this for the case where $V_{\text{c,flat}}=210 \rm{\ km\ s^{-1}}$, and $i = 70^\circ$. We do not show the results for every combination of velocity and inclination that we tested, but rather select this case as an illustrative example. We do however summarise the results for each function tested in the four panels of Figure~\ref{fig:biasmaps}. This shows how self-consistent each of the four tested functions is as a function of $V_{\text{c,flat}}$ and $i$, where
self-consistency is taken here to mean both that the measured width is similar to its true value and that it does not vary systematically with decreasing $A/N$. This is judged by measuring the bias $b$, defined as the average fractional width difference over all $A/N$ sampled. The bias for each function, and for each of the selected values of $V_{\text{c,flat}}$ and $i$, is shown in the panels of Figure~\ref{fig:biasmaps}. It is clear that both Double Peak functions overestimate to varying degrees the line width (with respect to $W_{50,\rm{true}}$) for low values of inclination ($i \lesssim 20^\circ$), but are otherwise only slightly negatively biased. The Busy function displays the same trends, but is negatively biased to a greater degree than the Double Peak functions for high values of inclination ($i \gtrsim 20^\circ$). The Gaussian function, however, is negatively biased in most cases, and to a greater degree than each of the three other functions.

\begin{figure}
  \includegraphics[scale=0.37,trim= 35 5 20 25, clip=True]{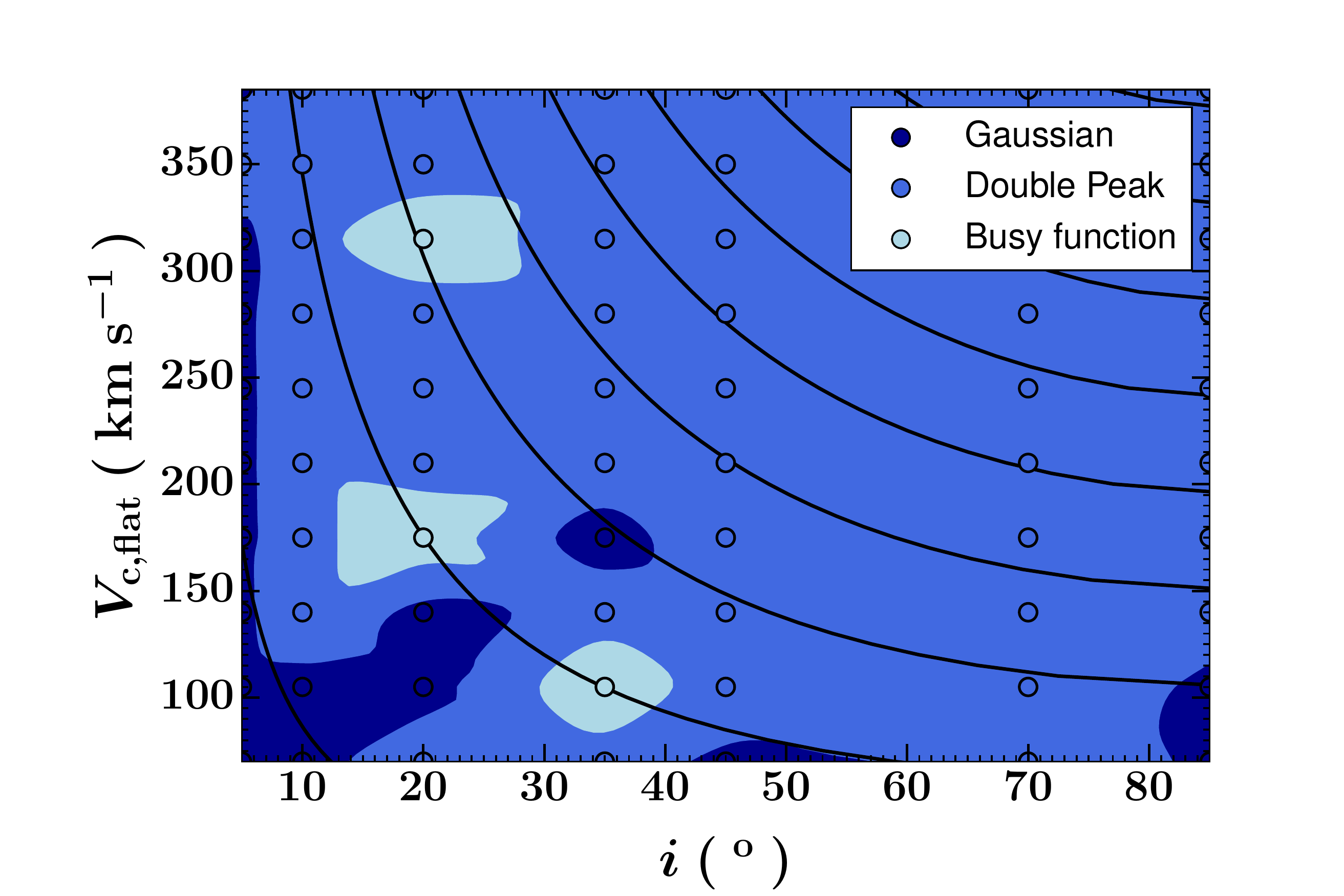}
  \caption{Self-consistency (as defined in the text) of the four
    analytical functions considered, as a function of the circular
    velocity $V_{\text{c,flat}}$ and the inclination $i$. The colour coding shows
    which function is most self-consistent (dark blue: Gaussian function;
    blue: either of the Gaussian Double Peak or Exponential Double
    Peak function; light blue: Busy function). The data points (open circles) indicate
    $V_{\text{c,flat}}$ -- $i$ pairs where a measurement was made. The
    underlying colour scale shows tri-tonal contours of the same
    measurements, for ease of interpretation. Black lines show curves
    of constant $V_{\text{c,flat}}\sin{i}$. The Double Peak functions are the most self-consistent in the majority of the cases tested.}
  \label{fig:selfconsistency}
\end{figure}

Figure~\ref{fig:selfconsistency} attempts to distill further the information contained in Figure~\ref{fig:biasmaps}. It shows which function is most
self-consistent as a function of $V_{\text{c,flat}}$ and $i$, where
self-consistency is again judged by the bias $b$. We found that the Gaussian Double Peak
and Exponential Double Peak functions yield very similar results in
terms of self-consistency, recovering the true width to a similar accuracy at a given circular velocity, inclination and amplitude-to-noise ratio. This is clear in the panels of Figure~\ref{fig:biasmaps}. Given the restricted
velocity resolution of the COLD GASS spectra, and since the only real
difference between the Gaussian Double Peak and the Exponential Double
Peak function is the shape of the edges, it was decided to group these
two functions together when considering their self-consistency with
respect to that of the Gaussian and Busy functions.

What is immediately obvious from Figure~\ref{fig:selfconsistency} is
that the Busy function features very little in the plot; it was
rarely the most self-consistent function. The main result is that the double-peaked functions are the most self-consistent in the majority of cases, be it the Gaussian
Double Peak or the Exponential Double Peak function. The single
Gaussian is the most self-consistent only at small inclinations (i.e.\
face-on discs) or small circular velocities. These two extremes are of
course degenerate observationally, and this result is easily
understood. Indeed, since any galaxy spectrum has a finite spectral
resolution, the integrated velocity profile will appear Gaussian
regardless of its intrinsic shape if the inclination or the circular
velocity is small enough.

\begin{figure*}
  \includegraphics[scale=0.47, trim= 40 0 50 0, clip=True]{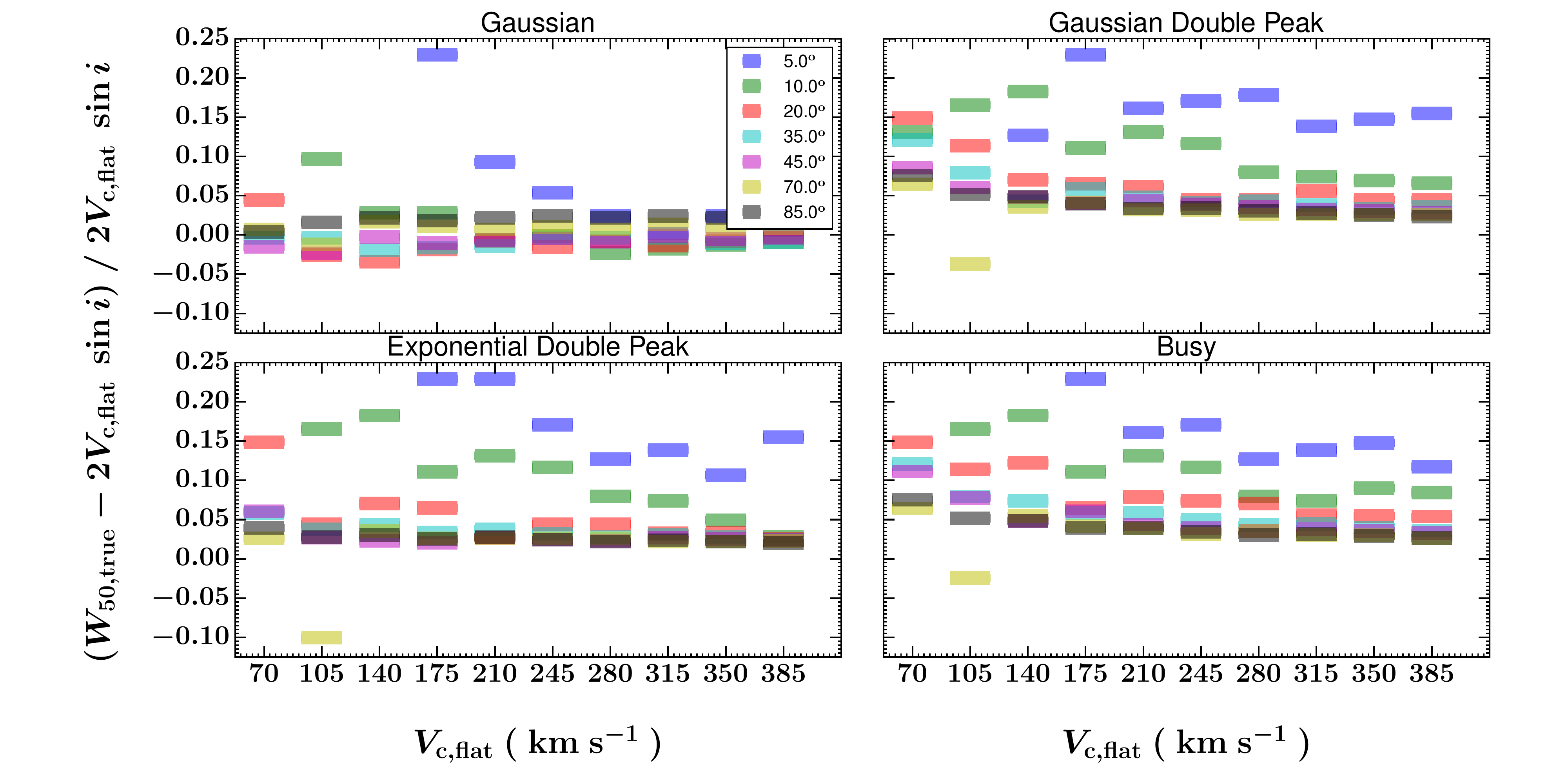}
  \caption{Fractional difference between the true width
    recovered ($W_{50,\rm{true}}$, $W_{50}$ for $A/N=10,000)$ and twice the projected input
    circular velocity $2V_{\text{c,flat}}\sin{i}$, this as a function of
    $V_{\text{c,flat}}$ itself for different inclinations $i$. Each panel shows a
    different fitting function. In general, measurements of $W_{50,\rm{true}}$ are reasonably constant (as a function of $V_{\text{c,flat}}$) for $i\geq35^\circ$ only.}
  \label{fig:incselfconsistency}
\end{figure*}

In addition to self-consistency, one should also consider how accurately each function recovers the circular velocity. For each of the
four functions tested, Figure~\ref{fig:incselfconsistency} thus shows the fractional difference between the true width ($W_{50,\rm{true}}$) and (twice) the projected circular velocity $2V_{\text{c,flat}}\sin{i}$, this as a function of $V_{\text{c,flat}}$ itself and for different values of the inclination $i$. In other words, Figure~\ref{fig:incselfconsistency} summarises the position of the dashed black line, with respect to $2V_{\text{c,flat}}\sin{i}$, in each of the panels of Figure~\ref{fig:fracdiff} (but now for every velocity and inclination tested).

We recall here that a constant fractional offset is unimportant for
TFR work, but that useful functions will minimise any systematic
variation with circular velocity (and
$A/N$). Figure~\ref{fig:incselfconsistency} thus clearly shows that no
function yields constant $W_{50,\rm{true}}$ measurements (as a function
of $V_{\text{c,flat}}$) at inclinations $i\leq10^\circ$, only the
Gaussian Double Peak function does a reasonable job at $i=20^\circ$,
and all functions are satisfactory at $i\geq35^\circ$.

Given that the cases in which the Gaussian function is most
self-consistent cover only a small area in the parameter space of
$V_{\text{c,flat}}$ and $i$ (see Fig.~\ref{fig:selfconsistency}), and
given that no function yields constant measurements at small $i$
(Fig.~\ref{fig:incselfconsistency}), two conclusions must be
drawn. First, galaxies observed at small inclinations ($i \le 30^\circ$) should be
avoided. Second, the Gaussian Double Peak function is the most
appropriate function with which to measure the CO(1-0) line widths of
the COLD GASS galaxies (the Gaussian form of the profile's edge is
also better justified physically than the exponential form). It should
however be stressed that these conclusions are drawn here exclusively
from simulated spectra with an intrinsic regular double-horned shape.

\section{Comparison with H{\small I}}
\label{sec:HIcomparison}

In \S~\ref{subsec:motivation} we discussed the relative advantages of constructing the TFR using CO(1-0) as a dynamical tracer rather than H{\small I}, the use of the latter being well established. In this section we compare the values of $W_{50}$ derived from COLD GASS CO(1-0) integrated profiles  to those derived from GASS H{\small I} integrated profiles of the same galaxies. We measure the latter by fitting the Double Peak Gaussian function to the H{\small I} spectra in the same manner as described in \S~\ref{subsec:measurew50}. 

The majority of the GASS H{\small I} galaxy spectra were obtained with the Arecibo 305 m telescope (see \S~\ref{subsec:CO} for more detail). The spectra are available as part of the GASS data releases\footnote{http://wwwmpa.mpa-garching.mpg.de/GASS/data.php} \citep{Catinella:2009aa,Catinella:2012,Catinella:2013}. However, as described in \citet{Catinella:2009aa}, GASS galaxies were not re-observed with Arecibo if a suitable H{\small I} detection was already available from the ALFALFA survey or the Cornell H{\small I} archive \citep{Haynes:2011}. As such, the comparison conducted in this section draws on H{\small I} $W_{50}$ values measured from spectra from all three sources. We denote all measurements of the width of a galaxy's H{\small I} integrated profile at $50$\% of its maximum as $W_{50, \rm{H{\small I}}}$. For clarity, in this section only, we denote the CO(1-0) $W_{50}$ values described in \S~\ref{subsec:measurew50} as $W_{50, \rm{CO}}$.      

Of the 207 COLD GASS galaxies in the initial sample defined in \S~\ref{sec:data}, 155 originate from the COLD GASS DR3, with the remaining 52 from the low mass extension of COLD GASS. H{\small I} spectra for galaxies in the COLD GASS low mass extension are not publicly available, so we exclude these galaxies from our comparison. Of the remaining 155 galaxies, 140 have a corresponding H{\small I} spectrum that is publicly available. We detect a signal (i.e. $A/N > 1$) in 136 of these (76 observed directly by GASS, 38 from ALFALFA and 22 from the Cornell archive). As all 136 of these spectra obey the selection criteria of the initial sample as defined in \S~\ref{sec:data}, we include them all in our comparison. Of the 83 COLD GASS galaxies comprising the sub-sample defined in \S~\ref{subsec:subsample}, we exclude 17 galaxies that originate from the COLD GASS low mass extension (for the reason discussed). The remaining 66 galaxies originate from the COLD GASS DR3. 59 of these have a corresponding H{\small I} spectrum that is publicly available. To maintain consistency in our comparison, we apply the same selection criteria to the derived H{\small I} line widths as those used to define the sub-sample in \S~\ref{subsec:subsample}. To this end we exclude two galaxies with $A/N < 1.5$, leaving 57 spectra (30 from GASS, 16 from ALFALFA and 11 from the Cornell archive). 

\begin{figure}
  \includegraphics[scale=0.42,trim= 15 2 10 10, clip=True]{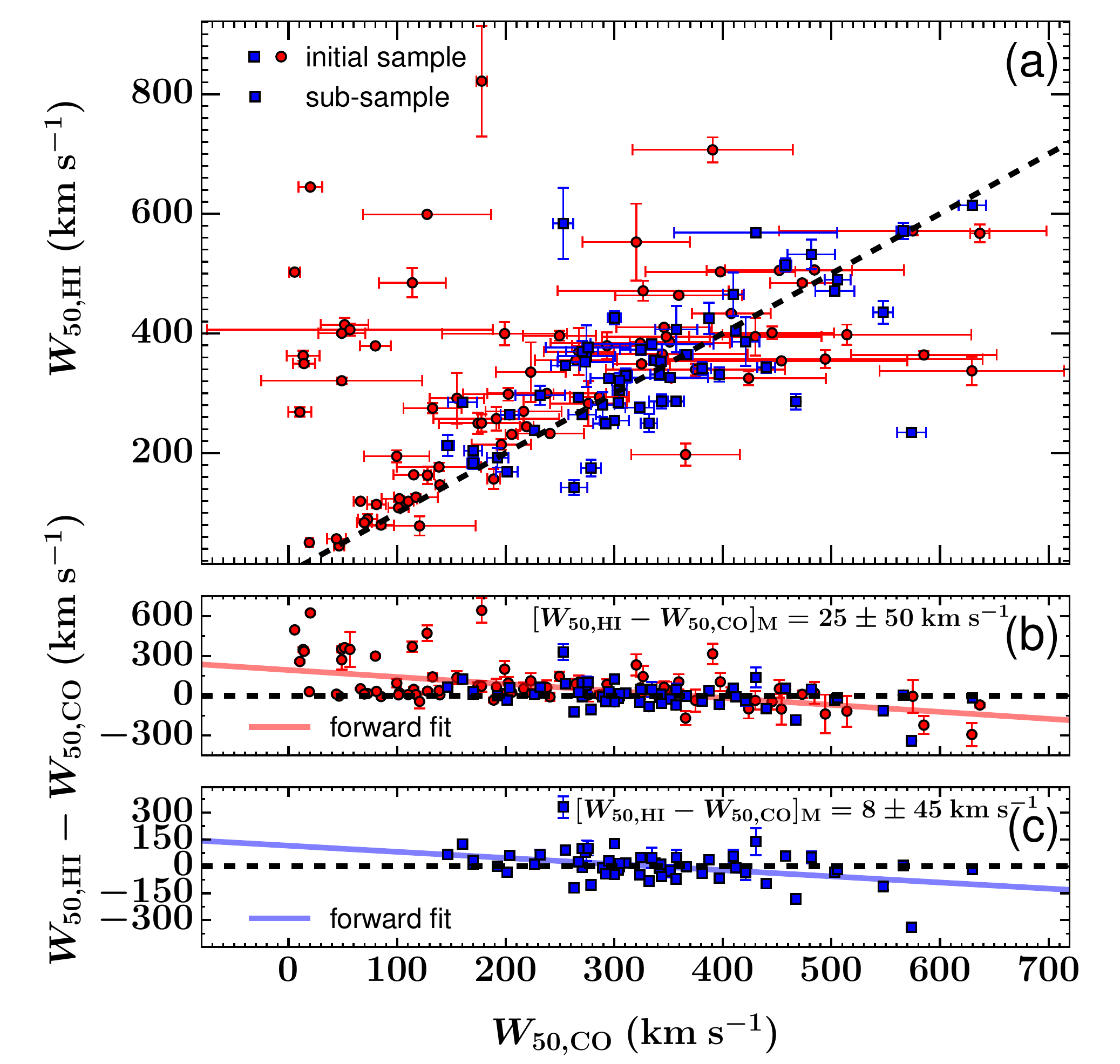}
  \caption{Comparison between $W_{50, \rm{CO}}$ and $W_{50, \rm{H{\small I}}}$ for galaxies in the initial sample (see \S~\ref{sec:data}; red circles and blue squares), and the final sub-sample (see \S~\ref{subsec:subsample}; blue squares) that have a corresponding H{\small I} detection in GASS. (a) Correlation between the two measures of line width for both the initial sample and sub-sample. The dashed black line represents the 1:1 relation. (b) and (c) Line width difference $W_{50, \rm{HI}} - W_{50, \rm{CO}}$ for the initial sample and sub-sample, respectively. The red solid line ($W_{50, \rm{HI}} - W_{50, \rm{CO}} =(-0.53\pm0.07) [W_{50, \rm{CO}} - 277.4] + 48 \pm 10$) and blue solid line ($W_{50, \rm{HI}} - W_{50, \rm{CO}}=(-0.3\pm0.1 )[W_{50, \rm{CO}} - 323.5] + 5 \pm 11$) represent a free forward fit to the data points in respectively (b) and (c). The median difference is displayed in the top right of both (b) and (c), along with its uncertainty (the median absolute deviation from the median itself). $W_{50, \rm{CO}}$ and $W_{50, \rm{H{\small I}}}$ differ much less for galaxies in the final sub-sample than for galaxies only included in the initial sample.}
  \label{fig:w50comp}
\end{figure}

We compare the values of $W_{50, \rm{CO}}$ and $W_{50, \rm{H{\small I}}}$ for both the initial sample and sub-sample in Figure~\ref{fig:w50comp}. This shows that, for both the sample and sub-sample, the two measures of line width generally correlate well with one another. However, the scatter around the 1:1 line is much larger for the initial sample than for the sub-sample. The increased scatter in the initial sample is a consequence of the inclusion of galaxies with CO(1-0) integrated profiles that do not display a double-horned or boxy shape, i.e. profiles for which we cannot be sure the CO sufficiently probes the outer parts of the galaxy's rotation curve (see \S~\ref{subsec:subsample}). Considering only the sub-sample, the two measures of line width are in good agreement ($[W_{50, \rm{H{\small I}}}-W_{50, \rm{CO}}]_{\text{M}} = 8 \pm 45$ km s$^{-1}$, where the uncertainty is the median absolute deviation with respect to the median itself).  

The H{\small I} velocity widths are typically larger than the CO velocity widths at lowest values of $W_{50, \rm{CO}}$, as is apparent from the free fit (solid red) line in panel (b) of  Figure~\ref{fig:w50comp}. Conversely, for the highest values of $W_{50, \rm{CO}}$, $W_{50, \rm{H{\small I}}}$ tends to be smaller. This trend is weaker but still present in the sub-sample, as is shown by the free fit (solid blue) line in panel (c) of  Figure~\ref{fig:w50comp}.   

\begin{figure*}
\centering
\begin{minipage}[]{1\textwidth}
\label{fig:HIMKall}
\centering
\includegraphics[width=0.5\textwidth, trim= 0 10 75 55, clip=True]{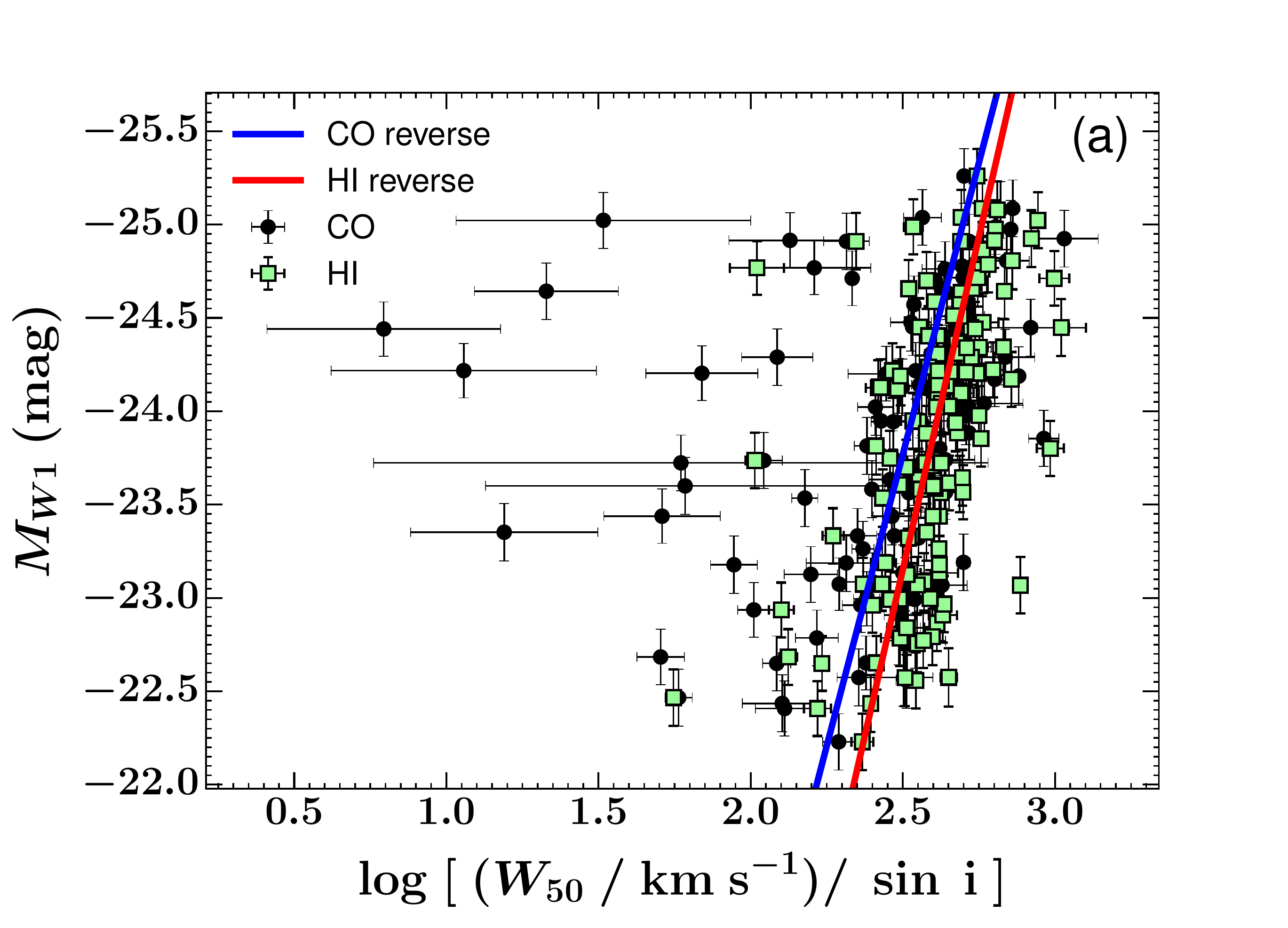}\includegraphics[width=0.5\textwidth, trim= 0 10 75 55, clip=True]{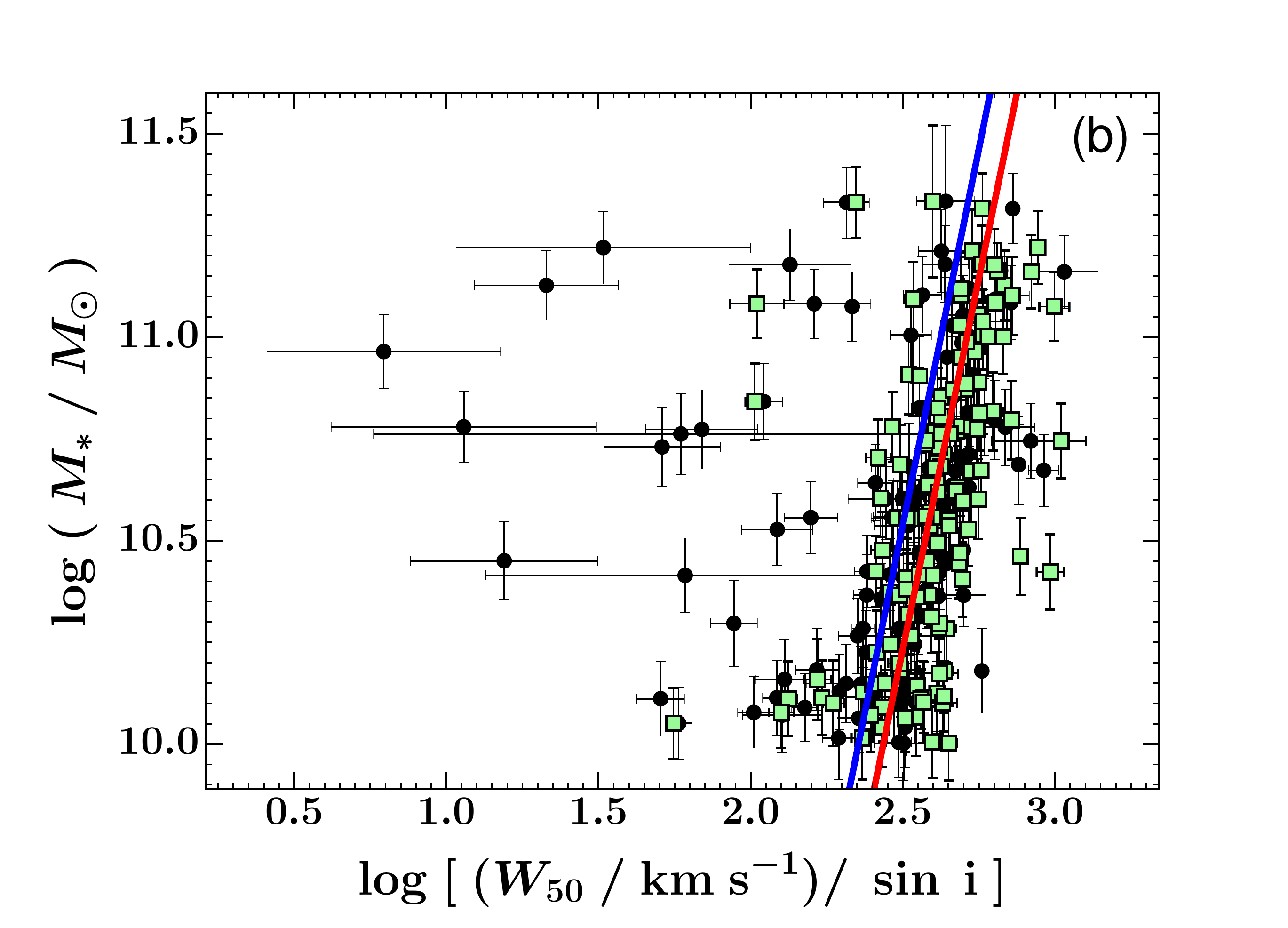}
\end{minipage}
\begin{minipage}[]{1\textwidth}
\label{fig:HIMKsub}
\centering
\includegraphics[width=0.5\textwidth,trim= 0 10 75 55, clip=True]{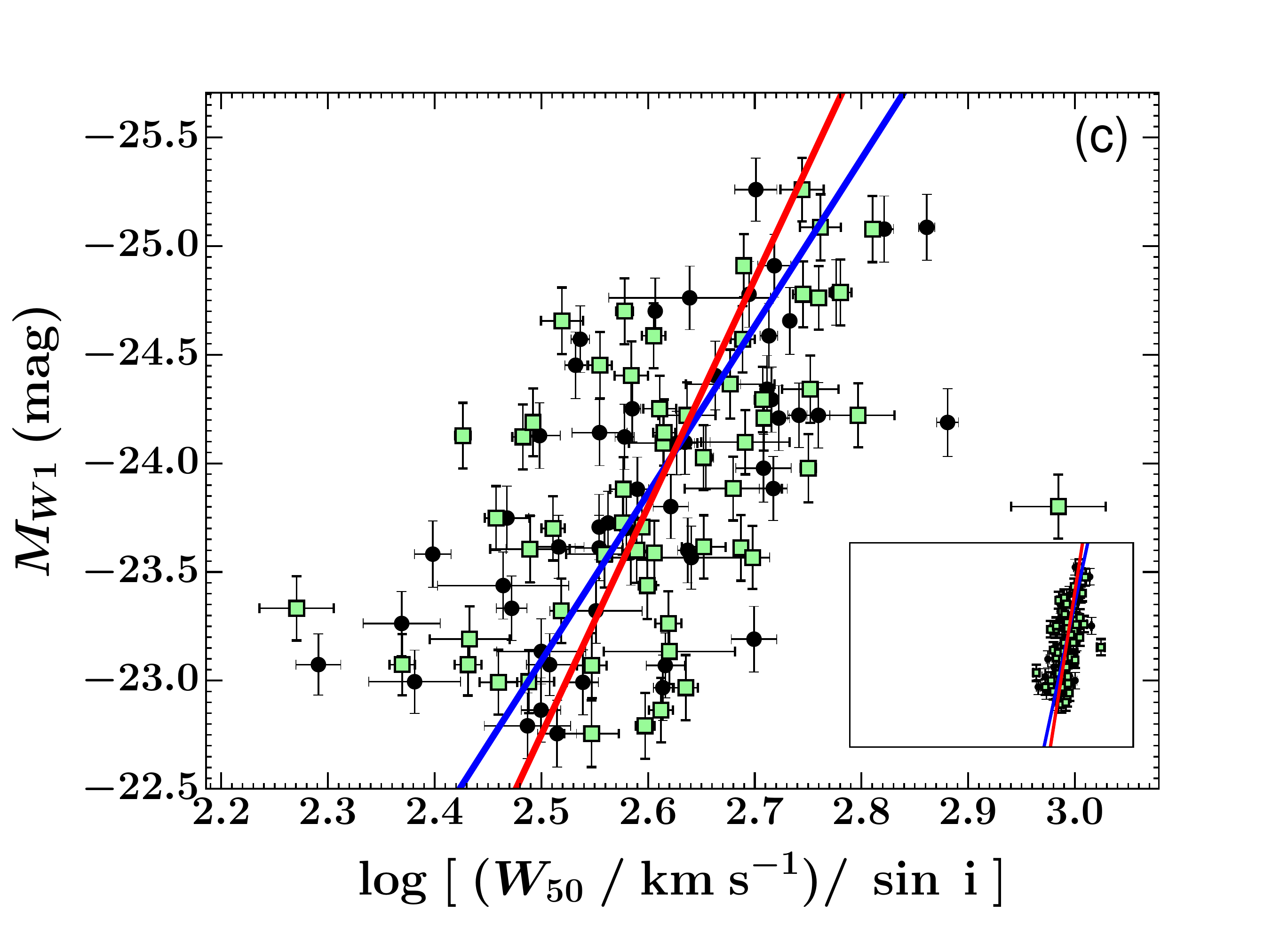}\includegraphics[width=0.5\textwidth, trim= 0 10 75 55, clip=True]{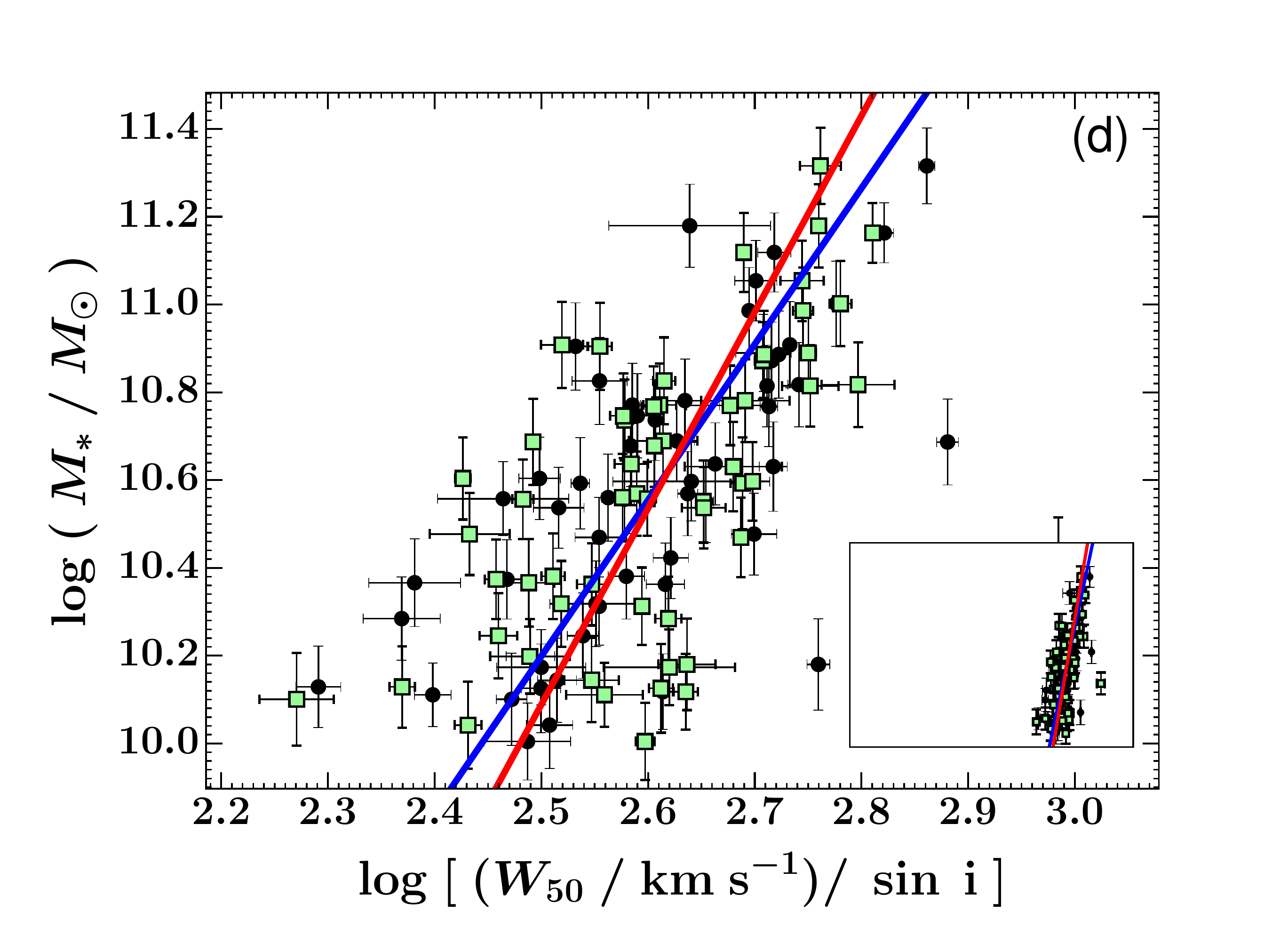}
\end{minipage}
  \caption{Comparison of the $W1$-band and stellar mass Tully-Fisher relations using $W_{50}$ values derived from COLD GASS CO(1-0) integrated profiles ($W_{50, \rm{CO}}$, black points) and GASS H{\small I} integrated profiles ($W_{50, \rm{HI}}$, pale green points), this for the same galaxies with both CO and H{\small I} data. (a) and (c)
    show the absolute $W1$-band TFR of those galaxies from respectively the initial COLD GASS sample and our final
    sub-sample. The x-axis is the width of the
    integrated profile (CO(1-0) or H{\small I}) at $50\%$ of the peak, corrected for
    the effect of inclination. For (a) and (c), the y-axis is the absolute {\it WISE}
    Band~1 magnitude $W1$ ($\approx3.4$~$\mu$m). The solid blue line show the reverse fit to the CO(1-0) data points. The reverse fit to the H{\small I} data points is shown as the solid red line. To demonstrate the reduction in scatter between the initial sample and final sub-sample, the embedded panel in (c) shows the TFR for the final sub-sample but over the same axis ranges as (a). (b) and (d) are as (a) and (c), but for the stellar mass TFRs.}
  \label{fig:HICOTFR}
\end{figure*}

To investigate whether this bias translates into a significant difference between TFRs constructed using CO(1-0) line widths and H{\small I} line widths, we plot in Figure~\ref{fig:HICOTFR} both the $W1$-band and the stellar mass TFR of both the initial sample and the final sub-sample, this using both $W_{50, \rm{CO}}$ and $W_{50, \rm{H{\small I}}}$ values. For consistency, we only consider here galaxies with both CO and H{\small I} data. This allows to directly compare how the use of either measure of line width affects the relations. The corresponding TFR fit parameters are listed in Table~\ref{tab:COHITFR}. For reasons discussed in \S~\ref{subsec:TFRs}, the reverse fit is preferred to the conventional forward fit (see \S~\ref{subsec:fitting} for a description of both). In addition, it is more informative here to examine how the scatter in the velocity width changes when considering either $W_{50, \rm{CO}}$ or $W_{50, \rm{H{\small I}}}$ than the scatter in the $W1$ magnitudes or stellar masses. For both these reasons, we examine only the reverse fits to the TFRs, as shown in Figure~\ref{fig:HICOTFR}.  

Whilst in all cases the slopes of both the $W_{50, \rm{CO}}$ and $W_{50, \rm{HI}}$ TFRs agree within uncertainties, the slopes of both H{\small I} TFRs of the sub-sample are steeper than those of the CO(1-0) TFRs for the same galaxies. The intercepts agree within the uncertainties for all the TFRs of the sub-sample. However, the intercepts of both the stellar mass and $W1$ CO(1-0) TFRs of the initial sample are offset to higher values along the ordinate compared to those of the corresponding H{\small I} relations. For both the stellar mass and $W1$-band TFRs of the initial sample, the total and intrinsic scatter of the $W_{50, \rm{HI}}$ relations are significantly less than those of the $W_{50, \rm{CO}}$ relations. Conversely, considering the TFRs of the sub-sample, the total and intrinsic scatter of the CO relations are significantly smaller. It is clear then that TFRs constructed using CO and H{\sc I} line widths are comparable in terms of slope and intercept, but differ in terms of intrinsic and total scatter. Considering the H{\small I} TFRs (both the $W1$-band and stellar mass relations), there is only a small reduction in the intrinsic and total scatter between the initial sample and the final sub-sample. The reduction in scatter for the CO TFRs is much greater - primarily as a result of excluding from the sub-sample those galaxies with CO(1-0) integrated profiles that do not display a boxy or double-horned shape (see \S~\ref{subsec:subsample} and \S~\ref{subsec:sample}). We therefore conclude that, assuming the selection criteria of the sub-sample are applied, the CO and H{\small I} line widths, and thus the resultant TFRs, are comparable, but that the scatter is greatly reduced by using CO. 

\begin{table*}
  \caption{Reverse fit parameters of the $W1$-band and stellar mass Tully-Fisher relations presented in Figure~\ref{fig:HICOTFR}. The pivot value is 2.6 for all relations (see \S~\ref{subsec:fitting}).}
  \centering
  \begin{tabular}{llllrrr}
    \hline
    TFR & Tracer & Sample & Slope & Intercept & Intrinsic Scatter & Total Scatter \\
    &  &  &  & (dex) & (dex) & (dex) \\
    \hline
    $M_{*}$ & CO & Inital & $\phantom{-}4\phantom{.0}\pm2$ & $\phantom{-}10.9\phantom{0}\pm0.2\phantom{0}$ & $0.93\pm0.09$ & $0.985\pm0.007$ \\
             & & Sub-sample & $\phantom{-}3.6\pm0.6$ & $\phantom{-}10.55\pm0.04$ & $0.28\pm0.03$ & $0.310\pm0.007$ \\
             & H {\small I} & Initial & $\phantom{-}3.6\pm0.6$ & $\phantom{-}10.60\pm0.05$ & $0.56\pm0.04$ & $0.580\pm0.005$ \\
              & & Sub-sample & $\phantom{-}4.5\pm0.9$ & $\phantom{-}10.53\pm0.06$ & $0.43\pm0.05$ & $0.445\pm0.009$ \\
    \hline
    &  &  &  & (mag) & (mag) & (mag) \\
    \hline
    $M_{W1}$ & CO & Inital & $-6\phantom{.0}\pm1$ & $-24.4\phantom{0}\pm0.2$ & $1.6\phantom{0}\pm0.1\phantom{0}$ & $1.61\phantom{0}\pm0.01\phantom{0}$ \\
             & & Sub-sample & $-8\phantom{.0}\pm1$ & $-23.9\phantom{0}\pm0.1$ & $0.68\pm0.07$ & $0.69\phantom{0}\pm0.01\phantom{0}$ \\
             & H {\small I} & Initial & $-7\phantom{.0}\pm1$ & $-23.9\phantom{0}\pm0.1$ & $1.07\pm0.07$ & $1.104\pm0.009$ \\
             & & Sub-sample & $-10\phantom{.}\pm3$ & $-23.8\phantom{0}\pm0.2$ & $1.1\phantom{0}\pm0.1\phantom{0}$ & $1.09\phantom{0}\pm0.02\phantom{0}$ \\
    \hline
  \end{tabular}
  \label{tab:COHITFR}
\end{table*}

\section[]{Galaxy Spectra} \label{sec:galfits}

Figure~\ref{fig:w50fits} shows the Gaussian Double Peak function fits to each of the COLD GASS galaxy spectra in our final sub-sample, as described in \S~\ref{subsec:measurew50} and \S~\ref{subsec:subsample}. The data are plotted in blue and the best fit function in green. The solid red line indicates the best fit central velocity ($v_{0}$), whilst the dashed red lines show the width at $50\%$ of the peak ($W_{50}$). 

\begin{landscape}
\includegraphics[scale=0.23, trim= 0 60 50 20, clip=true]{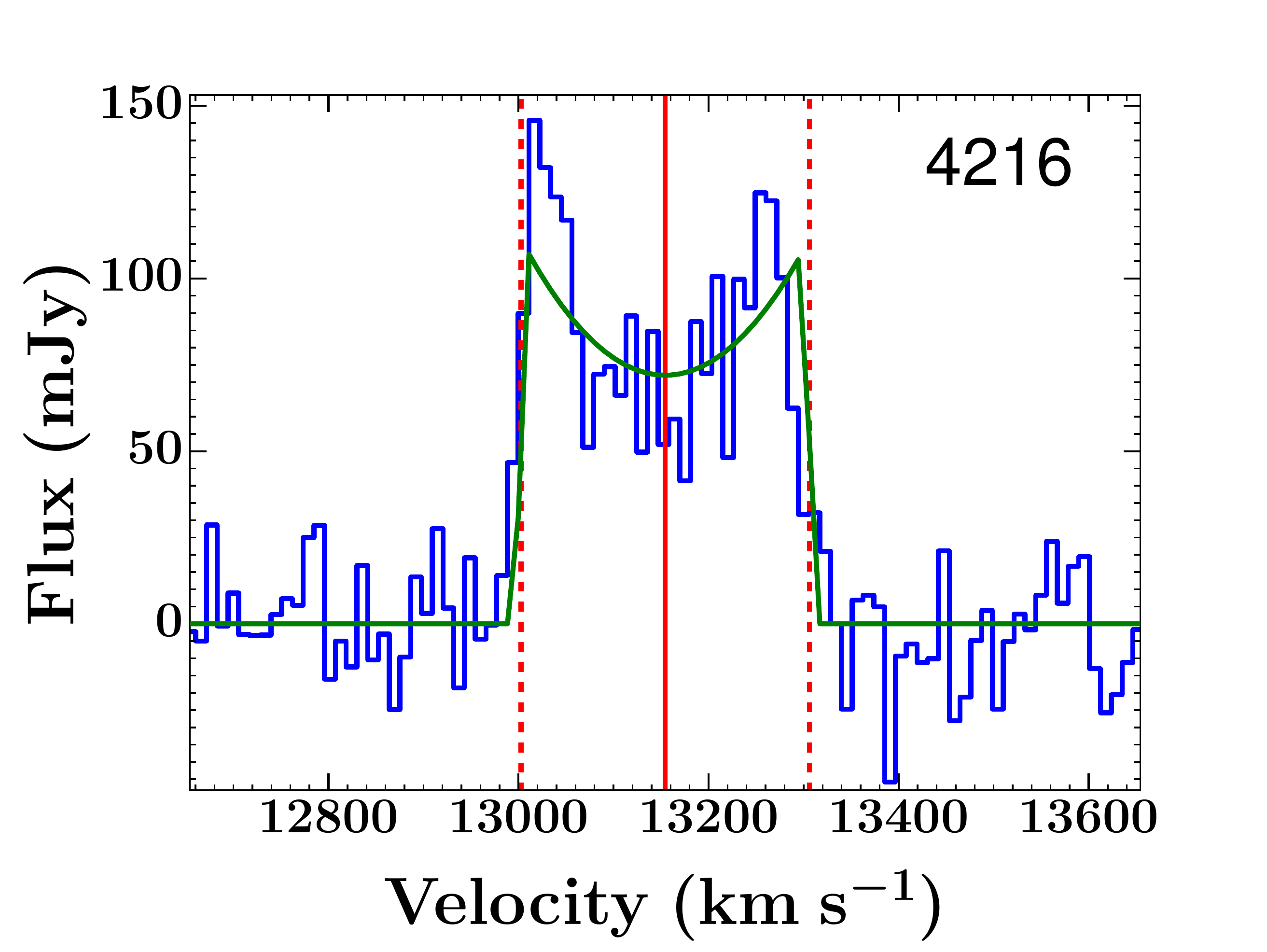}\includegraphics[scale=0.23, trim= 60 60 50 20, clip=true]{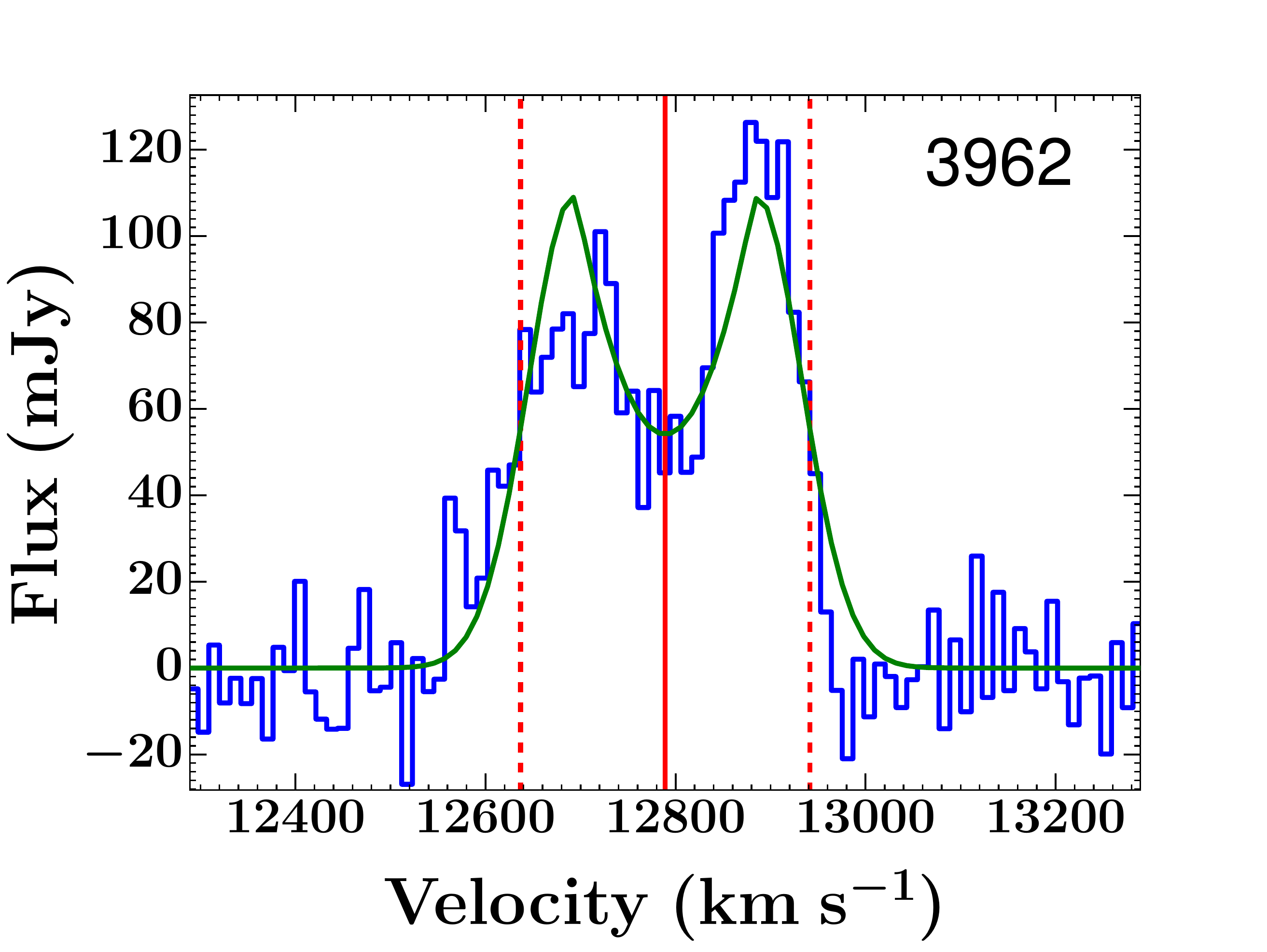}\includegraphics[scale=0.23, trim= 60 60 50 20, clip=true]{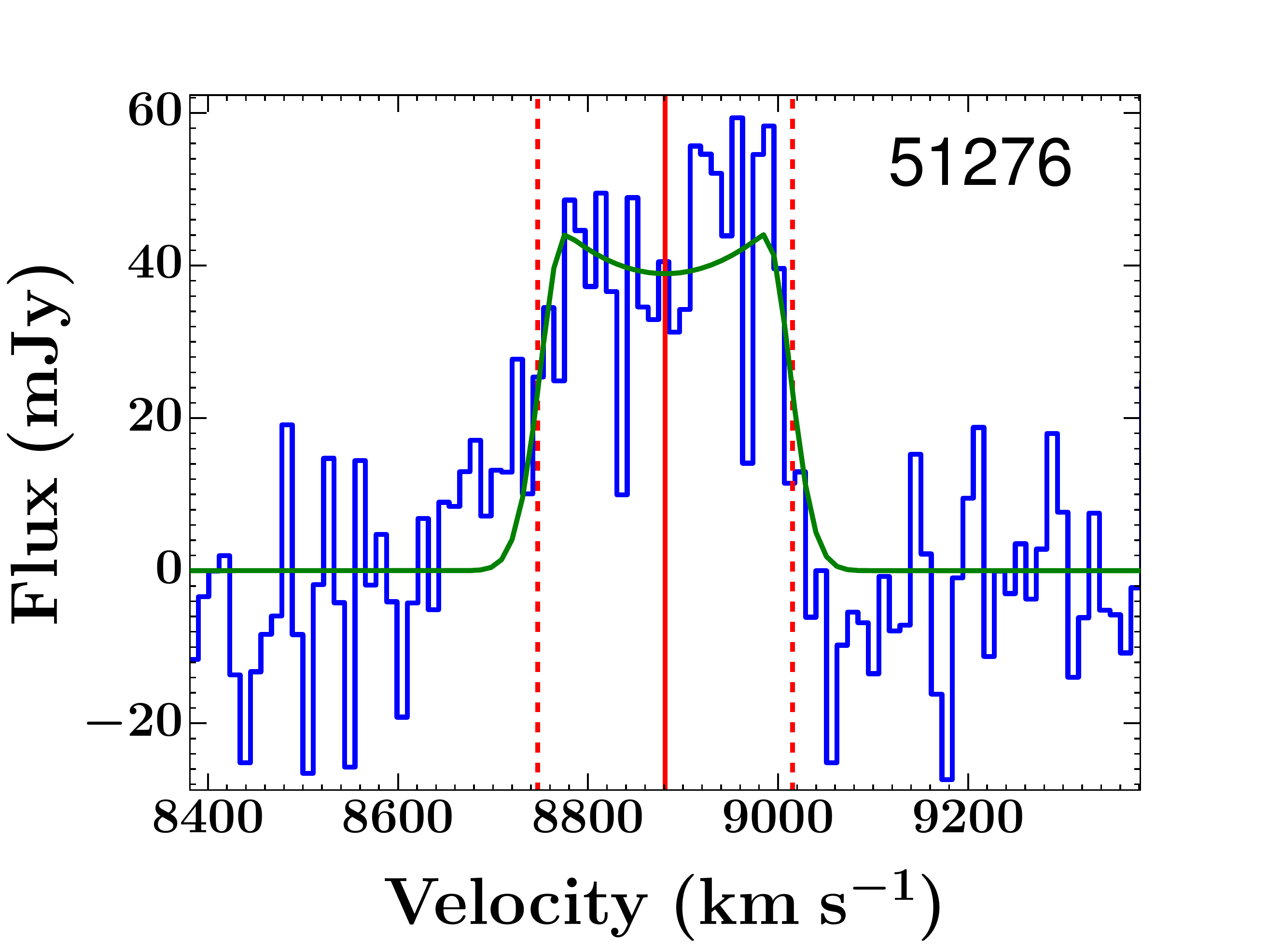}\includegraphics[scale=0.23, trim= 60 60 50 20, clip=true]{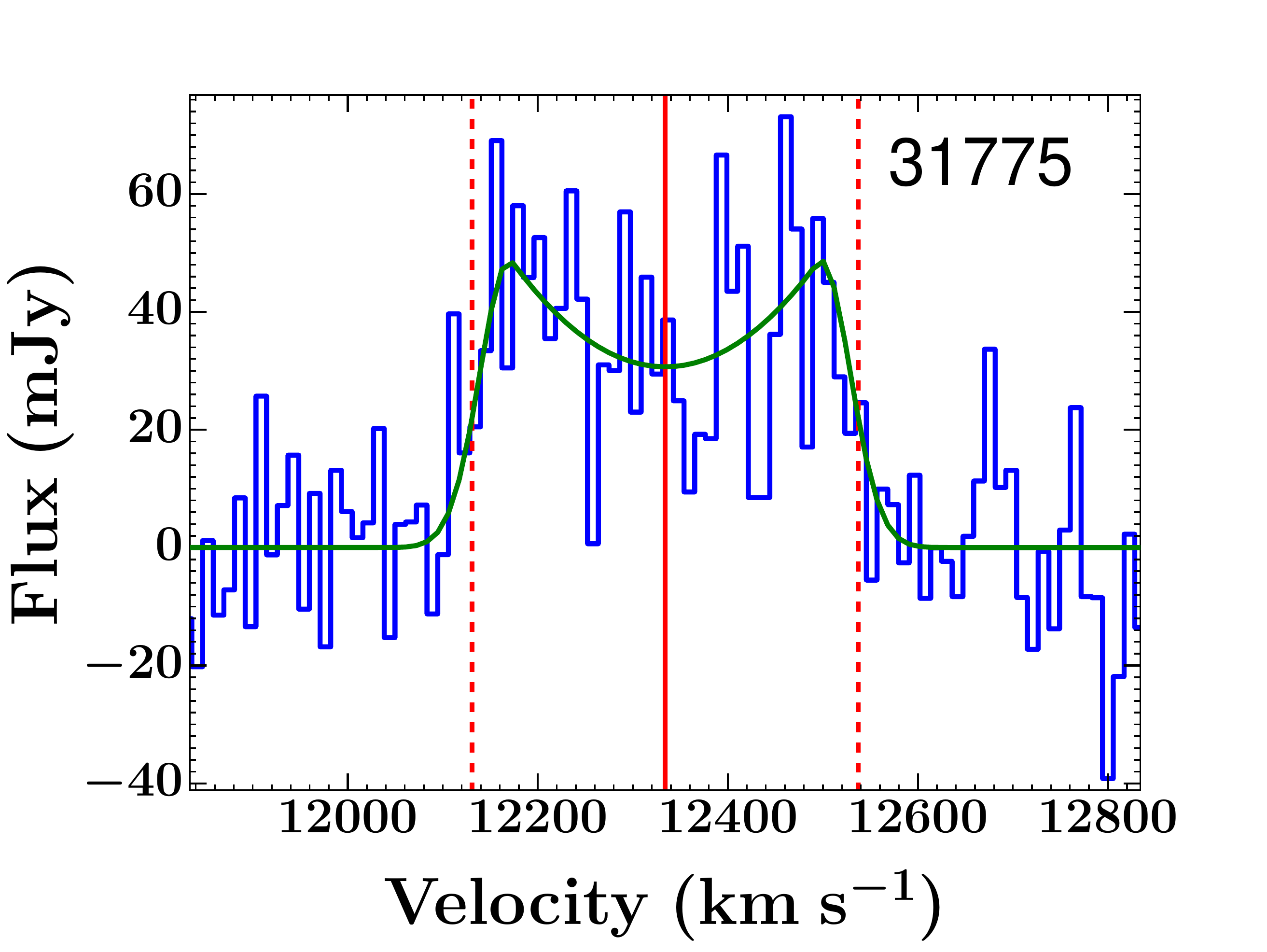}

\includegraphics[scale=0.23, trim= 0 60 50 20, clip=true]{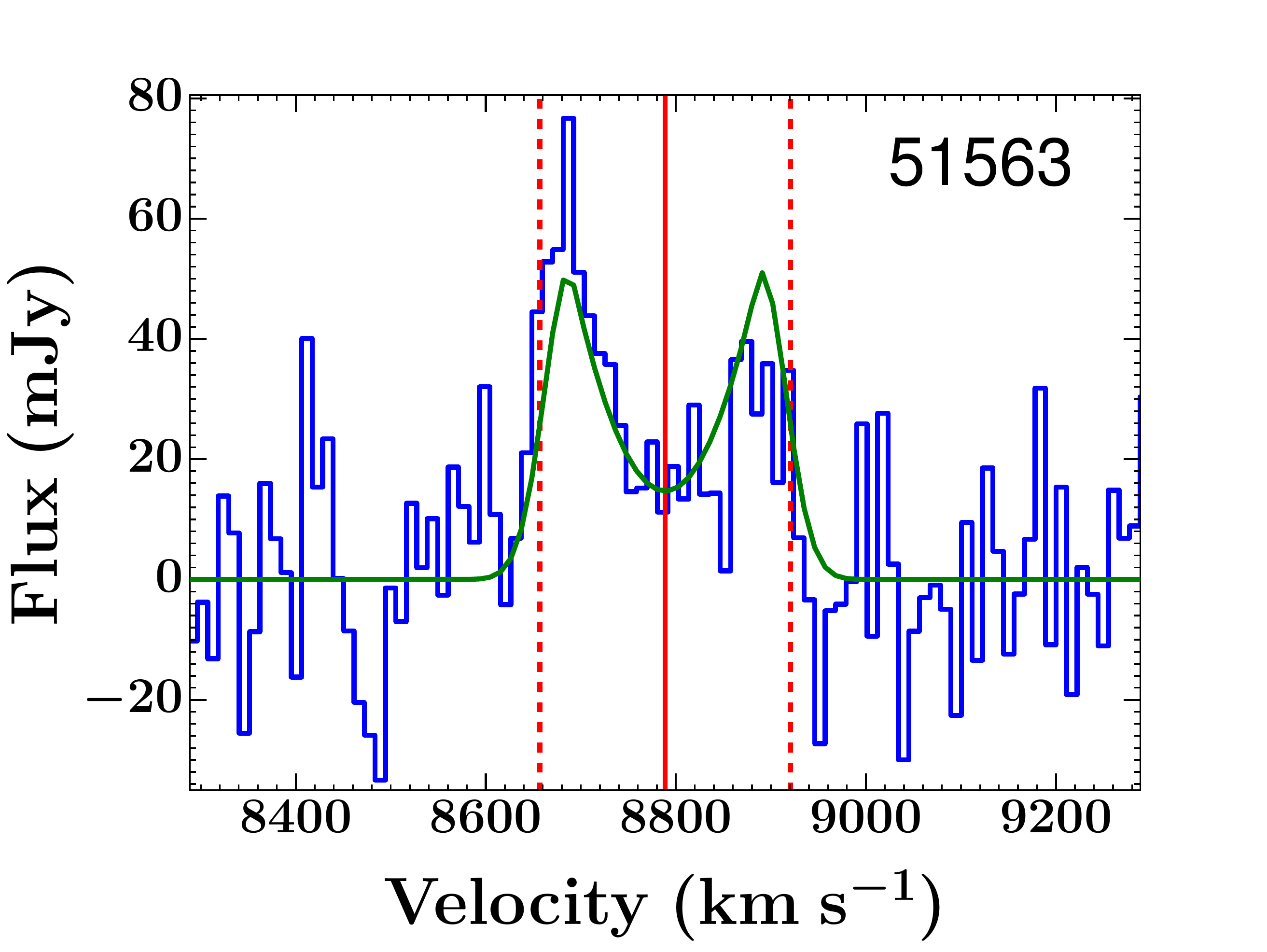}\includegraphics[scale=0.23, trim= 60 60 50 20, clip=true]{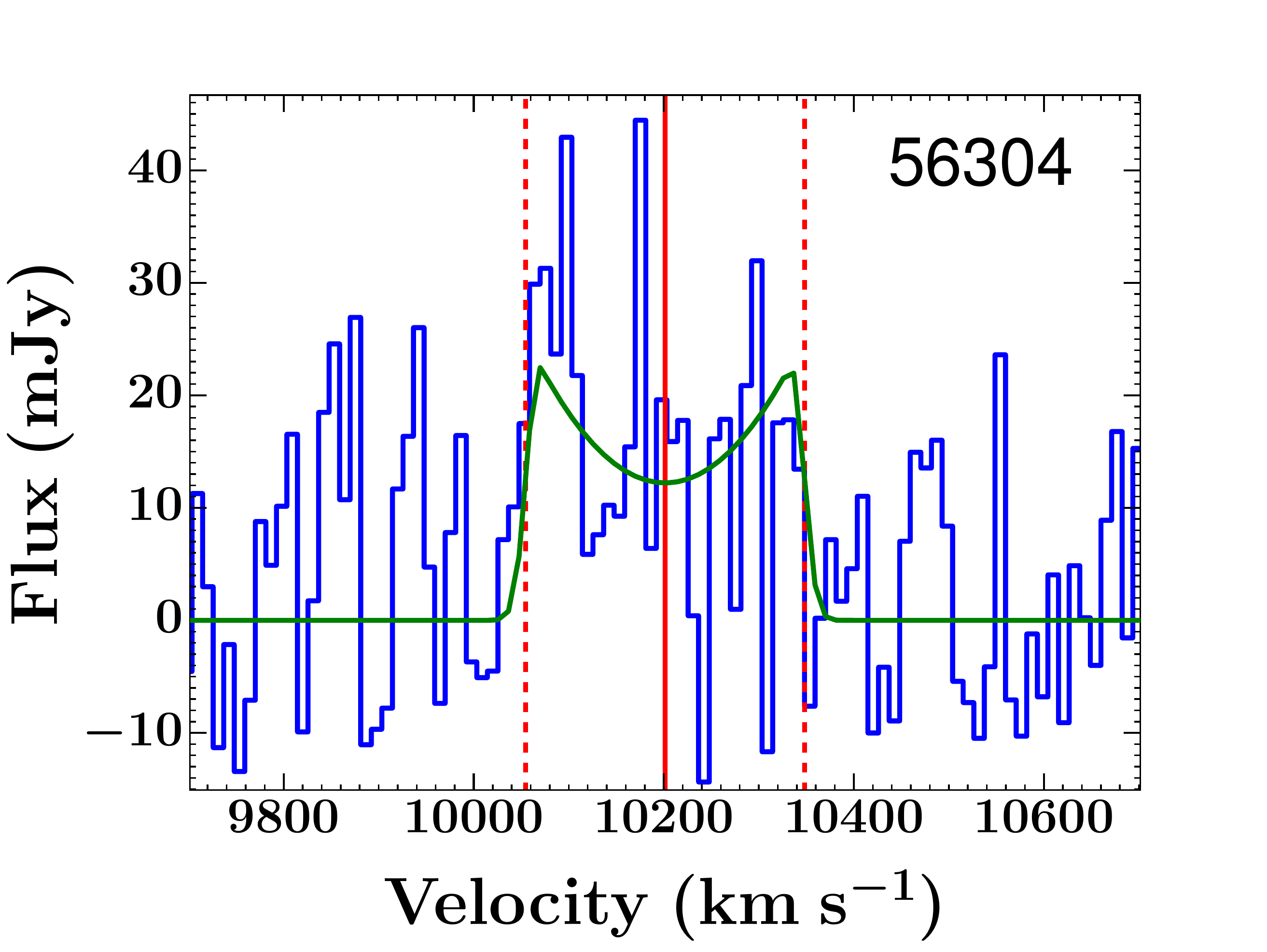}\includegraphics[scale=0.23, trim= 60 60 50 20, clip=true]{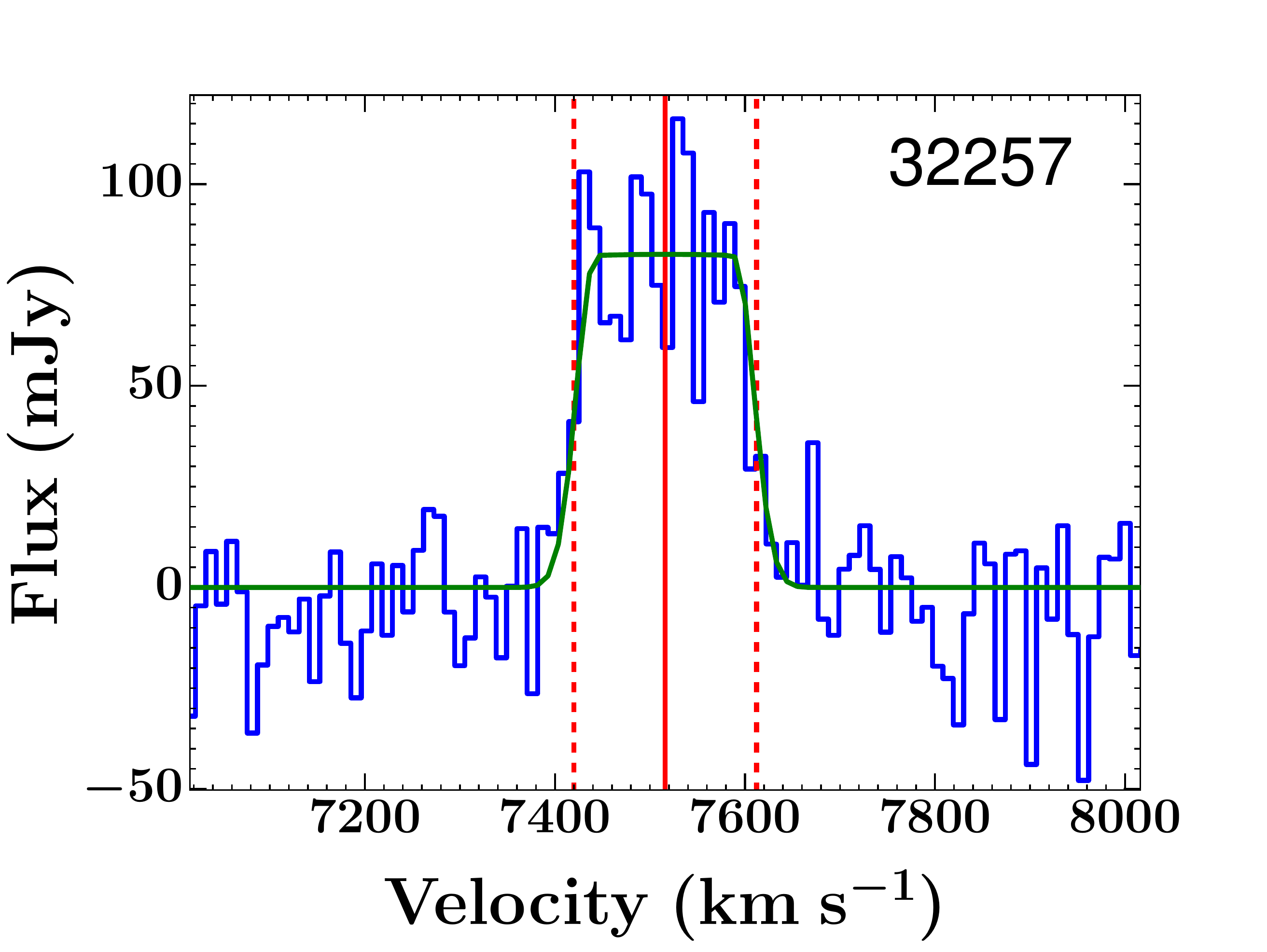}\includegraphics[scale=0.23, trim= 60 60 50 20, clip=true]{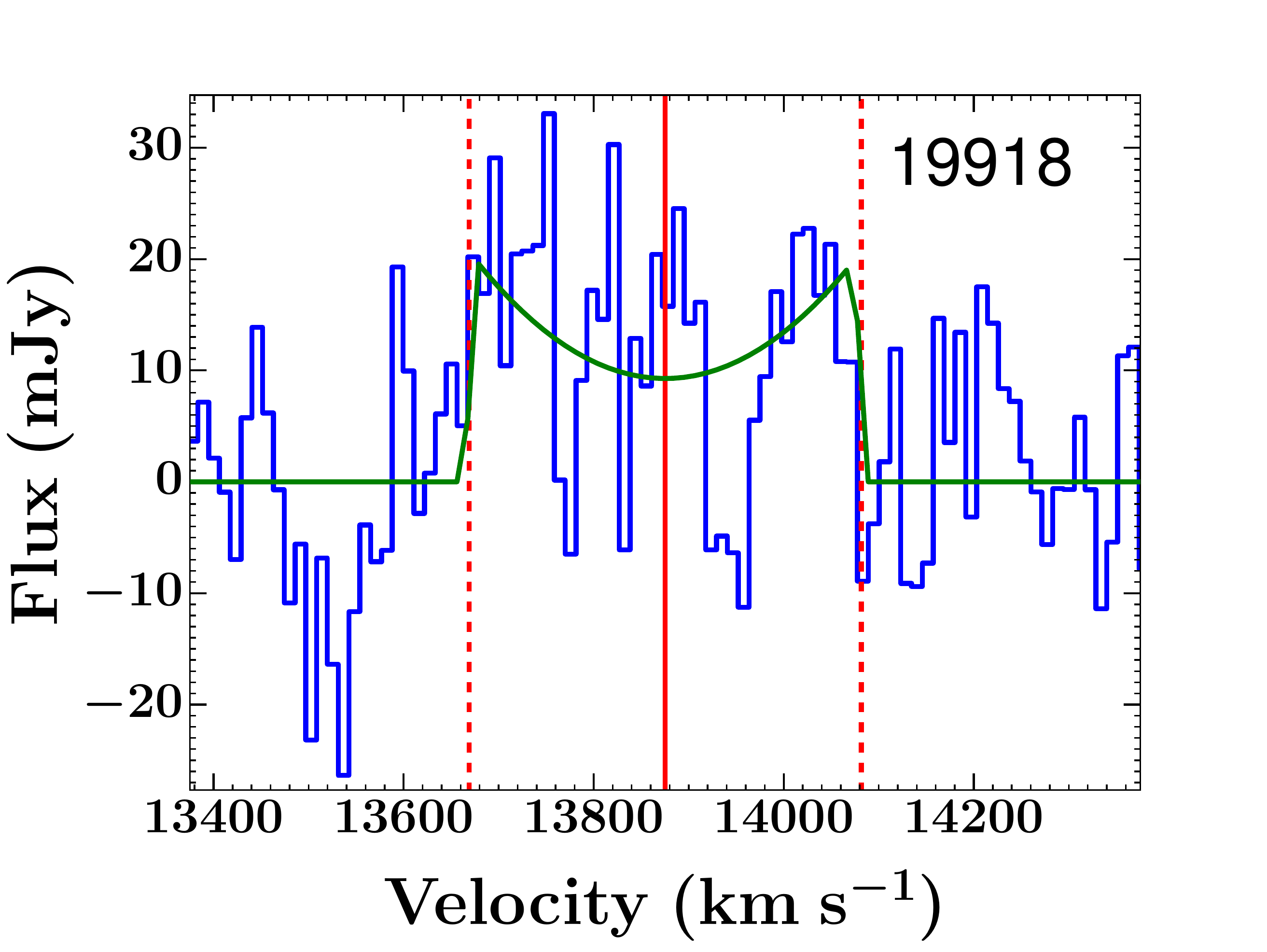}

\includegraphics[scale=0.23, trim= 0 60 50 20, clip=true]{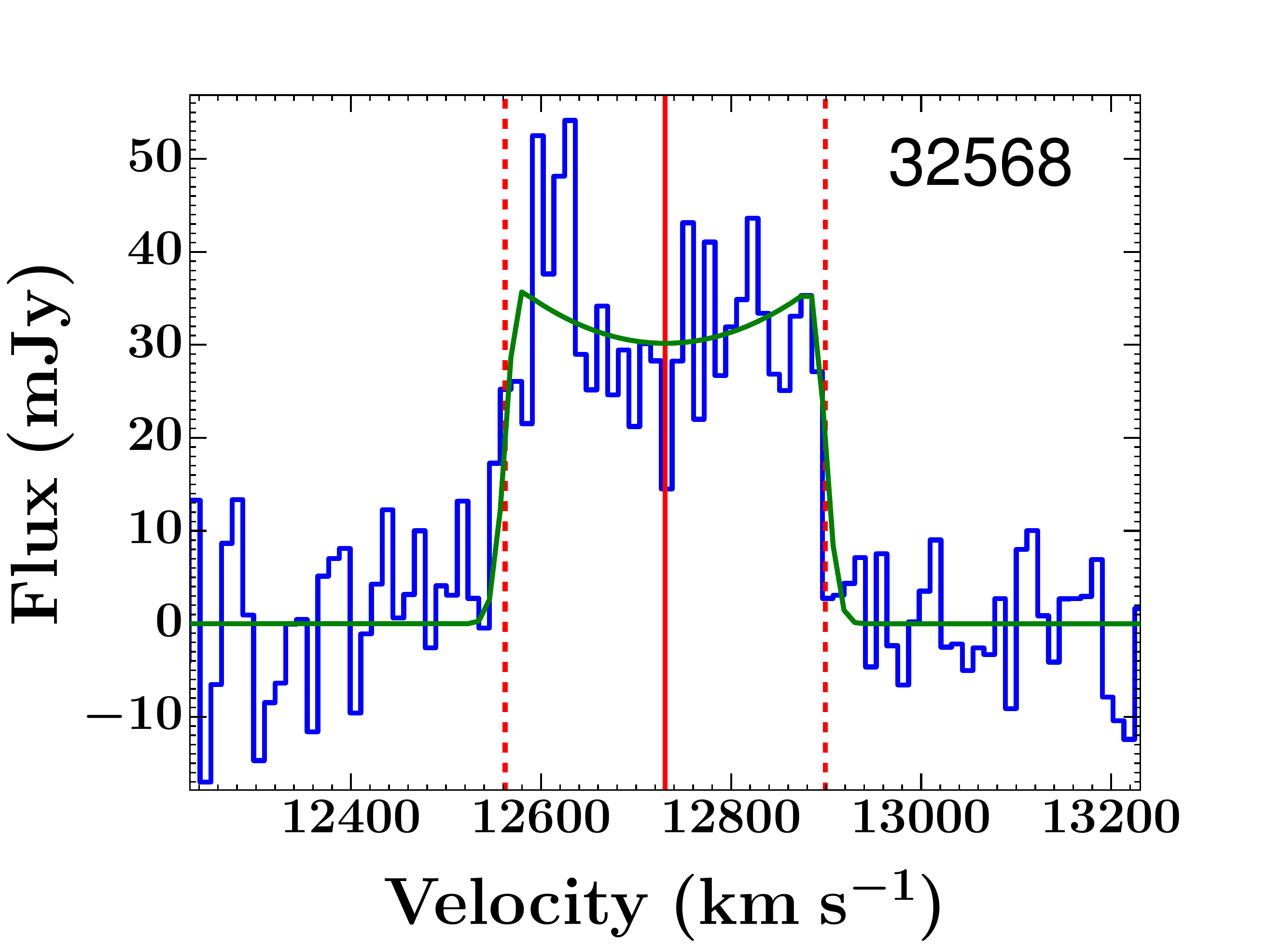}\includegraphics[scale=0.23, trim= 60 60 50 20, clip=true]{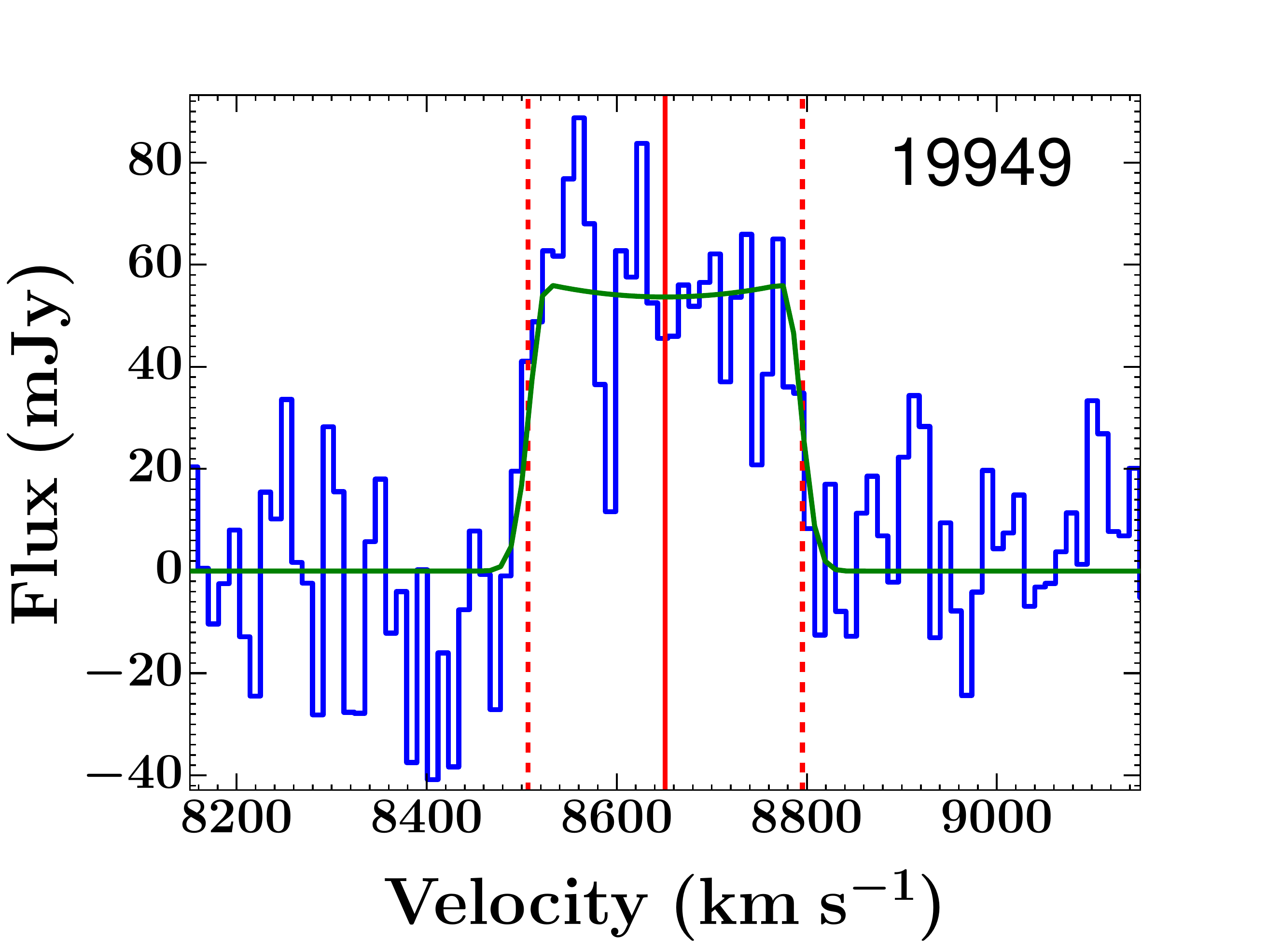}\includegraphics[scale=0.23, trim= 60 60 50 20, clip=true]{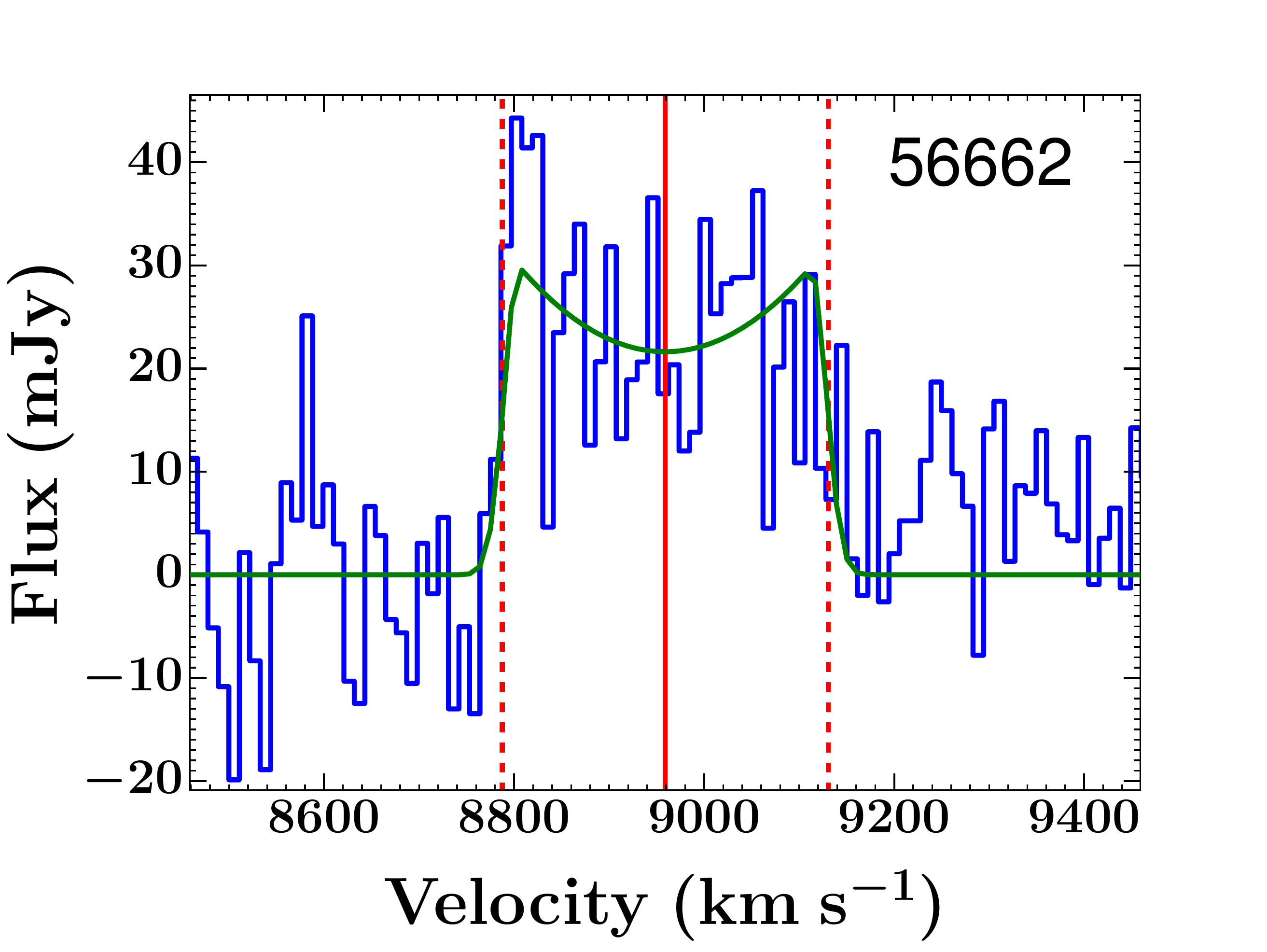}\includegraphics[scale=0.23, trim= 60 60 50 20, clip=true]{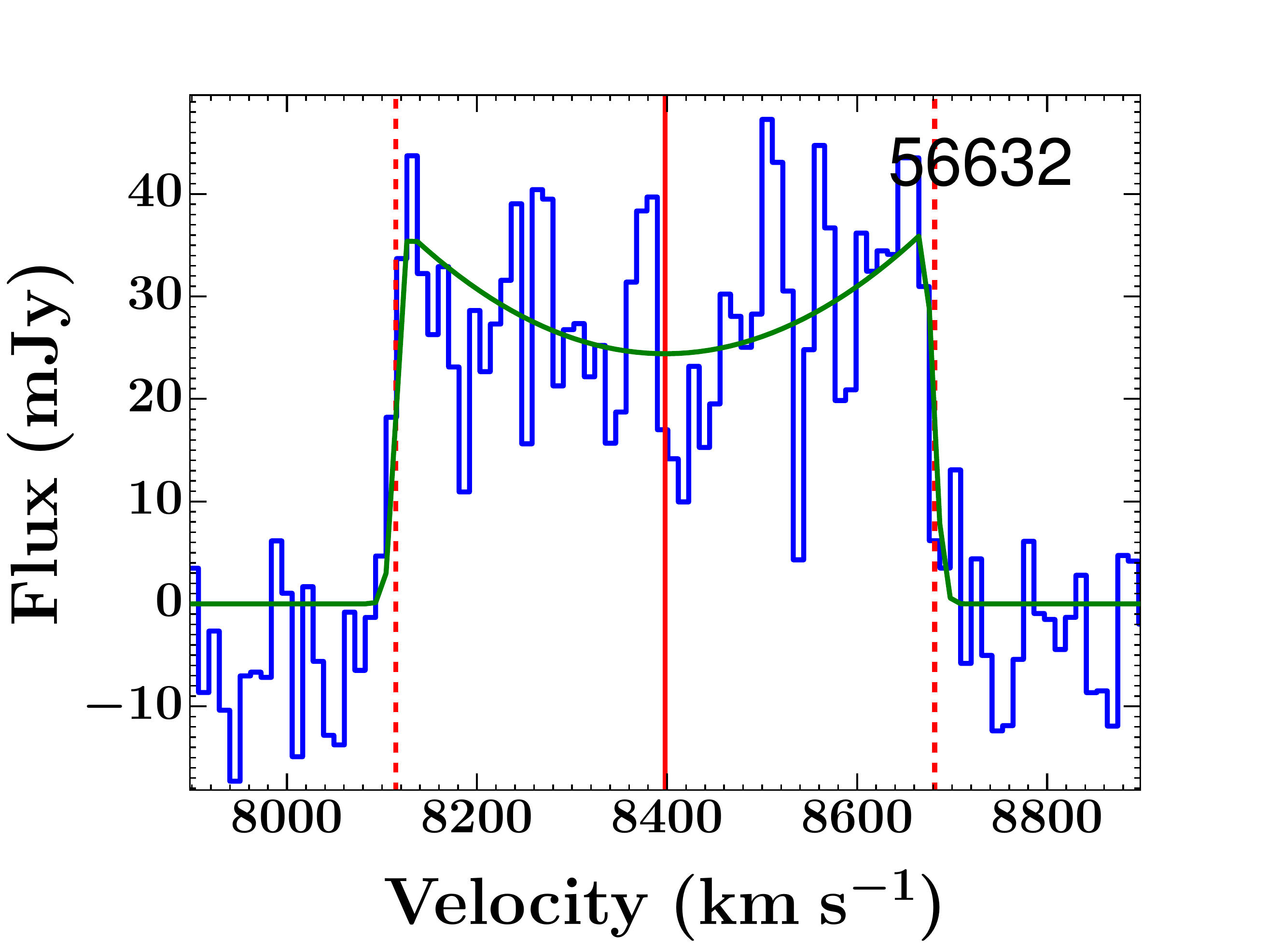}

\includegraphics[scale=0.23, trim= 0 5 50 20, clip=true]{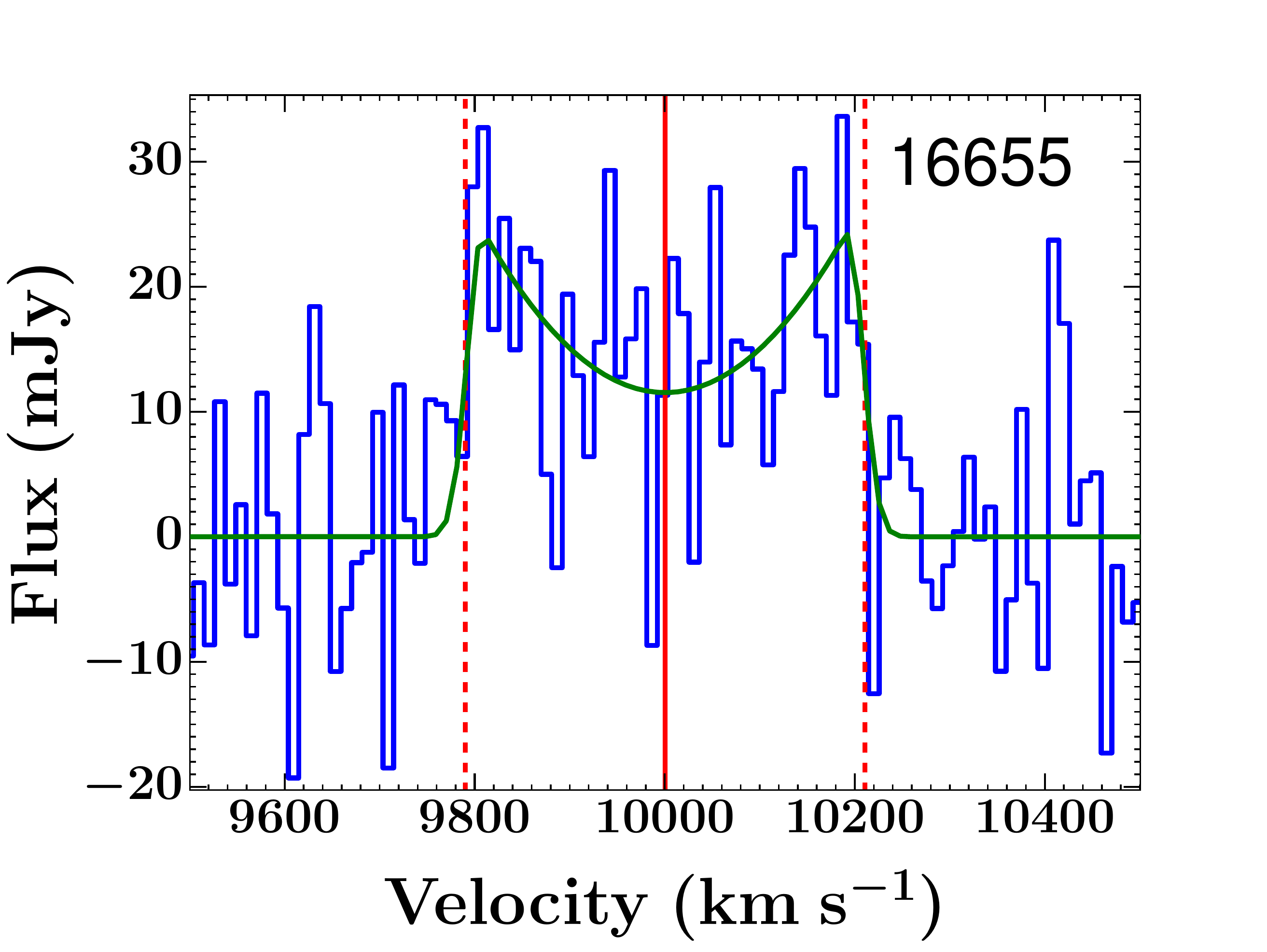}\includegraphics[scale=0.23, trim= 60 5 50 20, clip=true]{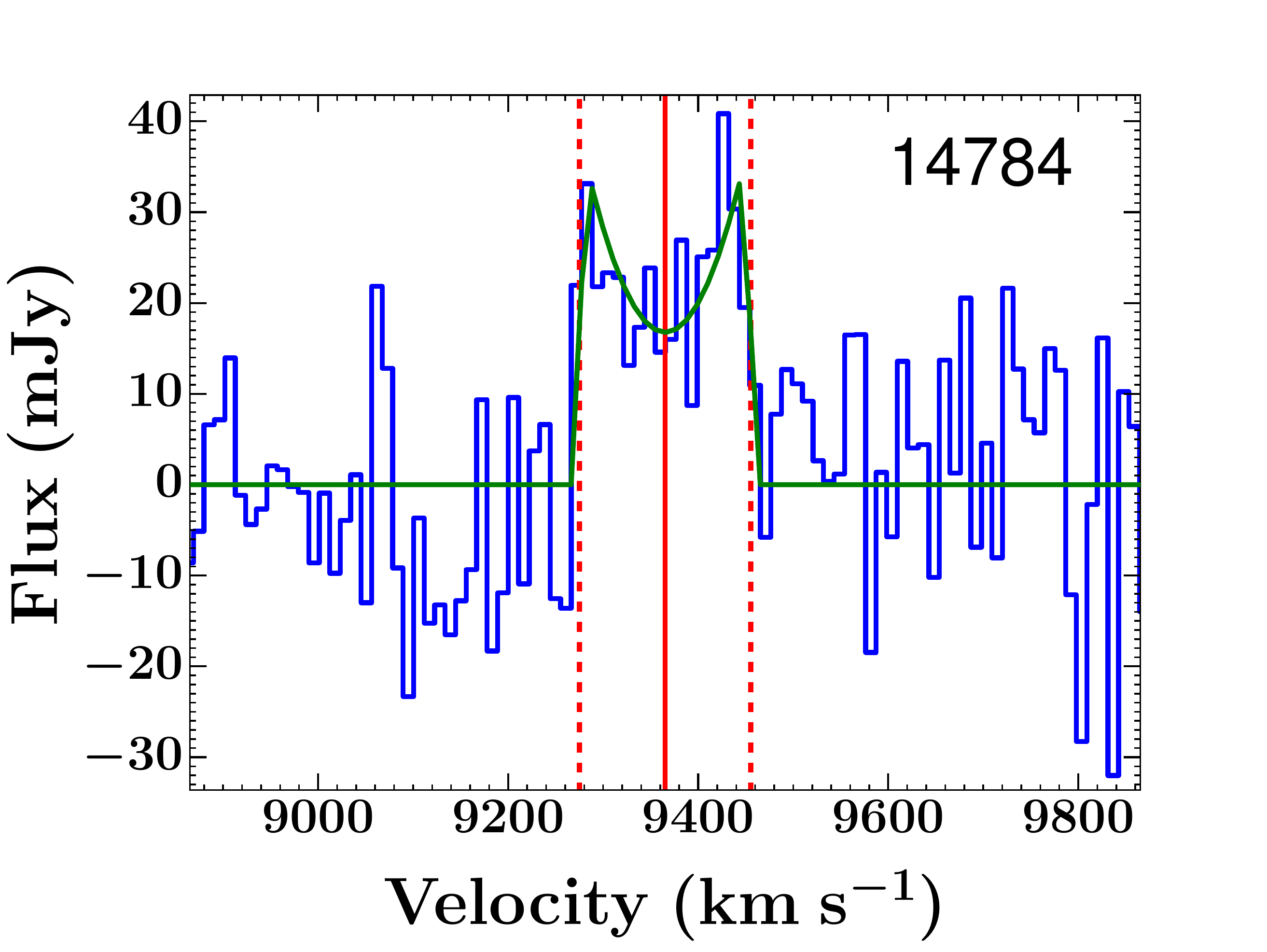}\includegraphics[scale=0.23, trim= 60 5 50 20, clip=true]{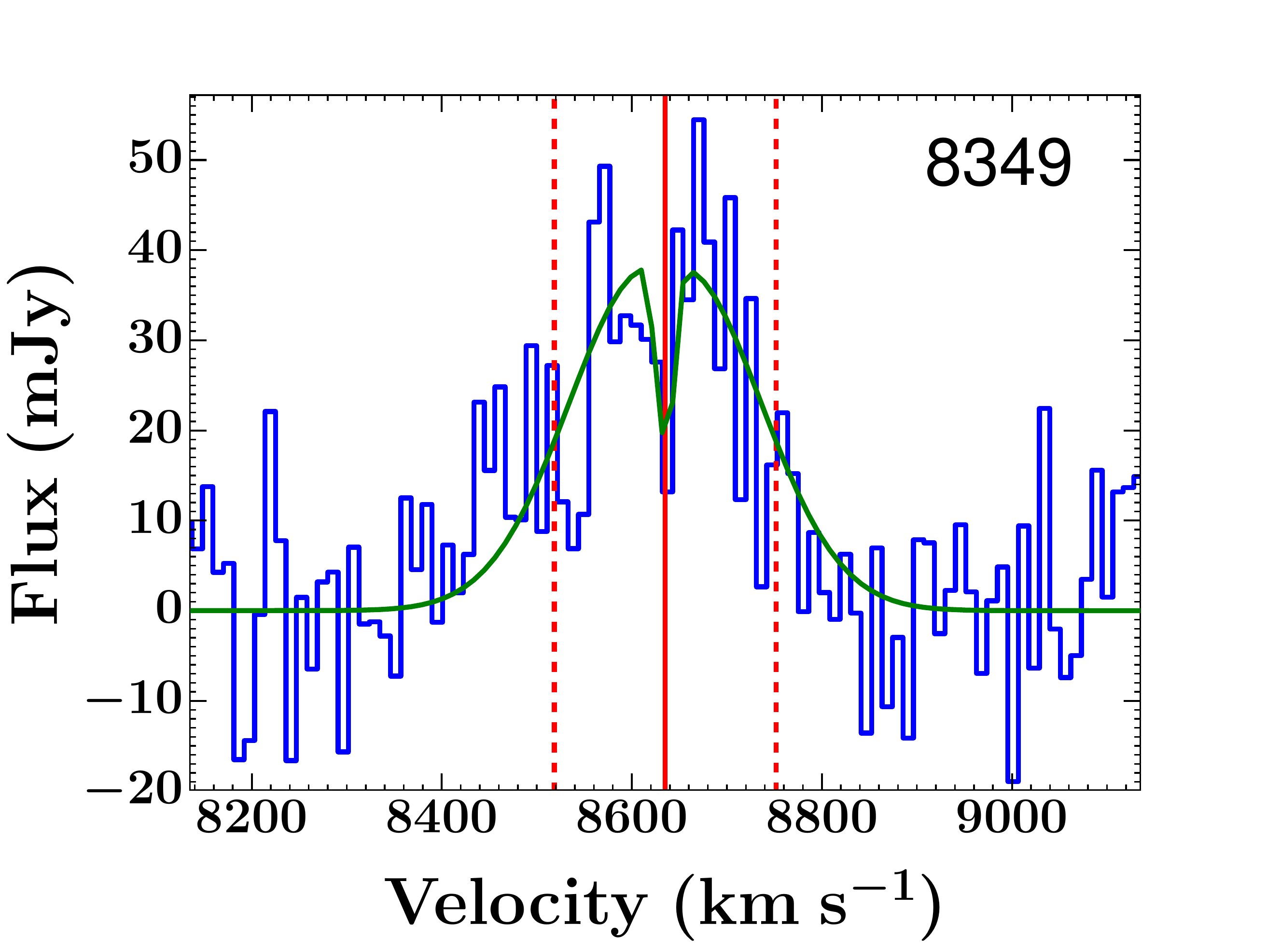}\includegraphics[scale=0.23, trim= 60 5 50 20, clip=true]{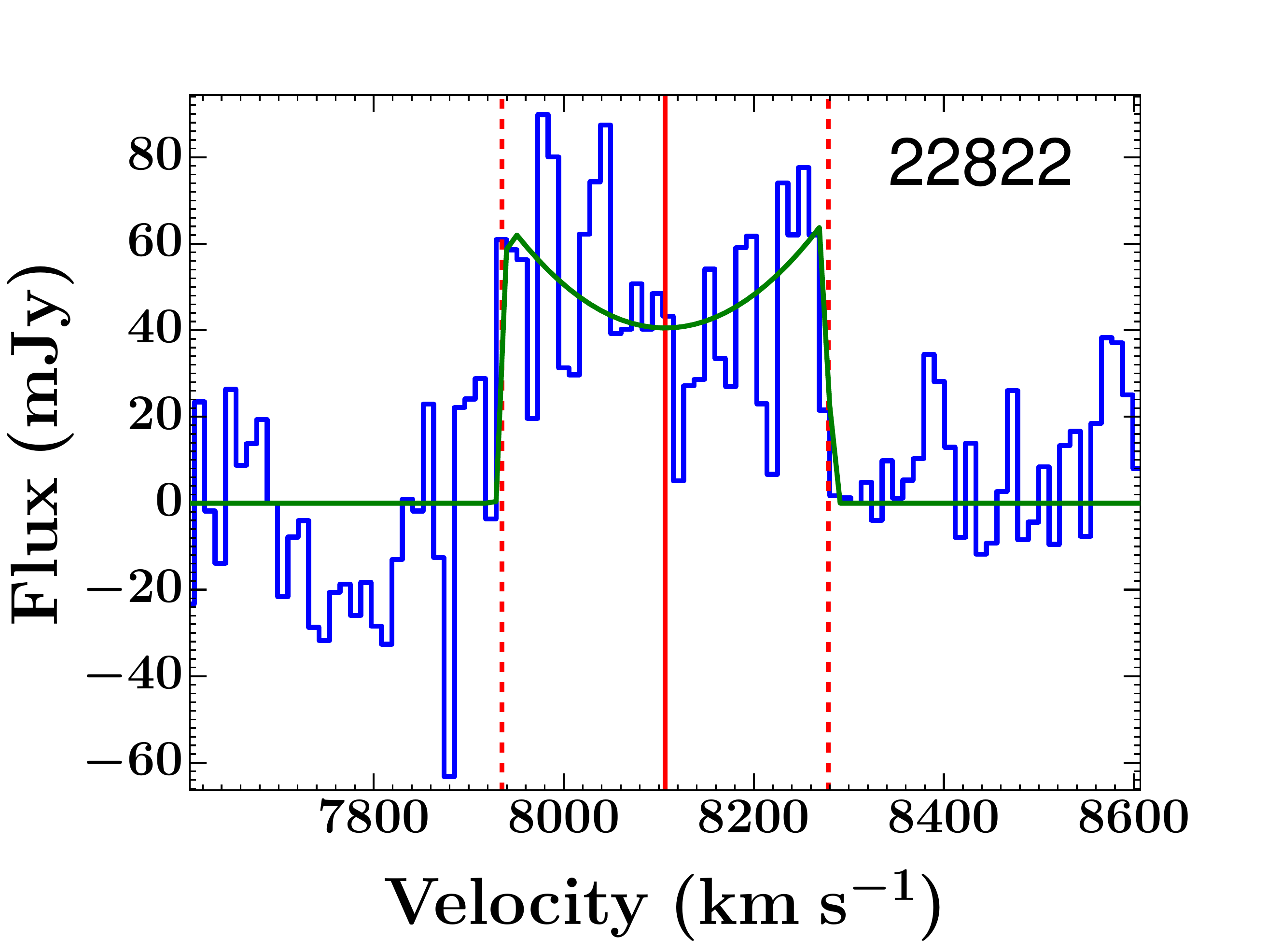}
\end{landscape}
\begin{landscape}
\includegraphics[scale=0.23, trim= 0 60 50 20, clip=true]{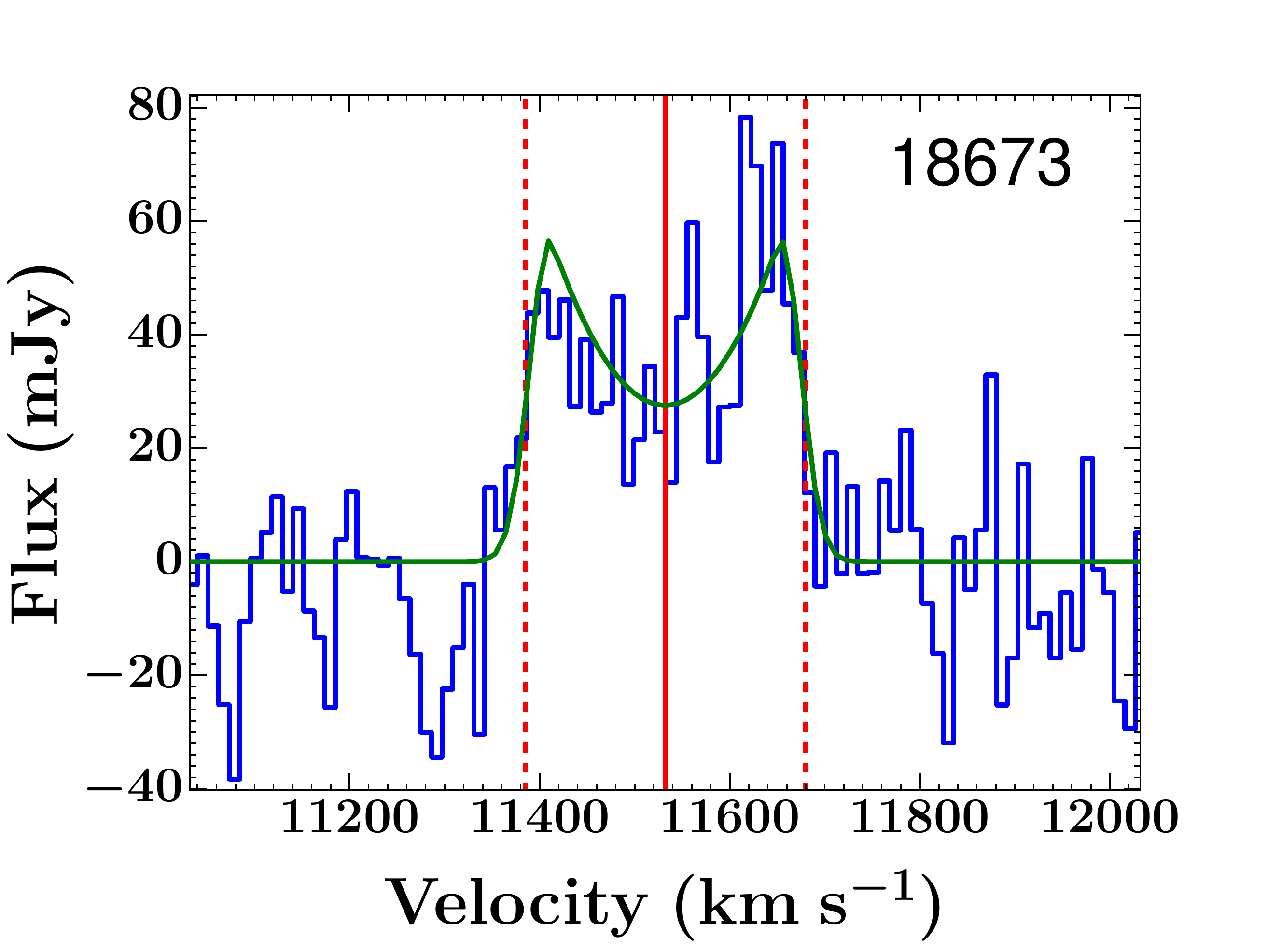}\includegraphics[scale=0.23, trim= 60 60 50 20, clip=true]{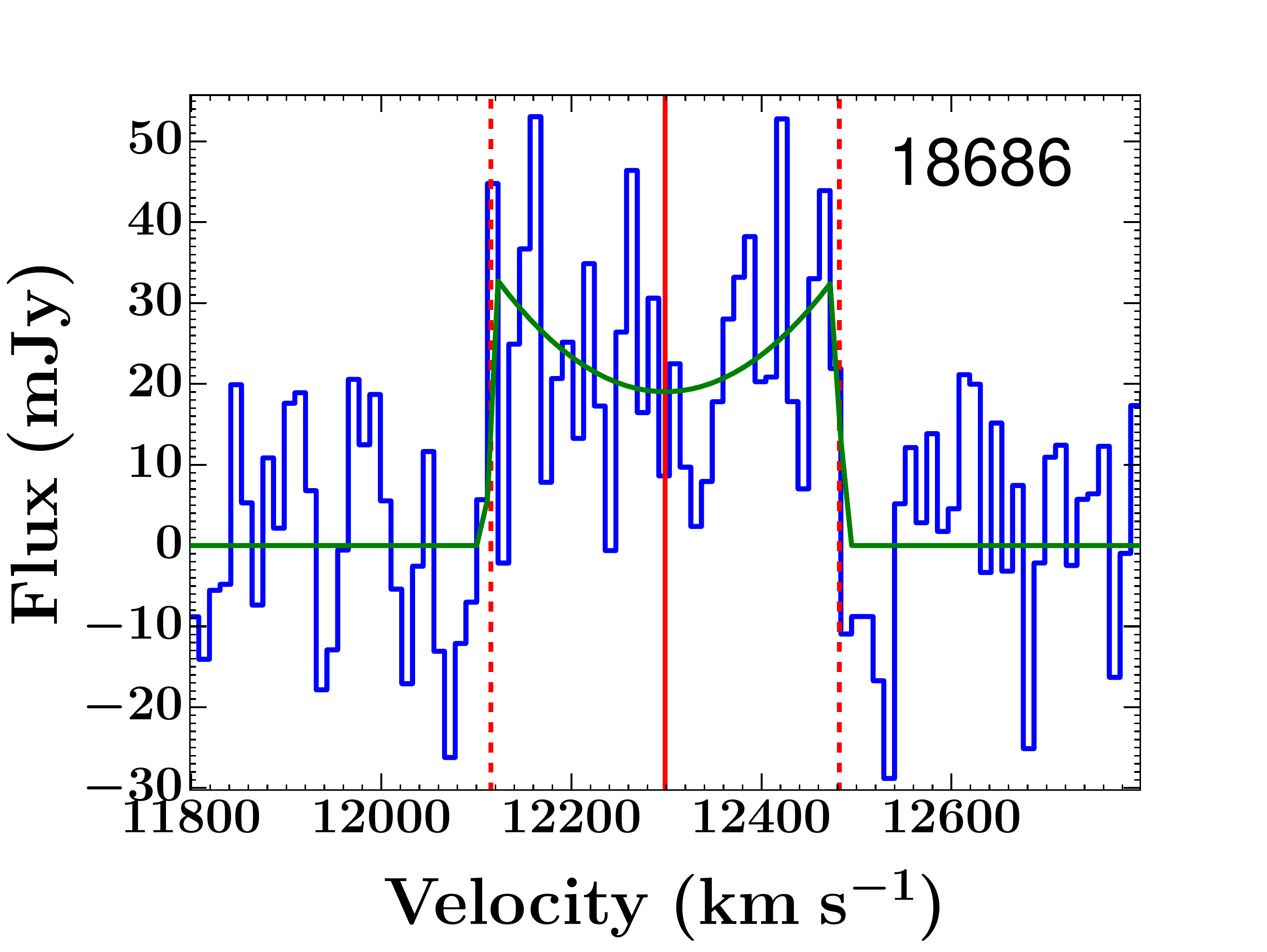}\includegraphics[scale=0.23, trim= 60 60 50 20, clip=true]{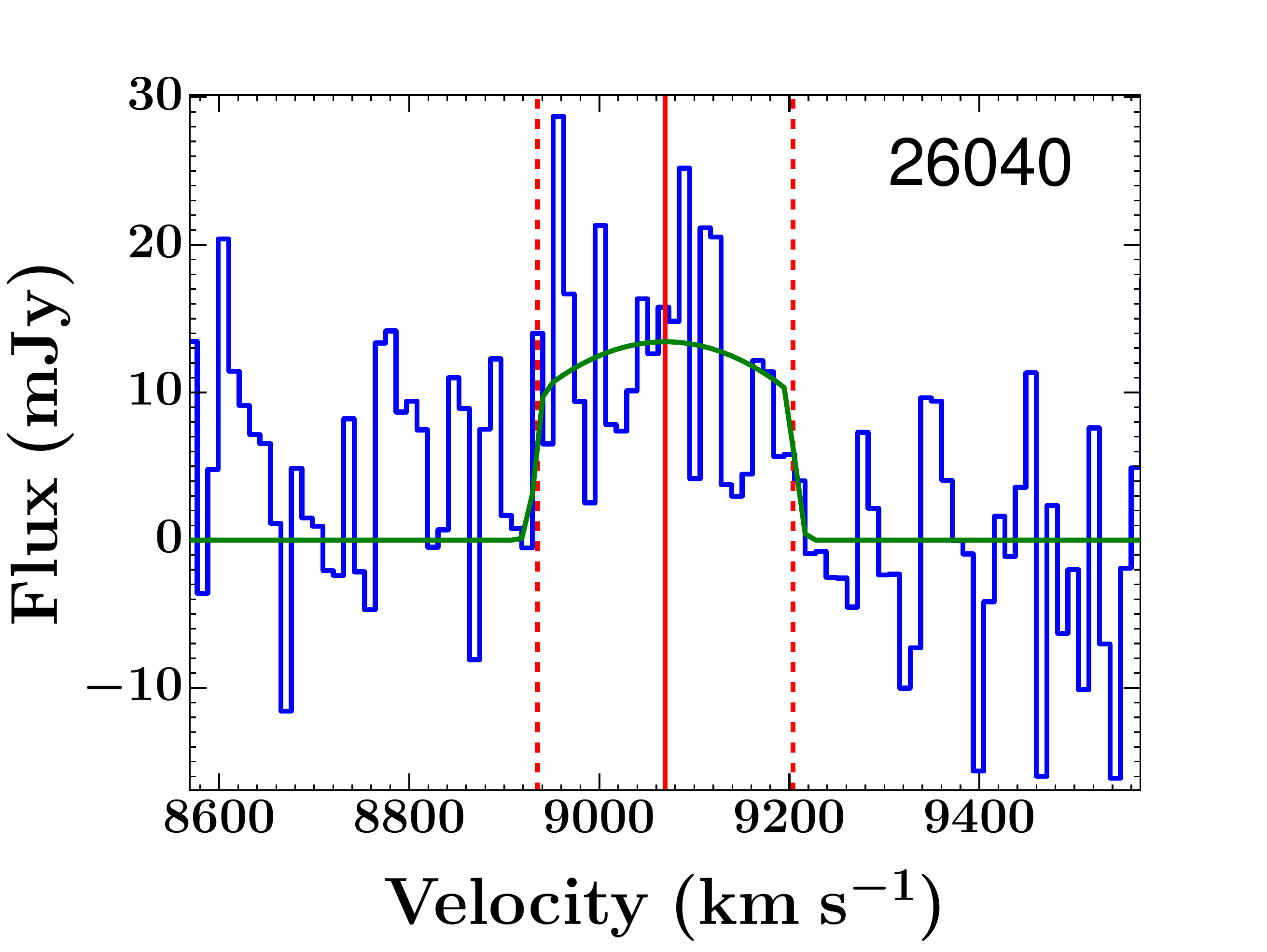}\includegraphics[scale=0.23, trim= 60 60 50 20, clip=true]{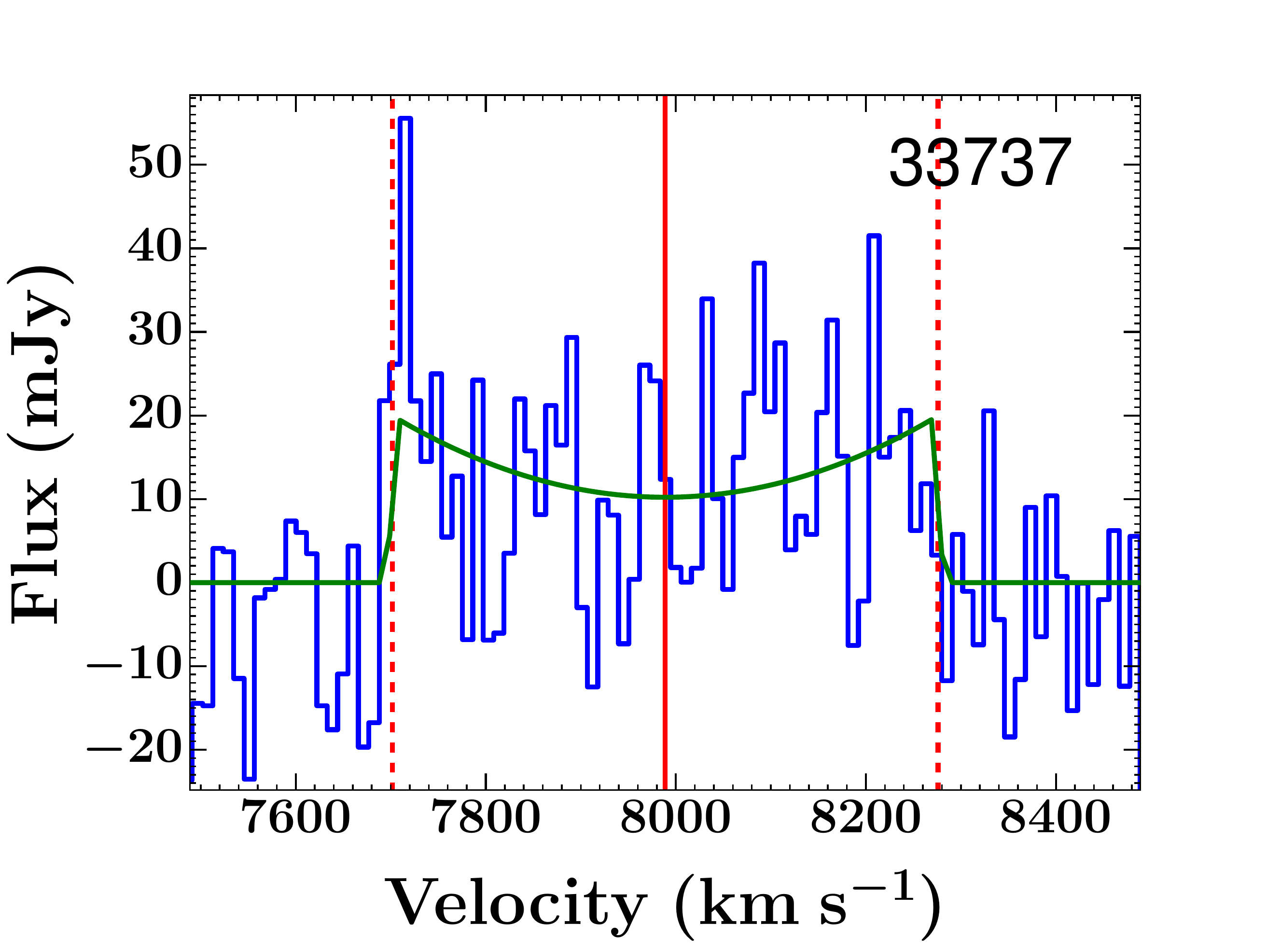}

\includegraphics[scale=0.23, trim= 0 60 50 20, clip=true]{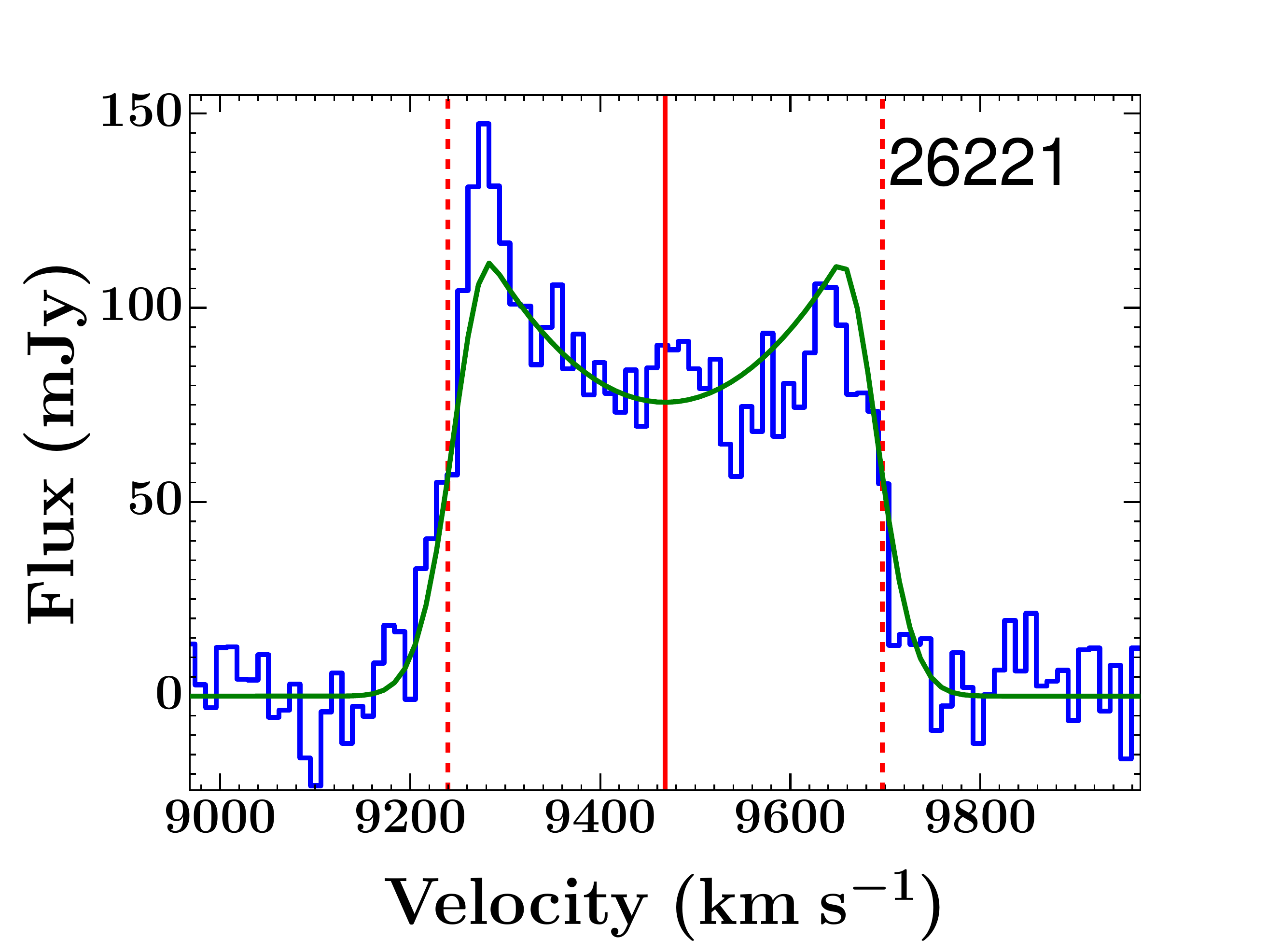}\includegraphics[scale=0.23, trim= 60 60 50 20, clip=true]{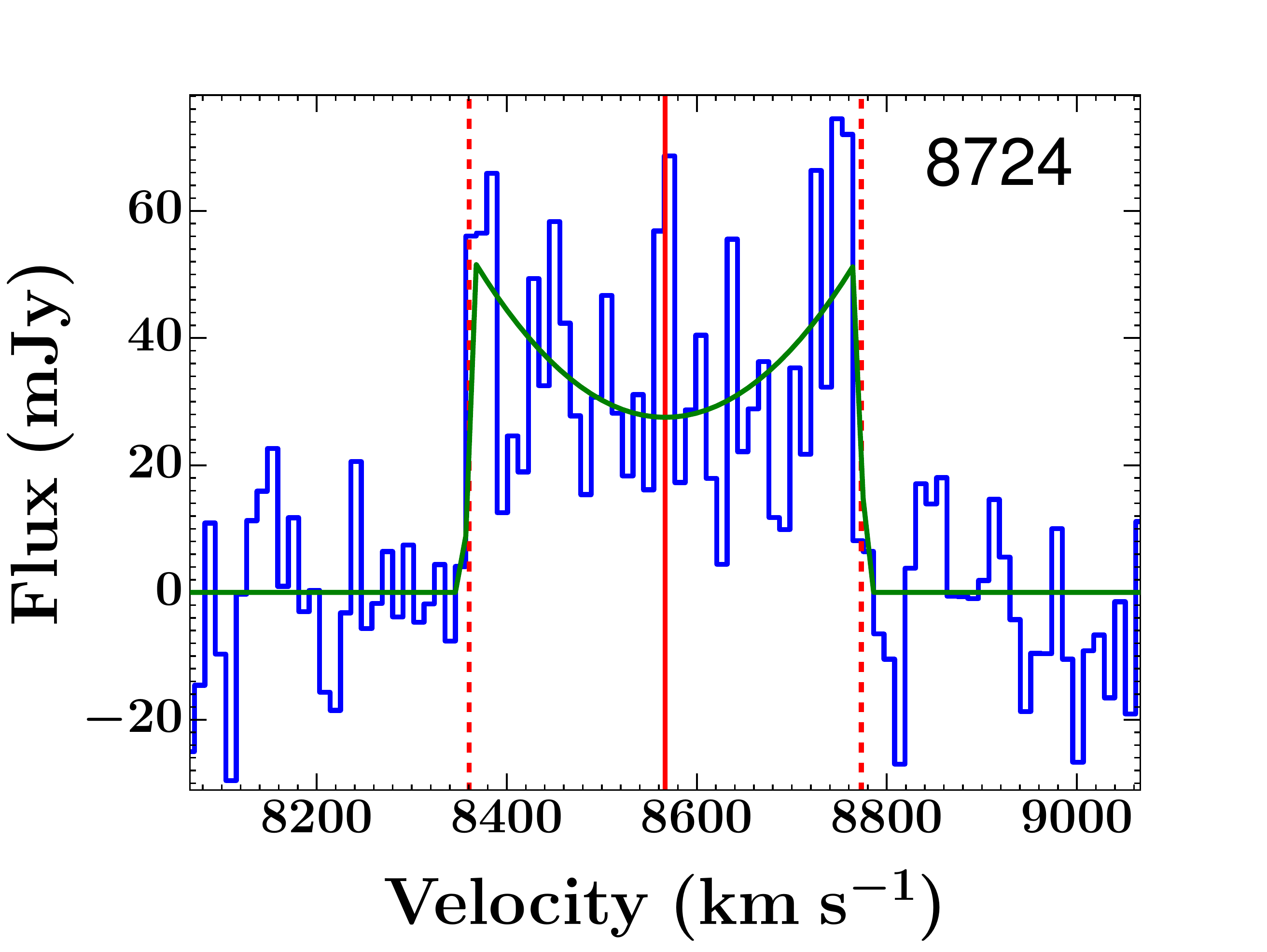}\includegraphics[scale=0.23, trim= 60 60 50 20, clip=true]{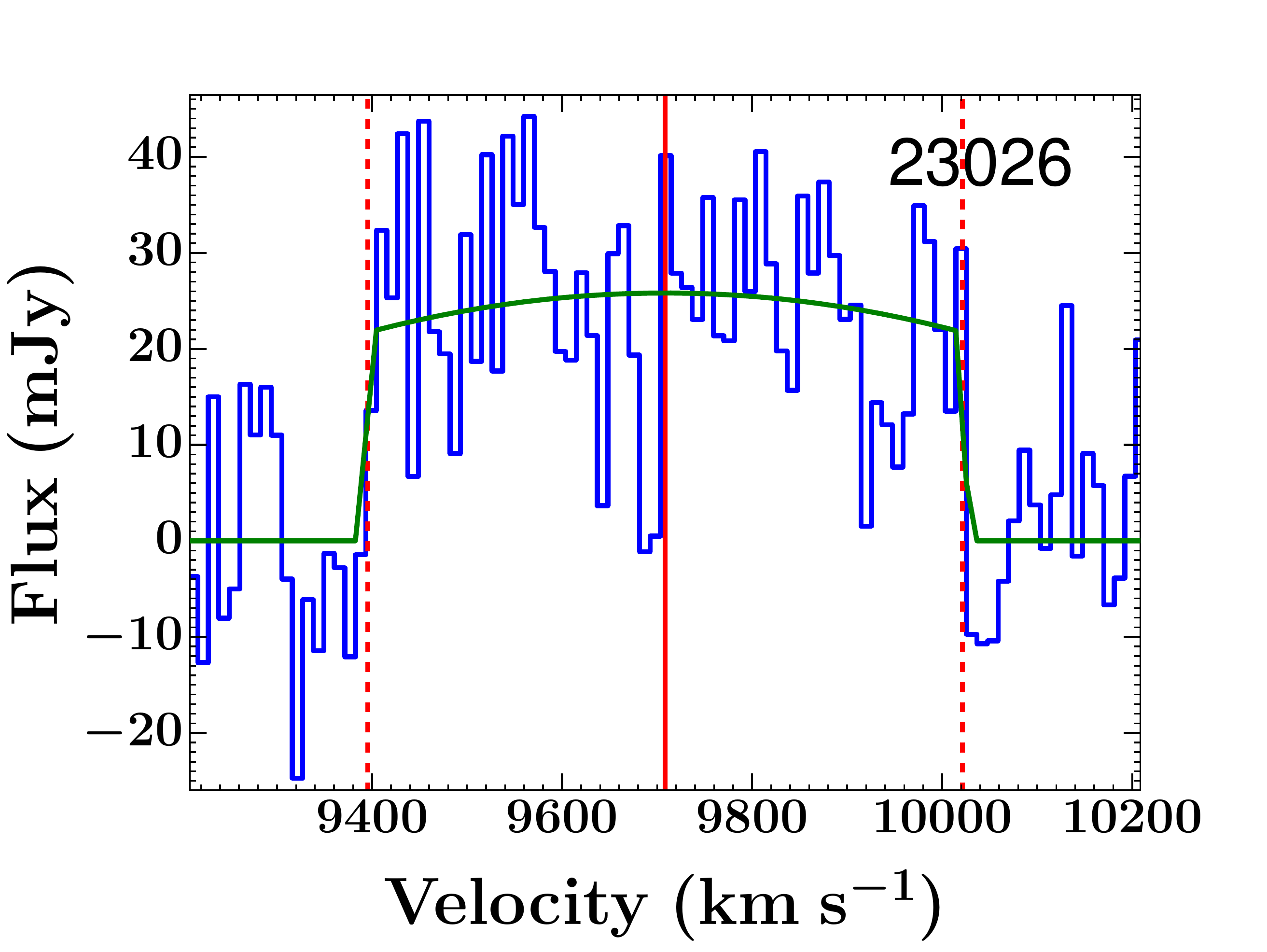}\includegraphics[scale=0.23, trim= 60 60 50 20, clip=true]{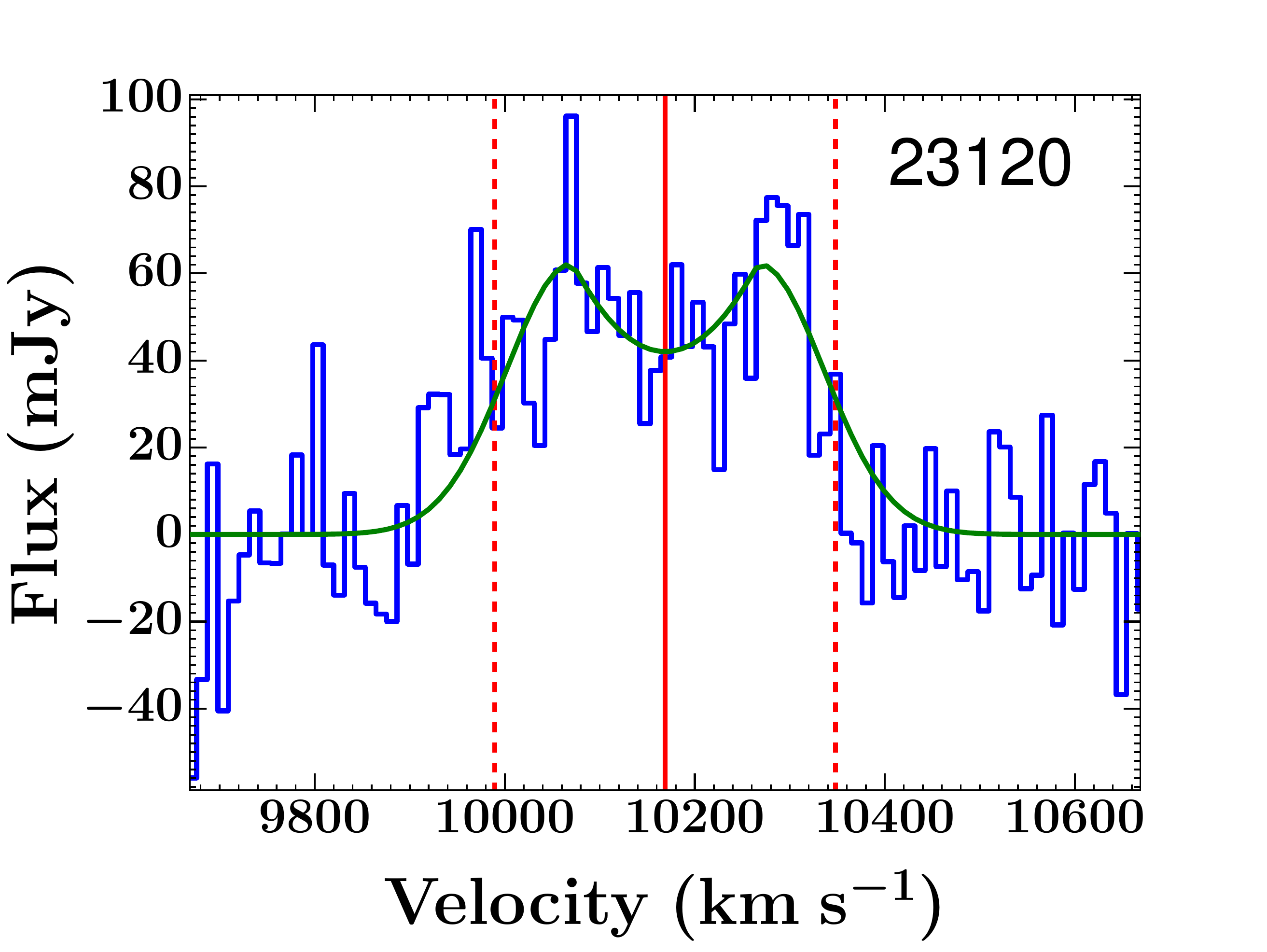}

\includegraphics[scale=0.23, trim= 0 60 50 20, clip=true]{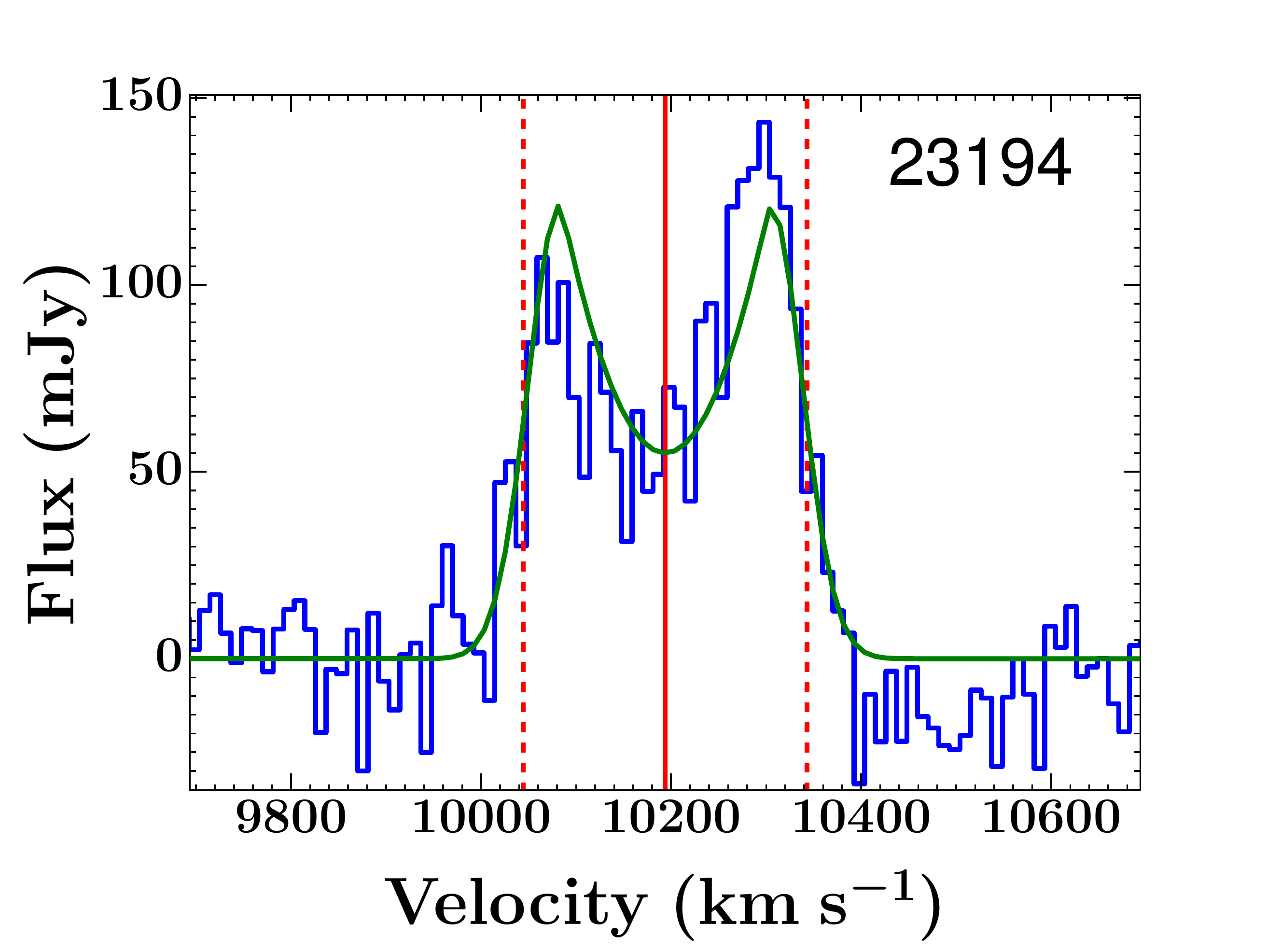}\includegraphics[scale=0.23, trim= 60 60 50 20, clip=true]{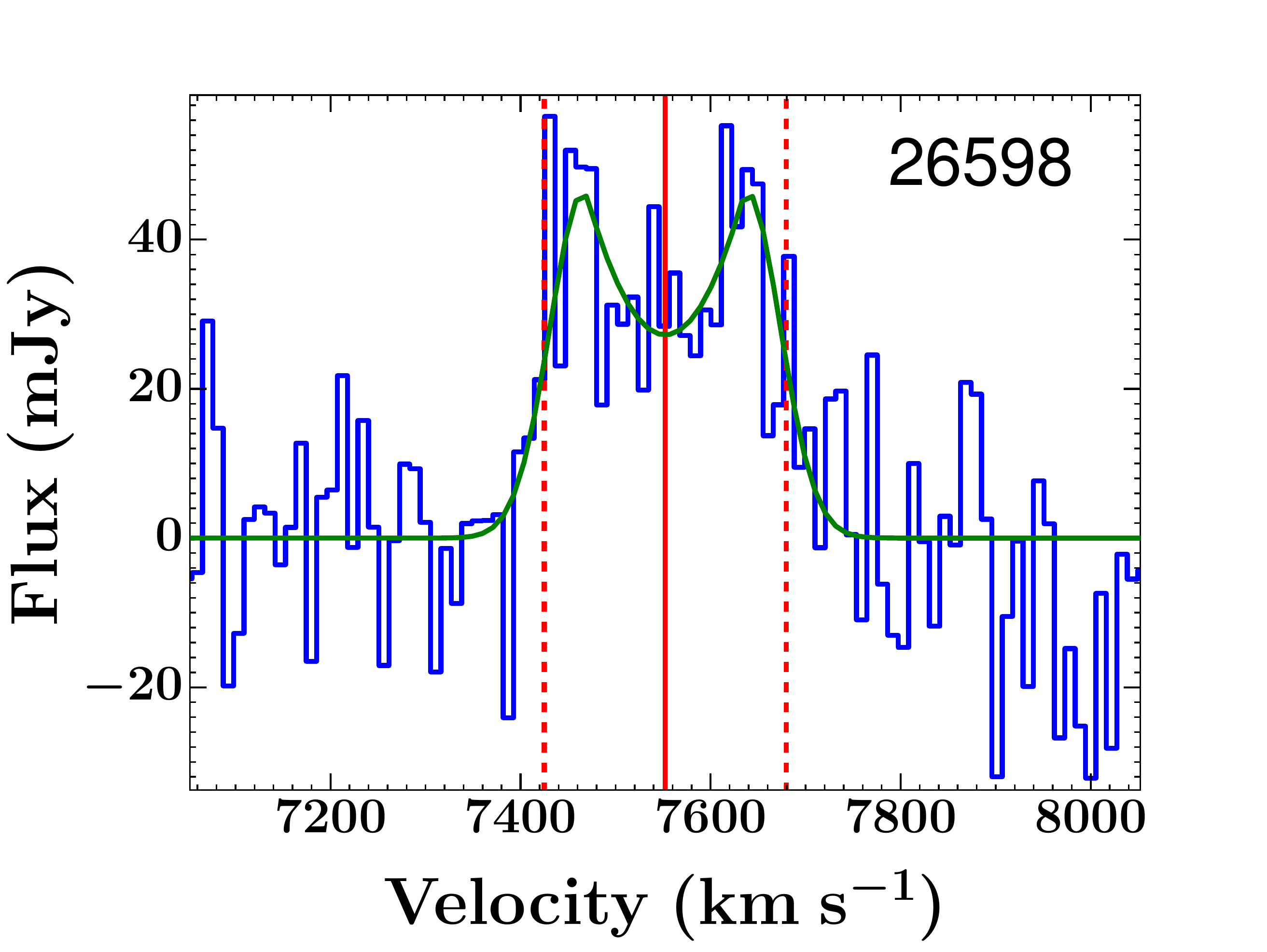}\includegraphics[scale=0.23, trim= 60 60 50 20, clip=true]{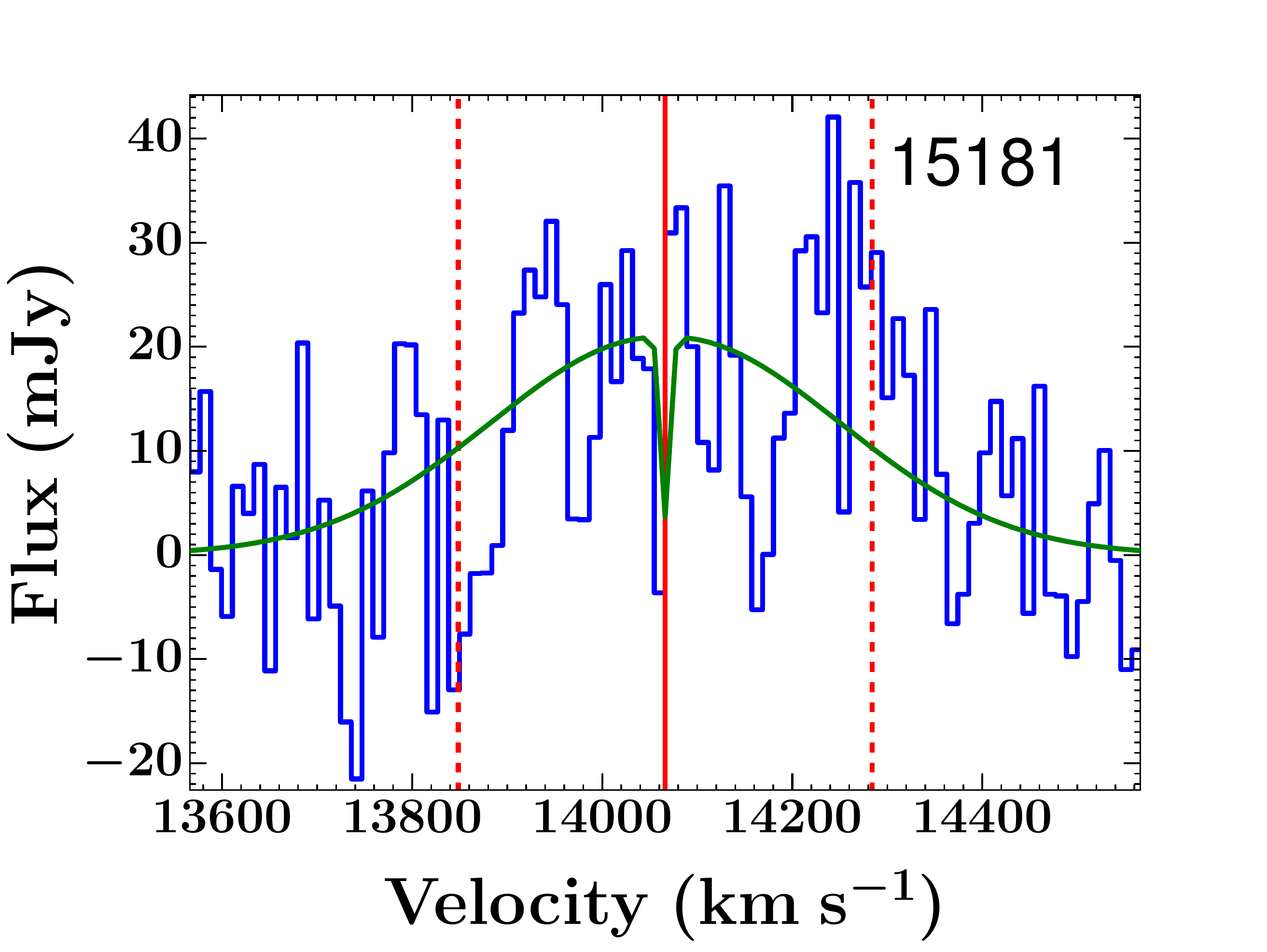}\includegraphics[scale=0.23, trim= 60 60 50 20, clip=true]{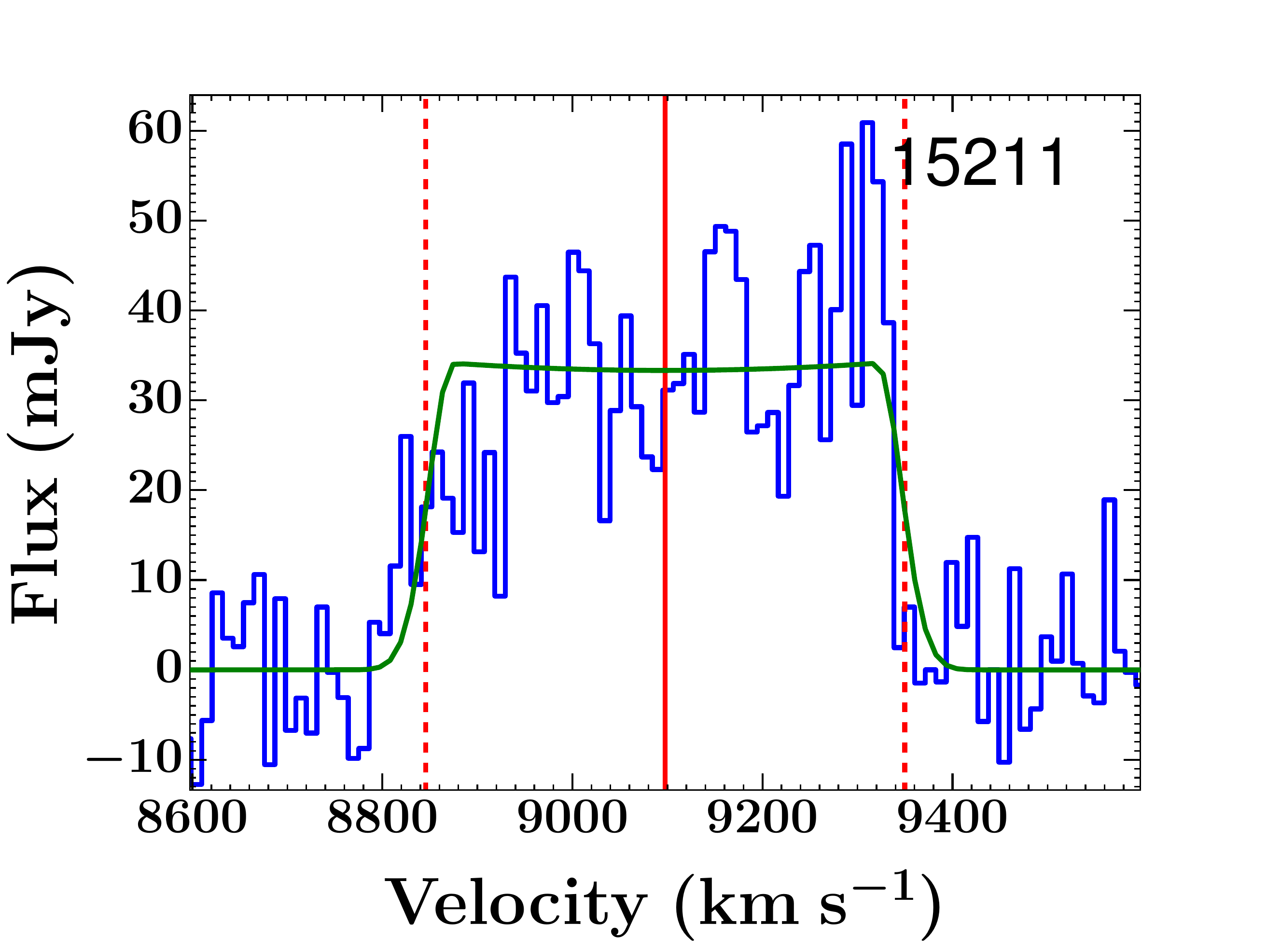}

\includegraphics[scale=0.23, trim= 0 5 50 20, clip=true]{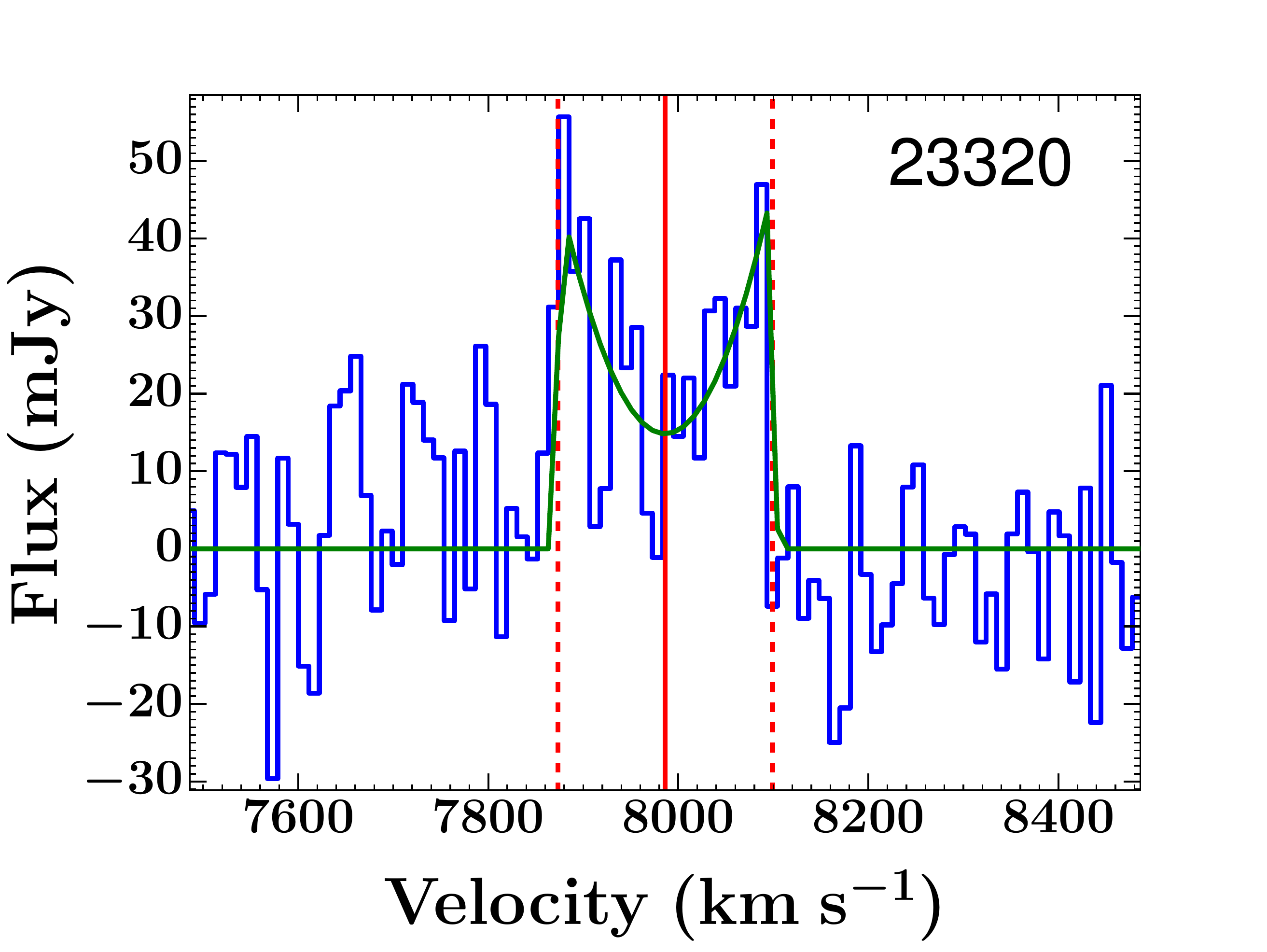}\includegraphics[scale=0.23, trim= 60 5 50 20, clip=true]{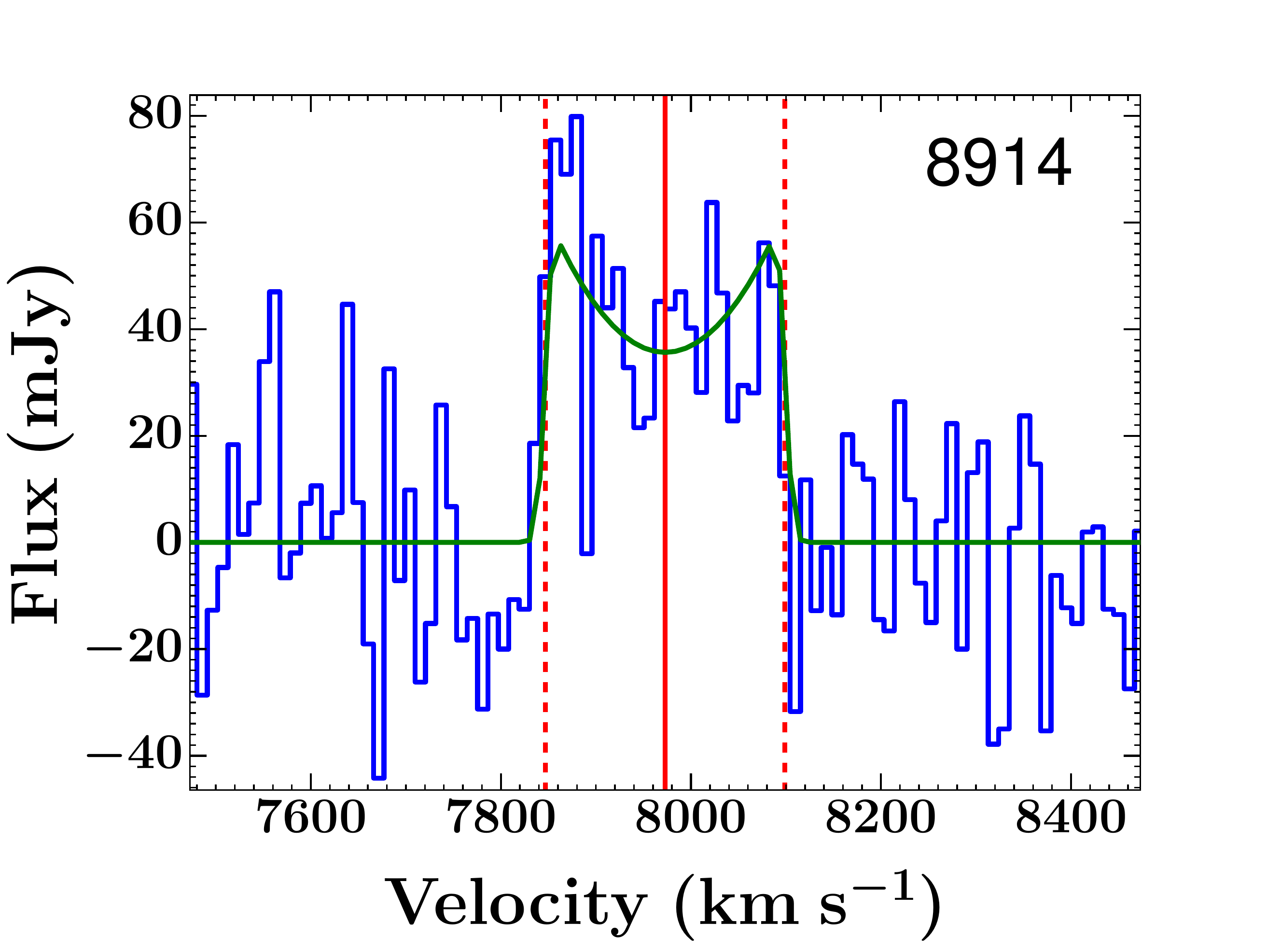}\includegraphics[scale=0.23, trim= 60 5 50 20, clip=true]{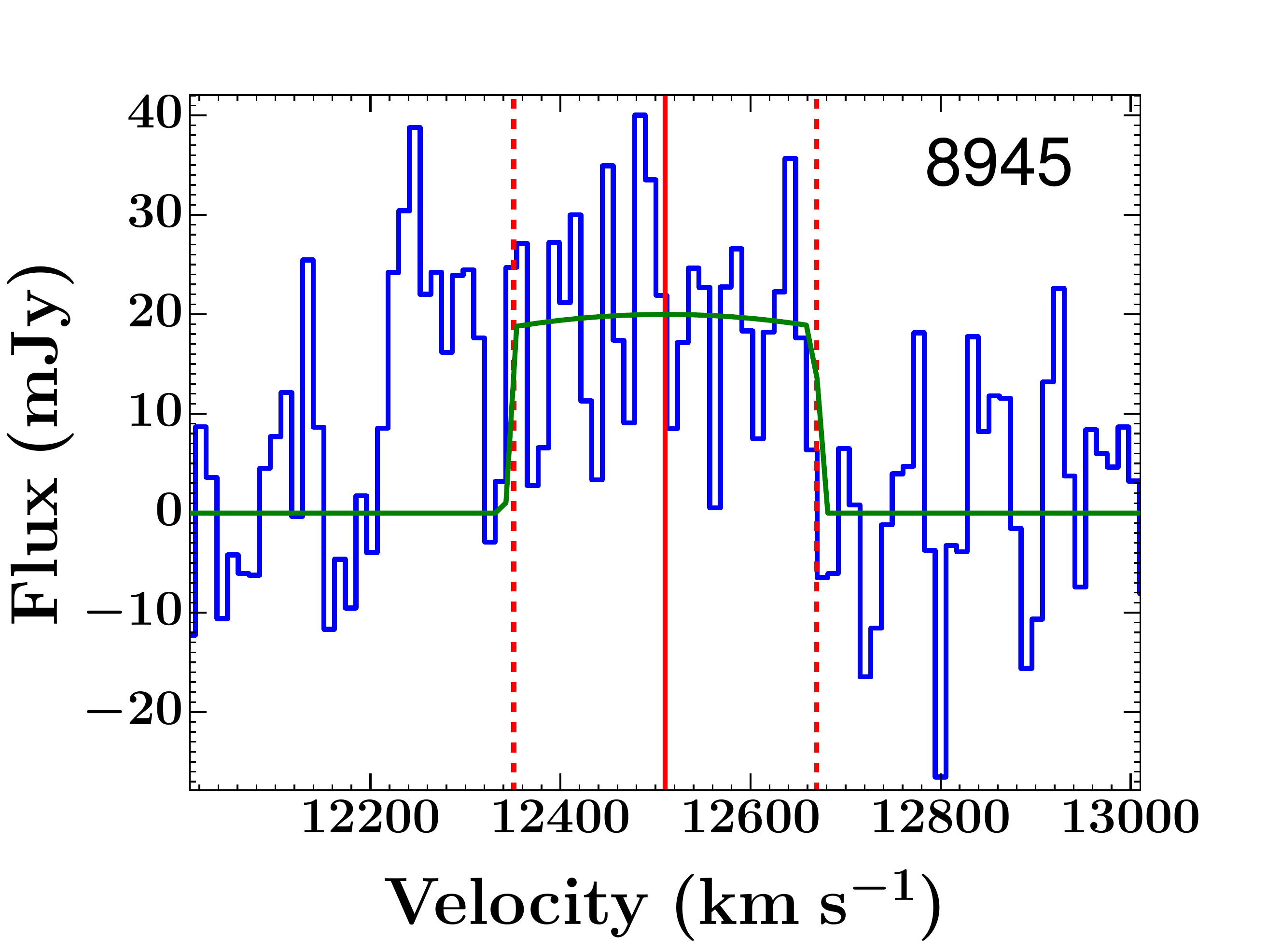}\includegraphics[scale=0.23, trim= 60 5 50 20, clip=true]{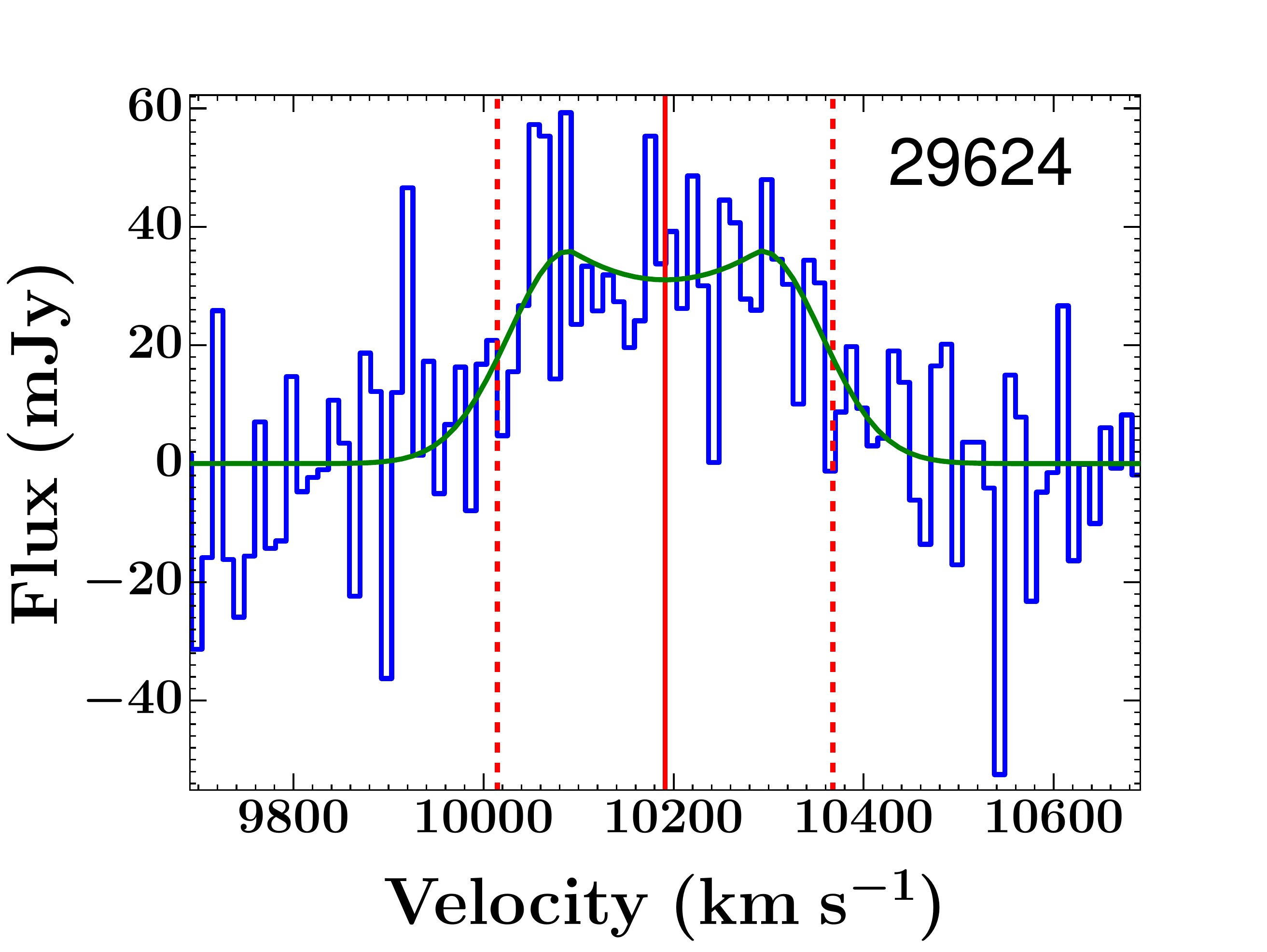}
\end{landscape}
\begin{landscape}
\includegraphics[scale=0.23, trim= 0 60 50 20, clip=true]{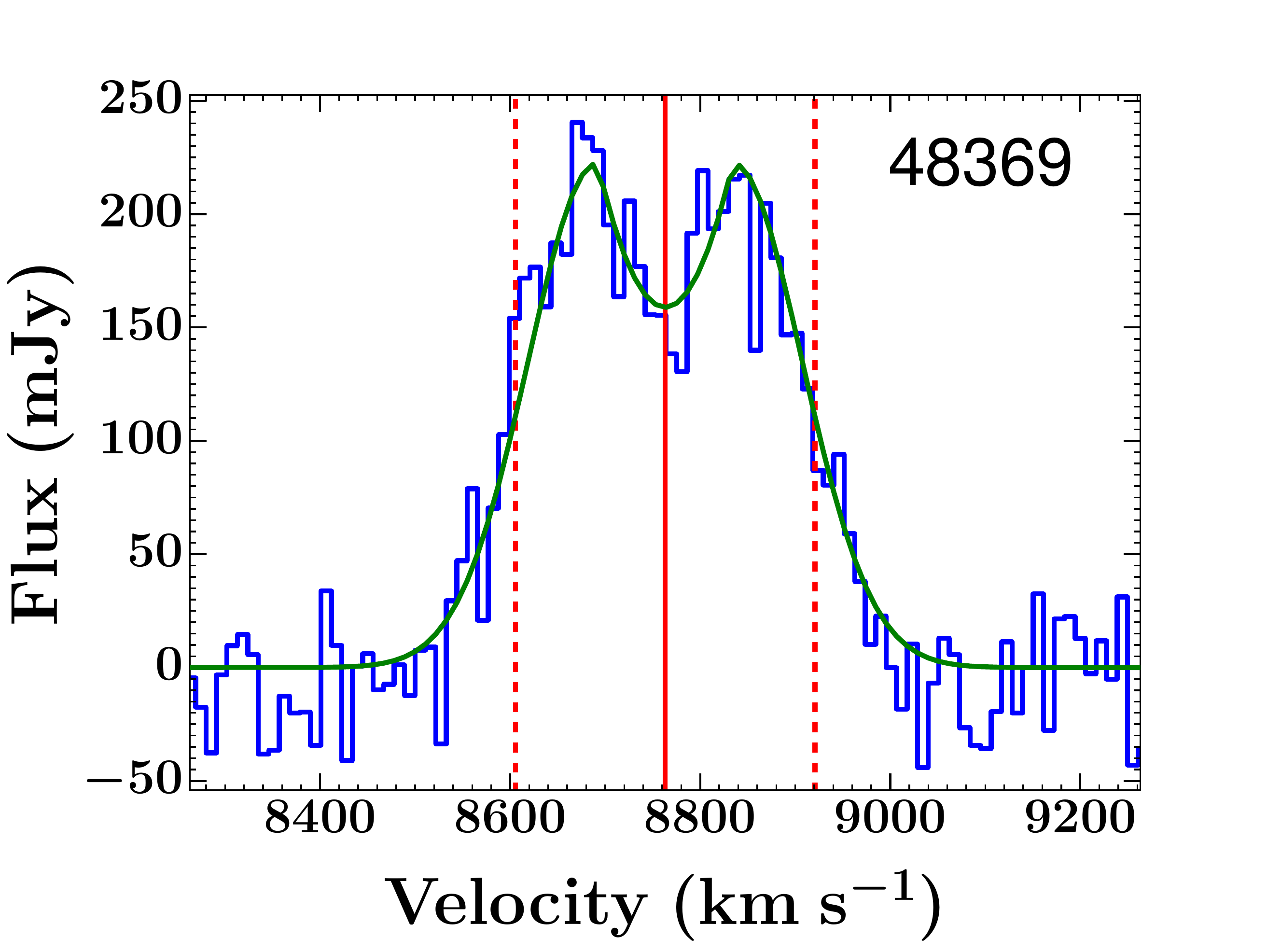}\includegraphics[scale=0.23, trim= 60 60 50 20, clip=true]{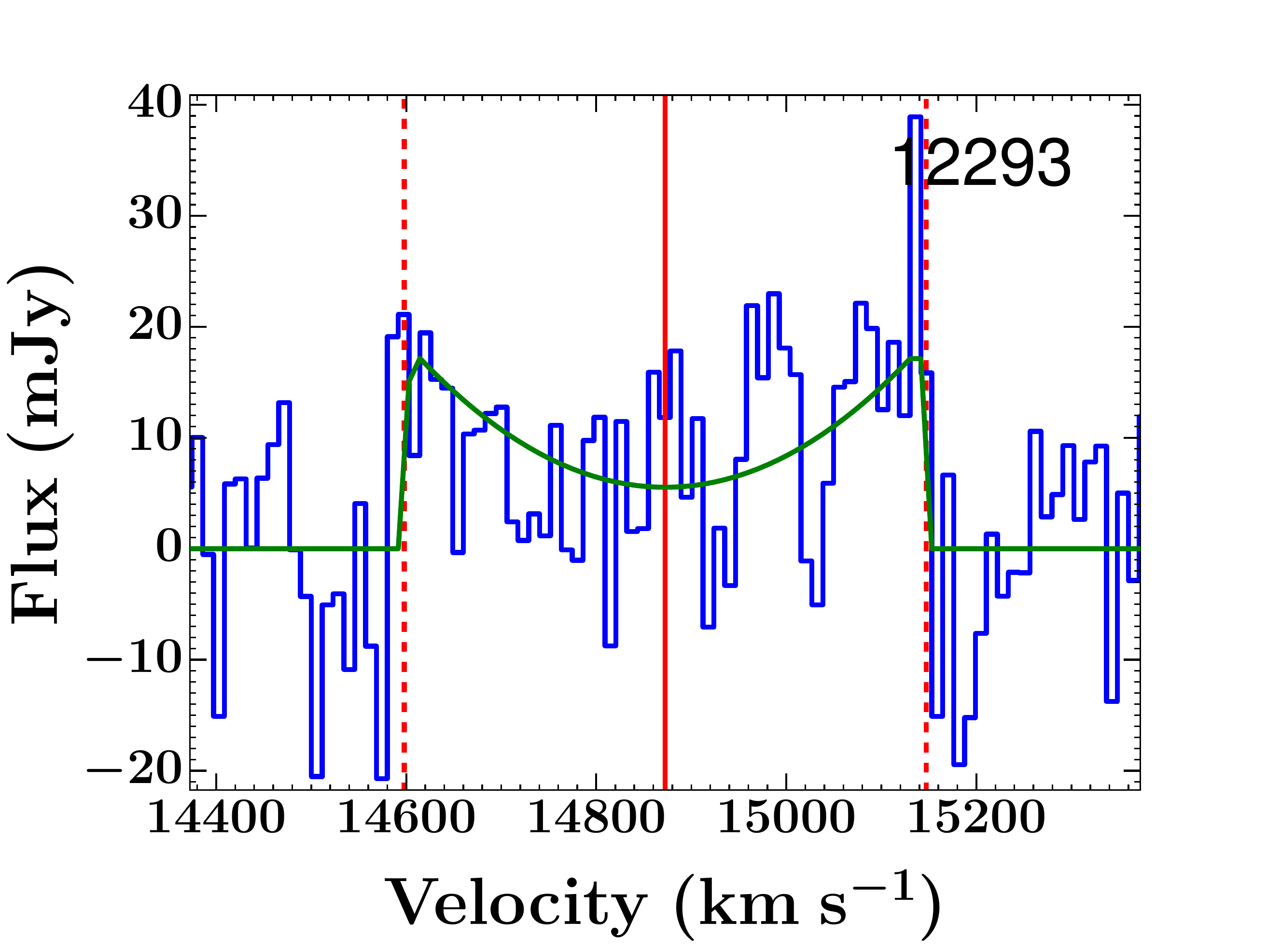}\includegraphics[scale=0.23, trim= 60 60 50 20, clip=true]{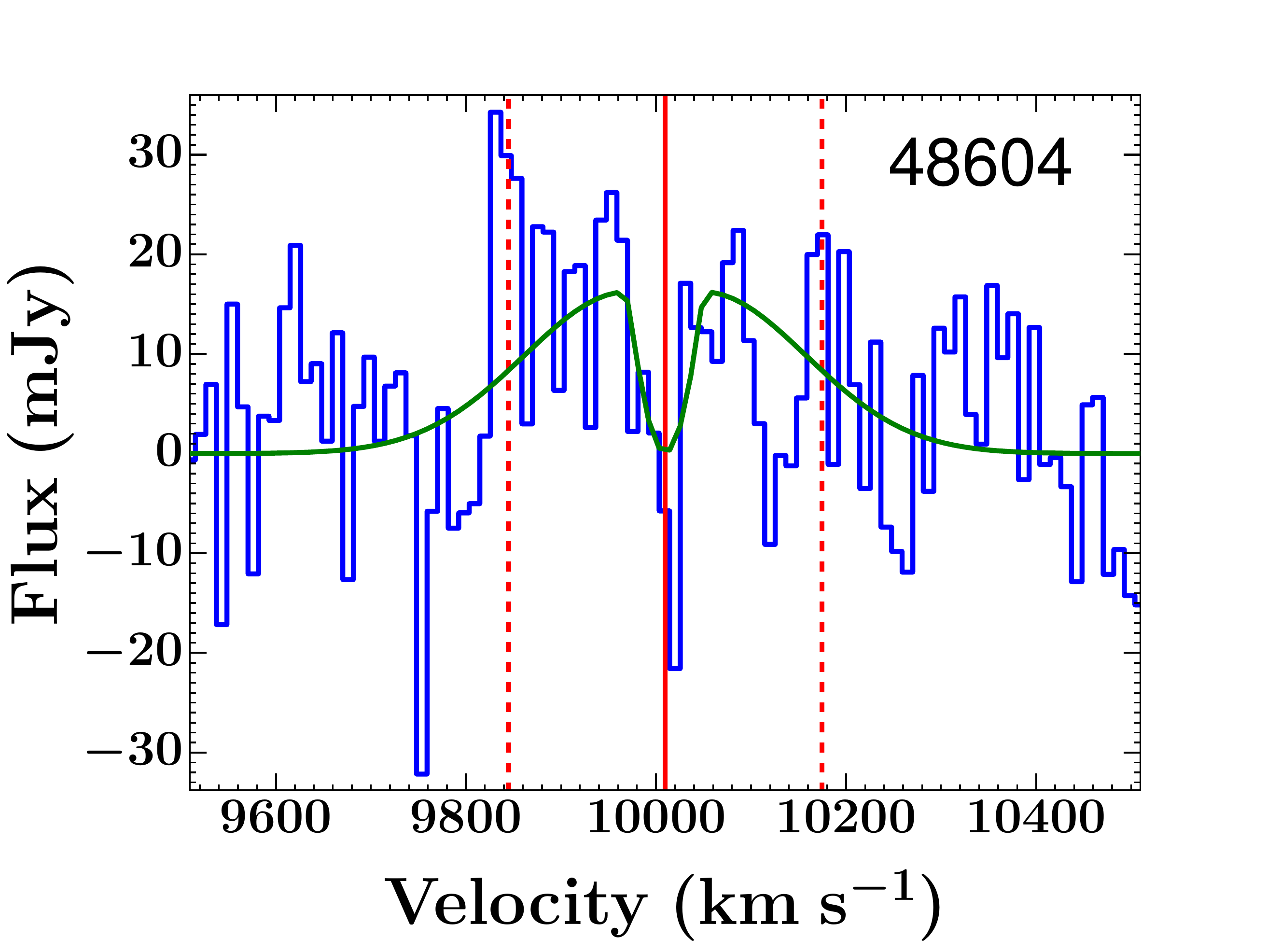}\includegraphics[scale=0.23, trim= 60 60 50 20, clip=true]{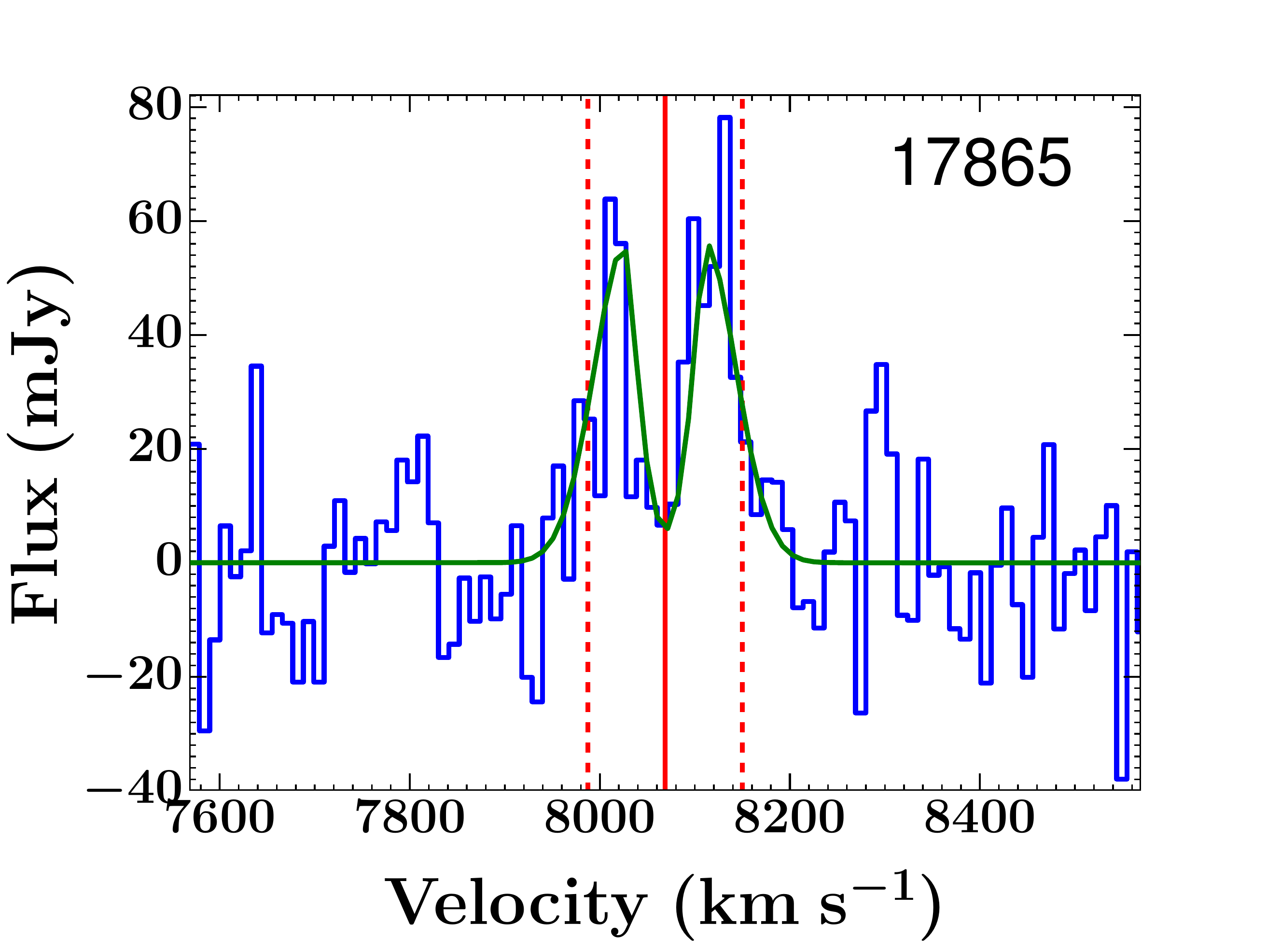}

\includegraphics[scale=0.23, trim= 0 60 50 20, clip=true]{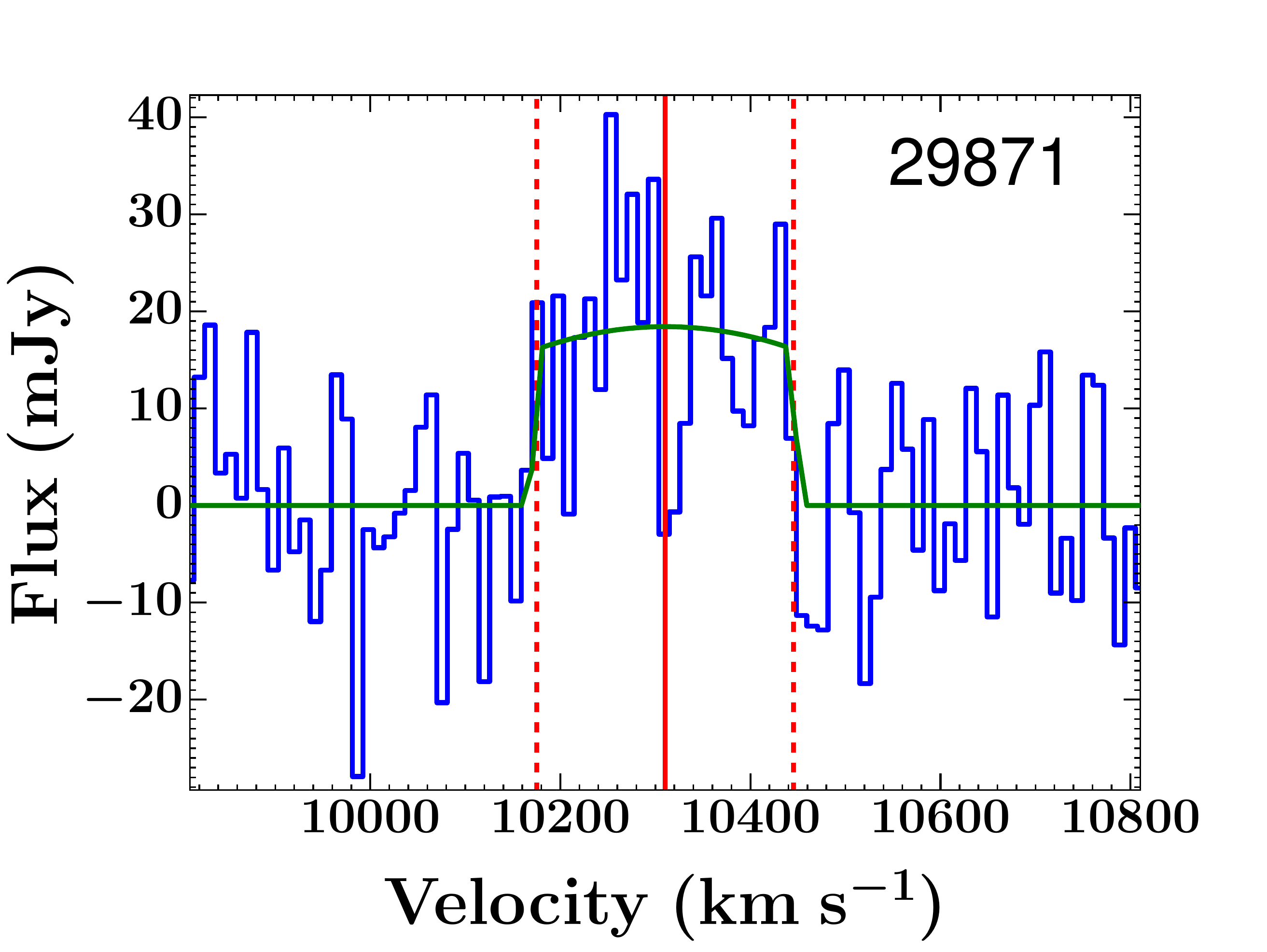}\includegraphics[scale=0.23, trim= 60 60 50 20, clip=true]{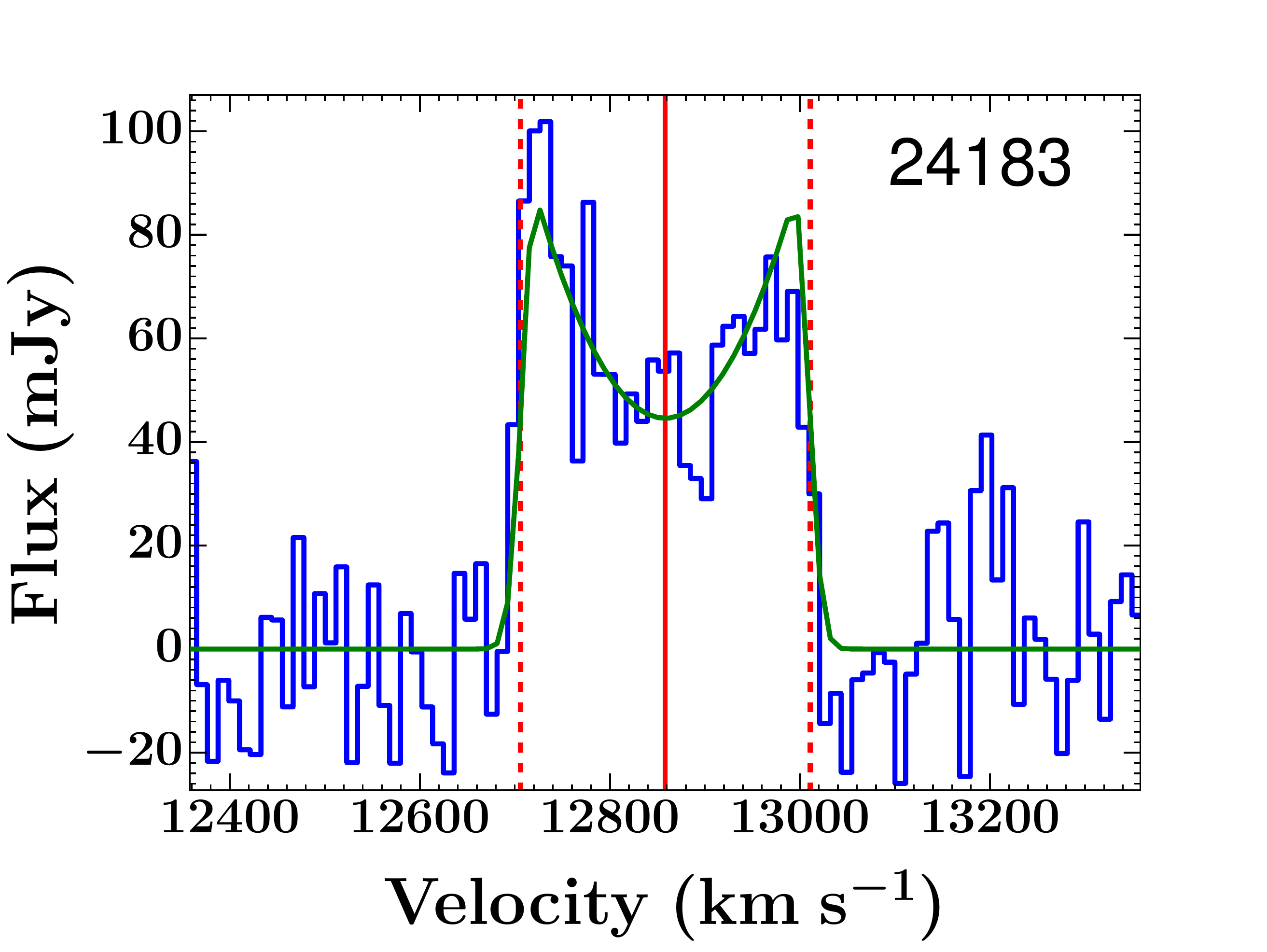}\includegraphics[scale=0.23, trim= 60 60 50 20, clip=true]{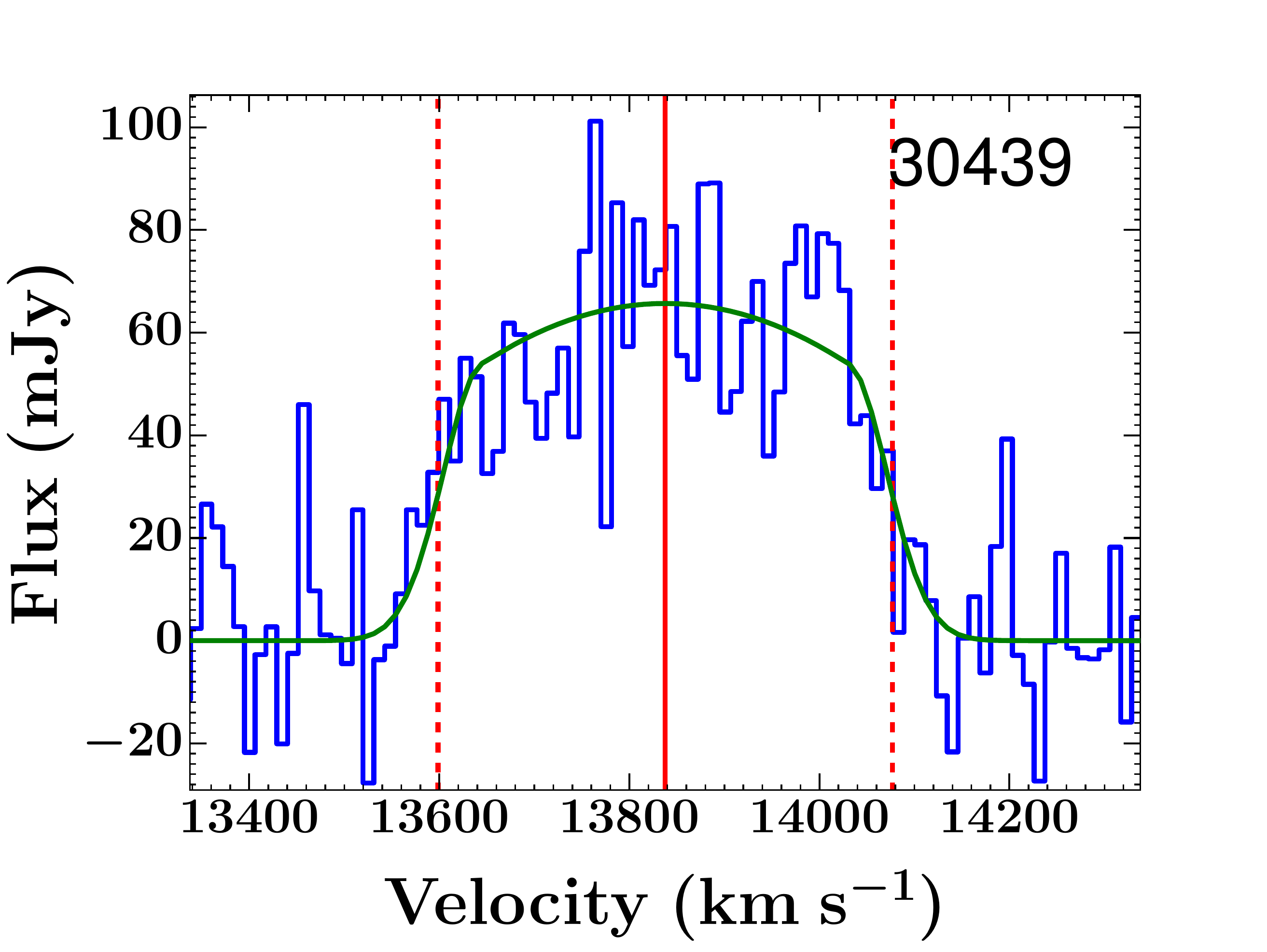}\includegraphics[scale=0.23, trim= 60 60 50 20, clip=true]{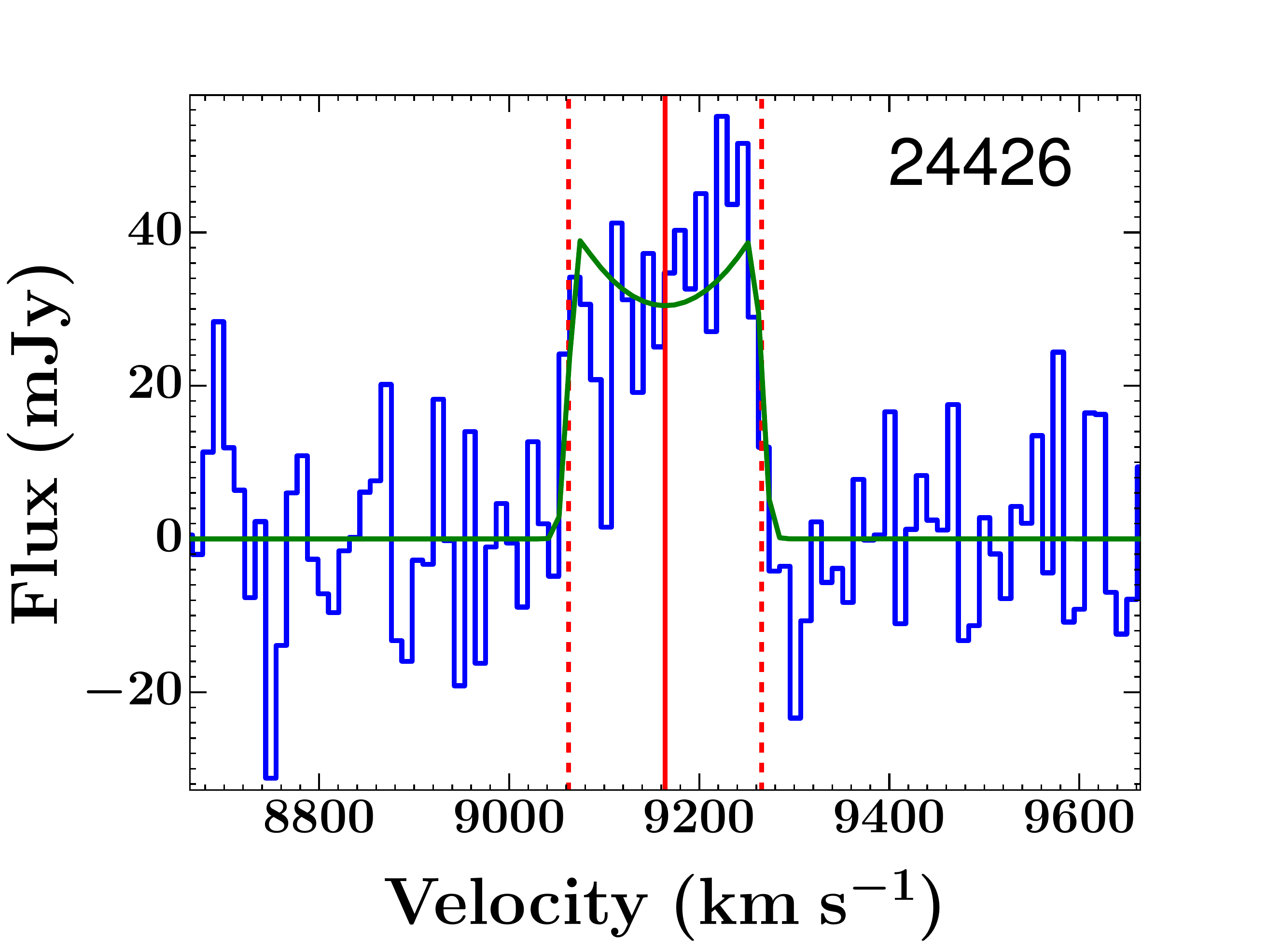}

\includegraphics[scale=0.23, trim= 0 60 50 20, clip=true]{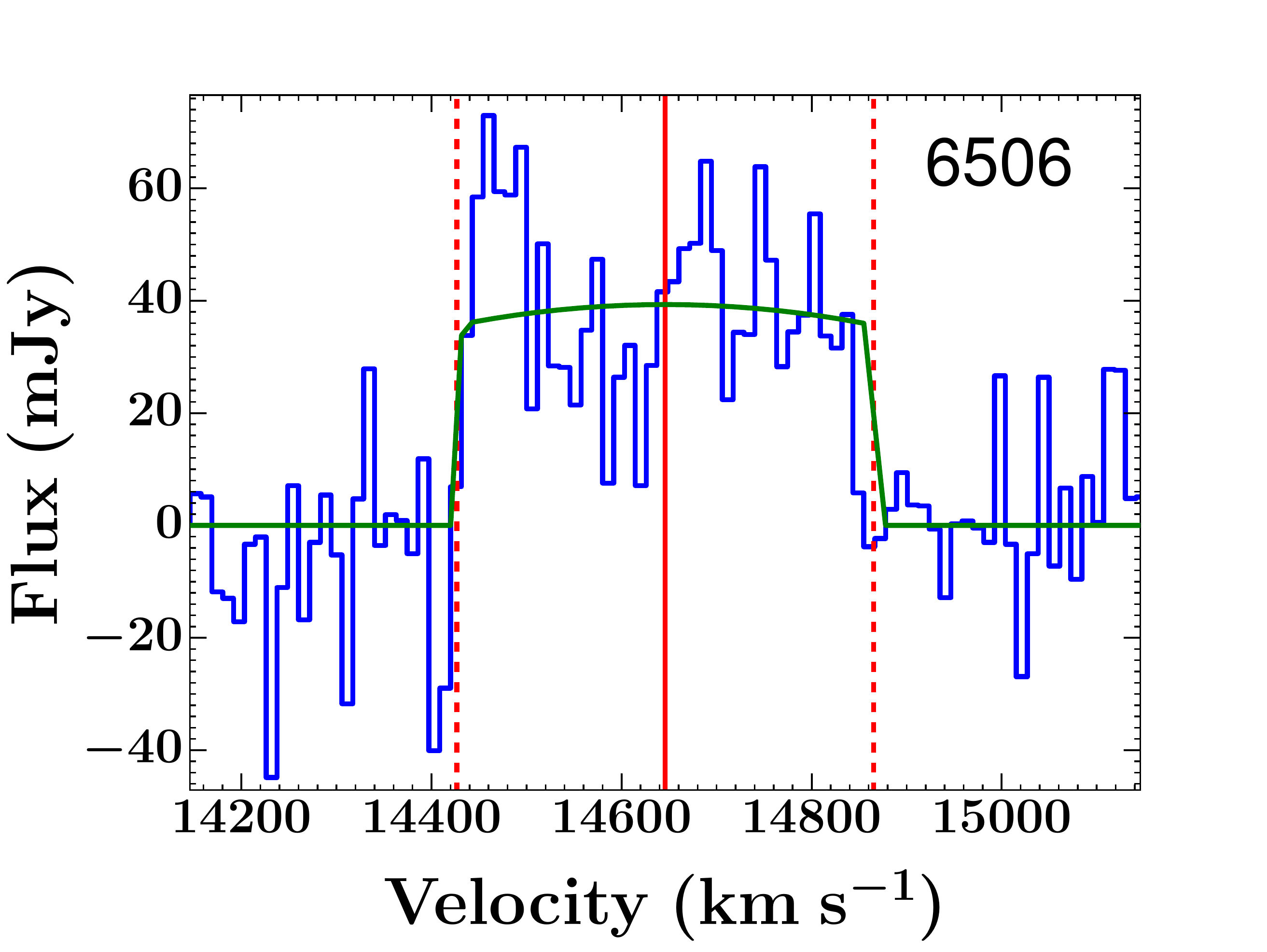}\includegraphics[scale=0.23, trim= 60 60 50 20, clip=true]{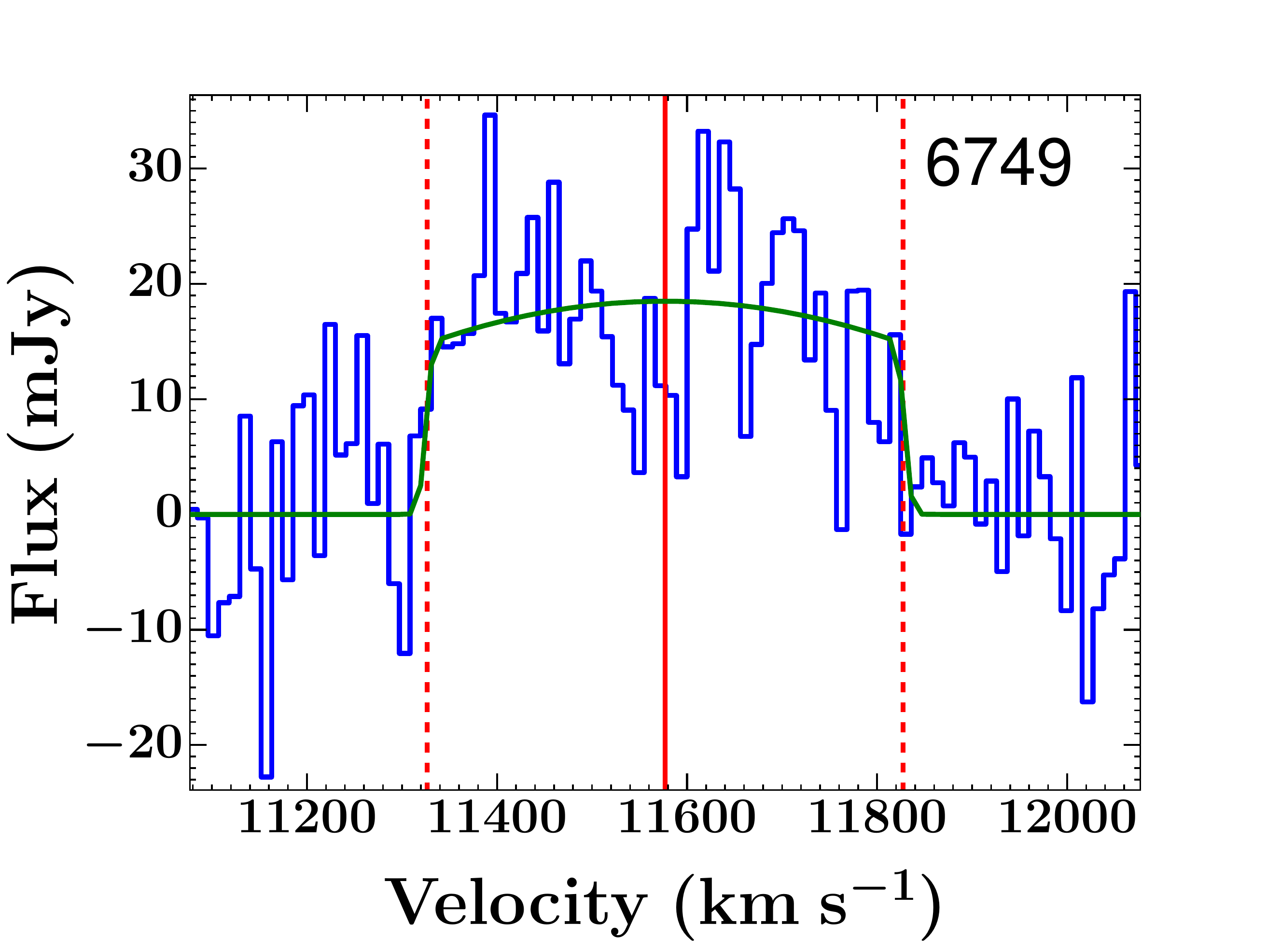}\includegraphics[scale=0.23, trim= 60 60 50 20, clip=true]{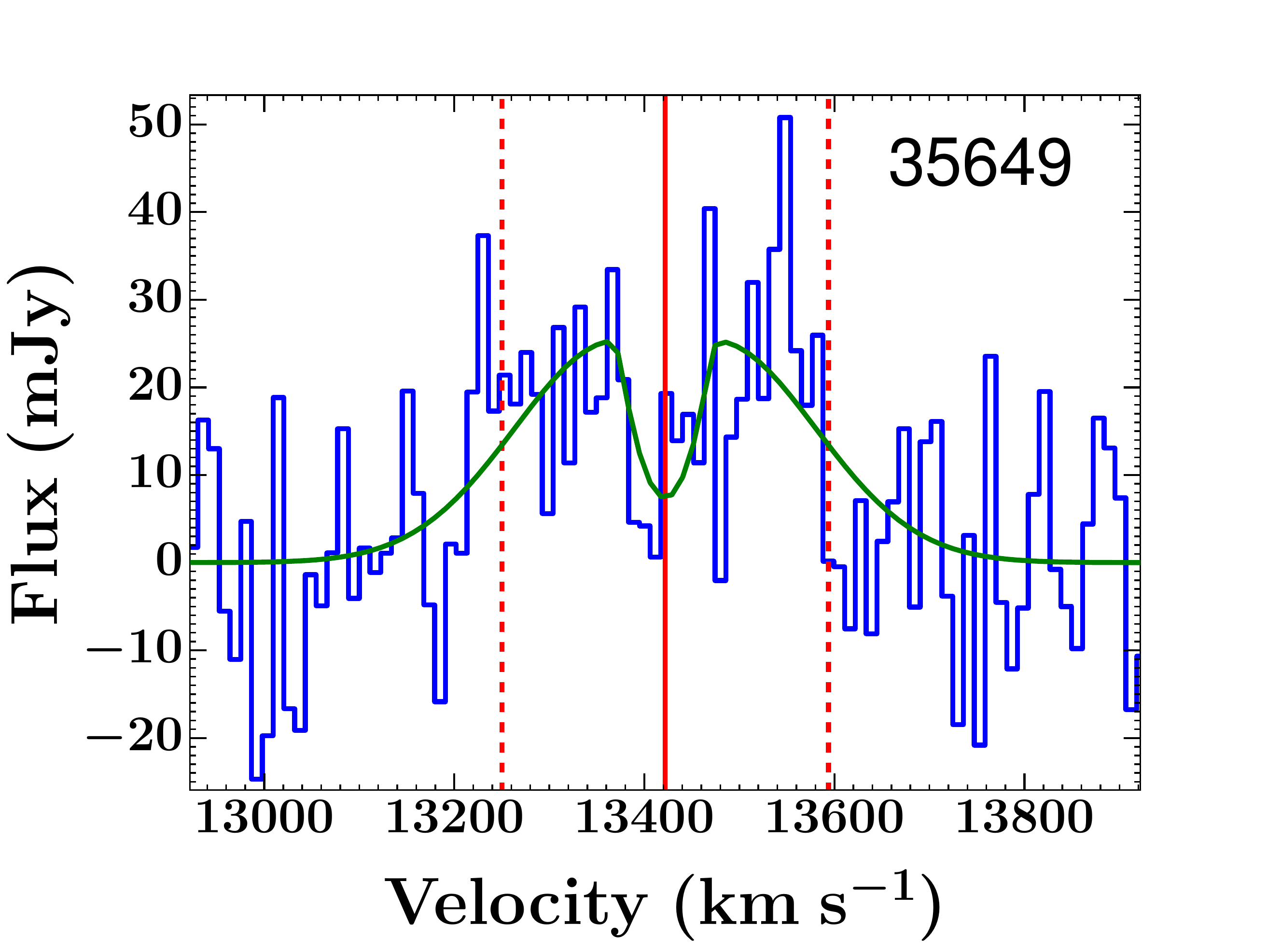}\includegraphics[scale=0.23, trim= 60 60 50 20, clip=true]{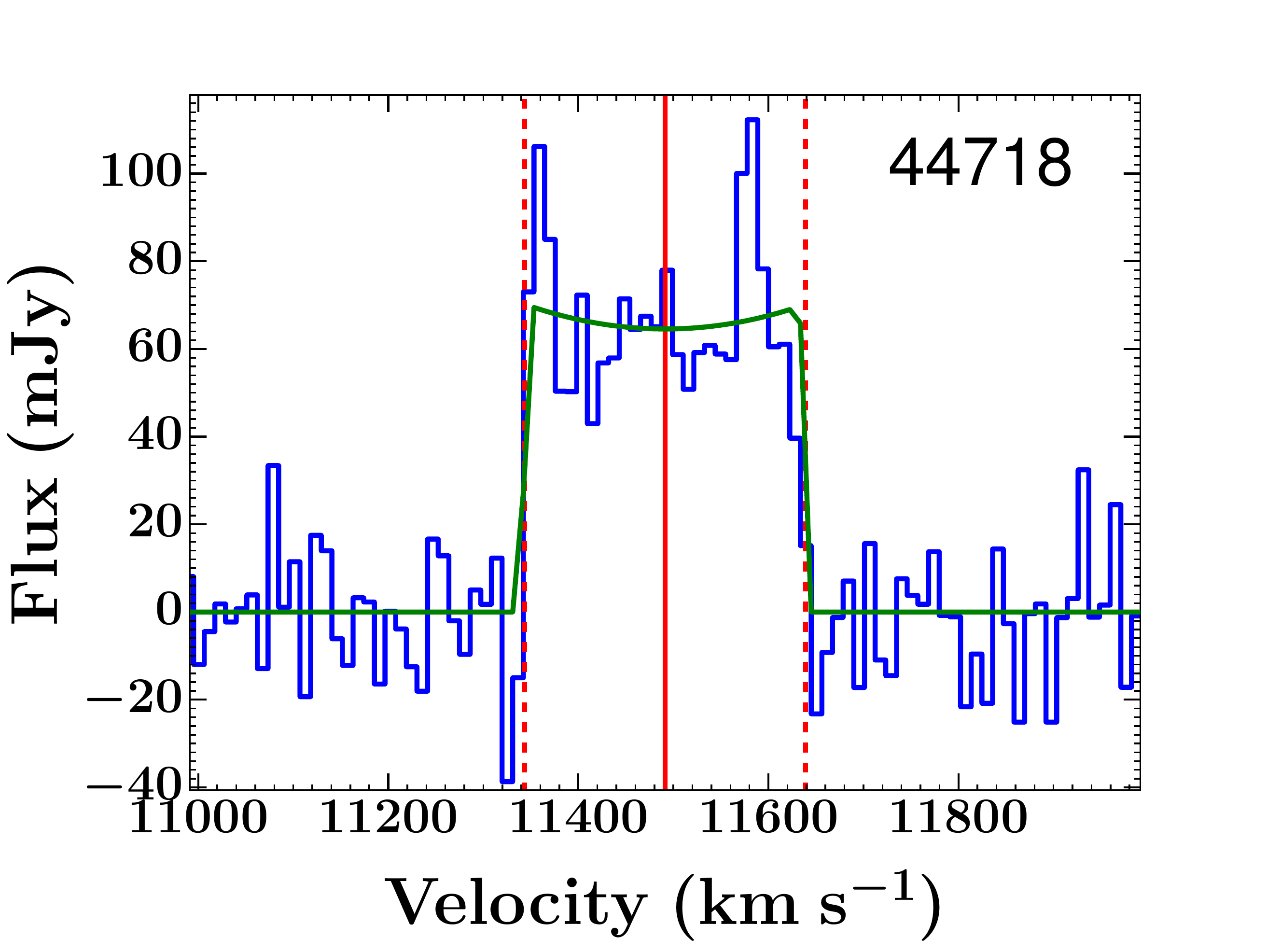}

\includegraphics[scale=0.23, trim= 0 5 50 20, clip=true]{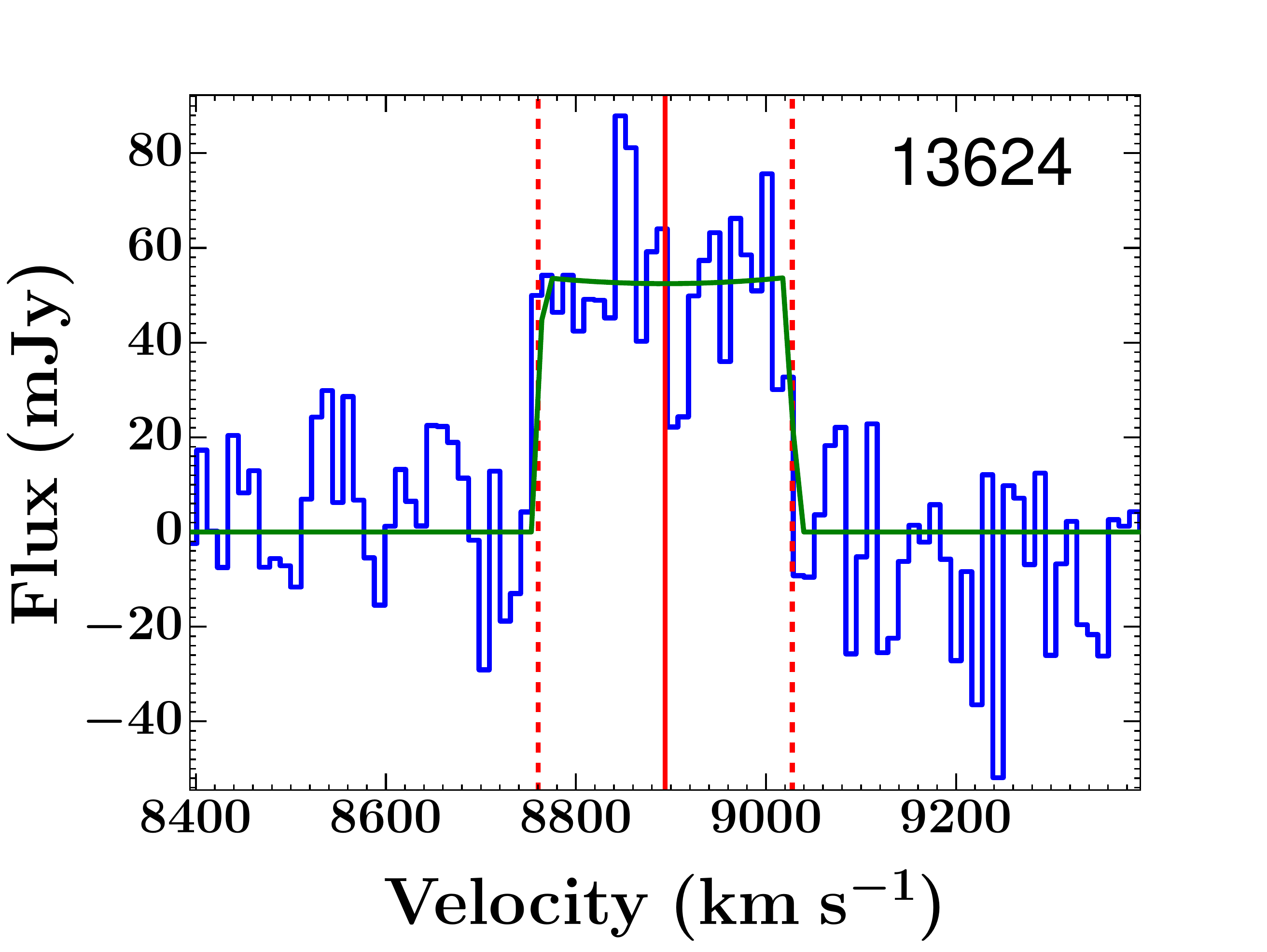}\includegraphics[scale=0.23, trim= 60 5 50 20, clip=true]{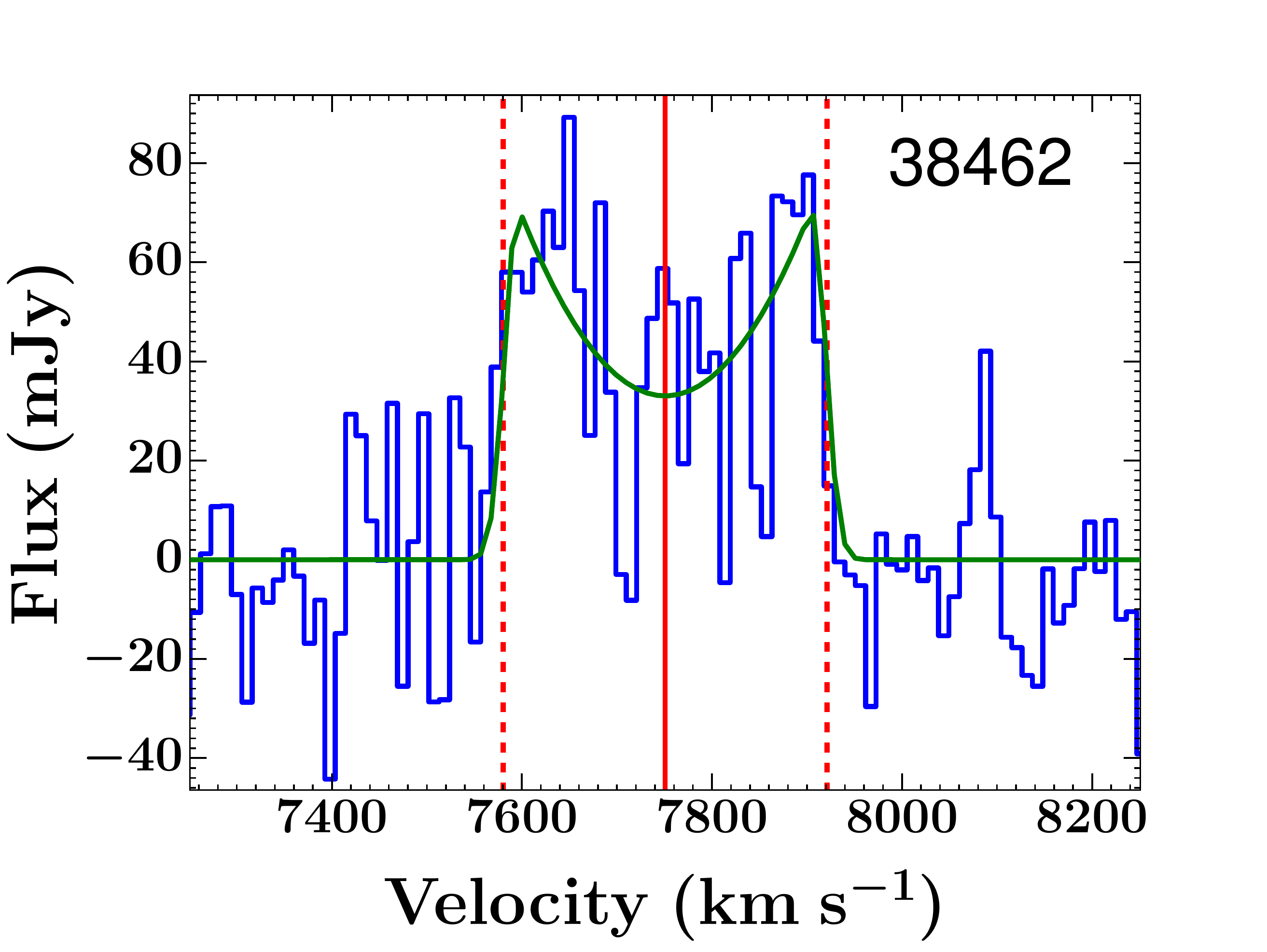}\includegraphics[scale=0.23, trim= 60 5 50 20, clip=true]{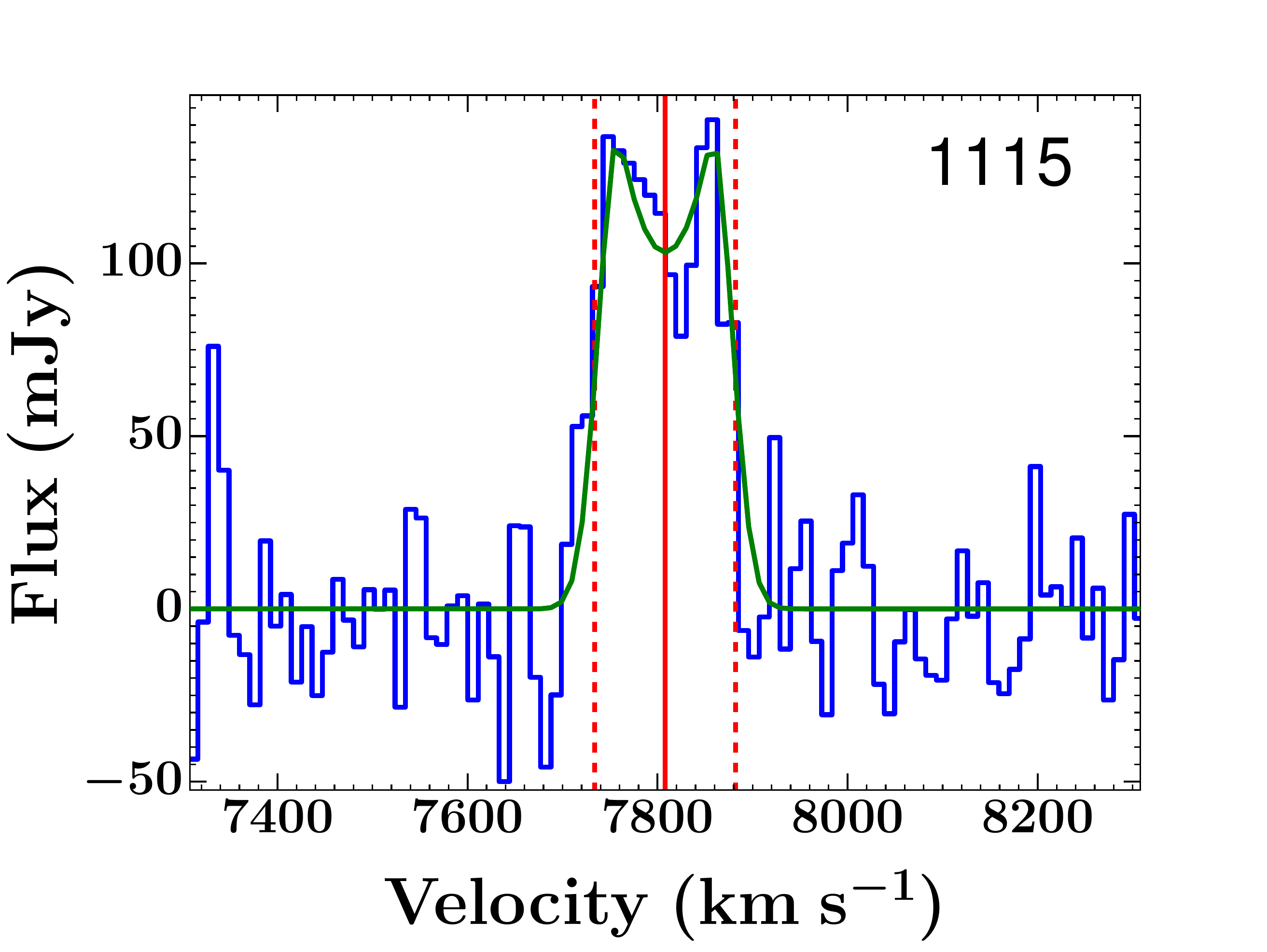}\includegraphics[scale=0.23, trim= 60 5 50 20, clip=true]{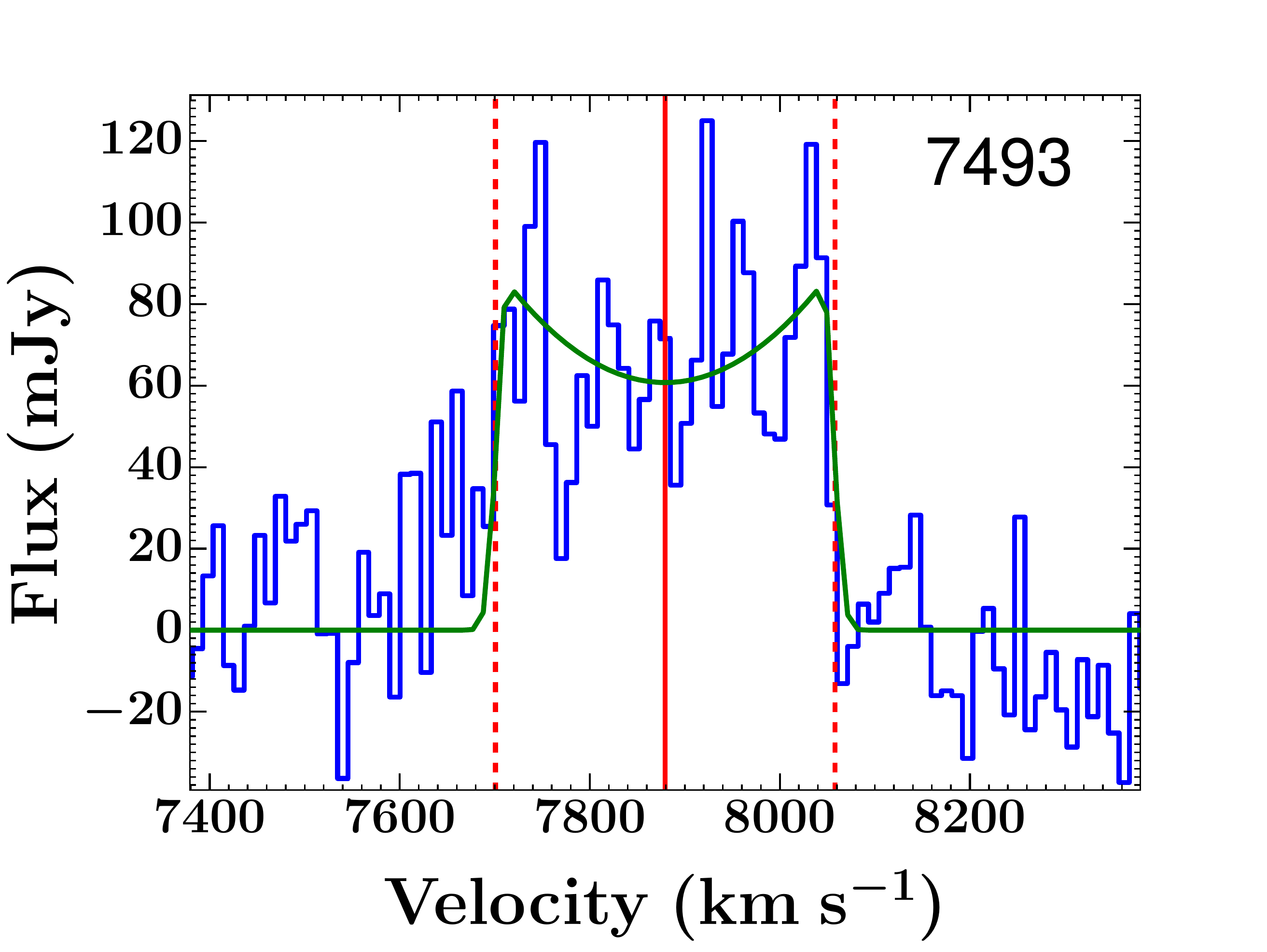}
\end{landscape}
\begin{landscape}
\includegraphics[scale=0.23, trim= 0 60 50 20, clip=true]{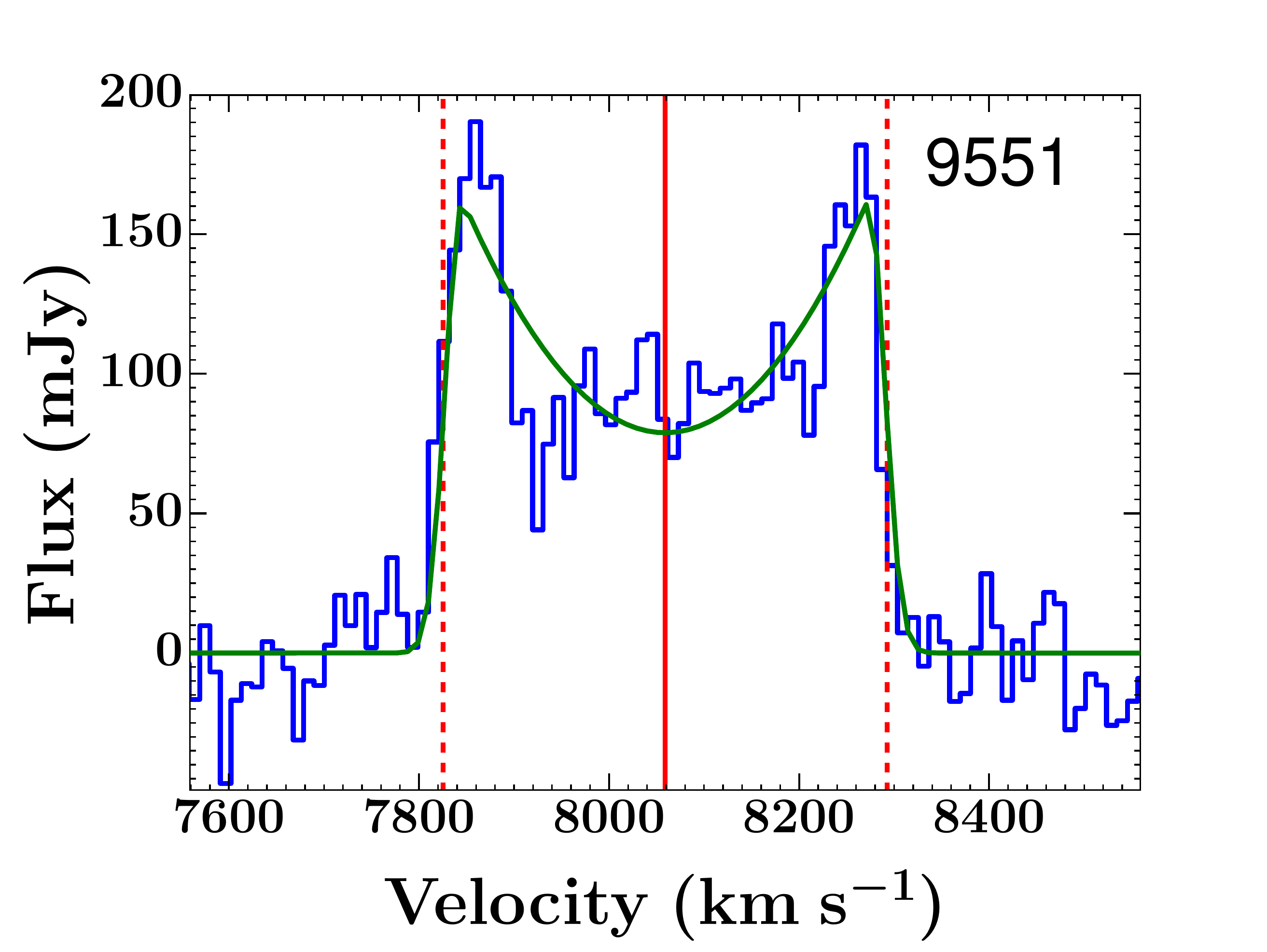}\includegraphics[scale=0.23, trim= 60 60 50 20, clip=true]{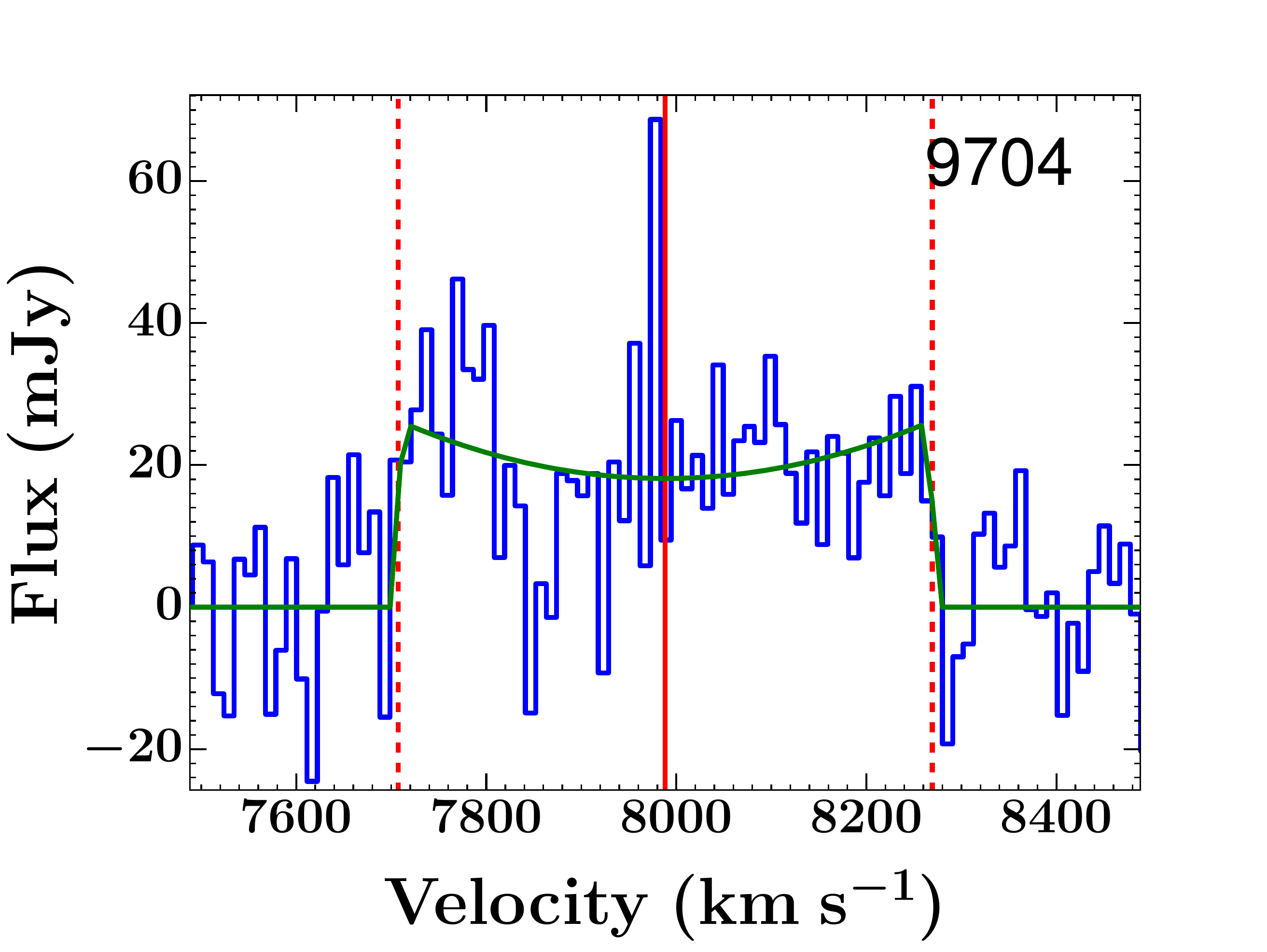}\includegraphics[scale=0.23, trim= 60 60 50 20, clip=true]{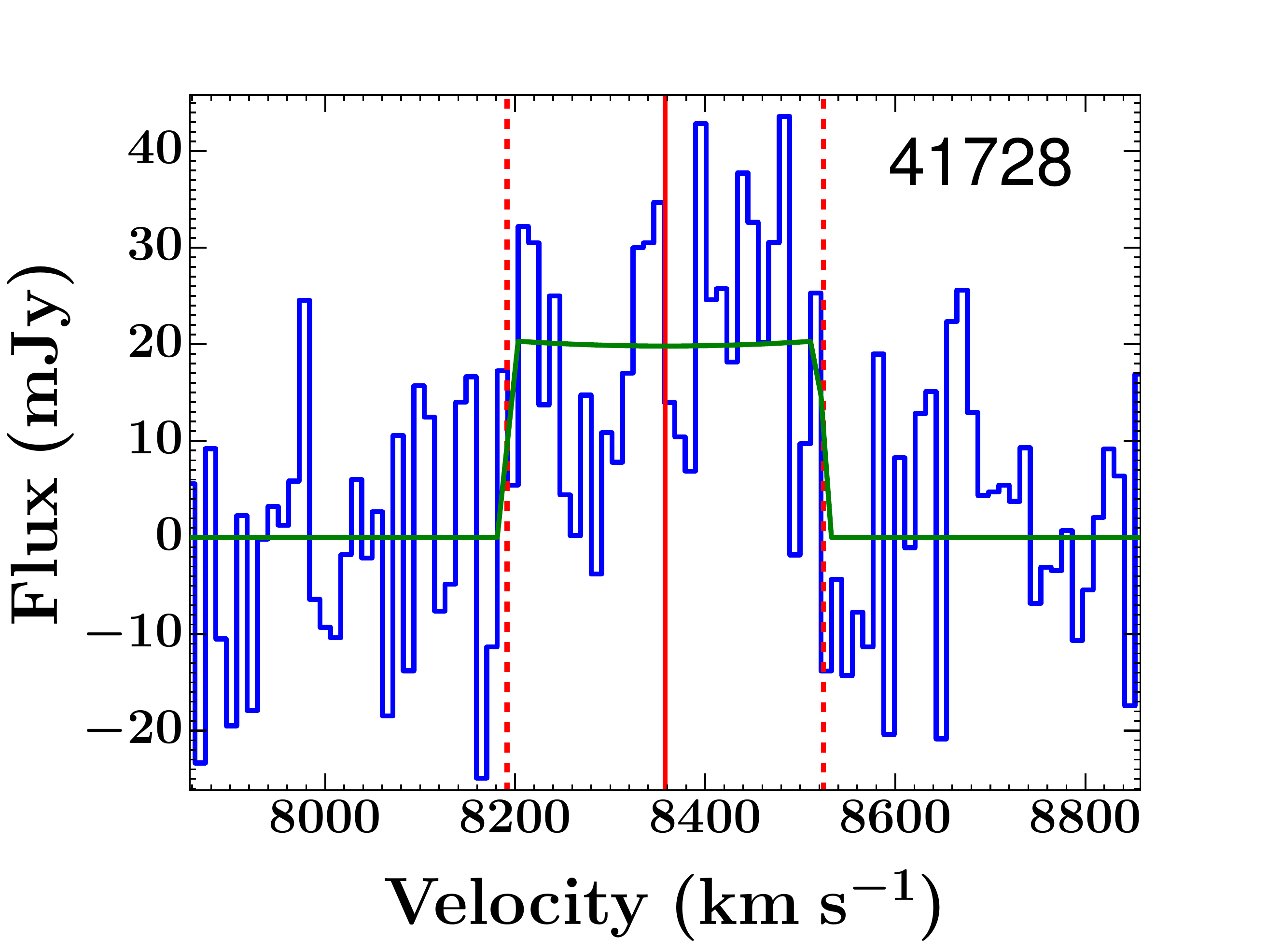}\includegraphics[scale=0.23, trim= 60 60 50 20, clip=true]{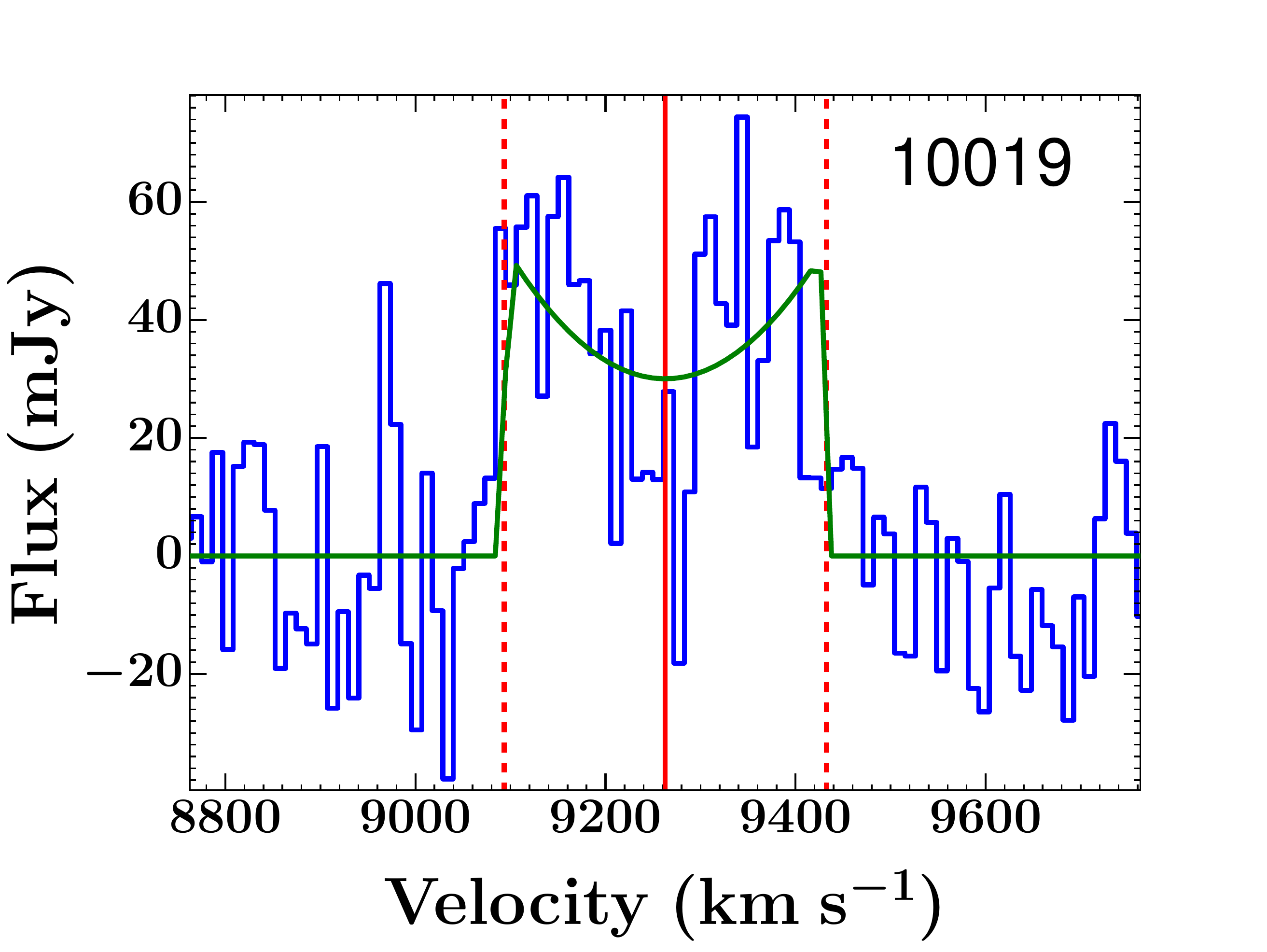}

\includegraphics[scale=0.23, trim= 0 60 50 20, clip=true]{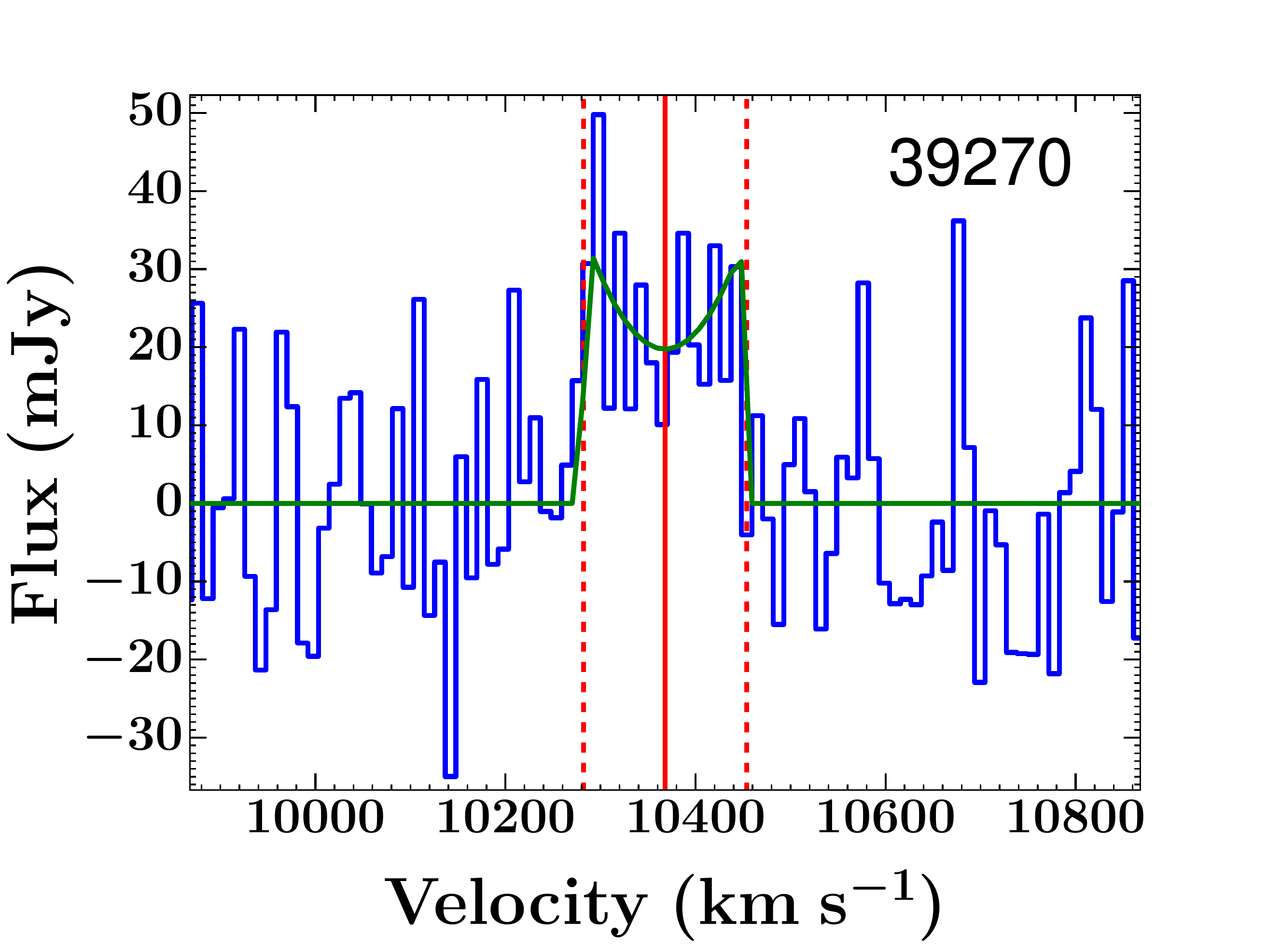}\includegraphics[scale=0.23, trim= 60 60 50 20, clip=true]{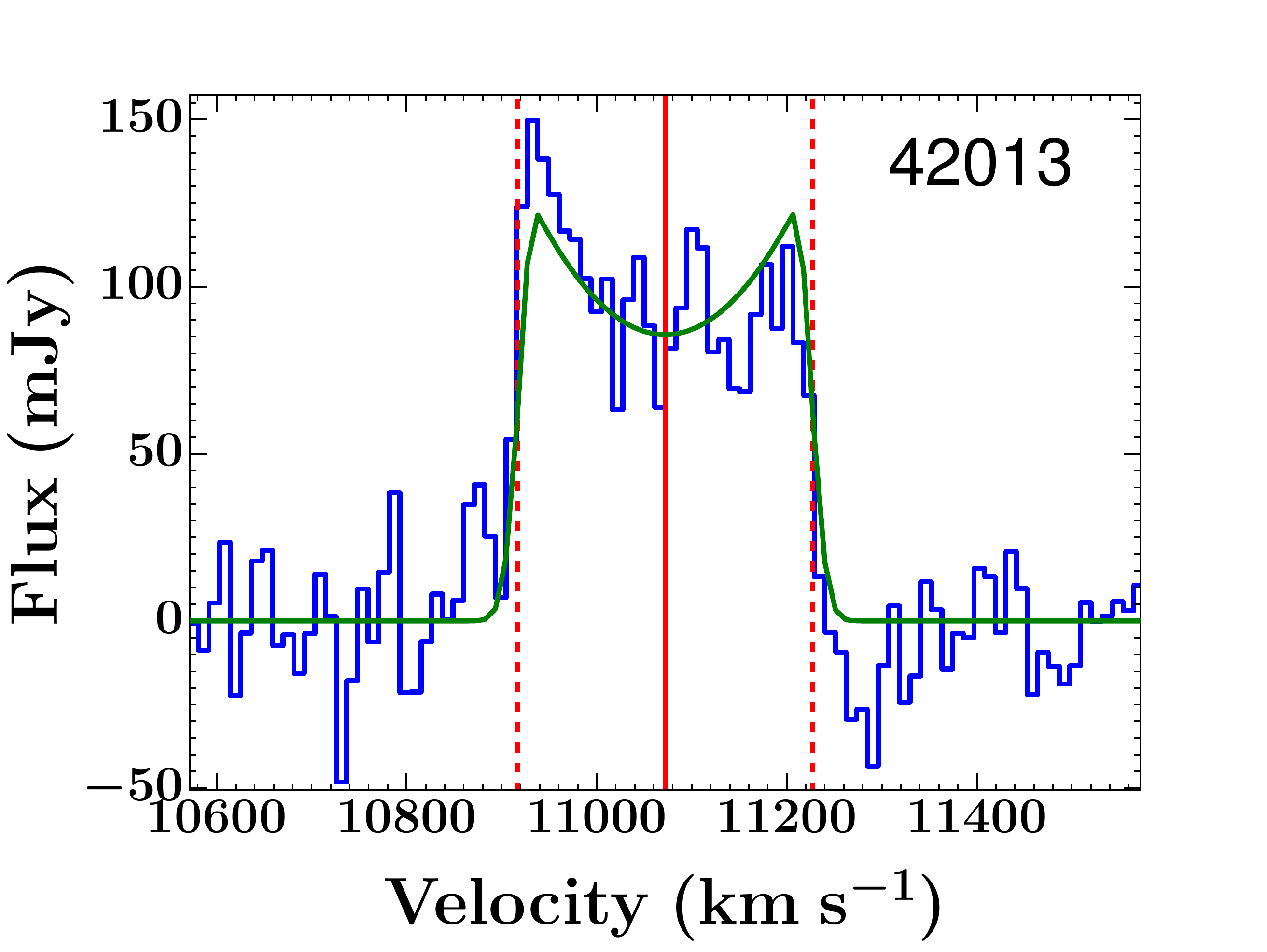}\includegraphics[scale=0.23, trim= 60 60 50 20, clip=true]{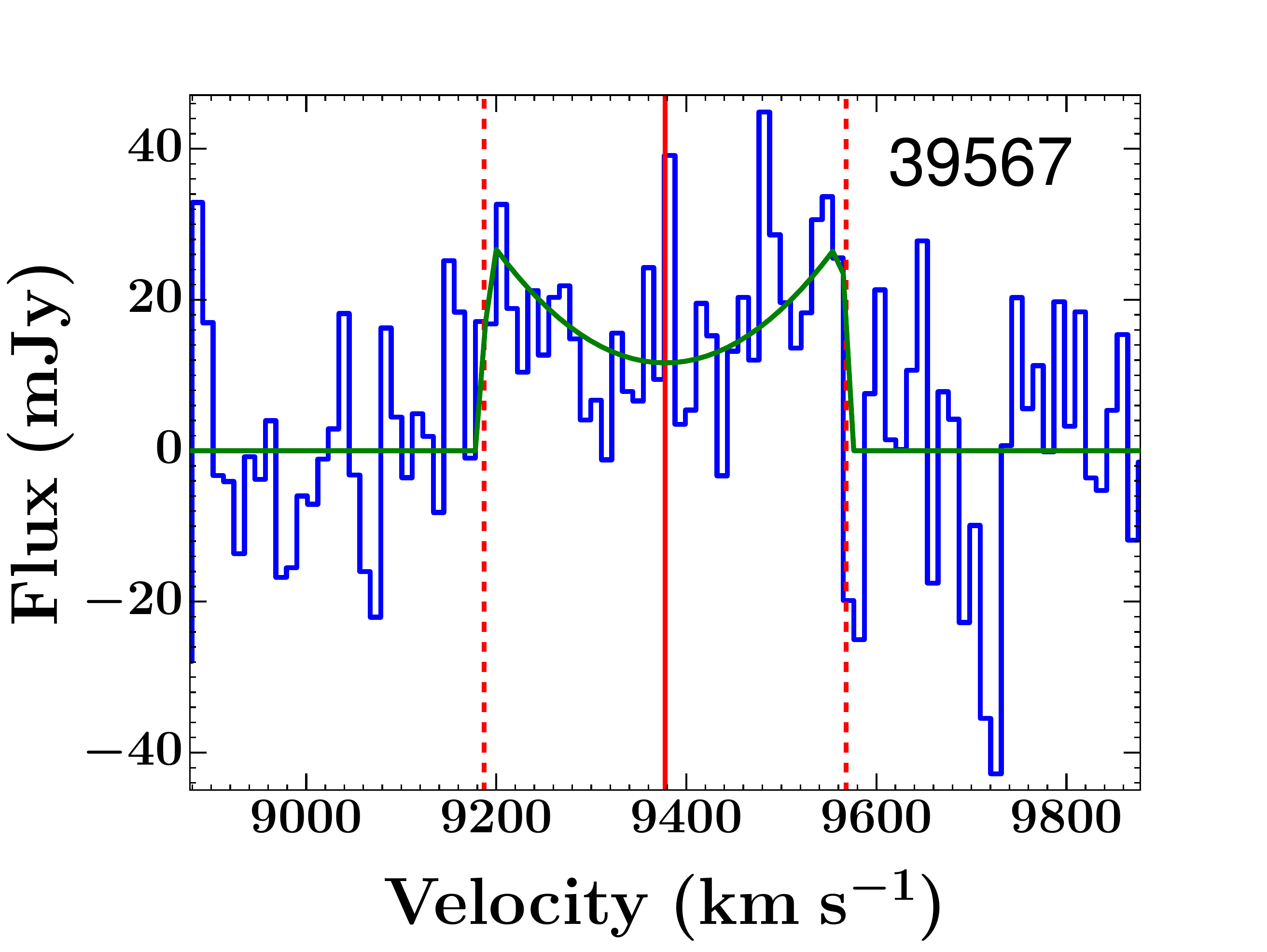}\includegraphics[scale=0.23, trim= 60 60 50 20, clip=true]{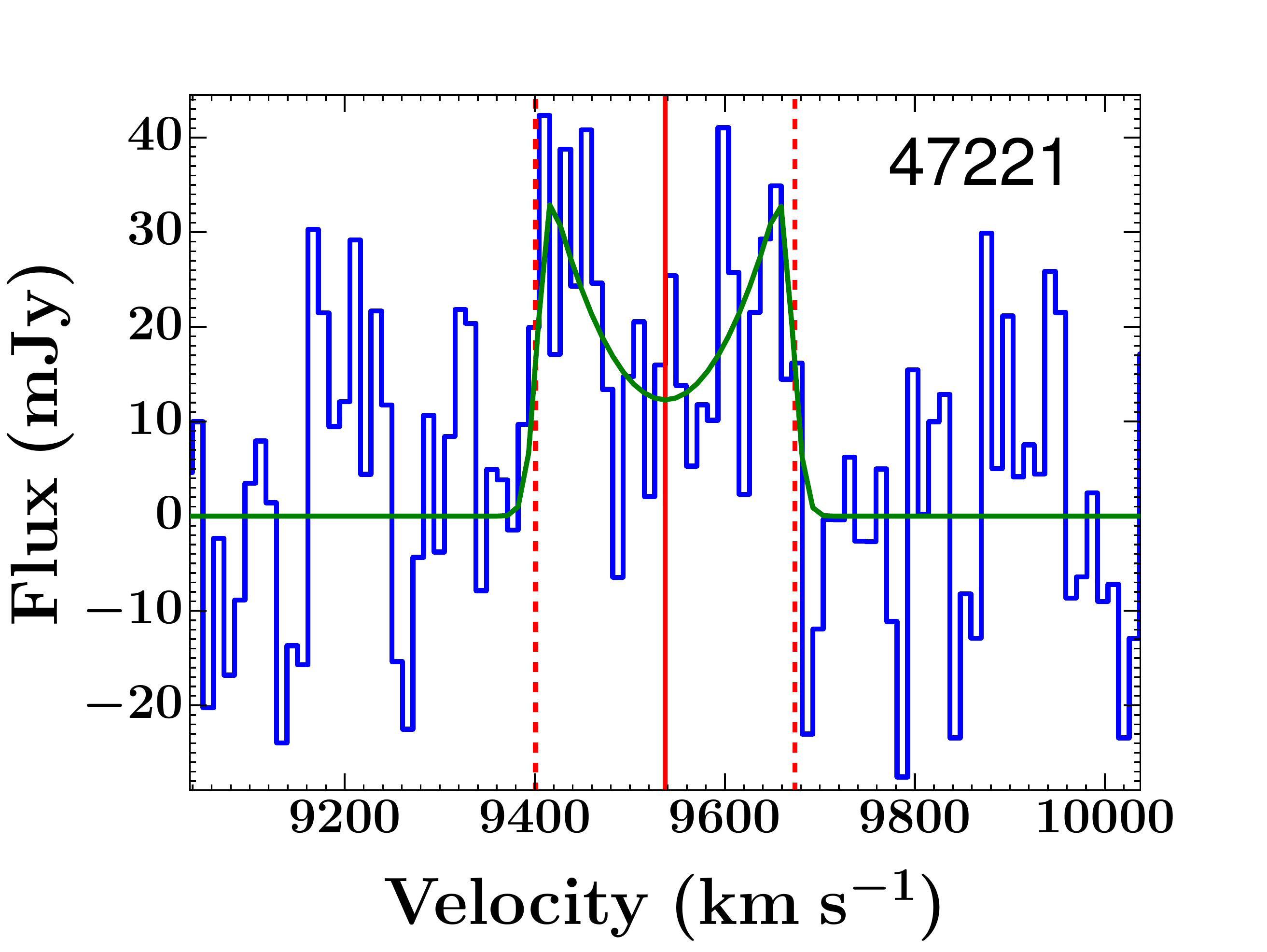}

\includegraphics[scale=0.23, trim= 0 60 50 20, clip=true]{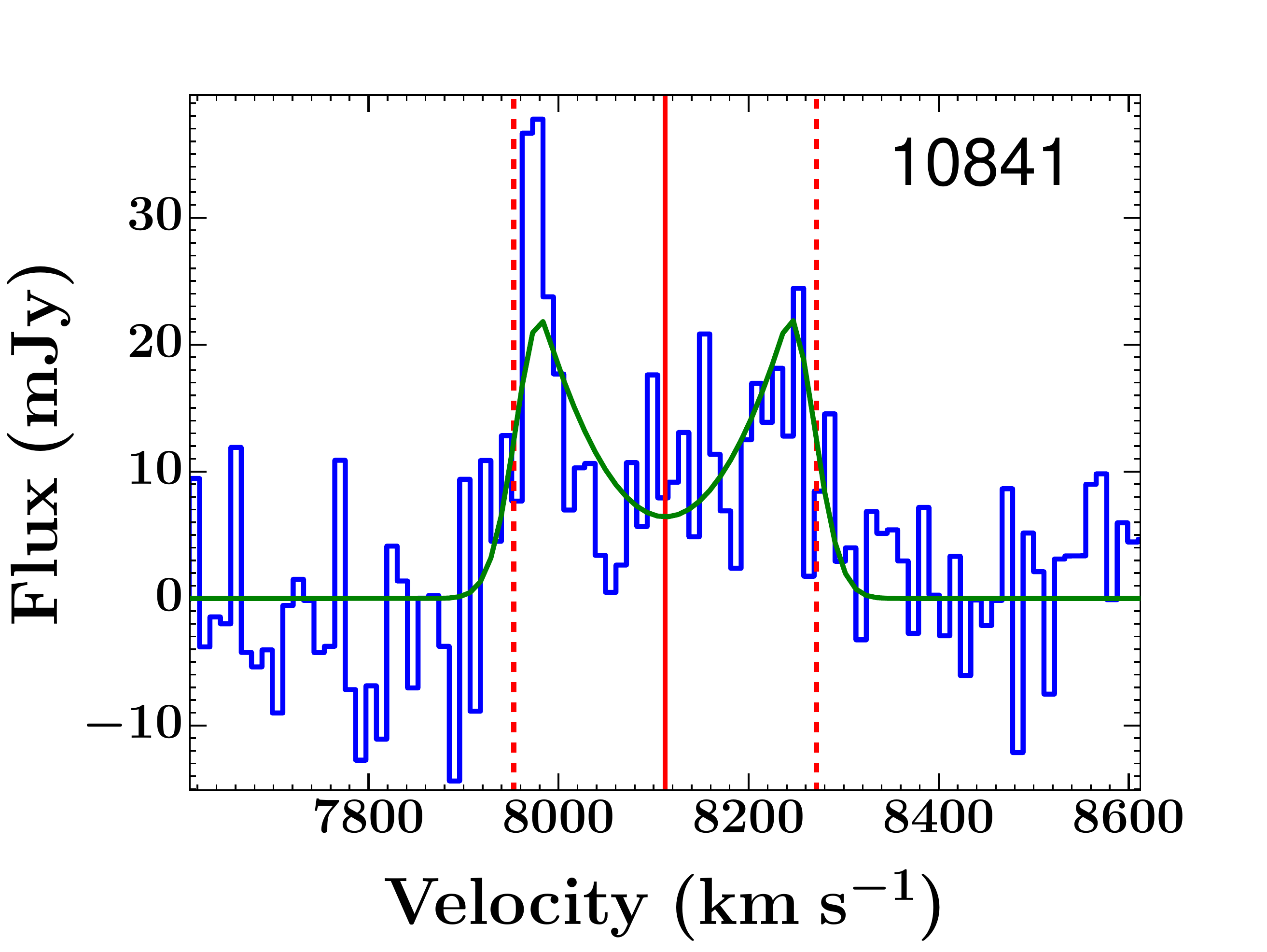}\includegraphics[scale=0.23, trim= 60 60 50 20, clip=true]{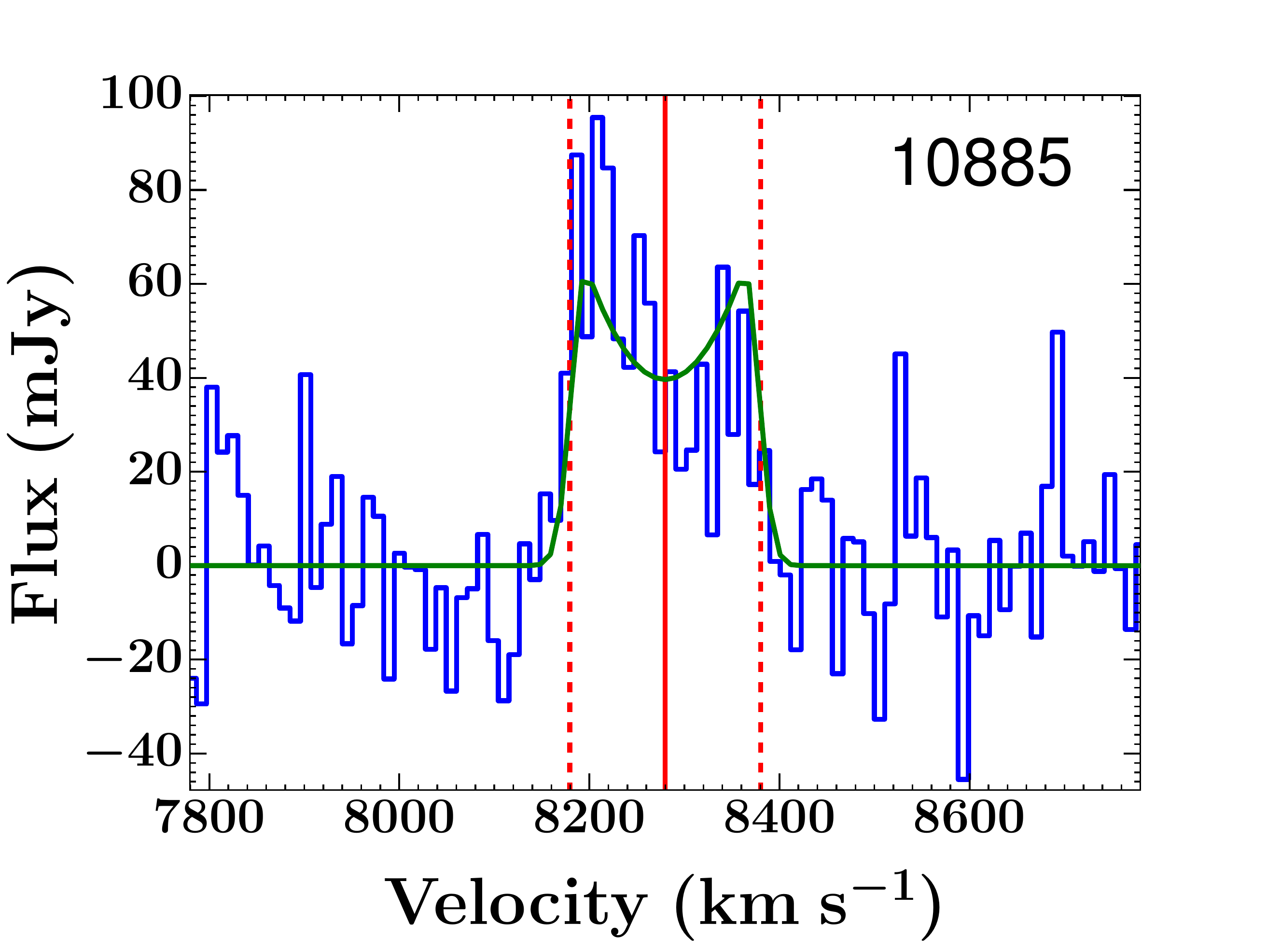}\includegraphics[scale=0.23, trim= 60 60 50 20, clip=true]{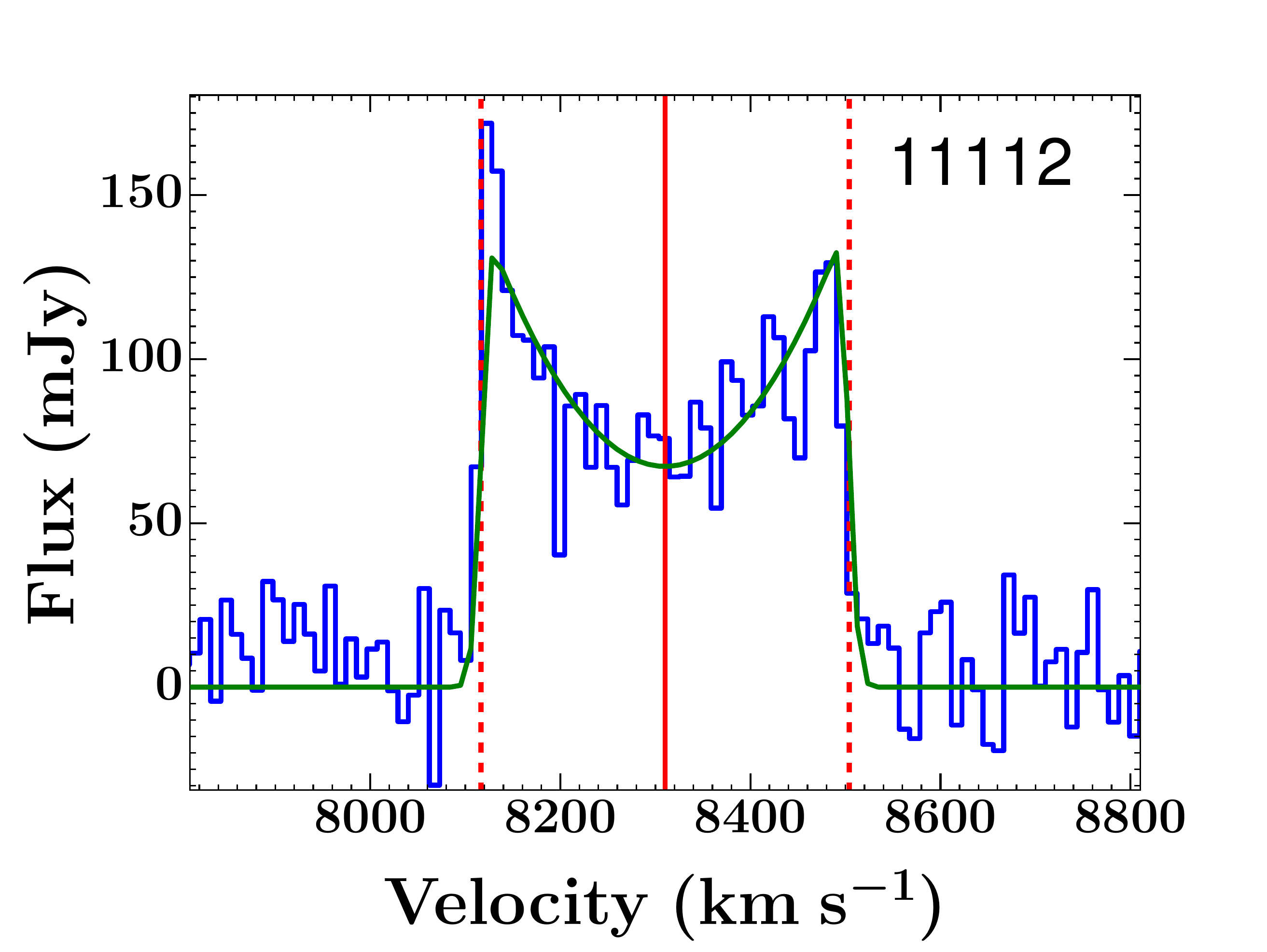}\includegraphics[scale=0.23, trim= 60 60 50 20, clip=true]{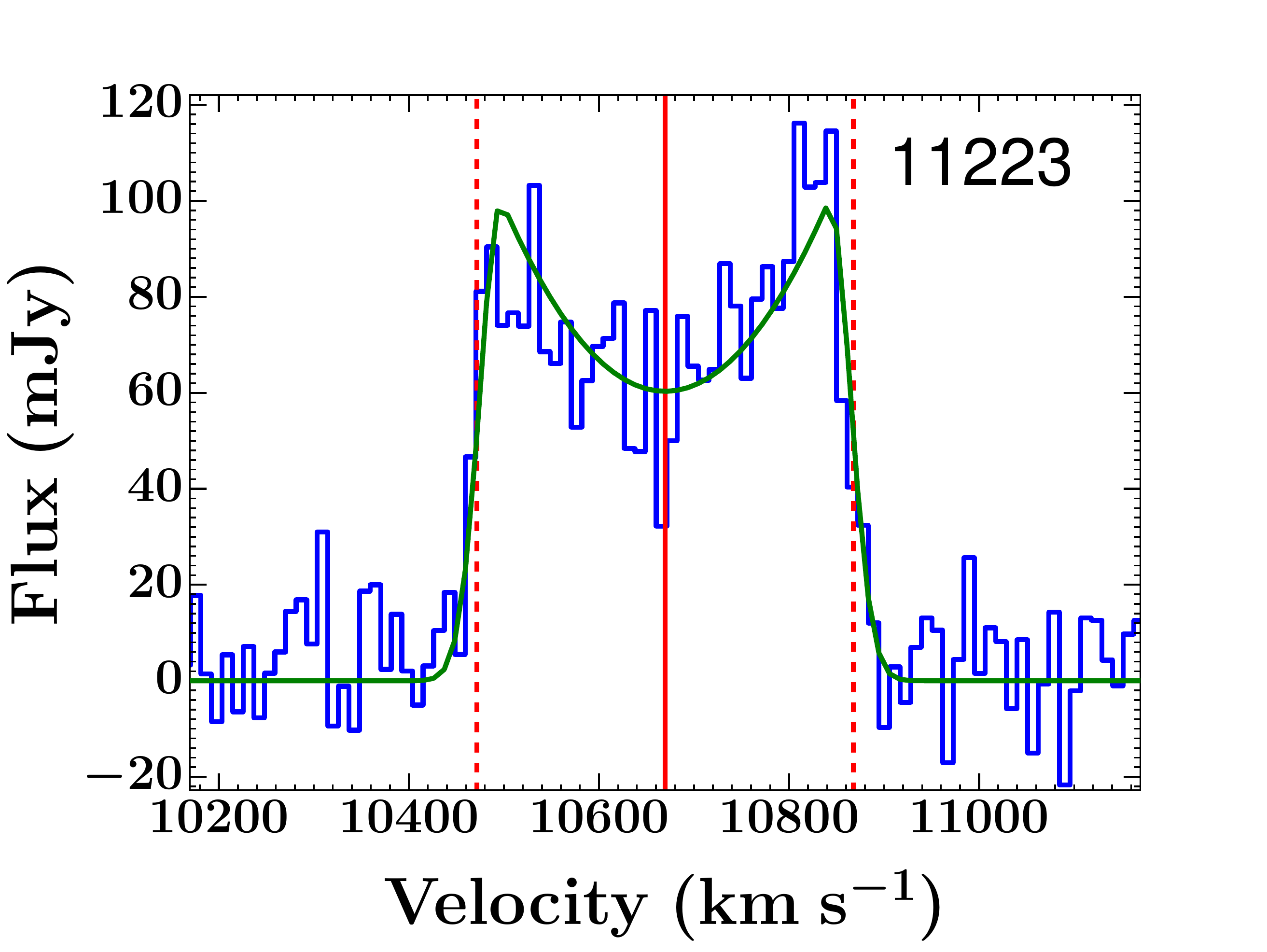}

\includegraphics[scale=0.23, trim= 0 5 50 20, clip=true]{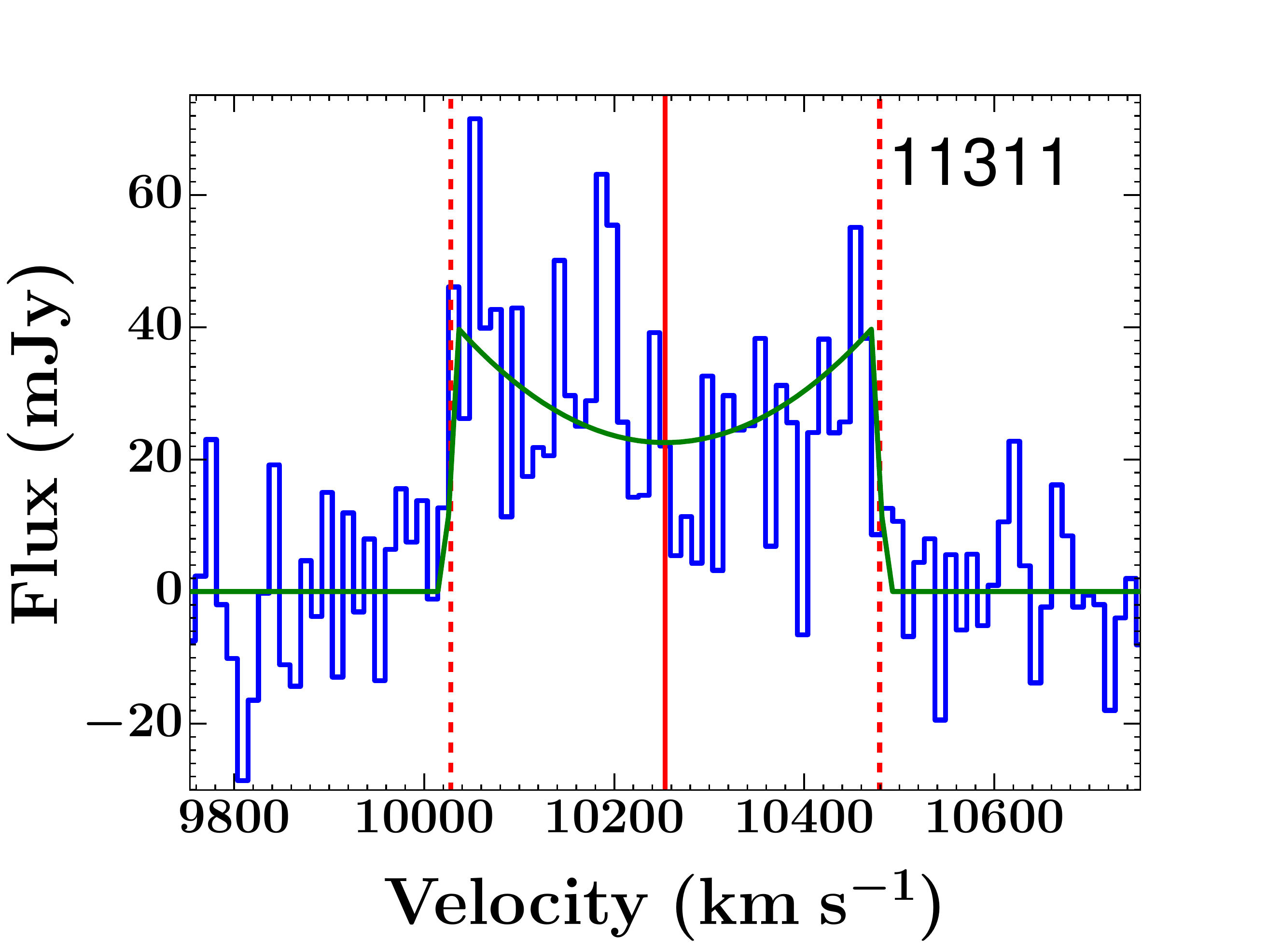}\includegraphics[scale=0.23, trim= 60 5 50 20, clip=true]{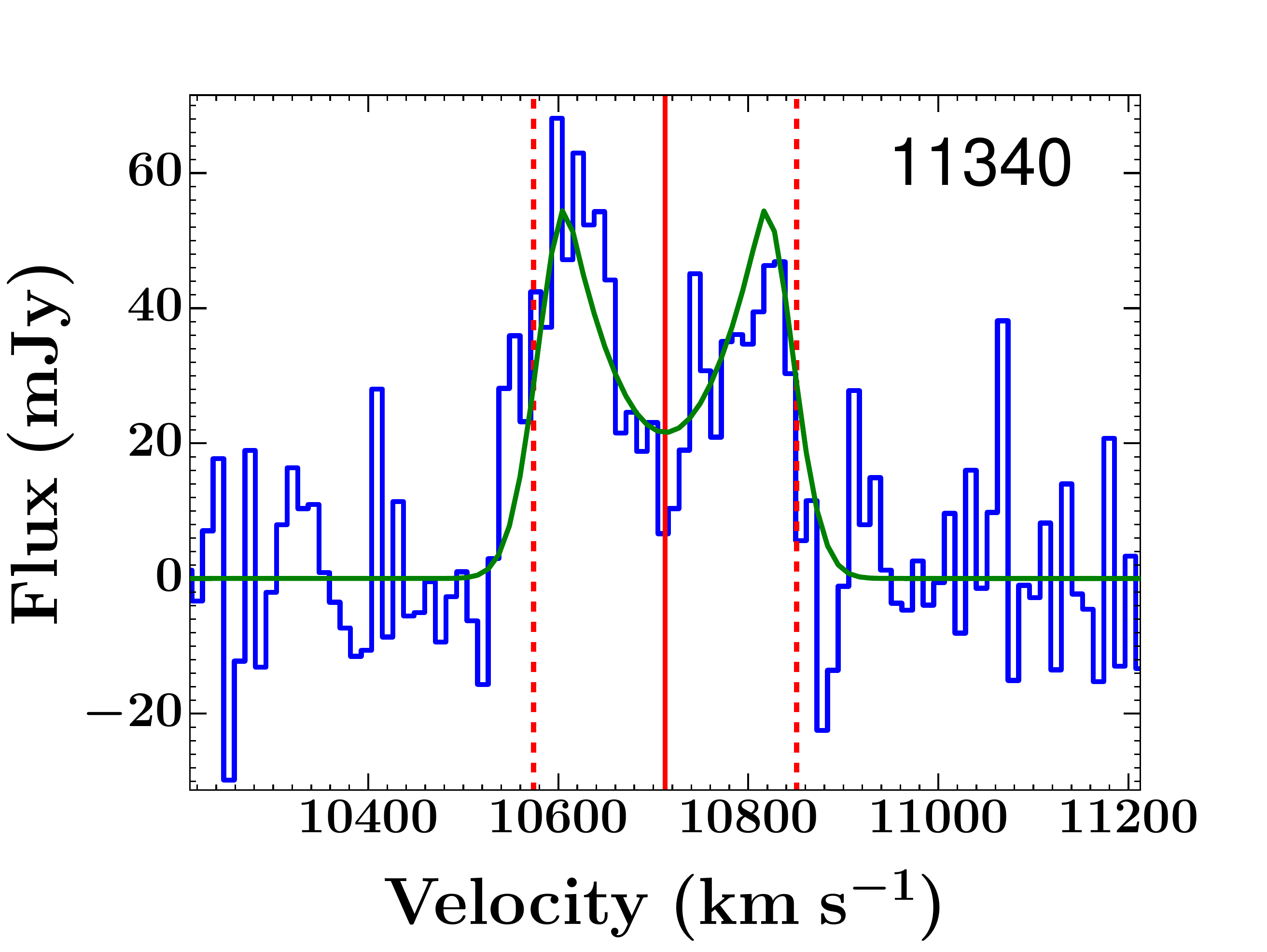}\includegraphics[scale=0.23, trim= 60 5 50 20, clip=true]{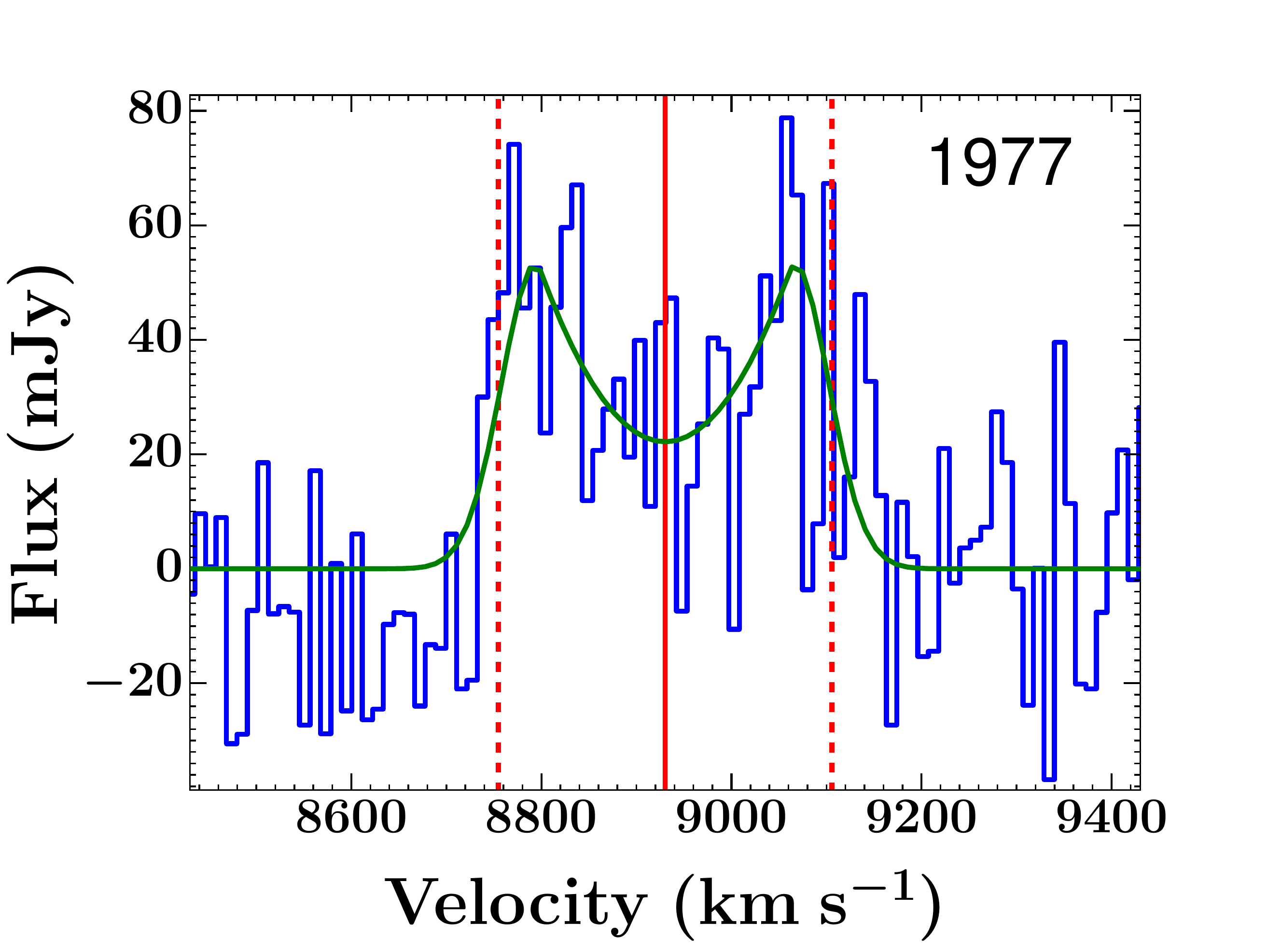}\includegraphics[scale=0.23, trim= 60 5 50 20, clip=true]{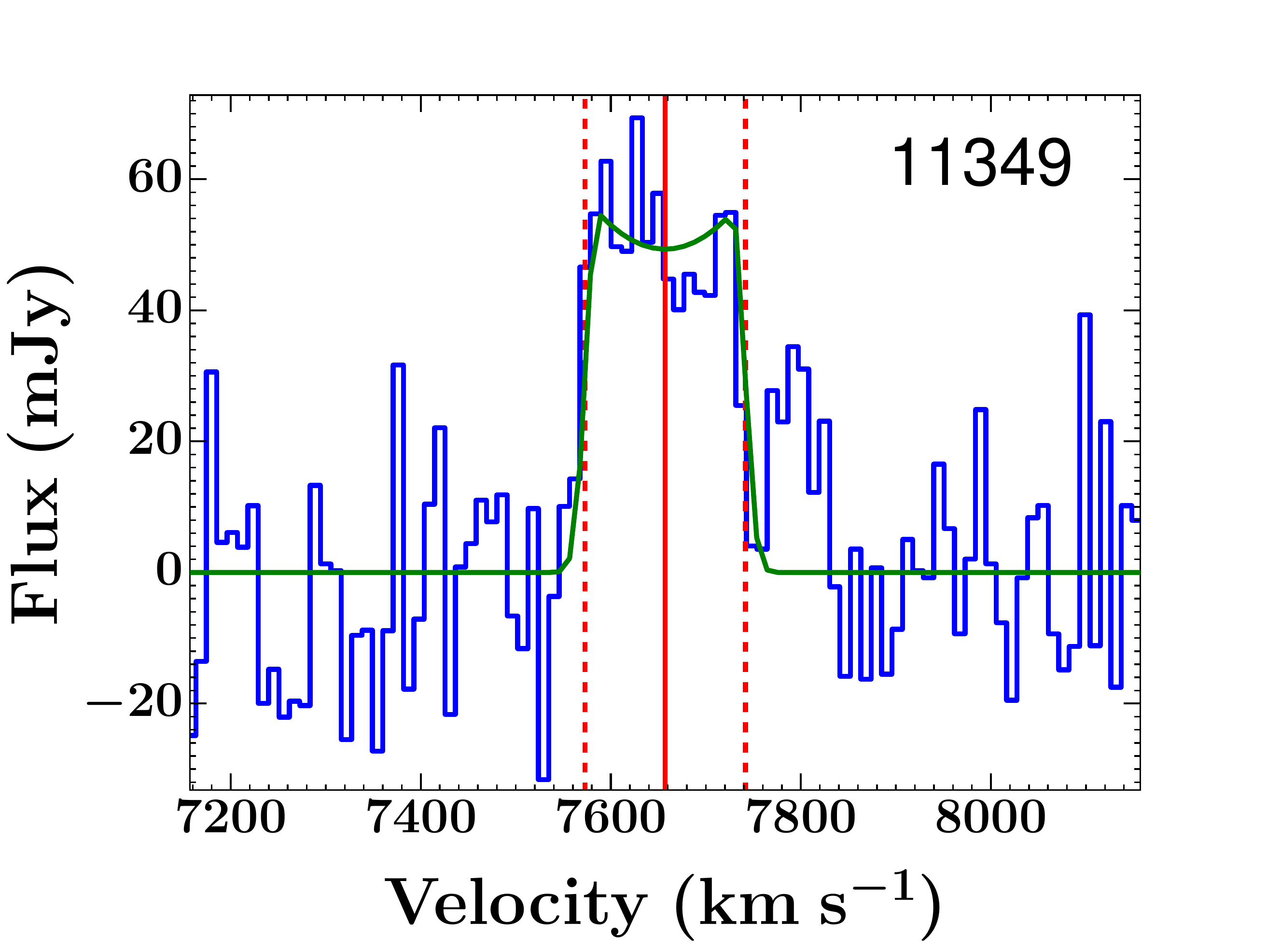}
\end{landscape}
\begin{landscape}
\includegraphics[scale=0.23, trim= 0 60 50 20, clip=true]{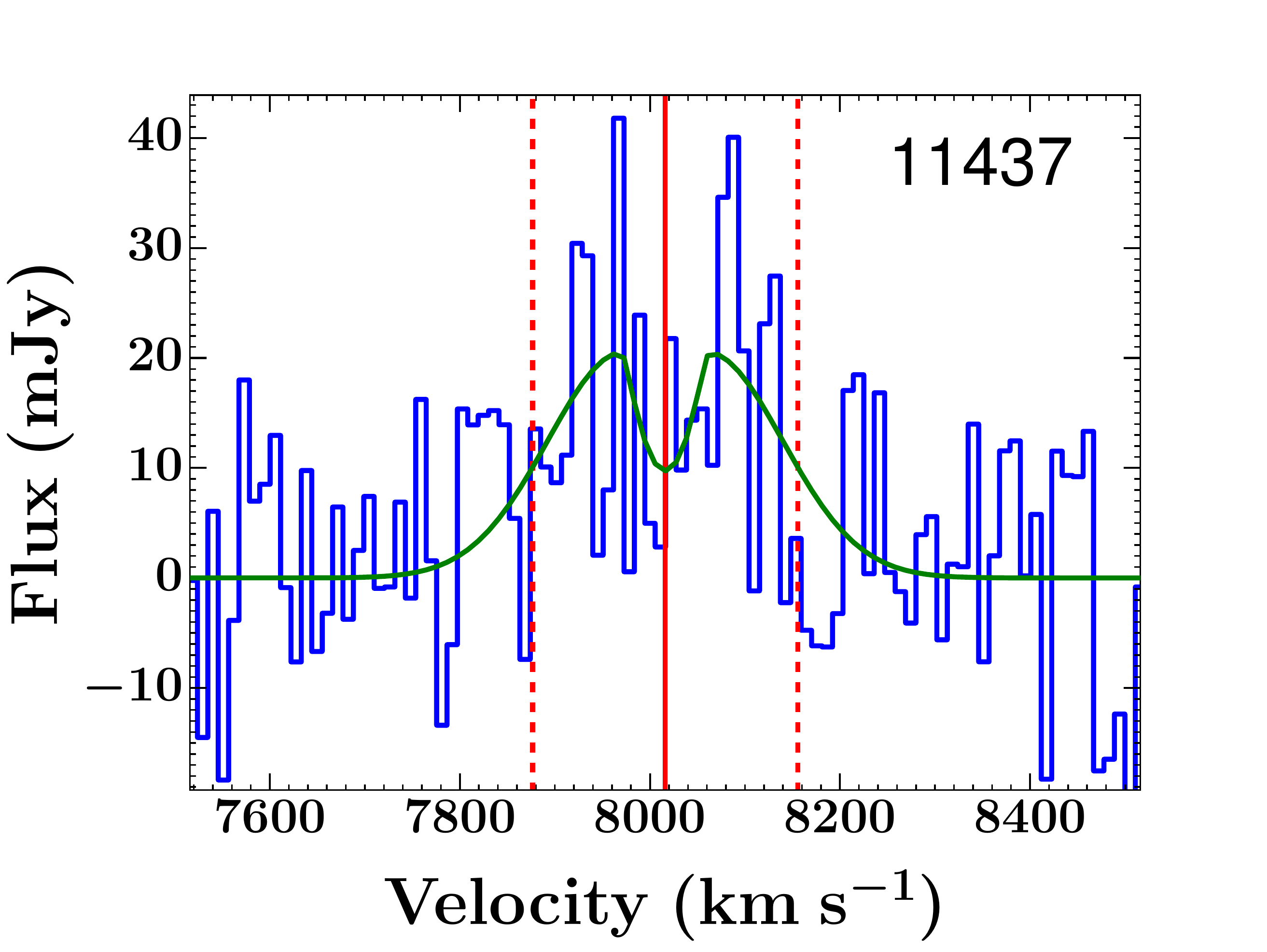}\includegraphics[scale=0.23, trim= 60 60 50 20, clip=true]{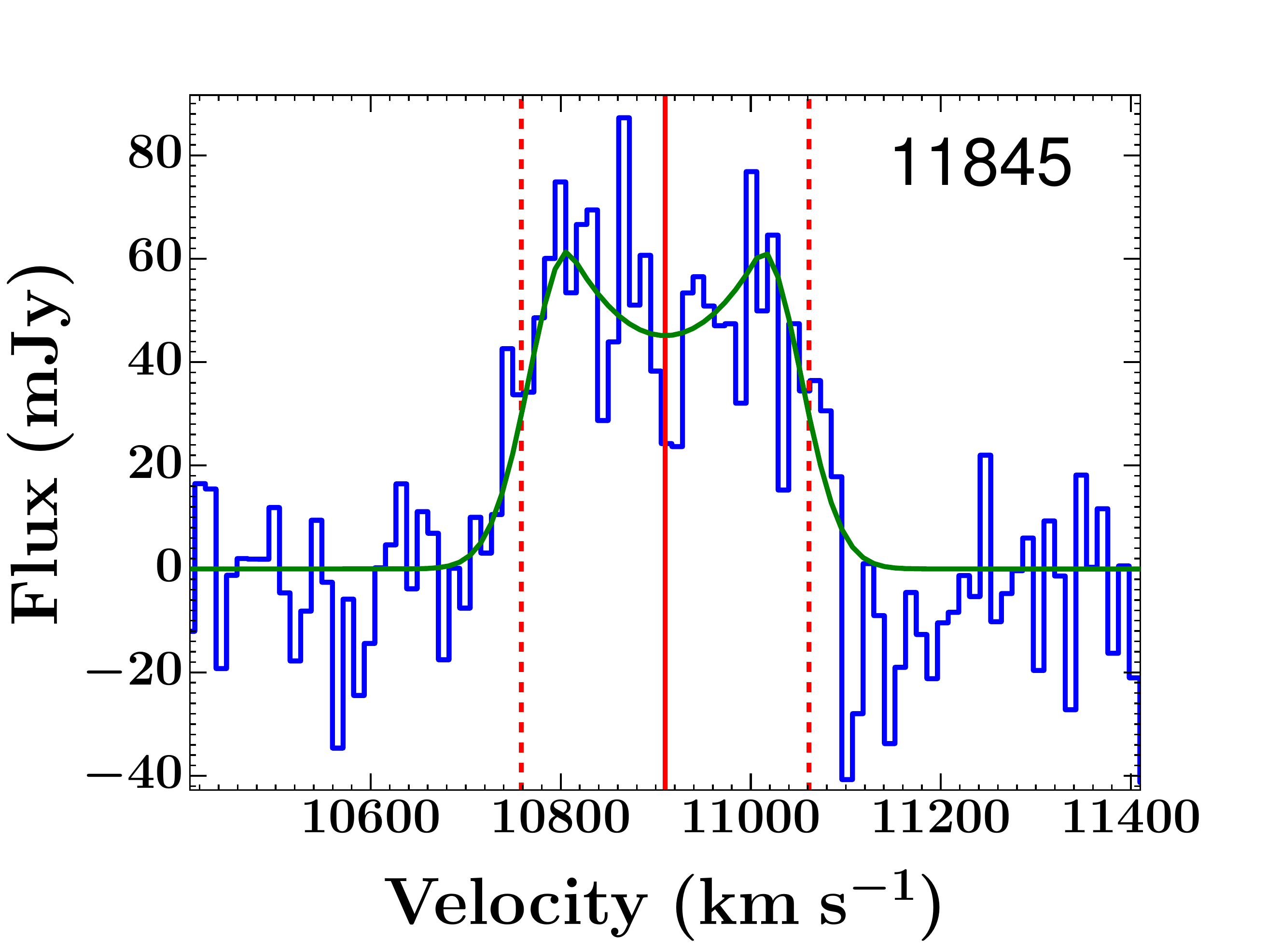}\includegraphics[scale=0.23, trim= 60 60 50 20, clip=true]{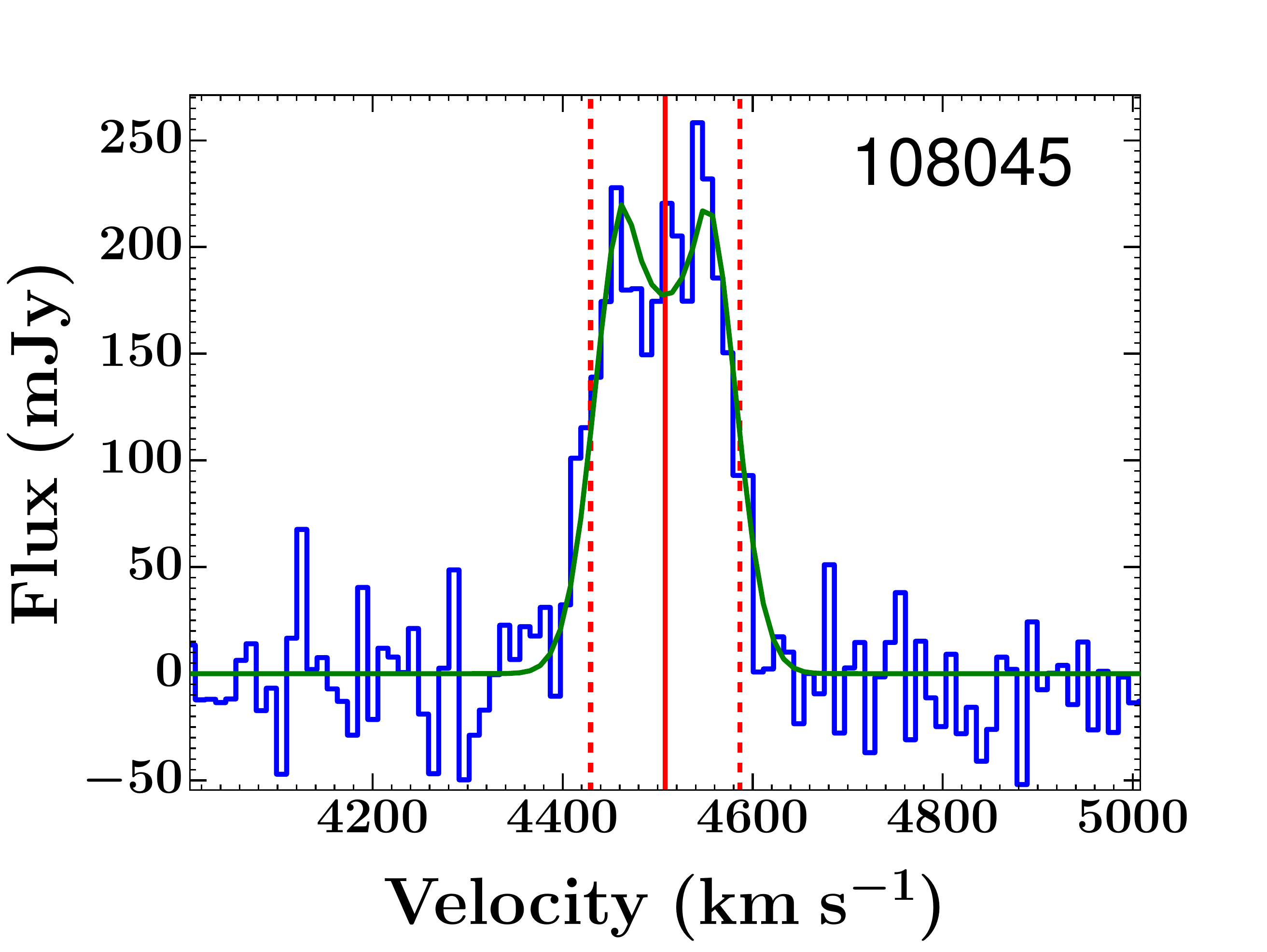}\includegraphics[scale=0.23, trim= 60 60 50 20, clip=true]{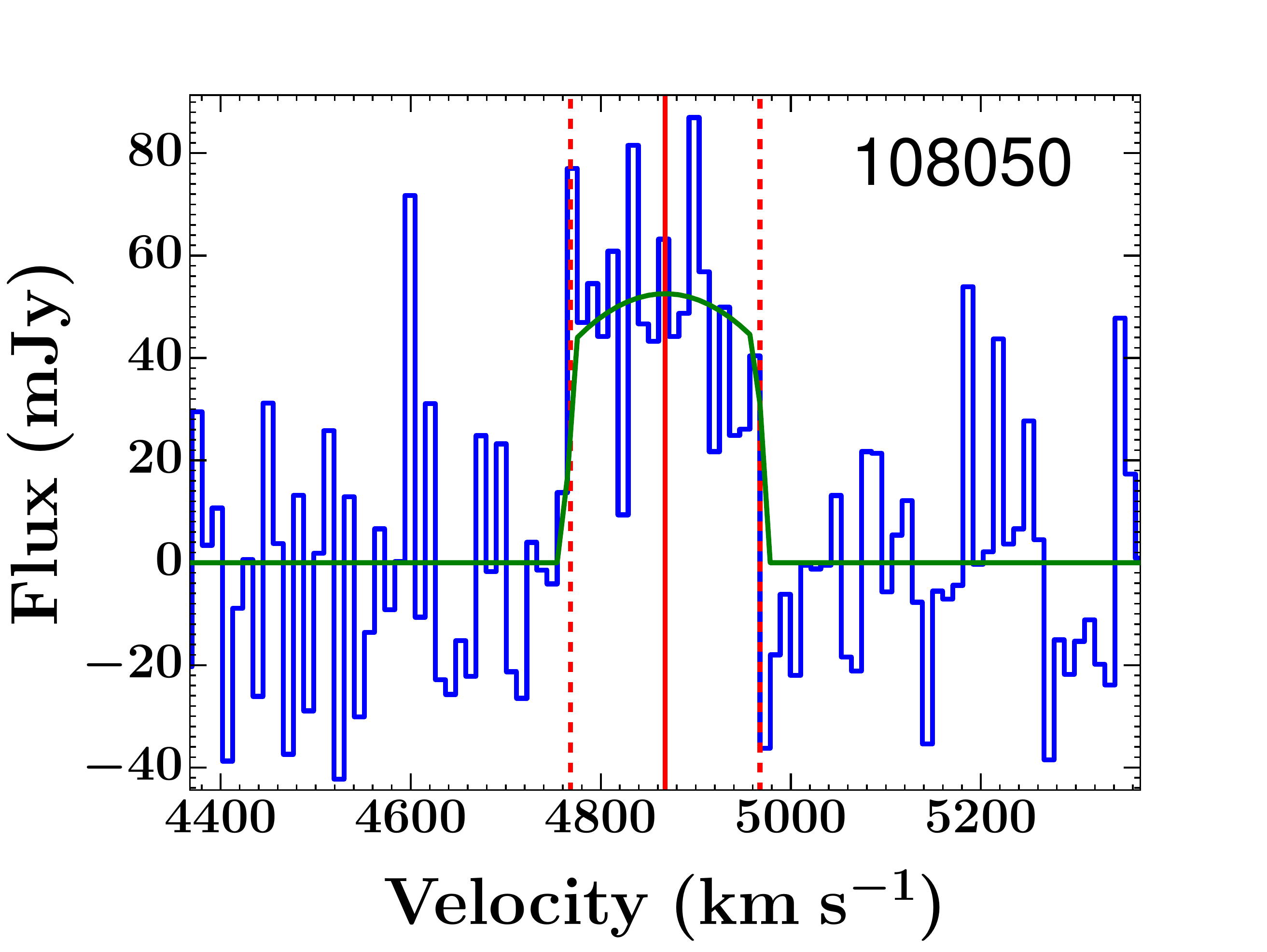}

\includegraphics[scale=0.23, trim= 0 60 50 20, clip=true]{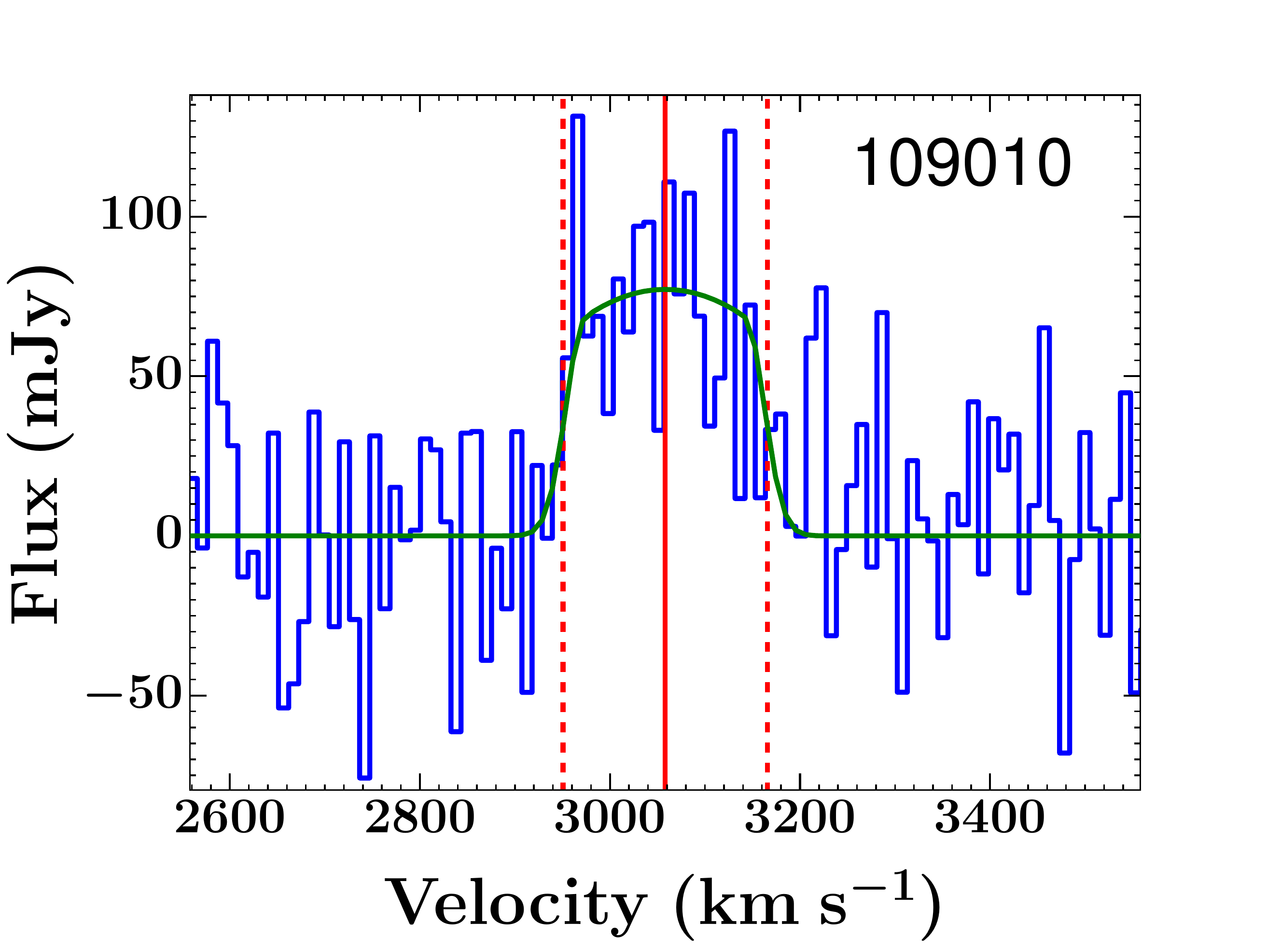}\includegraphics[scale=0.23, trim= 60 60 50 20, clip=true]{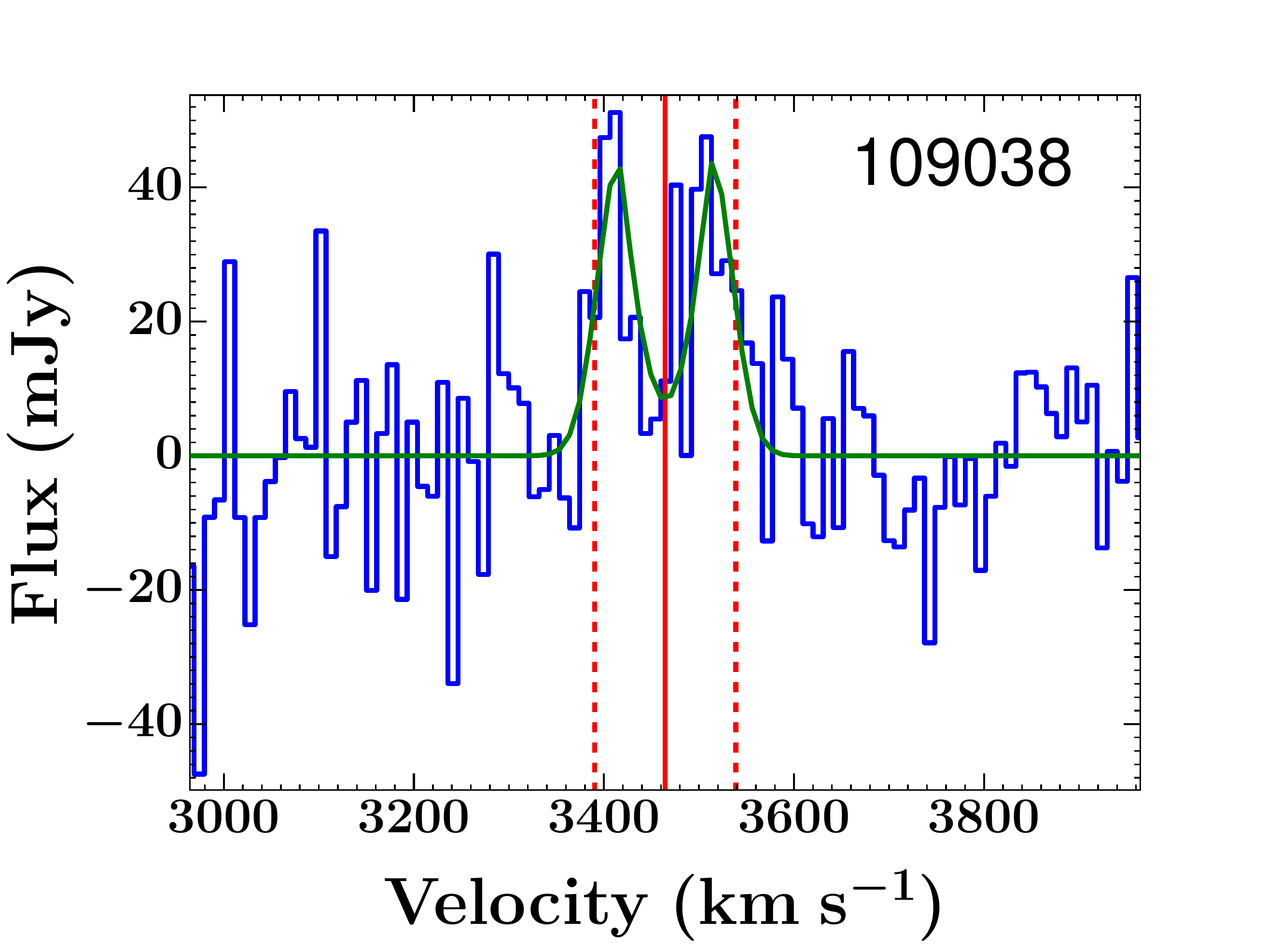}\includegraphics[scale=0.23, trim= 60 60 50 20, clip=true]{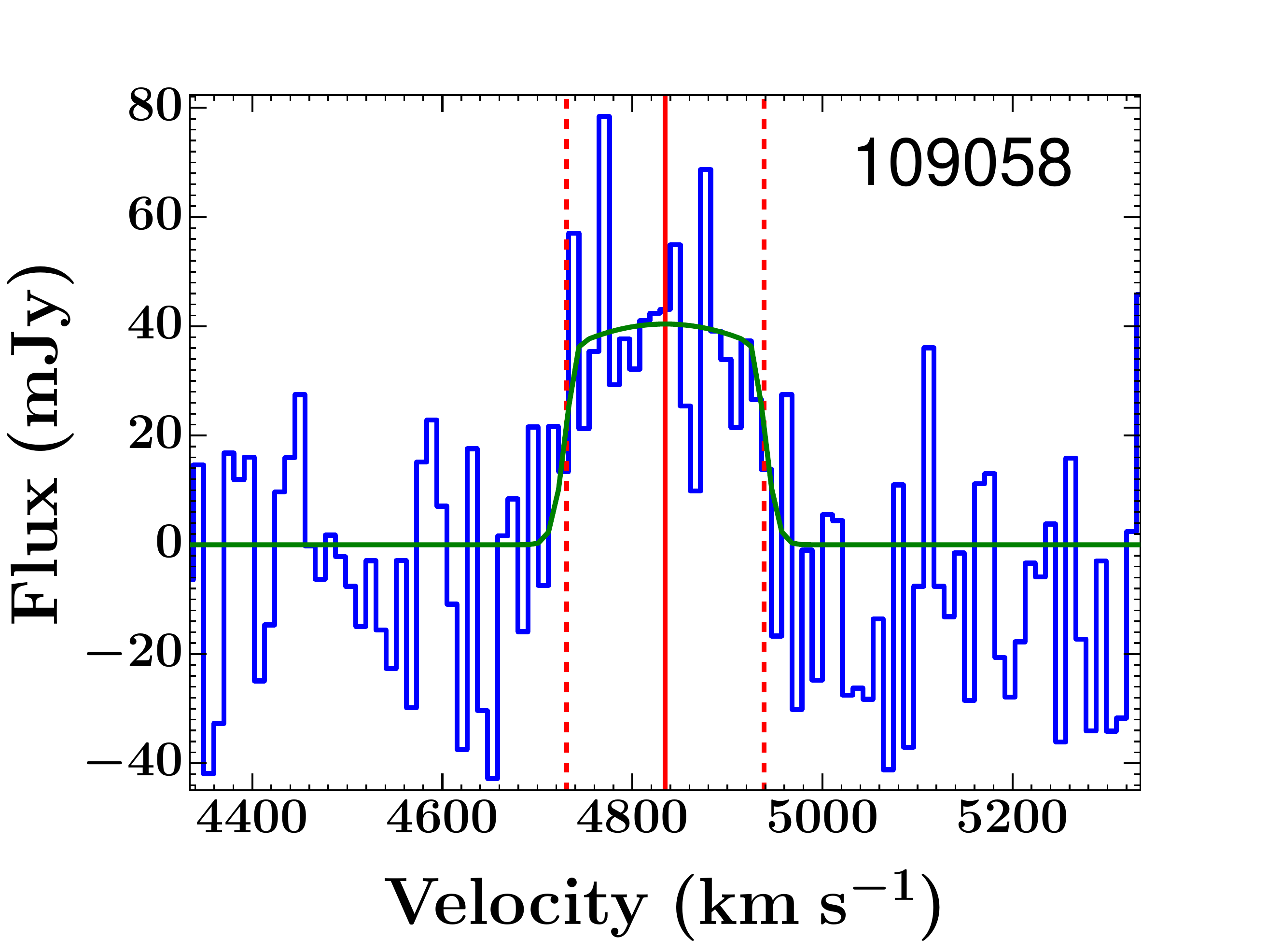}\includegraphics[scale=0.23, trim= 60 60 50 20, clip=true]{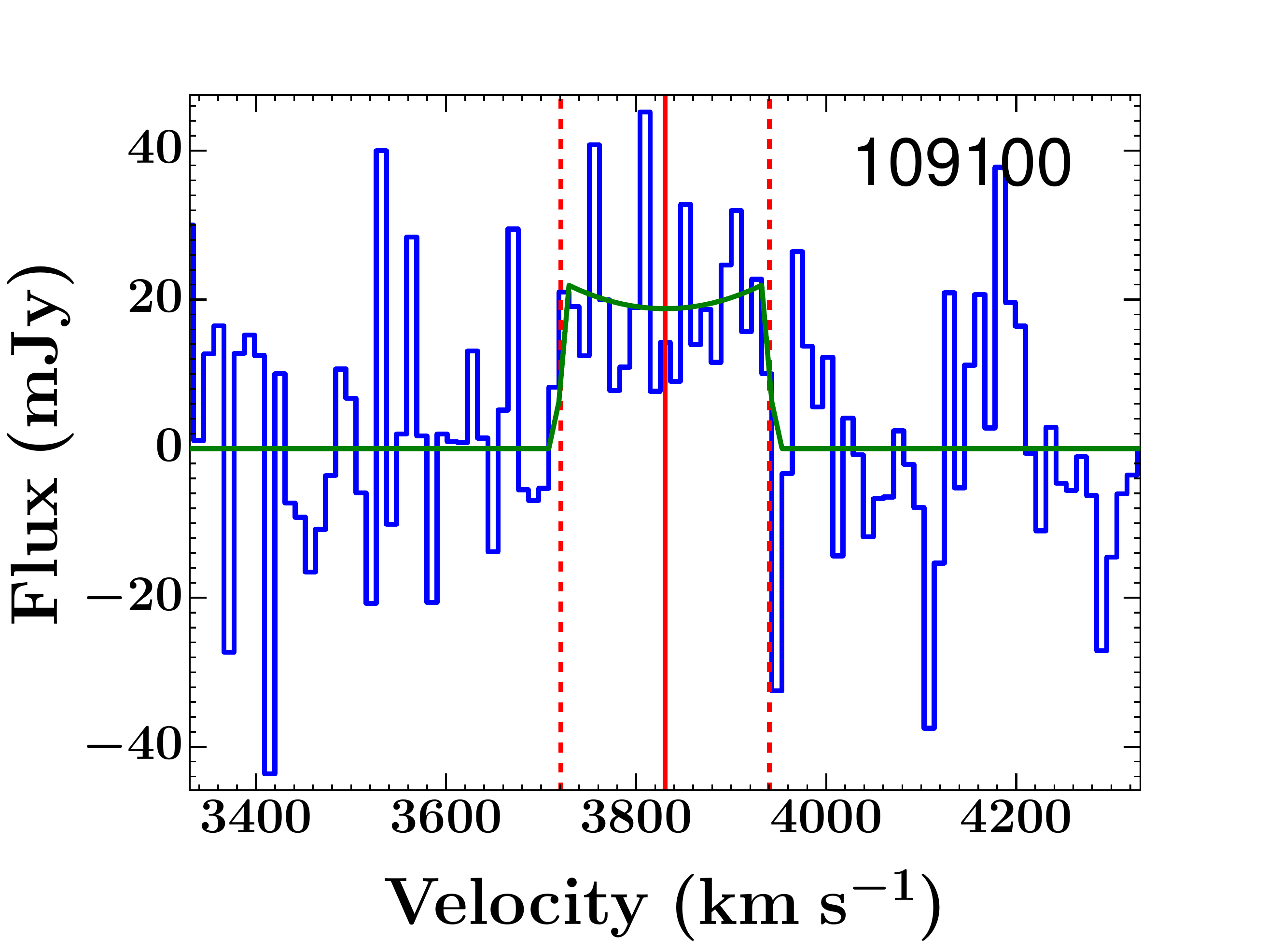}

\includegraphics[scale=0.23, trim= 0 60 50 20, clip=true]{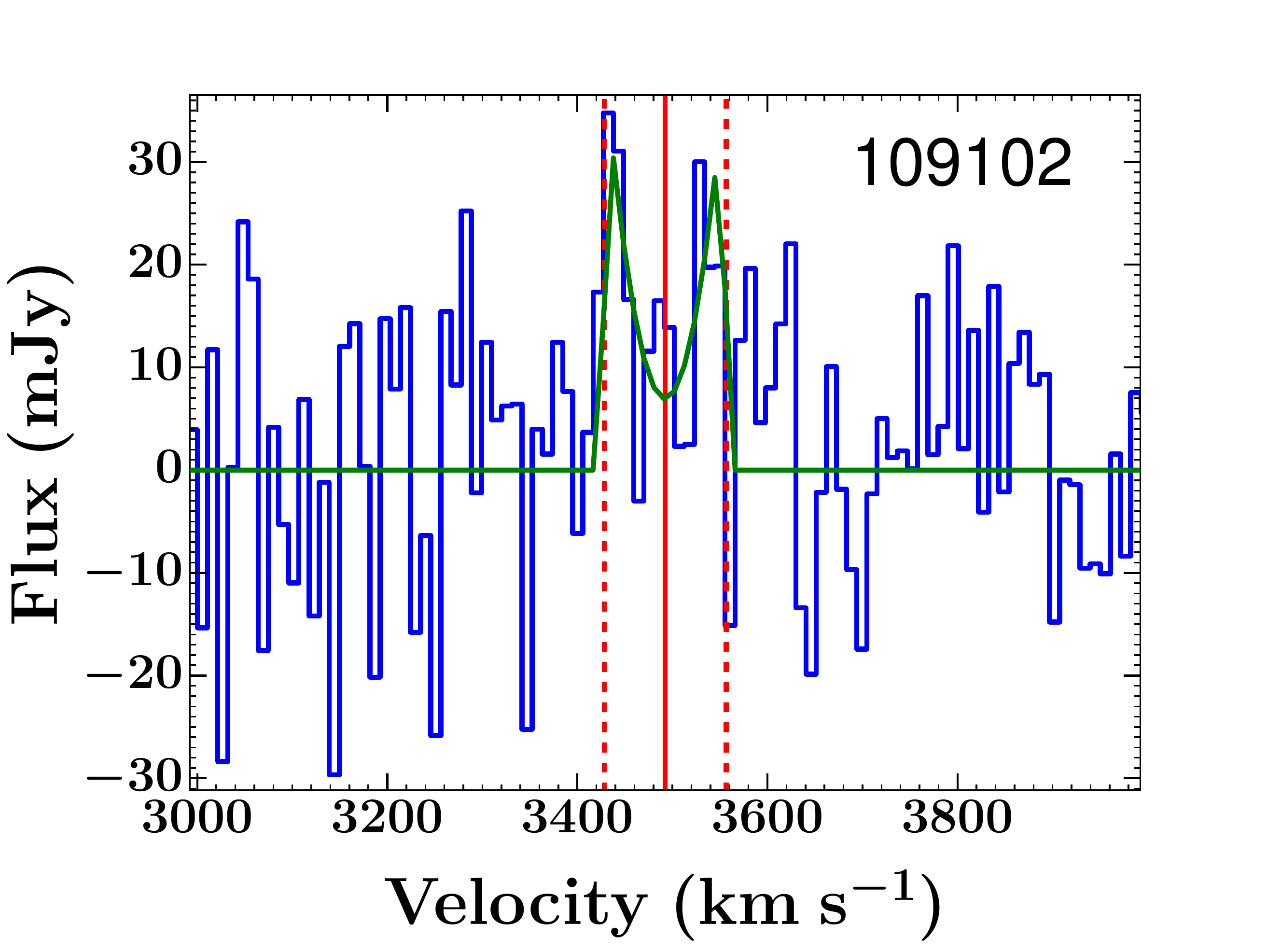}\includegraphics[scale=0.23, trim= 60 60 50 20, clip=true]{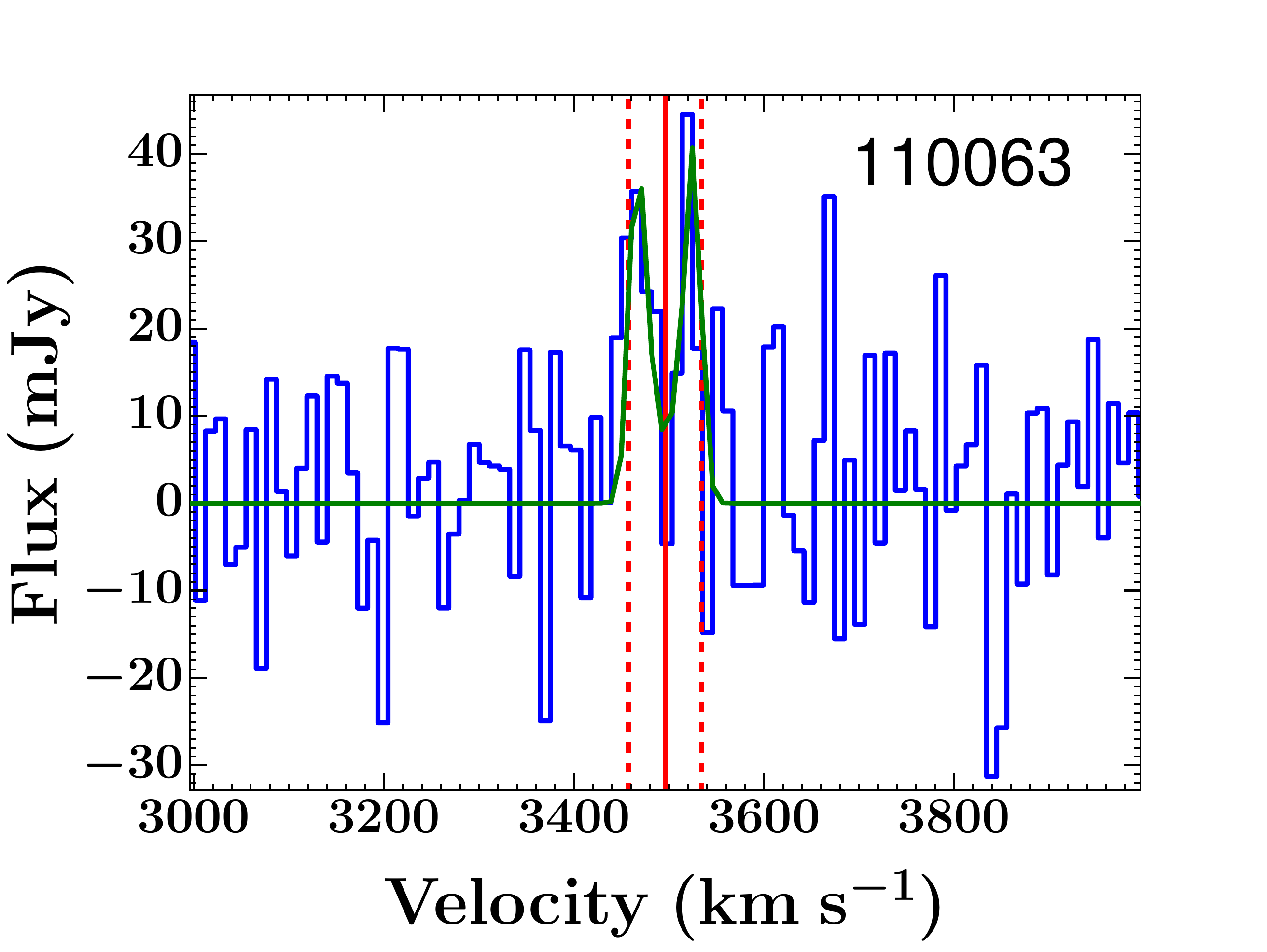}\includegraphics[scale=0.23, trim= 60 60 50 20, clip=true]{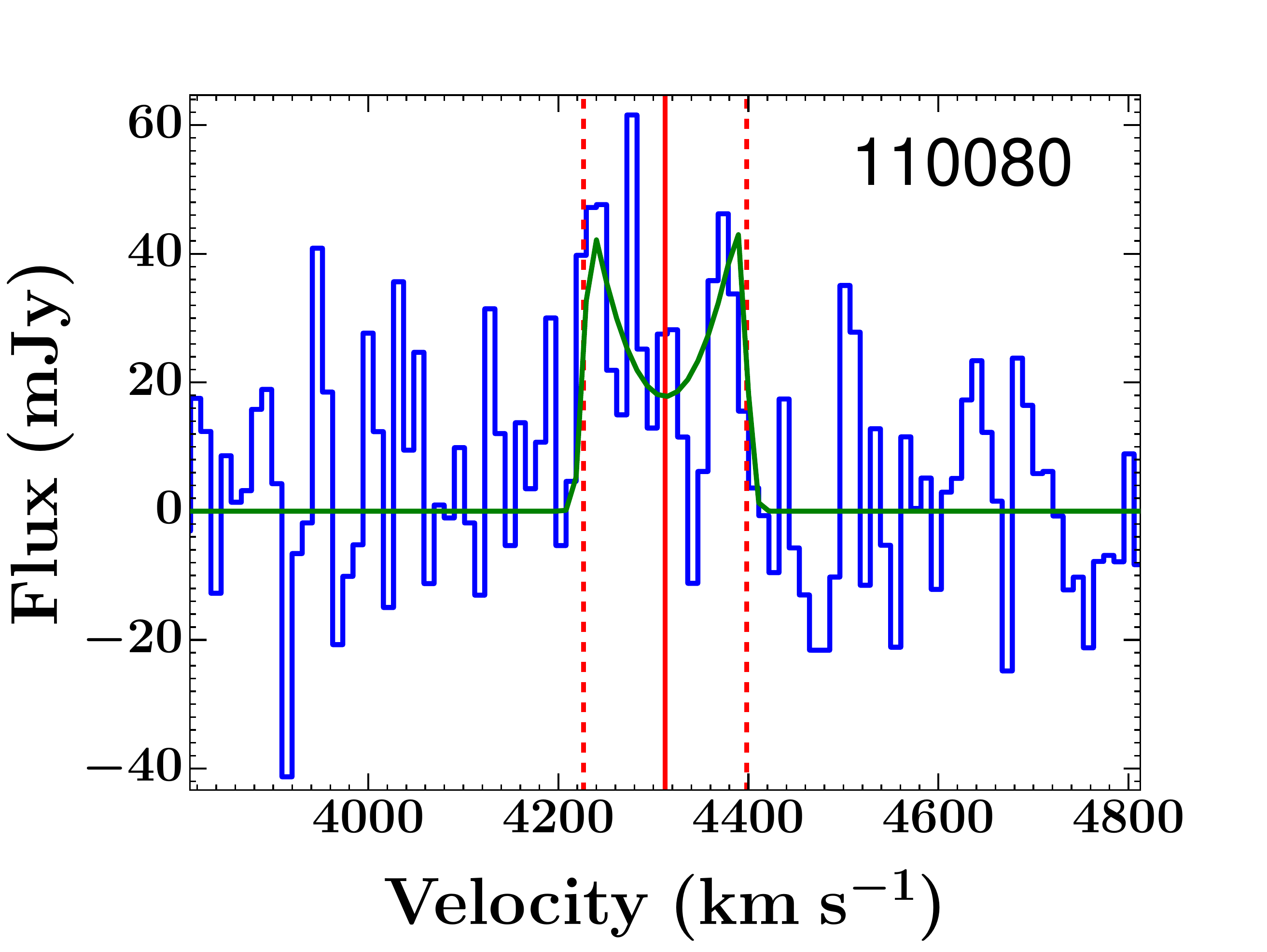}\includegraphics[scale=0.23, trim= 60 60 50 20, clip=true]{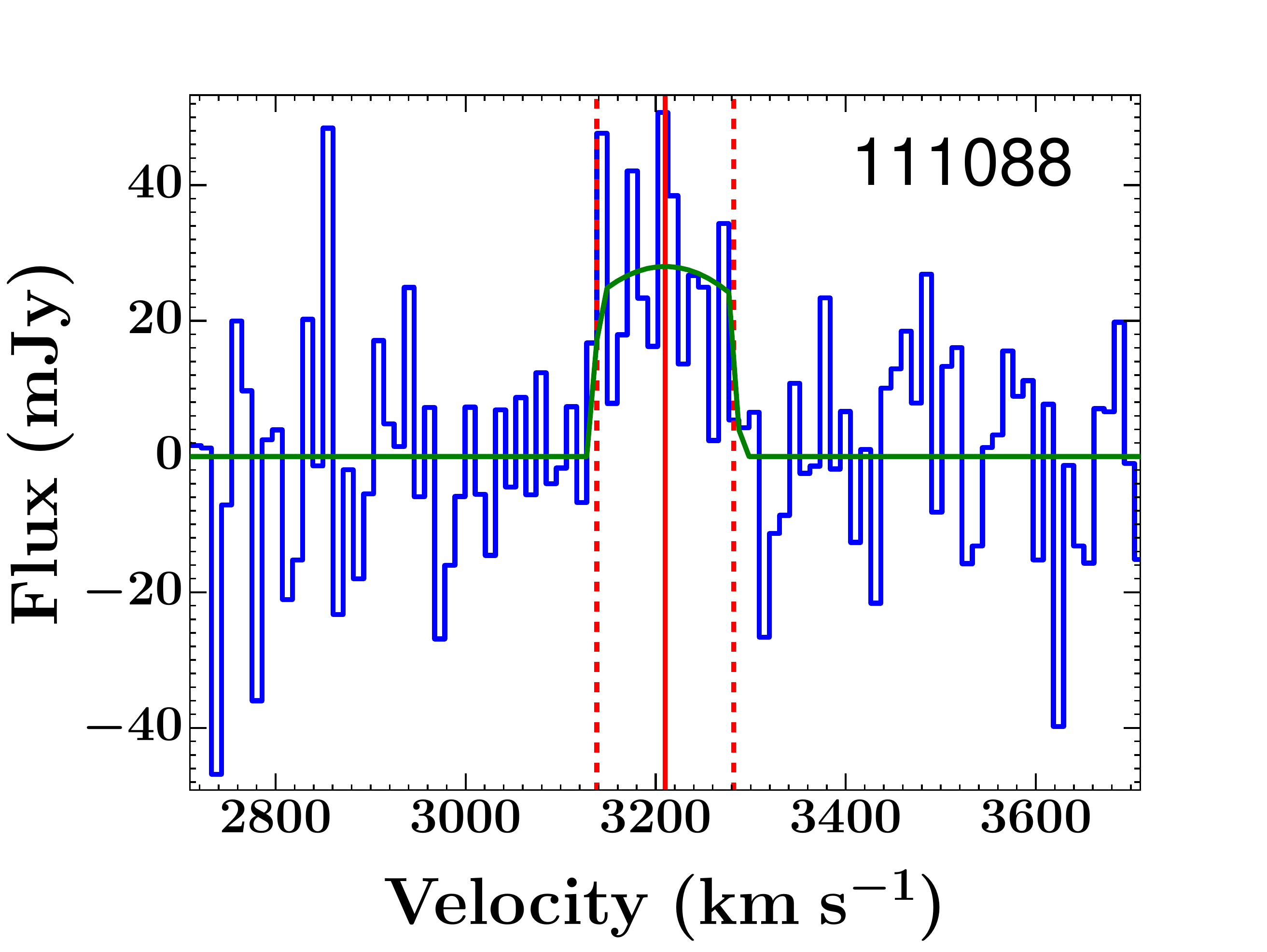}

\includegraphics[scale=0.23, trim= 0 5 50 20, clip=true]{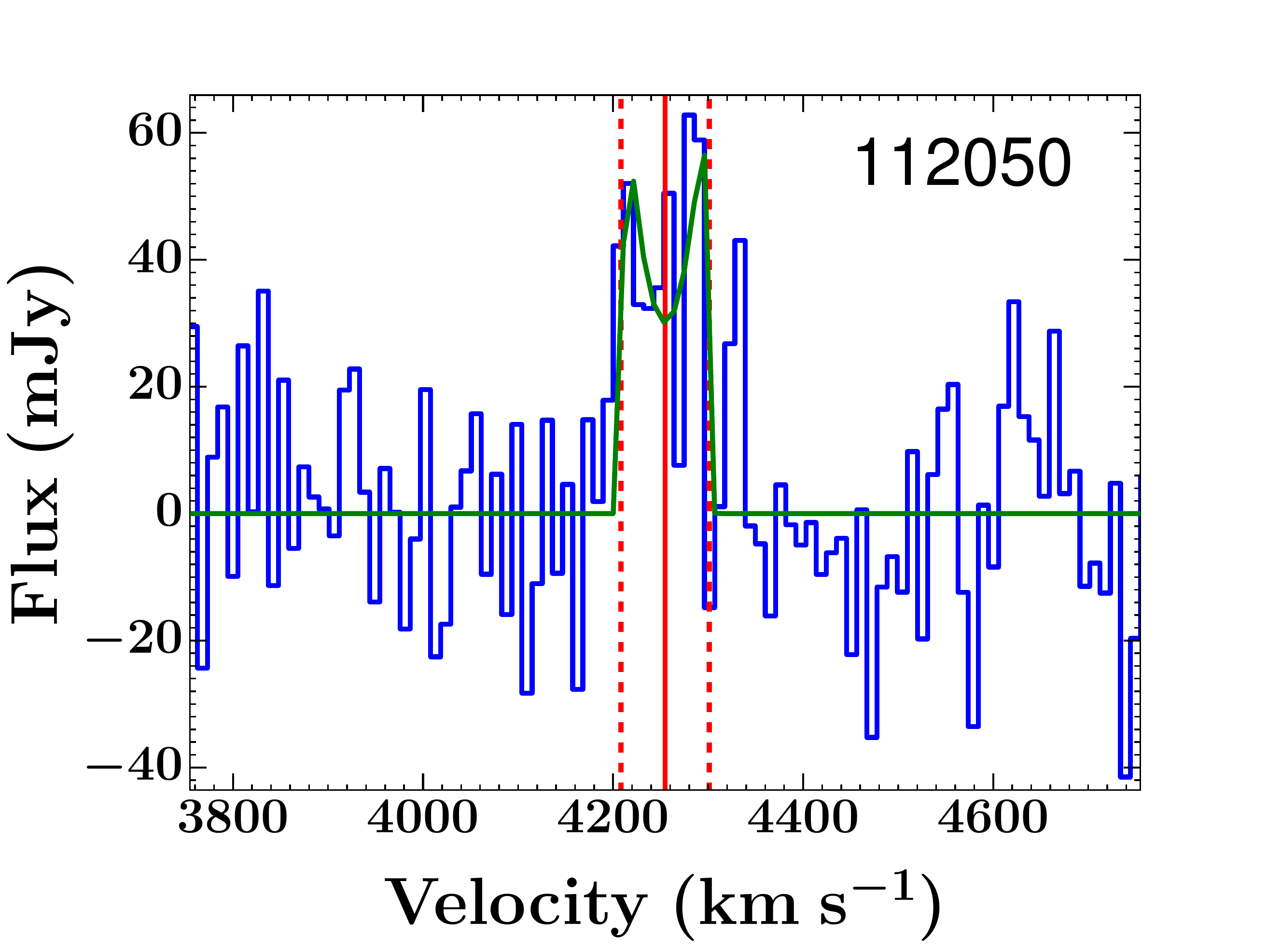}\includegraphics[scale=0.23, trim= 60 5 50 20, clip=true]{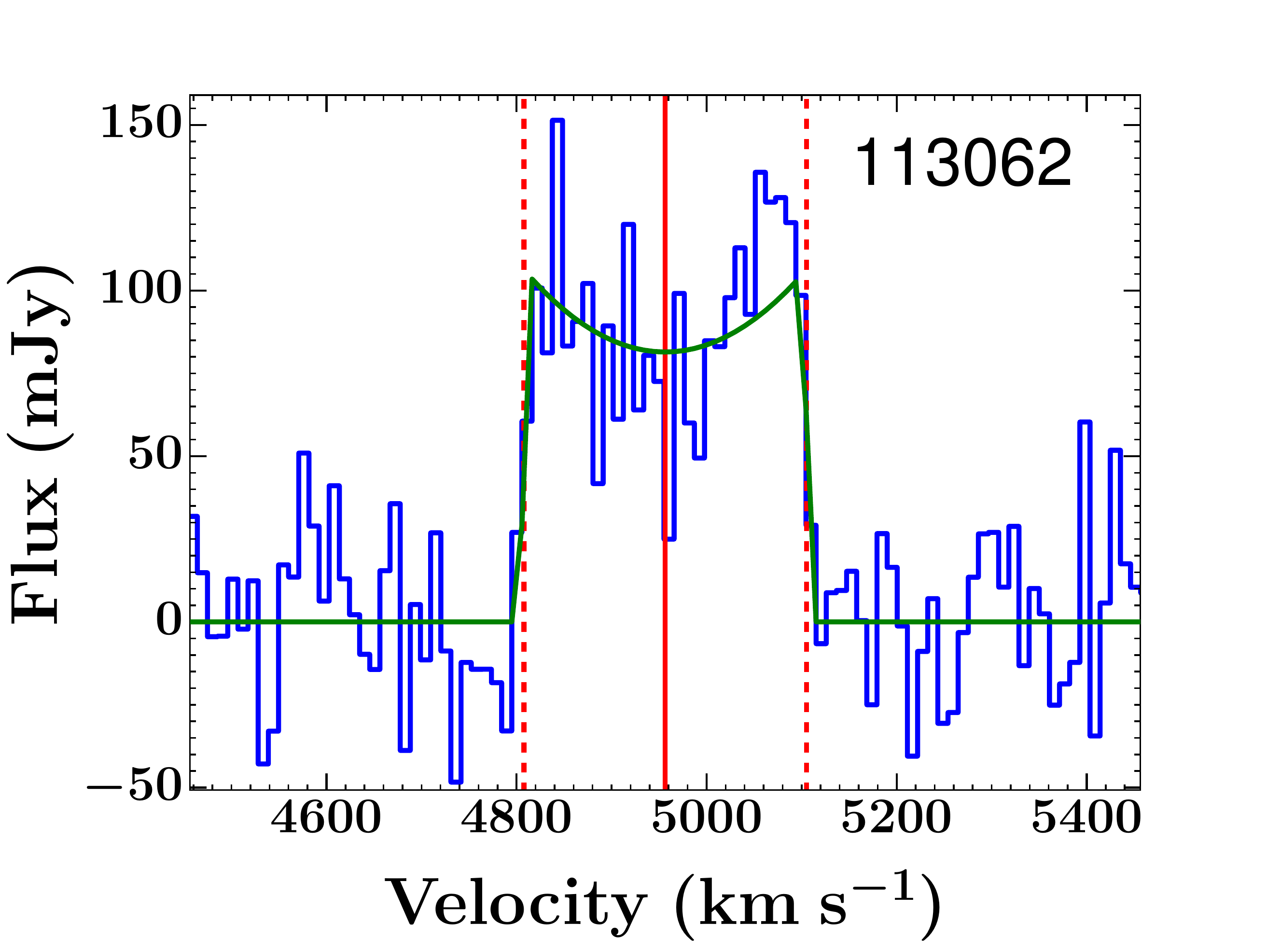}\includegraphics[scale=0.23, trim= 60 5 50 20, clip=true]{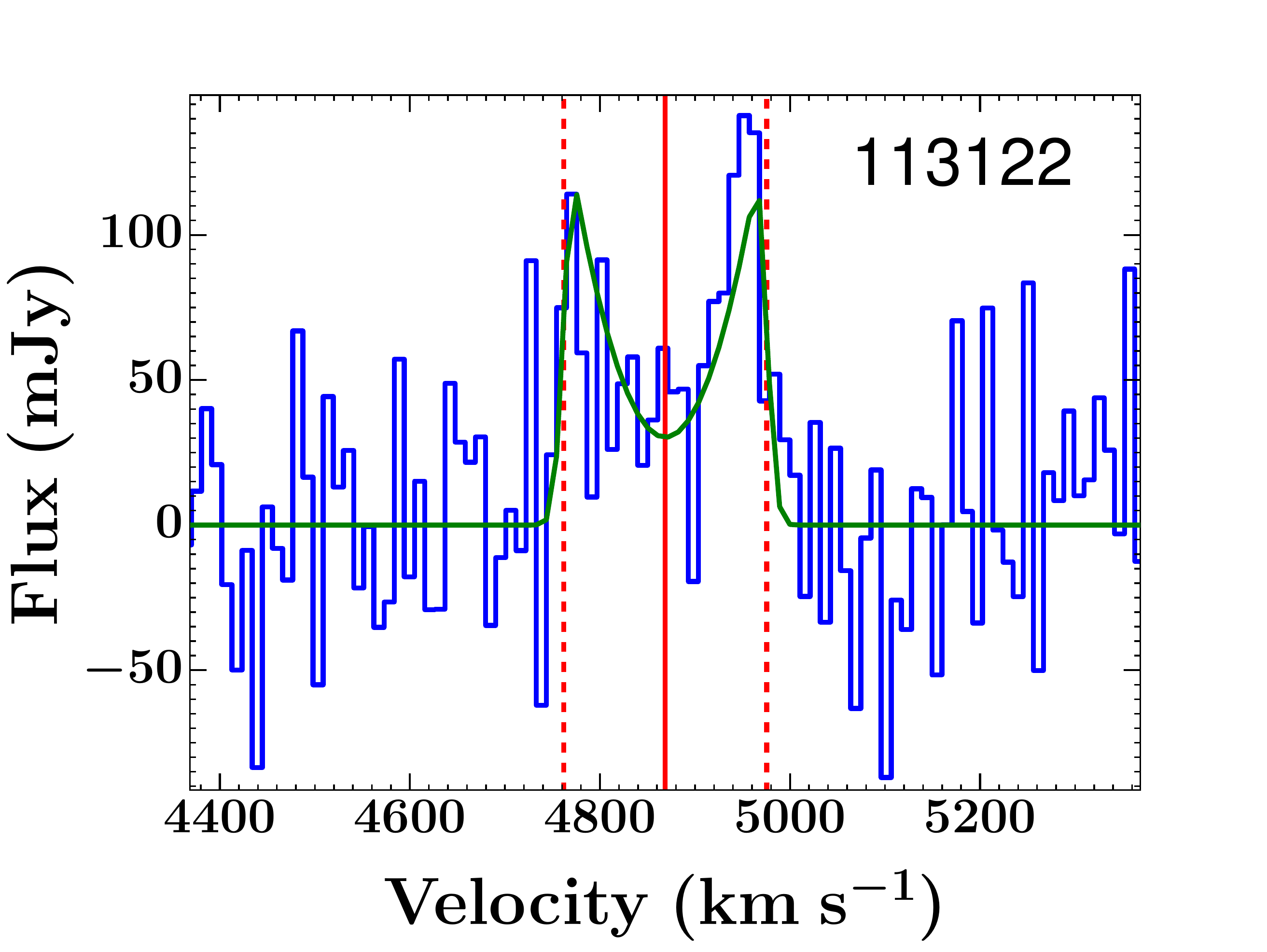}\includegraphics[scale=0.23, trim= 60 5 50 20, clip=true]{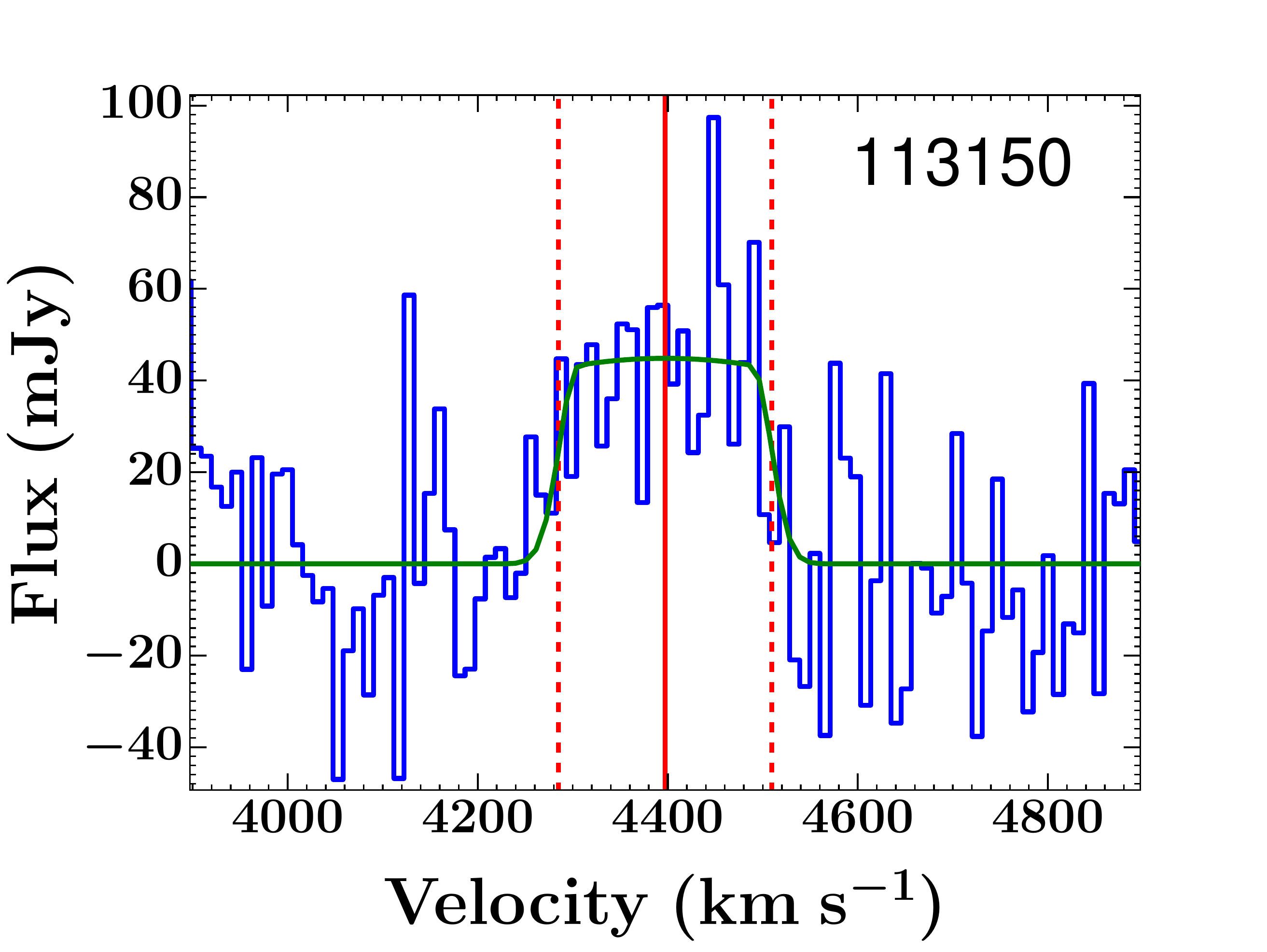}
\end{landscape}
\begin{landscape}
\includegraphics[scale=0.23, trim= 0 5 50 20, clip=true]{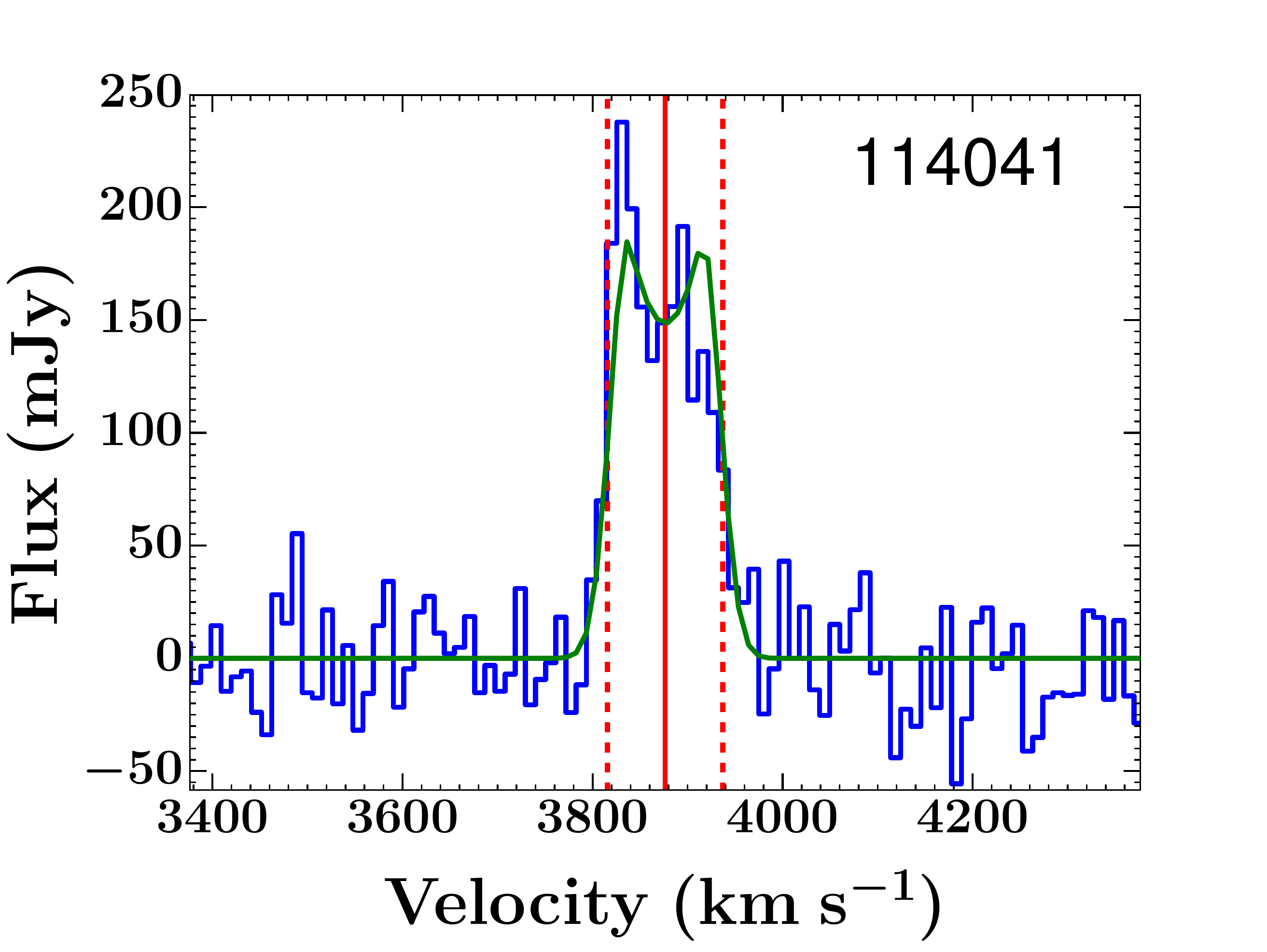}\includegraphics[scale=0.23, trim= 60 5 50 20, clip=true]{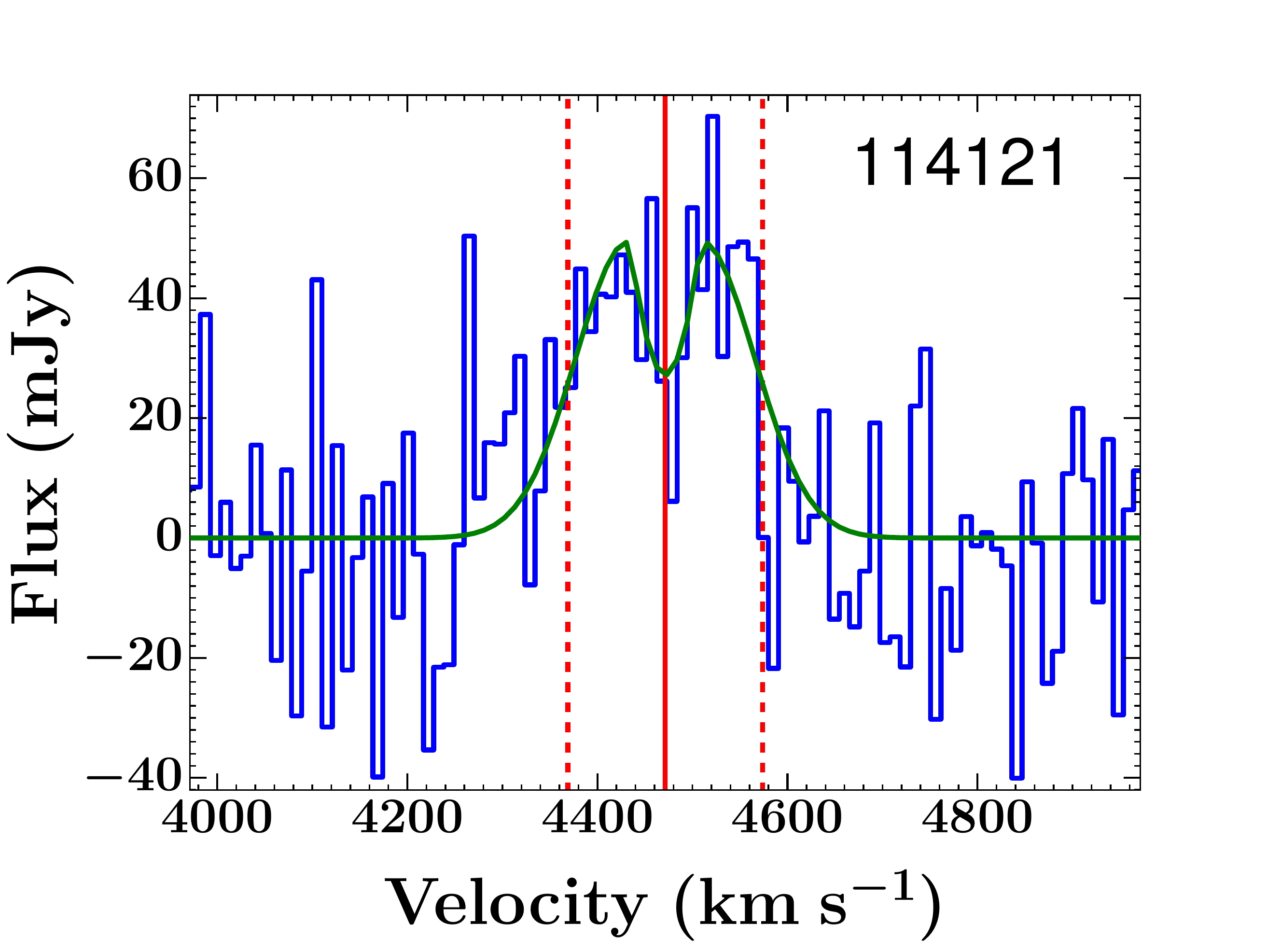}\includegraphics[scale=0.23, trim= 60 5 50 20, clip=true]{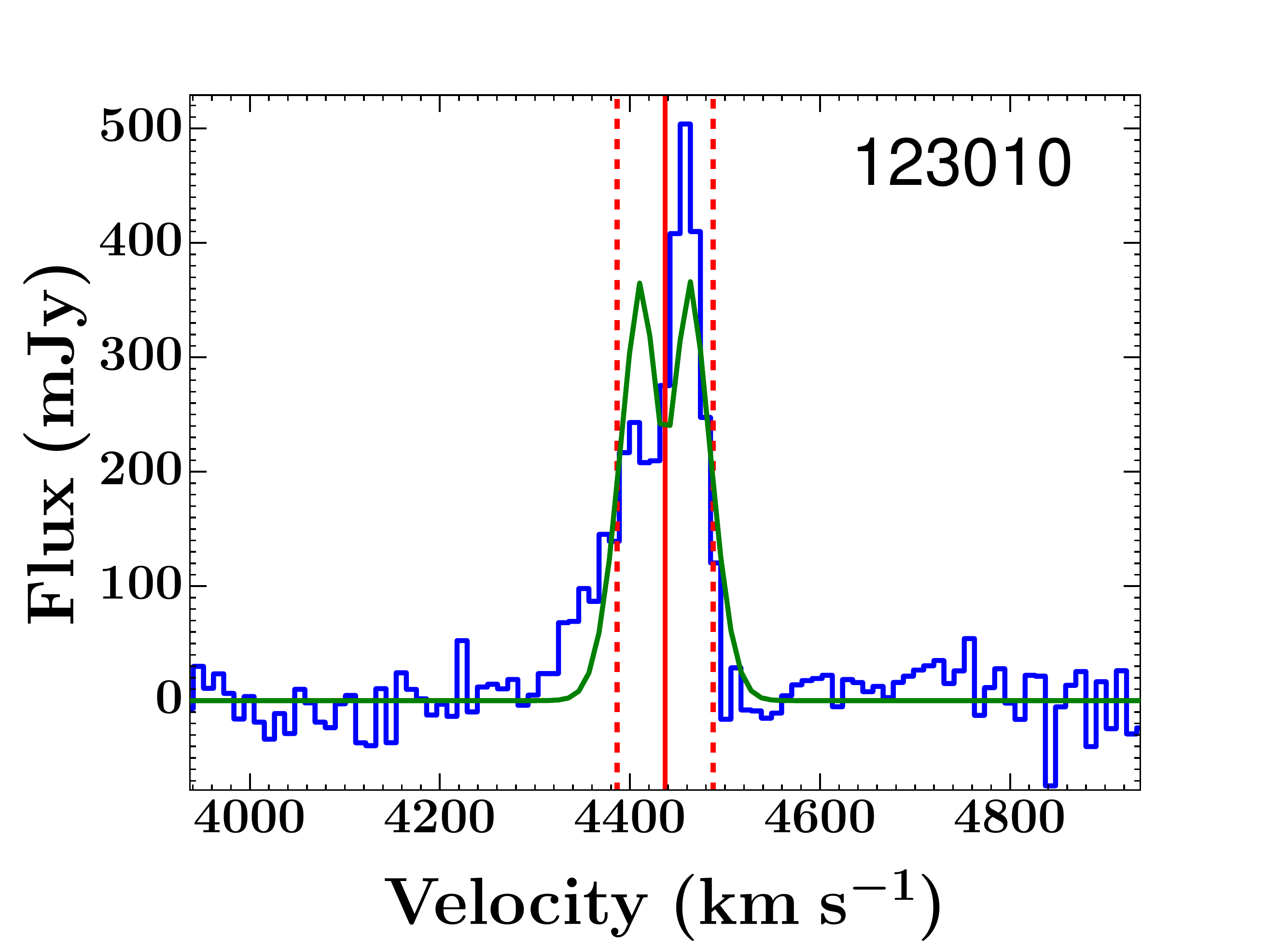}
\captionof{figure}{The Gaussian Double Peak function fits to each of the COLD GASS galaxy spectra in our final sub-sample, as described in \S~\ref{subsec:measurew50} and \S~\ref{subsec:subsample}. The data are plotted in blue and the best fit function in green. The solid red line indicates the best fit central velocity ($v_{0}$), whilst the dashed red lines show the width at $50\%$ of the peak ($W_{50}$). The COLD GASS ID is displayed in the top right of each panel.}\label{fig:w50fits}
\end{landscape}

\bsp	
\label{lastpage}
\end{document}